\documentclass[letterpaper,twocolumn,10pt]{article}
\usepackage[letterpaper,textwidth=7in,textheight=9in,columnsep=0.33in,centering]{geometry}
\usepackage{mathptmx}               %
\usepackage[T1]{fontenc}
\usepackage[utf8]{inputenc}
\usepackage[english]{babel}
\ifdefined\pdfgentounicode   %
\pdfglyphtounicode{A}{0041}
\pdfglyphtounicode{AE}{00C6}
\pdfglyphtounicode{AEacute}{01FC}
\pdfglyphtounicode{AEmacron}{01E2}
\pdfglyphtounicode{AEsmall}{00E6}
\pdfglyphtounicode{Aacute}{00C1}
\pdfglyphtounicode{Aacutesmall}{00E1}
\pdfglyphtounicode{Abreve}{0102}
\pdfglyphtounicode{Abreveacute}{1EAE}
\pdfglyphtounicode{Abrevecyrillic}{04D0}
\pdfglyphtounicode{Abrevedotbelow}{1EB6}
\pdfglyphtounicode{Abrevegrave}{1EB0}
\pdfglyphtounicode{Abrevehookabove}{1EB2}
\pdfglyphtounicode{Abrevetilde}{1EB4}
\pdfglyphtounicode{Acaron}{01CD}
\pdfglyphtounicode{Acircle}{24B6}
\pdfglyphtounicode{Acircumflex}{00C2}
\pdfglyphtounicode{Acircumflexacute}{1EA4}
\pdfglyphtounicode{Acircumflexdotbelow}{1EAC}
\pdfglyphtounicode{Acircumflexgrave}{1EA6}
\pdfglyphtounicode{Acircumflexhookabove}{1EA8}
\pdfglyphtounicode{Acircumflexsmall}{00E2}
\pdfglyphtounicode{Acircumflextilde}{1EAA}
\pdfglyphtounicode{Acute}{00B4}
\pdfglyphtounicode{Acutesmall}{00B4}
\pdfglyphtounicode{Acyrillic}{0410}
\pdfglyphtounicode{Adblgrave}{0200}
\pdfglyphtounicode{Adieresis}{00C4}
\pdfglyphtounicode{Adieresiscyrillic}{04D2}
\pdfglyphtounicode{Adieresismacron}{01DE}
\pdfglyphtounicode{Adieresissmall}{00E4}
\pdfglyphtounicode{Adotbelow}{1EA0}
\pdfglyphtounicode{Adotmacron}{01E0}
\pdfglyphtounicode{Agrave}{00C0}
\pdfglyphtounicode{Agravesmall}{00E0}
\pdfglyphtounicode{Ahookabove}{1EA2}
\pdfglyphtounicode{Aiecyrillic}{04D4}
\pdfglyphtounicode{Ainvertedbreve}{0202}
\pdfglyphtounicode{Alpha}{0391}
\pdfglyphtounicode{Alphatonos}{0386}
\pdfglyphtounicode{Amacron}{0100}
\pdfglyphtounicode{Amonospace}{FF21}
\pdfglyphtounicode{Aogonek}{0104}
\pdfglyphtounicode{Aring}{00C5}
\pdfglyphtounicode{Aringacute}{01FA}
\pdfglyphtounicode{Aringbelow}{1E00}
\pdfglyphtounicode{Aringsmall}{00E5}
\pdfglyphtounicode{Asmall}{0061}
\pdfglyphtounicode{Atilde}{00C3}
\pdfglyphtounicode{Atildesmall}{00E3}
\pdfglyphtounicode{Aybarmenian}{0531}
\pdfglyphtounicode{B}{0042}
\pdfglyphtounicode{Bcircle}{24B7}
\pdfglyphtounicode{Bdotaccent}{1E02}
\pdfglyphtounicode{Bdotbelow}{1E04}
\pdfglyphtounicode{Becyrillic}{0411}
\pdfglyphtounicode{Benarmenian}{0532}
\pdfglyphtounicode{Beta}{0392}
\pdfglyphtounicode{Bhook}{0181}
\pdfglyphtounicode{Blinebelow}{1E06}
\pdfglyphtounicode{Bmonospace}{FF22}
\pdfglyphtounicode{Brevesmall}{02D8}
\pdfglyphtounicode{Bsmall}{0062}
\pdfglyphtounicode{Btopbar}{0182}
\pdfglyphtounicode{C}{0043}
\pdfglyphtounicode{Caarmenian}{053E}
\pdfglyphtounicode{Cacute}{0106}
\pdfglyphtounicode{Caron}{02C7}
\pdfglyphtounicode{Caronsmall}{02C7}
\pdfglyphtounicode{Ccaron}{010C}
\pdfglyphtounicode{Ccedilla}{00C7}
\pdfglyphtounicode{Ccedillaacute}{1E08}
\pdfglyphtounicode{Ccedillasmall}{00E7}
\pdfglyphtounicode{Ccircle}{24B8}
\pdfglyphtounicode{Ccircumflex}{0108}
\pdfglyphtounicode{Cdot}{010A}
\pdfglyphtounicode{Cdotaccent}{010A}
\pdfglyphtounicode{Cedillasmall}{00B8}
\pdfglyphtounicode{Chaarmenian}{0549}
\pdfglyphtounicode{Cheabkhasiancyrillic}{04BC}
\pdfglyphtounicode{Checyrillic}{0427}
\pdfglyphtounicode{Chedescenderabkhasiancyrillic}{04BE}
\pdfglyphtounicode{Chedescendercyrillic}{04B6}
\pdfglyphtounicode{Chedieresiscyrillic}{04F4}
\pdfglyphtounicode{Cheharmenian}{0543}
\pdfglyphtounicode{Chekhakassiancyrillic}{04CB}
\pdfglyphtounicode{Cheverticalstrokecyrillic}{04B8}
\pdfglyphtounicode{Chi}{03A7}
\pdfglyphtounicode{Chook}{0187}
\pdfglyphtounicode{Circumflexsmall}{02C6}
\pdfglyphtounicode{Cmonospace}{FF23}
\pdfglyphtounicode{Coarmenian}{0551}
\pdfglyphtounicode{Csmall}{0063}
\pdfglyphtounicode{D}{0044}
\pdfglyphtounicode{DZ}{01F1}
\pdfglyphtounicode{DZcaron}{01C4}
\pdfglyphtounicode{Daarmenian}{0534}
\pdfglyphtounicode{Dafrican}{0189}
\pdfglyphtounicode{Dbar}{0110}
\pdfglyphtounicode{Dcaron}{010E}
\pdfglyphtounicode{Dcedilla}{1E10}
\pdfglyphtounicode{Dcircle}{24B9}
\pdfglyphtounicode{Dcircumflexbelow}{1E12}
\pdfglyphtounicode{Dcroat}{0110}
\pdfglyphtounicode{Ddotaccent}{1E0A}
\pdfglyphtounicode{Ddotbelow}{1E0C}
\pdfglyphtounicode{Decyrillic}{0414}
\pdfglyphtounicode{Deicoptic}{03EE}
\pdfglyphtounicode{Delta}{2206}
\pdfglyphtounicode{Deltagreek}{0394}
\pdfglyphtounicode{Dhook}{018A}
\pdfglyphtounicode{Dieresis}{00A8}
\pdfglyphtounicode{DieresisAcute}{F6CC}
\pdfglyphtounicode{DieresisGrave}{F6CD}
\pdfglyphtounicode{Dieresissmall}{00A8}
\pdfglyphtounicode{Digamma}{D875 DFCB}
\pdfglyphtounicode{Digammagreek}{03DC}
\pdfglyphtounicode{Djecyrillic}{0402}
\pdfglyphtounicode{Dlinebelow}{1E0E}
\pdfglyphtounicode{Dmonospace}{FF24}
\pdfglyphtounicode{Dotaccentsmall}{02D9}
\pdfglyphtounicode{Dslash}{0110}
\pdfglyphtounicode{Dsmall}{0064}
\pdfglyphtounicode{Dtopbar}{018B}
\pdfglyphtounicode{Dz}{01F2}
\pdfglyphtounicode{Dzcaron}{01C5}
\pdfglyphtounicode{Dzeabkhasiancyrillic}{04E0}
\pdfglyphtounicode{Dzecyrillic}{0405}
\pdfglyphtounicode{Dzhecyrillic}{040F}
\pdfglyphtounicode{E}{0045}
\pdfglyphtounicode{Eacute}{00C9}
\pdfglyphtounicode{Eacutesmall}{00E9}
\pdfglyphtounicode{Ebreve}{0114}
\pdfglyphtounicode{Ecaron}{011A}
\pdfglyphtounicode{Ecedillabreve}{1E1C}
\pdfglyphtounicode{Echarmenian}{0535}
\pdfglyphtounicode{Ecircle}{24BA}
\pdfglyphtounicode{Ecircumflex}{00CA}
\pdfglyphtounicode{Ecircumflexacute}{1EBE}
\pdfglyphtounicode{Ecircumflexbelow}{1E18}
\pdfglyphtounicode{Ecircumflexdotbelow}{1EC6}
\pdfglyphtounicode{Ecircumflexgrave}{1EC0}
\pdfglyphtounicode{Ecircumflexhookabove}{1EC2}
\pdfglyphtounicode{Ecircumflexsmall}{00EA}
\pdfglyphtounicode{Ecircumflextilde}{1EC4}
\pdfglyphtounicode{Ecyrillic}{0404}
\pdfglyphtounicode{Edblgrave}{0204}
\pdfglyphtounicode{Edieresis}{00CB}
\pdfglyphtounicode{Edieresissmall}{00EB}
\pdfglyphtounicode{Edot}{0116}
\pdfglyphtounicode{Edotaccent}{0116}
\pdfglyphtounicode{Edotbelow}{1EB8}
\pdfglyphtounicode{Efcyrillic}{0424}
\pdfglyphtounicode{Egrave}{00C8}
\pdfglyphtounicode{Egravesmall}{00E8}
\pdfglyphtounicode{Eharmenian}{0537}
\pdfglyphtounicode{Ehookabove}{1EBA}
\pdfglyphtounicode{Eightroman}{2167}
\pdfglyphtounicode{Einvertedbreve}{0206}
\pdfglyphtounicode{Eiotifiedcyrillic}{0464}
\pdfglyphtounicode{Elcyrillic}{041B}
\pdfglyphtounicode{Elevenroman}{216A}
\pdfglyphtounicode{Emacron}{0112}
\pdfglyphtounicode{Emacronacute}{1E16}
\pdfglyphtounicode{Emacrongrave}{1E14}
\pdfglyphtounicode{Emcyrillic}{041C}
\pdfglyphtounicode{Emonospace}{FF25}
\pdfglyphtounicode{Encyrillic}{041D}
\pdfglyphtounicode{Endescendercyrillic}{04A2}
\pdfglyphtounicode{Eng}{014A}
\pdfglyphtounicode{Enghecyrillic}{04A4}
\pdfglyphtounicode{Enhookcyrillic}{04C7}
\pdfglyphtounicode{Eogonek}{0118}
\pdfglyphtounicode{Eopen}{0190}
\pdfglyphtounicode{Epsilon}{0395}
\pdfglyphtounicode{Epsilontonos}{0388}
\pdfglyphtounicode{Ercyrillic}{0420}
\pdfglyphtounicode{Ereversed}{018E}
\pdfglyphtounicode{Ereversedcyrillic}{042D}
\pdfglyphtounicode{Escyrillic}{0421}
\pdfglyphtounicode{Esdescendercyrillic}{04AA}
\pdfglyphtounicode{Esh}{01A9}
\pdfglyphtounicode{Esmall}{0065}
\pdfglyphtounicode{Eta}{0397}
\pdfglyphtounicode{Etarmenian}{0538}
\pdfglyphtounicode{Etatonos}{0389}
\pdfglyphtounicode{Eth}{00D0}
\pdfglyphtounicode{Ethsmall}{00F0}
\pdfglyphtounicode{Etilde}{1EBC}
\pdfglyphtounicode{Etildebelow}{1E1A}
\pdfglyphtounicode{Euro}{20AC}
\pdfglyphtounicode{Ezh}{01B7}
\pdfglyphtounicode{Ezhcaron}{01EE}
\pdfglyphtounicode{Ezhreversed}{01B8}
\pdfglyphtounicode{F}{0046}
\pdfglyphtounicode{FFIsmall}{0066 0066 0069}
\pdfglyphtounicode{FFLsmall}{0066 0066 006C}
\pdfglyphtounicode{FFsmall}{0066 0066}
\pdfglyphtounicode{FIsmall}{0066 0069}
\pdfglyphtounicode{FLsmall}{0066 006C}
\pdfglyphtounicode{Fcircle}{24BB}
\pdfglyphtounicode{Fdotaccent}{1E1E}
\pdfglyphtounicode{Feharmenian}{0556}
\pdfglyphtounicode{Feicoptic}{03E4}
\pdfglyphtounicode{Fhook}{0191}
\pdfglyphtounicode{Finv}{2132}
\pdfglyphtounicode{Fitacyrillic}{0472}
\pdfglyphtounicode{Fiveroman}{2164}
\pdfglyphtounicode{Fmonospace}{FF26}
\pdfglyphtounicode{Fourroman}{2163}
\pdfglyphtounicode{Fsmall}{0066}
\pdfglyphtounicode{G}{0047}
\pdfglyphtounicode{GBsquare}{3387}
\pdfglyphtounicode{Gacute}{01F4}
\pdfglyphtounicode{Gamma}{0393}
\pdfglyphtounicode{Gammaafrican}{0194}
\pdfglyphtounicode{Gangiacoptic}{03EA}
\pdfglyphtounicode{Gbreve}{011E}
\pdfglyphtounicode{Gcaron}{01E6}
\pdfglyphtounicode{Gcedilla}{0122}
\pdfglyphtounicode{Gcircle}{24BC}
\pdfglyphtounicode{Gcircumflex}{011C}
\pdfglyphtounicode{Gcommaaccent}{0122}
\pdfglyphtounicode{Gdot}{0120}
\pdfglyphtounicode{Gdotaccent}{0120}
\pdfglyphtounicode{Gecyrillic}{0413}
\pdfglyphtounicode{Germandbls}{0053 0053}
\pdfglyphtounicode{Germandblssmall}{0073 0073}
\pdfglyphtounicode{Ghadarmenian}{0542}
\pdfglyphtounicode{Ghemiddlehookcyrillic}{0494}
\pdfglyphtounicode{Ghestrokecyrillic}{0492}
\pdfglyphtounicode{Gheupturncyrillic}{0490}
\pdfglyphtounicode{Ghook}{0193}
\pdfglyphtounicode{Gimarmenian}{0533}
\pdfglyphtounicode{Gjecyrillic}{0403}
\pdfglyphtounicode{Gmacron}{1E20}
\pdfglyphtounicode{Gmir}{2141}
\pdfglyphtounicode{Gmonospace}{FF27}
\pdfglyphtounicode{Grave}{0060}
\pdfglyphtounicode{Gravesmall}{0060}
\pdfglyphtounicode{Gsmall}{0067}
\pdfglyphtounicode{Gsmallhook}{029B}
\pdfglyphtounicode{Gstroke}{01E4}
\pdfglyphtounicode{H}{0048}
\pdfglyphtounicode{H18533}{25CF}
\pdfglyphtounicode{H18543}{25AA}
\pdfglyphtounicode{H18551}{25AB}
\pdfglyphtounicode{H22073}{25A1}
\pdfglyphtounicode{HPsquare}{33CB}
\pdfglyphtounicode{Haabkhasiancyrillic}{04A8}
\pdfglyphtounicode{Hadescendercyrillic}{04B2}
\pdfglyphtounicode{Hardsigncyrillic}{042A}
\pdfglyphtounicode{Hbar}{0126}
\pdfglyphtounicode{Hbrevebelow}{1E2A}
\pdfglyphtounicode{Hcedilla}{1E28}
\pdfglyphtounicode{Hcircle}{24BD}
\pdfglyphtounicode{Hcircumflex}{0124}
\pdfglyphtounicode{Hdieresis}{1E26}
\pdfglyphtounicode{Hdotaccent}{1E22}
\pdfglyphtounicode{Hdotbelow}{1E24}
\pdfglyphtounicode{Hmonospace}{FF28}
\pdfglyphtounicode{Hoarmenian}{0540}
\pdfglyphtounicode{Horicoptic}{03E8}
\pdfglyphtounicode{Hsmall}{0068}
\pdfglyphtounicode{Hungarumlaut}{02DD}
\pdfglyphtounicode{Hungarumlautsmall}{02DD}
\pdfglyphtounicode{Hzsquare}{3390}
\pdfglyphtounicode{I}{0049}
\pdfglyphtounicode{IAcyrillic}{042F}
\pdfglyphtounicode{IJ}{0132}
\pdfglyphtounicode{IUcyrillic}{042E}
\pdfglyphtounicode{Iacute}{00CD}
\pdfglyphtounicode{Iacutesmall}{00ED}
\pdfglyphtounicode{Ibreve}{012C}
\pdfglyphtounicode{Icaron}{01CF}
\pdfglyphtounicode{Icircle}{24BE}
\pdfglyphtounicode{Icircumflex}{00CE}
\pdfglyphtounicode{Icircumflexsmall}{00EE}
\pdfglyphtounicode{Icyrillic}{0406}
\pdfglyphtounicode{Idblgrave}{0208}
\pdfglyphtounicode{Idieresis}{00CF}
\pdfglyphtounicode{Idieresisacute}{1E2E}
\pdfglyphtounicode{Idieresiscyrillic}{04E4}
\pdfglyphtounicode{Idieresissmall}{00EF}
\pdfglyphtounicode{Idot}{0130}
\pdfglyphtounicode{Idotaccent}{0130}
\pdfglyphtounicode{Idotbelow}{1ECA}
\pdfglyphtounicode{Iebrevecyrillic}{04D6}
\pdfglyphtounicode{Iecyrillic}{0415}
\pdfglyphtounicode{Ifractur}{2111}
\pdfglyphtounicode{Ifraktur}{2111}
\pdfglyphtounicode{Igrave}{00CC}
\pdfglyphtounicode{Igravesmall}{00EC}
\pdfglyphtounicode{Ihookabove}{1EC8}
\pdfglyphtounicode{Iicyrillic}{0418}
\pdfglyphtounicode{Iinvertedbreve}{020A}
\pdfglyphtounicode{Iishortcyrillic}{0419}
\pdfglyphtounicode{Imacron}{012A}
\pdfglyphtounicode{Imacroncyrillic}{04E2}
\pdfglyphtounicode{Imonospace}{FF29}
\pdfglyphtounicode{Iniarmenian}{053B}
\pdfglyphtounicode{Iocyrillic}{0401}
\pdfglyphtounicode{Iogonek}{012E}
\pdfglyphtounicode{Iota}{0399}
\pdfglyphtounicode{Iotaafrican}{0196}
\pdfglyphtounicode{Iotadieresis}{03AA}
\pdfglyphtounicode{Iotatonos}{038A}
\pdfglyphtounicode{Ismall}{0069}
\pdfglyphtounicode{Istroke}{0197}
\pdfglyphtounicode{Itilde}{0128}
\pdfglyphtounicode{Itildebelow}{1E2C}
\pdfglyphtounicode{Izhitsacyrillic}{0474}
\pdfglyphtounicode{Izhitsadblgravecyrillic}{0476}
\pdfglyphtounicode{J}{004A}
\pdfglyphtounicode{Jaarmenian}{0541}
\pdfglyphtounicode{Jcircle}{24BF}
\pdfglyphtounicode{Jcircumflex}{0134}
\pdfglyphtounicode{Jecyrillic}{0408}
\pdfglyphtounicode{Jheharmenian}{054B}
\pdfglyphtounicode{Jmonospace}{FF2A}
\pdfglyphtounicode{Jsmall}{006A}
\pdfglyphtounicode{K}{004B}
\pdfglyphtounicode{KBsquare}{3385}
\pdfglyphtounicode{KKsquare}{33CD}
\pdfglyphtounicode{Kabashkircyrillic}{04A0}
\pdfglyphtounicode{Kacute}{1E30}
\pdfglyphtounicode{Kacyrillic}{041A}
\pdfglyphtounicode{Kadescendercyrillic}{049A}
\pdfglyphtounicode{Kahookcyrillic}{04C3}
\pdfglyphtounicode{Kappa}{039A}
\pdfglyphtounicode{Kastrokecyrillic}{049E}
\pdfglyphtounicode{Kaverticalstrokecyrillic}{049C}
\pdfglyphtounicode{Kcaron}{01E8}
\pdfglyphtounicode{Kcedilla}{0136}
\pdfglyphtounicode{Kcircle}{24C0}
\pdfglyphtounicode{Kcommaaccent}{0136}
\pdfglyphtounicode{Kdotbelow}{1E32}
\pdfglyphtounicode{Keharmenian}{0554}
\pdfglyphtounicode{Kenarmenian}{053F}
\pdfglyphtounicode{Khacyrillic}{0425}
\pdfglyphtounicode{Kheicoptic}{03E6}
\pdfglyphtounicode{Khook}{0198}
\pdfglyphtounicode{Kjecyrillic}{040C}
\pdfglyphtounicode{Klinebelow}{1E34}
\pdfglyphtounicode{Kmonospace}{FF2B}
\pdfglyphtounicode{Koppacyrillic}{0480}
\pdfglyphtounicode{Koppagreek}{03DE}
\pdfglyphtounicode{Ksicyrillic}{046E}
\pdfglyphtounicode{Ksmall}{006B}
\pdfglyphtounicode{L}{004C}
\pdfglyphtounicode{LJ}{01C7}
\pdfglyphtounicode{LL}{004C 004C}
\pdfglyphtounicode{Lacute}{0139}
\pdfglyphtounicode{Lambda}{039B}
\pdfglyphtounicode{Lcaron}{013D}
\pdfglyphtounicode{Lcedilla}{013B}
\pdfglyphtounicode{Lcircle}{24C1}
\pdfglyphtounicode{Lcircumflexbelow}{1E3C}
\pdfglyphtounicode{Lcommaaccent}{013B}
\pdfglyphtounicode{Ldot}{013F}
\pdfglyphtounicode{Ldotaccent}{013F}
\pdfglyphtounicode{Ldotbelow}{1E36}
\pdfglyphtounicode{Ldotbelowmacron}{1E38}
\pdfglyphtounicode{Liwnarmenian}{053C}
\pdfglyphtounicode{Lj}{01C8}
\pdfglyphtounicode{Ljecyrillic}{0409}
\pdfglyphtounicode{Llinebelow}{1E3A}
\pdfglyphtounicode{Lmonospace}{FF2C}
\pdfglyphtounicode{Lslash}{0141}
\pdfglyphtounicode{Lslashsmall}{0142}
\pdfglyphtounicode{Lsmall}{006C}
\pdfglyphtounicode{M}{004D}
\pdfglyphtounicode{MBsquare}{3386}
\pdfglyphtounicode{Macron}{00AF}
\pdfglyphtounicode{Macronsmall}{00AF}
\pdfglyphtounicode{Macute}{1E3E}
\pdfglyphtounicode{Mcircle}{24C2}
\pdfglyphtounicode{Mdotaccent}{1E40}
\pdfglyphtounicode{Mdotbelow}{1E42}
\pdfglyphtounicode{Menarmenian}{0544}
\pdfglyphtounicode{Mmonospace}{FF2D}
\pdfglyphtounicode{Msmall}{006D}
\pdfglyphtounicode{Mturned}{019C}
\pdfglyphtounicode{Mu}{039C}
\pdfglyphtounicode{N}{004E}
\pdfglyphtounicode{NJ}{01CA}
\pdfglyphtounicode{Nacute}{0143}
\pdfglyphtounicode{Ncaron}{0147}
\pdfglyphtounicode{Ncedilla}{0145}
\pdfglyphtounicode{Ncircle}{24C3}
\pdfglyphtounicode{Ncircumflexbelow}{1E4A}
\pdfglyphtounicode{Ncommaaccent}{0145}
\pdfglyphtounicode{Ndotaccent}{1E44}
\pdfglyphtounicode{Ndotbelow}{1E46}
\pdfglyphtounicode{Ng}{014A}
\pdfglyphtounicode{Nhookleft}{019D}
\pdfglyphtounicode{Nineroman}{2168}
\pdfglyphtounicode{Nj}{01CB}
\pdfglyphtounicode{Njecyrillic}{040A}
\pdfglyphtounicode{Nlinebelow}{1E48}
\pdfglyphtounicode{Nmonospace}{FF2E}
\pdfglyphtounicode{Nowarmenian}{0546}
\pdfglyphtounicode{Nsmall}{006E}
\pdfglyphtounicode{Ntilde}{00D1}
\pdfglyphtounicode{Ntildesmall}{00F1}
\pdfglyphtounicode{Nu}{039D}
\pdfglyphtounicode{O}{004F}
\pdfglyphtounicode{OE}{0152}
\pdfglyphtounicode{OEsmall}{0153}
\pdfglyphtounicode{Oacute}{00D3}
\pdfglyphtounicode{Oacutesmall}{00F3}
\pdfglyphtounicode{Obarredcyrillic}{04E8}
\pdfglyphtounicode{Obarreddieresiscyrillic}{04EA}
\pdfglyphtounicode{Obreve}{014E}
\pdfglyphtounicode{Ocaron}{01D1}
\pdfglyphtounicode{Ocenteredtilde}{019F}
\pdfglyphtounicode{Ocircle}{24C4}
\pdfglyphtounicode{Ocircumflex}{00D4}
\pdfglyphtounicode{Ocircumflexacute}{1ED0}
\pdfglyphtounicode{Ocircumflexdotbelow}{1ED8}
\pdfglyphtounicode{Ocircumflexgrave}{1ED2}
\pdfglyphtounicode{Ocircumflexhookabove}{1ED4}
\pdfglyphtounicode{Ocircumflexsmall}{00F4}
\pdfglyphtounicode{Ocircumflextilde}{1ED6}
\pdfglyphtounicode{Ocyrillic}{041E}
\pdfglyphtounicode{Odblacute}{0150}
\pdfglyphtounicode{Odblgrave}{020C}
\pdfglyphtounicode{Odieresis}{00D6}
\pdfglyphtounicode{Odieresiscyrillic}{04E6}
\pdfglyphtounicode{Odieresissmall}{00F6}
\pdfglyphtounicode{Odotbelow}{1ECC}
\pdfglyphtounicode{Ogoneksmall}{02DB}
\pdfglyphtounicode{Ograve}{00D2}
\pdfglyphtounicode{Ogravesmall}{00F2}
\pdfglyphtounicode{Oharmenian}{0555}
\pdfglyphtounicode{Ohm}{2126}
\pdfglyphtounicode{Ohookabove}{1ECE}
\pdfglyphtounicode{Ohorn}{01A0}
\pdfglyphtounicode{Ohornacute}{1EDA}
\pdfglyphtounicode{Ohorndotbelow}{1EE2}
\pdfglyphtounicode{Ohorngrave}{1EDC}
\pdfglyphtounicode{Ohornhookabove}{1EDE}
\pdfglyphtounicode{Ohorntilde}{1EE0}
\pdfglyphtounicode{Ohungarumlaut}{0150}
\pdfglyphtounicode{Oi}{01A2}
\pdfglyphtounicode{Oinvertedbreve}{020E}
\pdfglyphtounicode{Omacron}{014C}
\pdfglyphtounicode{Omacronacute}{1E52}
\pdfglyphtounicode{Omacrongrave}{1E50}
\pdfglyphtounicode{Omega}{2126}
\pdfglyphtounicode{Omegacyrillic}{0460}
\pdfglyphtounicode{Omegagreek}{03A9}
\pdfglyphtounicode{Omegainv}{2127}
\pdfglyphtounicode{Omegaroundcyrillic}{047A}
\pdfglyphtounicode{Omegatitlocyrillic}{047C}
\pdfglyphtounicode{Omegatonos}{038F}
\pdfglyphtounicode{Omicron}{039F}
\pdfglyphtounicode{Omicrontonos}{038C}
\pdfglyphtounicode{Omonospace}{FF2F}
\pdfglyphtounicode{Oneroman}{2160}
\pdfglyphtounicode{Oogonek}{01EA}
\pdfglyphtounicode{Oogonekmacron}{01EC}
\pdfglyphtounicode{Oopen}{0186}
\pdfglyphtounicode{Oslash}{00D8}
\pdfglyphtounicode{Oslashacute}{01FE}
\pdfglyphtounicode{Oslashsmall}{00F8}
\pdfglyphtounicode{Osmall}{006F}
\pdfglyphtounicode{Ostrokeacute}{01FE}
\pdfglyphtounicode{Otcyrillic}{047E}
\pdfglyphtounicode{Otilde}{00D5}
\pdfglyphtounicode{Otildeacute}{1E4C}
\pdfglyphtounicode{Otildedieresis}{1E4E}
\pdfglyphtounicode{Otildesmall}{00F5}
\pdfglyphtounicode{P}{0050}
\pdfglyphtounicode{Pacute}{1E54}
\pdfglyphtounicode{Pcircle}{24C5}
\pdfglyphtounicode{Pdotaccent}{1E56}
\pdfglyphtounicode{Pecyrillic}{041F}
\pdfglyphtounicode{Peharmenian}{054A}
\pdfglyphtounicode{Pemiddlehookcyrillic}{04A6}
\pdfglyphtounicode{Phi}{03A6}
\pdfglyphtounicode{Phook}{01A4}
\pdfglyphtounicode{Pi}{03A0}
\pdfglyphtounicode{Piwrarmenian}{0553}
\pdfglyphtounicode{Pmonospace}{FF30}
\pdfglyphtounicode{Psi}{03A8}
\pdfglyphtounicode{Psicyrillic}{0470}
\pdfglyphtounicode{Psmall}{0070}
\pdfglyphtounicode{Q}{0051}
\pdfglyphtounicode{Qcircle}{24C6}
\pdfglyphtounicode{Qmonospace}{FF31}
\pdfglyphtounicode{Qsmall}{0071}
\pdfglyphtounicode{R}{0052}
\pdfglyphtounicode{Raarmenian}{054C}
\pdfglyphtounicode{Racute}{0154}
\pdfglyphtounicode{Rcaron}{0158}
\pdfglyphtounicode{Rcedilla}{0156}
\pdfglyphtounicode{Rcircle}{24C7}
\pdfglyphtounicode{Rcommaaccent}{0156}
\pdfglyphtounicode{Rdblgrave}{0210}
\pdfglyphtounicode{Rdotaccent}{1E58}
\pdfglyphtounicode{Rdotbelow}{1E5A}
\pdfglyphtounicode{Rdotbelowmacron}{1E5C}
\pdfglyphtounicode{Reharmenian}{0550}
\pdfglyphtounicode{Rfractur}{211C}
\pdfglyphtounicode{Rfraktur}{211C}
\pdfglyphtounicode{Rho}{03A1}
\pdfglyphtounicode{Ringsmall}{02DA}
\pdfglyphtounicode{Rinvertedbreve}{0212}
\pdfglyphtounicode{Rlinebelow}{1E5E}
\pdfglyphtounicode{Rmonospace}{FF32}
\pdfglyphtounicode{Rsmall}{0072}
\pdfglyphtounicode{Rsmallinverted}{0281}
\pdfglyphtounicode{Rsmallinvertedsuperior}{02B6}
\pdfglyphtounicode{S}{0053}
\pdfglyphtounicode{SF010000}{250C}
\pdfglyphtounicode{SF020000}{2514}
\pdfglyphtounicode{SF030000}{2510}
\pdfglyphtounicode{SF040000}{2518}
\pdfglyphtounicode{SF050000}{253C}
\pdfglyphtounicode{SF060000}{252C}
\pdfglyphtounicode{SF070000}{2534}
\pdfglyphtounicode{SF080000}{251C}
\pdfglyphtounicode{SF090000}{2524}
\pdfglyphtounicode{SF100000}{2500}
\pdfglyphtounicode{SF110000}{2502}
\pdfglyphtounicode{SF190000}{2561}
\pdfglyphtounicode{SF200000}{2562}
\pdfglyphtounicode{SF210000}{2556}
\pdfglyphtounicode{SF220000}{2555}
\pdfglyphtounicode{SF230000}{2563}
\pdfglyphtounicode{SF240000}{2551}
\pdfglyphtounicode{SF250000}{2557}
\pdfglyphtounicode{SF260000}{255D}
\pdfglyphtounicode{SF270000}{255C}
\pdfglyphtounicode{SF280000}{255B}
\pdfglyphtounicode{SF360000}{255E}
\pdfglyphtounicode{SF370000}{255F}
\pdfglyphtounicode{SF380000}{255A}
\pdfglyphtounicode{SF390000}{2554}
\pdfglyphtounicode{SF400000}{2569}
\pdfglyphtounicode{SF410000}{2566}
\pdfglyphtounicode{SF420000}{2560}
\pdfglyphtounicode{SF430000}{2550}
\pdfglyphtounicode{SF440000}{256C}
\pdfglyphtounicode{SF450000}{2567}
\pdfglyphtounicode{SF460000}{2568}
\pdfglyphtounicode{SF470000}{2564}
\pdfglyphtounicode{SF480000}{2565}
\pdfglyphtounicode{SF490000}{2559}
\pdfglyphtounicode{SF500000}{2558}
\pdfglyphtounicode{SF510000}{2552}
\pdfglyphtounicode{SF520000}{2553}
\pdfglyphtounicode{SF530000}{256B}
\pdfglyphtounicode{SF540000}{256A}
\pdfglyphtounicode{SS}{0053 0053}
\pdfglyphtounicode{SSsmall}{0073 0073}
\pdfglyphtounicode{Sacute}{015A}
\pdfglyphtounicode{Sacutedotaccent}{1E64}
\pdfglyphtounicode{Sampigreek}{03E0}
\pdfglyphtounicode{Scaron}{0160}
\pdfglyphtounicode{Scarondotaccent}{1E66}
\pdfglyphtounicode{Scaronsmall}{0161}
\pdfglyphtounicode{Scedilla}{015E}
\pdfglyphtounicode{Schwa}{018F}
\pdfglyphtounicode{Schwacyrillic}{04D8}
\pdfglyphtounicode{Schwadieresiscyrillic}{04DA}
\pdfglyphtounicode{Scircle}{24C8}
\pdfglyphtounicode{Scircumflex}{015C}
\pdfglyphtounicode{Scommaaccent}{0218}
\pdfglyphtounicode{Sdotaccent}{1E60}
\pdfglyphtounicode{Sdotbelow}{1E62}
\pdfglyphtounicode{Sdotbelowdotaccent}{1E68}
\pdfglyphtounicode{Seharmenian}{054D}
\pdfglyphtounicode{Sevenroman}{2166}
\pdfglyphtounicode{Shaarmenian}{0547}
\pdfglyphtounicode{Shacyrillic}{0428}
\pdfglyphtounicode{Shchacyrillic}{0429}
\pdfglyphtounicode{Sheicoptic}{03E2}
\pdfglyphtounicode{Shhacyrillic}{04BA}
\pdfglyphtounicode{Shimacoptic}{03EC}
\pdfglyphtounicode{Sigma}{03A3}
\pdfglyphtounicode{Sixroman}{2165}
\pdfglyphtounicode{Smonospace}{FF33}
\pdfglyphtounicode{Softsigncyrillic}{042C}
\pdfglyphtounicode{Ssmall}{0073}
\pdfglyphtounicode{Stigmagreek}{03DA}
\pdfglyphtounicode{T}{0054}
\pdfglyphtounicode{Tau}{03A4}
\pdfglyphtounicode{Tbar}{0166}
\pdfglyphtounicode{Tcaron}{0164}
\pdfglyphtounicode{Tcedilla}{0162}
\pdfglyphtounicode{Tcircle}{24C9}
\pdfglyphtounicode{Tcircumflexbelow}{1E70}
\pdfglyphtounicode{Tcommaaccent}{0162}
\pdfglyphtounicode{Tdotaccent}{1E6A}
\pdfglyphtounicode{Tdotbelow}{1E6C}
\pdfglyphtounicode{Tecyrillic}{0422}
\pdfglyphtounicode{Tedescendercyrillic}{04AC}
\pdfglyphtounicode{Tenroman}{2169}
\pdfglyphtounicode{Tetsecyrillic}{04B4}
\pdfglyphtounicode{Theta}{0398}
\pdfglyphtounicode{Thook}{01AC}
\pdfglyphtounicode{Thorn}{00DE}
\pdfglyphtounicode{Thornsmall}{00FE}
\pdfglyphtounicode{Threeroman}{2162}
\pdfglyphtounicode{Tildesmall}{02DC}
\pdfglyphtounicode{Tiwnarmenian}{054F}
\pdfglyphtounicode{Tlinebelow}{1E6E}
\pdfglyphtounicode{Tmonospace}{FF34}
\pdfglyphtounicode{Toarmenian}{0539}
\pdfglyphtounicode{Tonefive}{01BC}
\pdfglyphtounicode{Tonesix}{0184}
\pdfglyphtounicode{Tonetwo}{01A7}
\pdfglyphtounicode{Tretroflexhook}{01AE}
\pdfglyphtounicode{Tsecyrillic}{0426}
\pdfglyphtounicode{Tshecyrillic}{040B}
\pdfglyphtounicode{Tsmall}{0074}
\pdfglyphtounicode{Twelveroman}{216B}
\pdfglyphtounicode{Tworoman}{2161}
\pdfglyphtounicode{U}{0055}
\pdfglyphtounicode{Uacute}{00DA}
\pdfglyphtounicode{Uacutesmall}{00FA}
\pdfglyphtounicode{Ubreve}{016C}
\pdfglyphtounicode{Ucaron}{01D3}
\pdfglyphtounicode{Ucircle}{24CA}
\pdfglyphtounicode{Ucircumflex}{00DB}
\pdfglyphtounicode{Ucircumflexbelow}{1E76}
\pdfglyphtounicode{Ucircumflexsmall}{00FB}
\pdfglyphtounicode{Ucyrillic}{0423}
\pdfglyphtounicode{Udblacute}{0170}
\pdfglyphtounicode{Udblgrave}{0214}
\pdfglyphtounicode{Udieresis}{00DC}
\pdfglyphtounicode{Udieresisacute}{01D7}
\pdfglyphtounicode{Udieresisbelow}{1E72}
\pdfglyphtounicode{Udieresiscaron}{01D9}
\pdfglyphtounicode{Udieresiscyrillic}{04F0}
\pdfglyphtounicode{Udieresisgrave}{01DB}
\pdfglyphtounicode{Udieresismacron}{01D5}
\pdfglyphtounicode{Udieresissmall}{00FC}
\pdfglyphtounicode{Udotbelow}{1EE4}
\pdfglyphtounicode{Ugrave}{00D9}
\pdfglyphtounicode{Ugravesmall}{00F9}
\pdfglyphtounicode{Uhookabove}{1EE6}
\pdfglyphtounicode{Uhorn}{01AF}
\pdfglyphtounicode{Uhornacute}{1EE8}
\pdfglyphtounicode{Uhorndotbelow}{1EF0}
\pdfglyphtounicode{Uhorngrave}{1EEA}
\pdfglyphtounicode{Uhornhookabove}{1EEC}
\pdfglyphtounicode{Uhorntilde}{1EEE}
\pdfglyphtounicode{Uhungarumlaut}{0170}
\pdfglyphtounicode{Uhungarumlautcyrillic}{04F2}
\pdfglyphtounicode{Uinvertedbreve}{0216}
\pdfglyphtounicode{Ukcyrillic}{0478}
\pdfglyphtounicode{Umacron}{016A}
\pdfglyphtounicode{Umacroncyrillic}{04EE}
\pdfglyphtounicode{Umacrondieresis}{1E7A}
\pdfglyphtounicode{Umonospace}{FF35}
\pdfglyphtounicode{Uogonek}{0172}
\pdfglyphtounicode{Upsilon}{03A5}
\pdfglyphtounicode{Upsilon1}{03D2}
\pdfglyphtounicode{Upsilonacutehooksymbolgreek}{03D3}
\pdfglyphtounicode{Upsilonafrican}{01B1}
\pdfglyphtounicode{Upsilondieresis}{03AB}
\pdfglyphtounicode{Upsilondieresishooksymbolgreek}{03D4}
\pdfglyphtounicode{Upsilonhooksymbol}{03D2}
\pdfglyphtounicode{Upsilontonos}{038E}
\pdfglyphtounicode{Uring}{016E}
\pdfglyphtounicode{Ushortcyrillic}{040E}
\pdfglyphtounicode{Usmall}{0075}
\pdfglyphtounicode{Ustraightcyrillic}{04AE}
\pdfglyphtounicode{Ustraightstrokecyrillic}{04B0}
\pdfglyphtounicode{Utilde}{0168}
\pdfglyphtounicode{Utildeacute}{1E78}
\pdfglyphtounicode{Utildebelow}{1E74}
\pdfglyphtounicode{V}{0056}
\pdfglyphtounicode{Vcircle}{24CB}
\pdfglyphtounicode{Vdotbelow}{1E7E}
\pdfglyphtounicode{Vecyrillic}{0412}
\pdfglyphtounicode{Vewarmenian}{054E}
\pdfglyphtounicode{Vhook}{01B2}
\pdfglyphtounicode{Vmonospace}{FF36}
\pdfglyphtounicode{Voarmenian}{0548}
\pdfglyphtounicode{Vsmall}{0076}
\pdfglyphtounicode{Vtilde}{1E7C}
\pdfglyphtounicode{W}{0057}
\pdfglyphtounicode{Wacute}{1E82}
\pdfglyphtounicode{Wcircle}{24CC}
\pdfglyphtounicode{Wcircumflex}{0174}
\pdfglyphtounicode{Wdieresis}{1E84}
\pdfglyphtounicode{Wdotaccent}{1E86}
\pdfglyphtounicode{Wdotbelow}{1E88}
\pdfglyphtounicode{Wgrave}{1E80}
\pdfglyphtounicode{Wmonospace}{FF37}
\pdfglyphtounicode{Wsmall}{0077}
\pdfglyphtounicode{X}{0058}
\pdfglyphtounicode{Xcircle}{24CD}
\pdfglyphtounicode{Xdieresis}{1E8C}
\pdfglyphtounicode{Xdotaccent}{1E8A}
\pdfglyphtounicode{Xeharmenian}{053D}
\pdfglyphtounicode{Xi}{039E}
\pdfglyphtounicode{Xmonospace}{FF38}
\pdfglyphtounicode{Xsmall}{0078}
\pdfglyphtounicode{Y}{0059}
\pdfglyphtounicode{Yacute}{00DD}
\pdfglyphtounicode{Yacutesmall}{00FD}
\pdfglyphtounicode{Yatcyrillic}{0462}
\pdfglyphtounicode{Ycircle}{24CE}
\pdfglyphtounicode{Ycircumflex}{0176}
\pdfglyphtounicode{Ydieresis}{0178}
\pdfglyphtounicode{Ydieresissmall}{00FF}
\pdfglyphtounicode{Ydotaccent}{1E8E}
\pdfglyphtounicode{Ydotbelow}{1EF4}
\pdfglyphtounicode{Yen}{00A5}
\pdfglyphtounicode{Yericyrillic}{042B}
\pdfglyphtounicode{Yerudieresiscyrillic}{04F8}
\pdfglyphtounicode{Ygrave}{1EF2}
\pdfglyphtounicode{Yhook}{01B3}
\pdfglyphtounicode{Yhookabove}{1EF6}
\pdfglyphtounicode{Yiarmenian}{0545}
\pdfglyphtounicode{Yicyrillic}{0407}
\pdfglyphtounicode{Yiwnarmenian}{0552}
\pdfglyphtounicode{Ymonospace}{FF39}
\pdfglyphtounicode{Ysmall}{0079}
\pdfglyphtounicode{Ytilde}{1EF8}
\pdfglyphtounicode{Yusbigcyrillic}{046A}
\pdfglyphtounicode{Yusbigiotifiedcyrillic}{046C}
\pdfglyphtounicode{Yuslittlecyrillic}{0466}
\pdfglyphtounicode{Yuslittleiotifiedcyrillic}{0468}
\pdfglyphtounicode{Z}{005A}
\pdfglyphtounicode{Zaarmenian}{0536}
\pdfglyphtounicode{Zacute}{0179}
\pdfglyphtounicode{Zcaron}{017D}
\pdfglyphtounicode{Zcaronsmall}{017E}
\pdfglyphtounicode{Zcircle}{24CF}
\pdfglyphtounicode{Zcircumflex}{1E90}
\pdfglyphtounicode{Zdot}{017B}
\pdfglyphtounicode{Zdotaccent}{017B}
\pdfglyphtounicode{Zdotbelow}{1E92}
\pdfglyphtounicode{Zecyrillic}{0417}
\pdfglyphtounicode{Zedescendercyrillic}{0498}
\pdfglyphtounicode{Zedieresiscyrillic}{04DE}
\pdfglyphtounicode{Zeta}{0396}
\pdfglyphtounicode{Zhearmenian}{053A}
\pdfglyphtounicode{Zhebrevecyrillic}{04C1}
\pdfglyphtounicode{Zhecyrillic}{0416}
\pdfglyphtounicode{Zhedescendercyrillic}{0496}
\pdfglyphtounicode{Zhedieresiscyrillic}{04DC}
\pdfglyphtounicode{Zlinebelow}{1E94}
\pdfglyphtounicode{Zmonospace}{FF3A}
\pdfglyphtounicode{Zsmall}{007A}
\pdfglyphtounicode{Zstroke}{01B5}
\pdfglyphtounicode{a}{0061}
\pdfglyphtounicode{aabengali}{0986}
\pdfglyphtounicode{aacute}{00E1}
\pdfglyphtounicode{aadeva}{0906}
\pdfglyphtounicode{aagujarati}{0A86}
\pdfglyphtounicode{aagurmukhi}{0A06}
\pdfglyphtounicode{aamatragurmukhi}{0A3E}
\pdfglyphtounicode{aarusquare}{3303}
\pdfglyphtounicode{aavowelsignbengali}{09BE}
\pdfglyphtounicode{aavowelsigndeva}{093E}
\pdfglyphtounicode{aavowelsigngujarati}{0ABE}
\pdfglyphtounicode{abbreviationmarkarmenian}{055F}
\pdfglyphtounicode{abbreviationsigndeva}{0970}
\pdfglyphtounicode{abengali}{0985}
\pdfglyphtounicode{abopomofo}{311A}
\pdfglyphtounicode{abreve}{0103}
\pdfglyphtounicode{abreveacute}{1EAF}
\pdfglyphtounicode{abrevecyrillic}{04D1}
\pdfglyphtounicode{abrevedotbelow}{1EB7}
\pdfglyphtounicode{abrevegrave}{1EB1}
\pdfglyphtounicode{abrevehookabove}{1EB3}
\pdfglyphtounicode{abrevetilde}{1EB5}
\pdfglyphtounicode{acaron}{01CE}
\pdfglyphtounicode{acircle}{24D0}
\pdfglyphtounicode{acircumflex}{00E2}
\pdfglyphtounicode{acircumflexacute}{1EA5}
\pdfglyphtounicode{acircumflexdotbelow}{1EAD}
\pdfglyphtounicode{acircumflexgrave}{1EA7}
\pdfglyphtounicode{acircumflexhookabove}{1EA9}
\pdfglyphtounicode{acircumflextilde}{1EAB}
\pdfglyphtounicode{acute}{00B4}
\pdfglyphtounicode{acutebelowcmb}{0317}
\pdfglyphtounicode{acutecmb}{0301}
\pdfglyphtounicode{acutecomb}{0301}
\pdfglyphtounicode{acutedeva}{0954}
\pdfglyphtounicode{acutelowmod}{02CF}
\pdfglyphtounicode{acutetonecmb}{0341}
\pdfglyphtounicode{acyrillic}{0430}
\pdfglyphtounicode{adblgrave}{0201}
\pdfglyphtounicode{addakgurmukhi}{0A71}
\pdfglyphtounicode{adeva}{0905}
\pdfglyphtounicode{adieresis}{00E4}
\pdfglyphtounicode{adieresiscyrillic}{04D3}
\pdfglyphtounicode{adieresismacron}{01DF}
\pdfglyphtounicode{adotbelow}{1EA1}
\pdfglyphtounicode{adotmacron}{01E1}
\pdfglyphtounicode{ae}{00E6}
\pdfglyphtounicode{aeacute}{01FD}
\pdfglyphtounicode{aekorean}{3150}
\pdfglyphtounicode{aemacron}{01E3}
\pdfglyphtounicode{afii00208}{2015}
\pdfglyphtounicode{afii08941}{20A4}
\pdfglyphtounicode{afii10017}{0410}
\pdfglyphtounicode{afii10018}{0411}
\pdfglyphtounicode{afii10019}{0412}
\pdfglyphtounicode{afii10020}{0413}
\pdfglyphtounicode{afii10021}{0414}
\pdfglyphtounicode{afii10022}{0415}
\pdfglyphtounicode{afii10023}{0401}
\pdfglyphtounicode{afii10024}{0416}
\pdfglyphtounicode{afii10025}{0417}
\pdfglyphtounicode{afii10026}{0418}
\pdfglyphtounicode{afii10027}{0419}
\pdfglyphtounicode{afii10028}{041A}
\pdfglyphtounicode{afii10029}{041B}
\pdfglyphtounicode{afii10030}{041C}
\pdfglyphtounicode{afii10031}{041D}
\pdfglyphtounicode{afii10032}{041E}
\pdfglyphtounicode{afii10033}{041F}
\pdfglyphtounicode{afii10034}{0420}
\pdfglyphtounicode{afii10035}{0421}
\pdfglyphtounicode{afii10036}{0422}
\pdfglyphtounicode{afii10037}{0423}
\pdfglyphtounicode{afii10038}{0424}
\pdfglyphtounicode{afii10039}{0425}
\pdfglyphtounicode{afii10040}{0426}
\pdfglyphtounicode{afii10041}{0427}
\pdfglyphtounicode{afii10042}{0428}
\pdfglyphtounicode{afii10043}{0429}
\pdfglyphtounicode{afii10044}{042A}
\pdfglyphtounicode{afii10045}{042B}
\pdfglyphtounicode{afii10046}{042C}
\pdfglyphtounicode{afii10047}{042D}
\pdfglyphtounicode{afii10048}{042E}
\pdfglyphtounicode{afii10049}{042F}
\pdfglyphtounicode{afii10050}{0490}
\pdfglyphtounicode{afii10051}{0402}
\pdfglyphtounicode{afii10052}{0403}
\pdfglyphtounicode{afii10053}{0404}
\pdfglyphtounicode{afii10054}{0405}
\pdfglyphtounicode{afii10055}{0406}
\pdfglyphtounicode{afii10056}{0407}
\pdfglyphtounicode{afii10057}{0408}
\pdfglyphtounicode{afii10058}{0409}
\pdfglyphtounicode{afii10059}{040A}
\pdfglyphtounicode{afii10060}{040B}
\pdfglyphtounicode{afii10061}{040C}
\pdfglyphtounicode{afii10062}{040E}
\pdfglyphtounicode{afii10063}{F6C4}
\pdfglyphtounicode{afii10064}{F6C5}
\pdfglyphtounicode{afii10065}{0430}
\pdfglyphtounicode{afii10066}{0431}
\pdfglyphtounicode{afii10067}{0432}
\pdfglyphtounicode{afii10068}{0433}
\pdfglyphtounicode{afii10069}{0434}
\pdfglyphtounicode{afii10070}{0435}
\pdfglyphtounicode{afii10071}{0451}
\pdfglyphtounicode{afii10072}{0436}
\pdfglyphtounicode{afii10073}{0437}
\pdfglyphtounicode{afii10074}{0438}
\pdfglyphtounicode{afii10075}{0439}
\pdfglyphtounicode{afii10076}{043A}
\pdfglyphtounicode{afii10077}{043B}
\pdfglyphtounicode{afii10078}{043C}
\pdfglyphtounicode{afii10079}{043D}
\pdfglyphtounicode{afii10080}{043E}
\pdfglyphtounicode{afii10081}{043F}
\pdfglyphtounicode{afii10082}{0440}
\pdfglyphtounicode{afii10083}{0441}
\pdfglyphtounicode{afii10084}{0442}
\pdfglyphtounicode{afii10085}{0443}
\pdfglyphtounicode{afii10086}{0444}
\pdfglyphtounicode{afii10087}{0445}
\pdfglyphtounicode{afii10088}{0446}
\pdfglyphtounicode{afii10089}{0447}
\pdfglyphtounicode{afii10090}{0448}
\pdfglyphtounicode{afii10091}{0449}
\pdfglyphtounicode{afii10092}{044A}
\pdfglyphtounicode{afii10093}{044B}
\pdfglyphtounicode{afii10094}{044C}
\pdfglyphtounicode{afii10095}{044D}
\pdfglyphtounicode{afii10096}{044E}
\pdfglyphtounicode{afii10097}{044F}
\pdfglyphtounicode{afii10098}{0491}
\pdfglyphtounicode{afii10099}{0452}
\pdfglyphtounicode{afii10100}{0453}
\pdfglyphtounicode{afii10101}{0454}
\pdfglyphtounicode{afii10102}{0455}
\pdfglyphtounicode{afii10103}{0456}
\pdfglyphtounicode{afii10104}{0457}
\pdfglyphtounicode{afii10105}{0458}
\pdfglyphtounicode{afii10106}{0459}
\pdfglyphtounicode{afii10107}{045A}
\pdfglyphtounicode{afii10108}{045B}
\pdfglyphtounicode{afii10109}{045C}
\pdfglyphtounicode{afii10110}{045E}
\pdfglyphtounicode{afii10145}{040F}
\pdfglyphtounicode{afii10146}{0462}
\pdfglyphtounicode{afii10147}{0472}
\pdfglyphtounicode{afii10148}{0474}
\pdfglyphtounicode{afii10192}{F6C6}
\pdfglyphtounicode{afii10193}{045F}
\pdfglyphtounicode{afii10194}{0463}
\pdfglyphtounicode{afii10195}{0473}
\pdfglyphtounicode{afii10196}{0475}
\pdfglyphtounicode{afii10831}{F6C7}
\pdfglyphtounicode{afii10832}{F6C8}
\pdfglyphtounicode{afii10846}{04D9}
\pdfglyphtounicode{afii299}{200E}
\pdfglyphtounicode{afii300}{200F}
\pdfglyphtounicode{afii301}{200D}
\pdfglyphtounicode{afii57381}{066A}
\pdfglyphtounicode{afii57388}{060C}
\pdfglyphtounicode{afii57392}{0660}
\pdfglyphtounicode{afii57393}{0661}
\pdfglyphtounicode{afii57394}{0662}
\pdfglyphtounicode{afii57395}{0663}
\pdfglyphtounicode{afii57396}{0664}
\pdfglyphtounicode{afii57397}{0665}
\pdfglyphtounicode{afii57398}{0666}
\pdfglyphtounicode{afii57399}{0667}
\pdfglyphtounicode{afii57400}{0668}
\pdfglyphtounicode{afii57401}{0669}
\pdfglyphtounicode{afii57403}{061B}
\pdfglyphtounicode{afii57407}{061F}
\pdfglyphtounicode{afii57409}{0621}
\pdfglyphtounicode{afii57410}{0622}
\pdfglyphtounicode{afii57411}{0623}
\pdfglyphtounicode{afii57412}{0624}
\pdfglyphtounicode{afii57413}{0625}
\pdfglyphtounicode{afii57414}{0626}
\pdfglyphtounicode{afii57415}{0627}
\pdfglyphtounicode{afii57416}{0628}
\pdfglyphtounicode{afii57417}{0629}
\pdfglyphtounicode{afii57418}{062A}
\pdfglyphtounicode{afii57419}{062B}
\pdfglyphtounicode{afii57420}{062C}
\pdfglyphtounicode{afii57421}{062D}
\pdfglyphtounicode{afii57422}{062E}
\pdfglyphtounicode{afii57423}{062F}
\pdfglyphtounicode{afii57424}{0630}
\pdfglyphtounicode{afii57425}{0631}
\pdfglyphtounicode{afii57426}{0632}
\pdfglyphtounicode{afii57427}{0633}
\pdfglyphtounicode{afii57428}{0634}
\pdfglyphtounicode{afii57429}{0635}
\pdfglyphtounicode{afii57430}{0636}
\pdfglyphtounicode{afii57431}{0637}
\pdfglyphtounicode{afii57432}{0638}
\pdfglyphtounicode{afii57433}{0639}
\pdfglyphtounicode{afii57434}{063A}
\pdfglyphtounicode{afii57440}{0640}
\pdfglyphtounicode{afii57441}{0641}
\pdfglyphtounicode{afii57442}{0642}
\pdfglyphtounicode{afii57443}{0643}
\pdfglyphtounicode{afii57444}{0644}
\pdfglyphtounicode{afii57445}{0645}
\pdfglyphtounicode{afii57446}{0646}
\pdfglyphtounicode{afii57448}{0648}
\pdfglyphtounicode{afii57449}{0649}
\pdfglyphtounicode{afii57450}{064A}
\pdfglyphtounicode{afii57451}{064B}
\pdfglyphtounicode{afii57452}{064C}
\pdfglyphtounicode{afii57453}{064D}
\pdfglyphtounicode{afii57454}{064E}
\pdfglyphtounicode{afii57455}{064F}
\pdfglyphtounicode{afii57456}{0650}
\pdfglyphtounicode{afii57457}{0651}
\pdfglyphtounicode{afii57458}{0652}
\pdfglyphtounicode{afii57470}{0647}
\pdfglyphtounicode{afii57505}{06A4}
\pdfglyphtounicode{afii57506}{067E}
\pdfglyphtounicode{afii57507}{0686}
\pdfglyphtounicode{afii57508}{0698}
\pdfglyphtounicode{afii57509}{06AF}
\pdfglyphtounicode{afii57511}{0679}
\pdfglyphtounicode{afii57512}{0688}
\pdfglyphtounicode{afii57513}{0691}
\pdfglyphtounicode{afii57514}{06BA}
\pdfglyphtounicode{afii57519}{06D2}
\pdfglyphtounicode{afii57534}{06D5}
\pdfglyphtounicode{afii57636}{20AA}
\pdfglyphtounicode{afii57645}{05BE}
\pdfglyphtounicode{afii57658}{05C3}
\pdfglyphtounicode{afii57664}{05D0}
\pdfglyphtounicode{afii57665}{05D1}
\pdfglyphtounicode{afii57666}{05D2}
\pdfglyphtounicode{afii57667}{05D3}
\pdfglyphtounicode{afii57668}{05D4}
\pdfglyphtounicode{afii57669}{05D5}
\pdfglyphtounicode{afii57670}{05D6}
\pdfglyphtounicode{afii57671}{05D7}
\pdfglyphtounicode{afii57672}{05D8}
\pdfglyphtounicode{afii57673}{05D9}
\pdfglyphtounicode{afii57674}{05DA}
\pdfglyphtounicode{afii57675}{05DB}
\pdfglyphtounicode{afii57676}{05DC}
\pdfglyphtounicode{afii57677}{05DD}
\pdfglyphtounicode{afii57678}{05DE}
\pdfglyphtounicode{afii57679}{05DF}
\pdfglyphtounicode{afii57680}{05E0}
\pdfglyphtounicode{afii57681}{05E1}
\pdfglyphtounicode{afii57682}{05E2}
\pdfglyphtounicode{afii57683}{05E3}
\pdfglyphtounicode{afii57684}{05E4}
\pdfglyphtounicode{afii57685}{05E5}
\pdfglyphtounicode{afii57686}{05E6}
\pdfglyphtounicode{afii57687}{05E7}
\pdfglyphtounicode{afii57688}{05E8}
\pdfglyphtounicode{afii57689}{05E9}
\pdfglyphtounicode{afii57690}{05EA}
\pdfglyphtounicode{afii57694}{FB2A}
\pdfglyphtounicode{afii57695}{FB2B}
\pdfglyphtounicode{afii57700}{FB4B}
\pdfglyphtounicode{afii57705}{FB1F}
\pdfglyphtounicode{afii57716}{05F0}
\pdfglyphtounicode{afii57717}{05F1}
\pdfglyphtounicode{afii57718}{05F2}
\pdfglyphtounicode{afii57723}{FB35}
\pdfglyphtounicode{afii57793}{05B4}
\pdfglyphtounicode{afii57794}{05B5}
\pdfglyphtounicode{afii57795}{05B6}
\pdfglyphtounicode{afii57796}{05BB}
\pdfglyphtounicode{afii57797}{05B8}
\pdfglyphtounicode{afii57798}{05B7}
\pdfglyphtounicode{afii57799}{05B0}
\pdfglyphtounicode{afii57800}{05B2}
\pdfglyphtounicode{afii57801}{05B1}
\pdfglyphtounicode{afii57802}{05B3}
\pdfglyphtounicode{afii57803}{05C2}
\pdfglyphtounicode{afii57804}{05C1}
\pdfglyphtounicode{afii57806}{05B9}
\pdfglyphtounicode{afii57807}{05BC}
\pdfglyphtounicode{afii57839}{05BD}
\pdfglyphtounicode{afii57841}{05BF}
\pdfglyphtounicode{afii57842}{05C0}
\pdfglyphtounicode{afii57929}{02BC}
\pdfglyphtounicode{afii61248}{2105}
\pdfglyphtounicode{afii61289}{2113}
\pdfglyphtounicode{afii61352}{2116}
\pdfglyphtounicode{afii61573}{202C}
\pdfglyphtounicode{afii61574}{202D}
\pdfglyphtounicode{afii61575}{202E}
\pdfglyphtounicode{afii61664}{200C}
\pdfglyphtounicode{afii63167}{066D}
\pdfglyphtounicode{afii64937}{02BD}
\pdfglyphtounicode{agrave}{00E0}
\pdfglyphtounicode{agujarati}{0A85}
\pdfglyphtounicode{agurmukhi}{0A05}
\pdfglyphtounicode{ahiragana}{3042}
\pdfglyphtounicode{ahookabove}{1EA3}
\pdfglyphtounicode{aibengali}{0990}
\pdfglyphtounicode{aibopomofo}{311E}
\pdfglyphtounicode{aideva}{0910}
\pdfglyphtounicode{aiecyrillic}{04D5}
\pdfglyphtounicode{aigujarati}{0A90}
\pdfglyphtounicode{aigurmukhi}{0A10}
\pdfglyphtounicode{aimatragurmukhi}{0A48}
\pdfglyphtounicode{ainarabic}{0639}
\pdfglyphtounicode{ainfinalarabic}{FECA}
\pdfglyphtounicode{aininitialarabic}{FECB}
\pdfglyphtounicode{ainmedialarabic}{FECC}
\pdfglyphtounicode{ainvertedbreve}{0203}
\pdfglyphtounicode{aivowelsignbengali}{09C8}
\pdfglyphtounicode{aivowelsigndeva}{0948}
\pdfglyphtounicode{aivowelsigngujarati}{0AC8}
\pdfglyphtounicode{akatakana}{30A2}
\pdfglyphtounicode{akatakanahalfwidth}{FF71}
\pdfglyphtounicode{akorean}{314F}
\pdfglyphtounicode{alef}{05D0}
\pdfglyphtounicode{alefarabic}{0627}
\pdfglyphtounicode{alefdageshhebrew}{FB30}
\pdfglyphtounicode{aleffinalarabic}{FE8E}
\pdfglyphtounicode{alefhamzaabovearabic}{0623}
\pdfglyphtounicode{alefhamzaabovefinalarabic}{FE84}
\pdfglyphtounicode{alefhamzabelowarabic}{0625}
\pdfglyphtounicode{alefhamzabelowfinalarabic}{FE88}
\pdfglyphtounicode{alefhebrew}{05D0}
\pdfglyphtounicode{aleflamedhebrew}{FB4F}
\pdfglyphtounicode{alefmaddaabovearabic}{0622}
\pdfglyphtounicode{alefmaddaabovefinalarabic}{FE82}
\pdfglyphtounicode{alefmaksuraarabic}{0649}
\pdfglyphtounicode{alefmaksurafinalarabic}{FEF0}
\pdfglyphtounicode{alefmaksurainitialarabic}{FEF3}
\pdfglyphtounicode{alefmaksuramedialarabic}{FEF4}
\pdfglyphtounicode{alefpatahhebrew}{FB2E}
\pdfglyphtounicode{alefqamatshebrew}{FB2F}
\pdfglyphtounicode{aleph}{2135}
\pdfglyphtounicode{allequal}{224C}
\pdfglyphtounicode{alpha}{03B1}
\pdfglyphtounicode{alphatonos}{03AC}
\pdfglyphtounicode{amacron}{0101}
\pdfglyphtounicode{amonospace}{FF41}
\pdfglyphtounicode{ampersand}{0026}
\pdfglyphtounicode{ampersandmonospace}{FF06}
\pdfglyphtounicode{ampersandsmall}{0026}
\pdfglyphtounicode{amsquare}{33C2}
\pdfglyphtounicode{anbopomofo}{3122}
\pdfglyphtounicode{angbopomofo}{3124}
\pdfglyphtounicode{angbracketleft}{27E8}
\pdfglyphtounicode{angbracketright}{27E9}
\pdfglyphtounicode{angkhankhuthai}{0E5A}
\pdfglyphtounicode{angle}{2220}
\pdfglyphtounicode{anglebracketleft}{3008}
\pdfglyphtounicode{anglebracketleftvertical}{FE3F}
\pdfglyphtounicode{anglebracketright}{3009}
\pdfglyphtounicode{anglebracketrightvertical}{FE40}
\pdfglyphtounicode{angleleft}{2329}
\pdfglyphtounicode{angleright}{232A}
\pdfglyphtounicode{angstrom}{212B}
\pdfglyphtounicode{anoteleia}{0387}
\pdfglyphtounicode{anticlockwise}{27F2}
\pdfglyphtounicode{anudattadeva}{0952}
\pdfglyphtounicode{anusvarabengali}{0982}
\pdfglyphtounicode{anusvaradeva}{0902}
\pdfglyphtounicode{anusvaragujarati}{0A82}
\pdfglyphtounicode{aogonek}{0105}
\pdfglyphtounicode{apaatosquare}{3300}
\pdfglyphtounicode{aparen}{249C}
\pdfglyphtounicode{apostrophearmenian}{055A}
\pdfglyphtounicode{apostrophemod}{02BC}
\pdfglyphtounicode{apple}{F8FF}
\pdfglyphtounicode{approaches}{2250}
\pdfglyphtounicode{approxequal}{2248}
\pdfglyphtounicode{approxequalorimage}{2252}
\pdfglyphtounicode{approximatelyequal}{2245}
\pdfglyphtounicode{approxorequal}{224A}
\pdfglyphtounicode{araeaekorean}{318E}
\pdfglyphtounicode{araeakorean}{318D}
\pdfglyphtounicode{arc}{2312}
\pdfglyphtounicode{archleftdown}{21B6}
\pdfglyphtounicode{archrightdown}{21B7}
\pdfglyphtounicode{arighthalfring}{1E9A}
\pdfglyphtounicode{aring}{00E5}
\pdfglyphtounicode{aringacute}{01FB}
\pdfglyphtounicode{aringbelow}{1E01}
\pdfglyphtounicode{arrowboth}{2194}
\pdfglyphtounicode{arrowbothv}{2195}
\pdfglyphtounicode{arrowdashdown}{21E3}
\pdfglyphtounicode{arrowdashleft}{21E0}
\pdfglyphtounicode{arrowdashright}{21E2}
\pdfglyphtounicode{arrowdashup}{21E1}
\pdfglyphtounicode{arrowdblboth}{21D4}
\pdfglyphtounicode{arrowdblbothv}{21D5}
\pdfglyphtounicode{arrowdbldown}{21D3}
\pdfglyphtounicode{arrowdblleft}{21D0}
\pdfglyphtounicode{arrowdblright}{21D2}
\pdfglyphtounicode{arrowdblup}{21D1}
\pdfglyphtounicode{arrowdown}{2193}
\pdfglyphtounicode{arrowdownleft}{2199}
\pdfglyphtounicode{arrowdownright}{2198}
\pdfglyphtounicode{arrowdownwhite}{21E9}
\pdfglyphtounicode{arrowheaddownmod}{02C5}
\pdfglyphtounicode{arrowheadleftmod}{02C2}
\pdfglyphtounicode{arrowheadrightmod}{02C3}
\pdfglyphtounicode{arrowheadupmod}{02C4}
\pdfglyphtounicode{arrowhorizex}{F8E7}
\pdfglyphtounicode{arrowleft}{2190}
\pdfglyphtounicode{arrowleftbothalf}{21BD}
\pdfglyphtounicode{arrowleftdbl}{21D0}
\pdfglyphtounicode{arrowleftdblstroke}{21CD}
\pdfglyphtounicode{arrowleftoverright}{21C6}
\pdfglyphtounicode{arrowlefttophalf}{21BC}
\pdfglyphtounicode{arrowleftwhite}{21E6}
\pdfglyphtounicode{arrownortheast}{2197}
\pdfglyphtounicode{arrownorthwest}{2196}
\pdfglyphtounicode{arrowparrleftright}{21C6}
\pdfglyphtounicode{arrowparrrightleft}{21C4}
\pdfglyphtounicode{arrowright}{2192}
\pdfglyphtounicode{arrowrightbothalf}{21C1}
\pdfglyphtounicode{arrowrightdblstroke}{21CF}
\pdfglyphtounicode{arrowrightheavy}{279E}
\pdfglyphtounicode{arrowrightoverleft}{21C4}
\pdfglyphtounicode{arrowrighttophalf}{21C0}
\pdfglyphtounicode{arrowrightwhite}{21E8}
\pdfglyphtounicode{arrowsoutheast}{2198}
\pdfglyphtounicode{arrowsouthwest}{2199}
\pdfglyphtounicode{arrowtableft}{21E4}
\pdfglyphtounicode{arrowtabright}{21E5}
\pdfglyphtounicode{arrowtailleft}{21A2}
\pdfglyphtounicode{arrowtailright}{21A3}
\pdfglyphtounicode{arrowtripleleft}{21DA}
\pdfglyphtounicode{arrowtripleright}{21DB}
\pdfglyphtounicode{arrowup}{2191}
\pdfglyphtounicode{arrowupdn}{2195}
\pdfglyphtounicode{arrowupdnbse}{21A8}
\pdfglyphtounicode{arrowupdownbase}{21A8}
\pdfglyphtounicode{arrowupleft}{2196}
\pdfglyphtounicode{arrowupleftofdown}{21C5}
\pdfglyphtounicode{arrowupright}{2197}
\pdfglyphtounicode{arrowupwhite}{21E7}
\pdfglyphtounicode{arrowvertex}{F8E6}
\pdfglyphtounicode{asciicircum}{005E}
\pdfglyphtounicode{asciicircummonospace}{FF3E}
\pdfglyphtounicode{asciitilde}{007E}
\pdfglyphtounicode{asciitildemonospace}{FF5E}
\pdfglyphtounicode{ascript}{0251}
\pdfglyphtounicode{ascriptturned}{0252}
\pdfglyphtounicode{asmallhiragana}{3041}
\pdfglyphtounicode{asmallkatakana}{30A1}
\pdfglyphtounicode{asmallkatakanahalfwidth}{FF67}
\pdfglyphtounicode{asterisk}{002A}
\pdfglyphtounicode{asteriskaltonearabic}{066D}
\pdfglyphtounicode{asteriskarabic}{066D}
\pdfglyphtounicode{asteriskcentered}{2217}
\pdfglyphtounicode{asteriskmath}{2217}
\pdfglyphtounicode{asteriskmonospace}{FF0A}
\pdfglyphtounicode{asterisksmall}{FE61}
\pdfglyphtounicode{asterism}{2042}
\pdfglyphtounicode{asuperior}{0061}
\pdfglyphtounicode{asymptoticallyequal}{2243}
\pdfglyphtounicode{at}{0040}
\pdfglyphtounicode{atilde}{00E3}
\pdfglyphtounicode{atmonospace}{FF20}
\pdfglyphtounicode{atsmall}{FE6B}
\pdfglyphtounicode{aturned}{0250}
\pdfglyphtounicode{aubengali}{0994}
\pdfglyphtounicode{aubopomofo}{3120}
\pdfglyphtounicode{audeva}{0914}
\pdfglyphtounicode{augujarati}{0A94}
\pdfglyphtounicode{augurmukhi}{0A14}
\pdfglyphtounicode{aulengthmarkbengali}{09D7}
\pdfglyphtounicode{aumatragurmukhi}{0A4C}
\pdfglyphtounicode{auvowelsignbengali}{09CC}
\pdfglyphtounicode{auvowelsigndeva}{094C}
\pdfglyphtounicode{auvowelsigngujarati}{0ACC}
\pdfglyphtounicode{avagrahadeva}{093D}
\pdfglyphtounicode{aybarmenian}{0561}
\pdfglyphtounicode{ayin}{05E2}
\pdfglyphtounicode{ayinaltonehebrew}{FB20}
\pdfglyphtounicode{ayinhebrew}{05E2}
\pdfglyphtounicode{b}{0062}
\pdfglyphtounicode{babengali}{09AC}
\pdfglyphtounicode{backslash}{005C}
\pdfglyphtounicode{backslashmonospace}{FF3C}
\pdfglyphtounicode{badeva}{092C}
\pdfglyphtounicode{bagujarati}{0AAC}
\pdfglyphtounicode{bagurmukhi}{0A2C}
\pdfglyphtounicode{bahiragana}{3070}
\pdfglyphtounicode{bahtthai}{0E3F}
\pdfglyphtounicode{bakatakana}{30D0}
\pdfglyphtounicode{bar}{007C}
\pdfglyphtounicode{bardbl}{2225}
\pdfglyphtounicode{barmonospace}{FF5C}
\pdfglyphtounicode{bbopomofo}{3105}
\pdfglyphtounicode{bcircle}{24D1}
\pdfglyphtounicode{bdotaccent}{1E03}
\pdfglyphtounicode{bdotbelow}{1E05}
\pdfglyphtounicode{beamedsixteenthnotes}{266C}
\pdfglyphtounicode{because}{2235}
\pdfglyphtounicode{becyrillic}{0431}
\pdfglyphtounicode{beharabic}{0628}
\pdfglyphtounicode{behfinalarabic}{FE90}
\pdfglyphtounicode{behinitialarabic}{FE91}
\pdfglyphtounicode{behiragana}{3079}
\pdfglyphtounicode{behmedialarabic}{FE92}
\pdfglyphtounicode{behmeeminitialarabic}{FC9F}
\pdfglyphtounicode{behmeemisolatedarabic}{FC08}
\pdfglyphtounicode{behnoonfinalarabic}{FC6D}
\pdfglyphtounicode{bekatakana}{30D9}
\pdfglyphtounicode{benarmenian}{0562}
\pdfglyphtounicode{bet}{05D1}
\pdfglyphtounicode{beta}{03B2}
\pdfglyphtounicode{betasymbolgreek}{03D0}
\pdfglyphtounicode{betdagesh}{FB31}
\pdfglyphtounicode{betdageshhebrew}{FB31}
\pdfglyphtounicode{beth}{2136}
\pdfglyphtounicode{bethebrew}{05D1}
\pdfglyphtounicode{betrafehebrew}{FB4C}
\pdfglyphtounicode{between}{226C}
\pdfglyphtounicode{bhabengali}{09AD}
\pdfglyphtounicode{bhadeva}{092D}
\pdfglyphtounicode{bhagujarati}{0AAD}
\pdfglyphtounicode{bhagurmukhi}{0A2D}
\pdfglyphtounicode{bhook}{0253}
\pdfglyphtounicode{bihiragana}{3073}
\pdfglyphtounicode{bikatakana}{30D3}
\pdfglyphtounicode{bilabialclick}{0298}
\pdfglyphtounicode{bindigurmukhi}{0A02}
\pdfglyphtounicode{birusquare}{3331}
\pdfglyphtounicode{blackcircle}{25CF}
\pdfglyphtounicode{blackdiamond}{25C6}
\pdfglyphtounicode{blackdownpointingtriangle}{25BC}
\pdfglyphtounicode{blackleftpointingpointer}{25C4}
\pdfglyphtounicode{blackleftpointingtriangle}{25C0}
\pdfglyphtounicode{blacklenticularbracketleft}{3010}
\pdfglyphtounicode{blacklenticularbracketleftvertical}{FE3B}
\pdfglyphtounicode{blacklenticularbracketright}{3011}
\pdfglyphtounicode{blacklenticularbracketrightvertical}{FE3C}
\pdfglyphtounicode{blacklowerlefttriangle}{25E3}
\pdfglyphtounicode{blacklowerrighttriangle}{25E2}
\pdfglyphtounicode{blackrectangle}{25AC}
\pdfglyphtounicode{blackrightpointingpointer}{25BA}
\pdfglyphtounicode{blackrightpointingtriangle}{25B6}
\pdfglyphtounicode{blacksmallsquare}{25AA}
\pdfglyphtounicode{blacksmilingface}{263B}
\pdfglyphtounicode{blacksquare}{25A0}
\pdfglyphtounicode{blackstar}{2605}
\pdfglyphtounicode{blackupperlefttriangle}{25E4}
\pdfglyphtounicode{blackupperrighttriangle}{25E5}
\pdfglyphtounicode{blackuppointingsmalltriangle}{25B4}
\pdfglyphtounicode{blackuppointingtriangle}{25B2}
\pdfglyphtounicode{blank}{2423}
\pdfglyphtounicode{blinebelow}{1E07}
\pdfglyphtounicode{block}{2588}
\pdfglyphtounicode{bmonospace}{FF42}
\pdfglyphtounicode{bobaimaithai}{0E1A}
\pdfglyphtounicode{bohiragana}{307C}
\pdfglyphtounicode{bokatakana}{30DC}
\pdfglyphtounicode{bparen}{249D}
\pdfglyphtounicode{bqsquare}{33C3}
\pdfglyphtounicode{braceex}{F8F4}
\pdfglyphtounicode{braceleft}{007B}
\pdfglyphtounicode{braceleftbt}{F8F3}
\pdfglyphtounicode{braceleftmid}{F8F2}
\pdfglyphtounicode{braceleftmonospace}{FF5B}
\pdfglyphtounicode{braceleftsmall}{FE5B}
\pdfglyphtounicode{bracelefttp}{F8F1}
\pdfglyphtounicode{braceleftvertical}{FE37}
\pdfglyphtounicode{braceright}{007D}
\pdfglyphtounicode{bracerightbt}{F8FE}
\pdfglyphtounicode{bracerightmid}{F8FD}
\pdfglyphtounicode{bracerightmonospace}{FF5D}
\pdfglyphtounicode{bracerightsmall}{FE5C}
\pdfglyphtounicode{bracerighttp}{F8FC}
\pdfglyphtounicode{bracerightvertical}{FE38}
\pdfglyphtounicode{bracketleft}{005B}
\pdfglyphtounicode{bracketleftbt}{F8F0}
\pdfglyphtounicode{bracketleftex}{F8EF}
\pdfglyphtounicode{bracketleftmonospace}{FF3B}
\pdfglyphtounicode{bracketlefttp}{F8EE}
\pdfglyphtounicode{bracketright}{005D}
\pdfglyphtounicode{bracketrightbt}{F8FB}
\pdfglyphtounicode{bracketrightex}{F8FA}
\pdfglyphtounicode{bracketrightmonospace}{FF3D}
\pdfglyphtounicode{bracketrighttp}{F8F9}
\pdfglyphtounicode{breve}{02D8}
\pdfglyphtounicode{brevebelowcmb}{032E}
\pdfglyphtounicode{brevecmb}{0306}
\pdfglyphtounicode{breveinvertedbelowcmb}{032F}
\pdfglyphtounicode{breveinvertedcmb}{0311}
\pdfglyphtounicode{breveinverteddoublecmb}{0361}
\pdfglyphtounicode{bridgebelowcmb}{032A}
\pdfglyphtounicode{bridgeinvertedbelowcmb}{033A}
\pdfglyphtounicode{brokenbar}{00A6}
\pdfglyphtounicode{bstroke}{0180}
\pdfglyphtounicode{bsuperior}{0062}
\pdfglyphtounicode{btopbar}{0183}
\pdfglyphtounicode{buhiragana}{3076}
\pdfglyphtounicode{bukatakana}{30D6}
\pdfglyphtounicode{bullet}{2022}
\pdfglyphtounicode{bulletinverse}{25D8}
\pdfglyphtounicode{bulletoperator}{2219}
\pdfglyphtounicode{bullseye}{25CE}
\pdfglyphtounicode{c}{0063}
\pdfglyphtounicode{caarmenian}{056E}
\pdfglyphtounicode{cabengali}{099A}
\pdfglyphtounicode{cacute}{0107}
\pdfglyphtounicode{cadeva}{091A}
\pdfglyphtounicode{cagujarati}{0A9A}
\pdfglyphtounicode{cagurmukhi}{0A1A}
\pdfglyphtounicode{calsquare}{3388}
\pdfglyphtounicode{candrabindubengali}{0981}
\pdfglyphtounicode{candrabinducmb}{0310}
\pdfglyphtounicode{candrabindudeva}{0901}
\pdfglyphtounicode{candrabindugujarati}{0A81}
\pdfglyphtounicode{capslock}{21EA}
\pdfglyphtounicode{careof}{2105}
\pdfglyphtounicode{caron}{02C7}
\pdfglyphtounicode{caronbelowcmb}{032C}
\pdfglyphtounicode{caroncmb}{030C}
\pdfglyphtounicode{carriagereturn}{21B5}
\pdfglyphtounicode{cbopomofo}{3118}
\pdfglyphtounicode{ccaron}{010D}
\pdfglyphtounicode{ccedilla}{00E7}
\pdfglyphtounicode{ccedillaacute}{1E09}
\pdfglyphtounicode{ccircle}{24D2}
\pdfglyphtounicode{ccircumflex}{0109}
\pdfglyphtounicode{ccurl}{0255}
\pdfglyphtounicode{cdot}{010B}
\pdfglyphtounicode{cdotaccent}{010B}
\pdfglyphtounicode{cdsquare}{33C5}
\pdfglyphtounicode{cedilla}{00B8}
\pdfglyphtounicode{cedillacmb}{0327}
\pdfglyphtounicode{ceilingleft}{2308}
\pdfglyphtounicode{ceilingright}{2309}
\pdfglyphtounicode{cent}{00A2}
\pdfglyphtounicode{centigrade}{2103}
\pdfglyphtounicode{centinferior}{00A2}
\pdfglyphtounicode{centmonospace}{FFE0}
\pdfglyphtounicode{centoldstyle}{00A2}
\pdfglyphtounicode{centsuperior}{00A2}
\pdfglyphtounicode{chaarmenian}{0579}
\pdfglyphtounicode{chabengali}{099B}
\pdfglyphtounicode{chadeva}{091B}
\pdfglyphtounicode{chagujarati}{0A9B}
\pdfglyphtounicode{chagurmukhi}{0A1B}
\pdfglyphtounicode{chbopomofo}{3114}
\pdfglyphtounicode{cheabkhasiancyrillic}{04BD}
\pdfglyphtounicode{check}{2713}
\pdfglyphtounicode{checkmark}{2713}
\pdfglyphtounicode{checyrillic}{0447}
\pdfglyphtounicode{chedescenderabkhasiancyrillic}{04BF}
\pdfglyphtounicode{chedescendercyrillic}{04B7}
\pdfglyphtounicode{chedieresiscyrillic}{04F5}
\pdfglyphtounicode{cheharmenian}{0573}
\pdfglyphtounicode{chekhakassiancyrillic}{04CC}
\pdfglyphtounicode{cheverticalstrokecyrillic}{04B9}
\pdfglyphtounicode{chi}{03C7}
\pdfglyphtounicode{chieuchacirclekorean}{3277}
\pdfglyphtounicode{chieuchaparenkorean}{3217}
\pdfglyphtounicode{chieuchcirclekorean}{3269}
\pdfglyphtounicode{chieuchkorean}{314A}
\pdfglyphtounicode{chieuchparenkorean}{3209}
\pdfglyphtounicode{chochangthai}{0E0A}
\pdfglyphtounicode{chochanthai}{0E08}
\pdfglyphtounicode{chochingthai}{0E09}
\pdfglyphtounicode{chochoethai}{0E0C}
\pdfglyphtounicode{chook}{0188}
\pdfglyphtounicode{cieucacirclekorean}{3276}
\pdfglyphtounicode{cieucaparenkorean}{3216}
\pdfglyphtounicode{cieuccirclekorean}{3268}
\pdfglyphtounicode{cieuckorean}{3148}
\pdfglyphtounicode{cieucparenkorean}{3208}
\pdfglyphtounicode{cieucuparenkorean}{321C}
\pdfglyphtounicode{circle}{25CB}
\pdfglyphtounicode{circleR}{00AE}
\pdfglyphtounicode{circleS}{24C8}
\pdfglyphtounicode{circleasterisk}{229B}
\pdfglyphtounicode{circlecopyrt}{20DD}
\pdfglyphtounicode{circledivide}{2298}
\pdfglyphtounicode{circledot}{2299}
\pdfglyphtounicode{circleequal}{229C}
\pdfglyphtounicode{circleminus}{2296}
\pdfglyphtounicode{circlemultiply}{2297}
\pdfglyphtounicode{circleot}{2299}
\pdfglyphtounicode{circleplus}{2295}
\pdfglyphtounicode{circlepostalmark}{3036}
\pdfglyphtounicode{circlering}{229A}
\pdfglyphtounicode{circlewithlefthalfblack}{25D0}
\pdfglyphtounicode{circlewithrighthalfblack}{25D1}
\pdfglyphtounicode{circumflex}{02C6}
\pdfglyphtounicode{circumflexbelowcmb}{032D}
\pdfglyphtounicode{circumflexcmb}{0302}
\pdfglyphtounicode{clear}{2327}
\pdfglyphtounicode{clickalveolar}{01C2}
\pdfglyphtounicode{clickdental}{01C0}
\pdfglyphtounicode{clicklateral}{01C1}
\pdfglyphtounicode{clickretroflex}{01C3}
\pdfglyphtounicode{clockwise}{27F3}
\pdfglyphtounicode{club}{2663}
\pdfglyphtounicode{clubsuitblack}{2663}
\pdfglyphtounicode{clubsuitwhite}{2667}
\pdfglyphtounicode{cmcubedsquare}{33A4}
\pdfglyphtounicode{cmonospace}{FF43}
\pdfglyphtounicode{cmsquaredsquare}{33A0}
\pdfglyphtounicode{coarmenian}{0581}
\pdfglyphtounicode{colon}{003A}
\pdfglyphtounicode{colonmonetary}{20A1}
\pdfglyphtounicode{colonmonospace}{FF1A}
\pdfglyphtounicode{colonsign}{20A1}
\pdfglyphtounicode{colonsmall}{FE55}
\pdfglyphtounicode{colontriangularhalfmod}{02D1}
\pdfglyphtounicode{colontriangularmod}{02D0}
\pdfglyphtounicode{comma}{002C}
\pdfglyphtounicode{commaabovecmb}{0313}
\pdfglyphtounicode{commaaboverightcmb}{0315}
\pdfglyphtounicode{commaaccent}{F6C3}
\pdfglyphtounicode{commaarabic}{060C}
\pdfglyphtounicode{commaarmenian}{055D}
\pdfglyphtounicode{commainferior}{002C}
\pdfglyphtounicode{commamonospace}{FF0C}
\pdfglyphtounicode{commareversedabovecmb}{0314}
\pdfglyphtounicode{commareversedmod}{02BD}
\pdfglyphtounicode{commasmall}{FE50}
\pdfglyphtounicode{commasuperior}{002C}
\pdfglyphtounicode{commaturnedabovecmb}{0312}
\pdfglyphtounicode{commaturnedmod}{02BB}
\pdfglyphtounicode{compass}{263C}
\pdfglyphtounicode{complement}{2201}
\pdfglyphtounicode{compwordmark}{200C}
\pdfglyphtounicode{congruent}{2245}
\pdfglyphtounicode{contourintegral}{222E}
\pdfglyphtounicode{control}{2303}
\pdfglyphtounicode{controlACK}{0006}
\pdfglyphtounicode{controlBEL}{0007}
\pdfglyphtounicode{controlBS}{0008}
\pdfglyphtounicode{controlCAN}{0018}
\pdfglyphtounicode{controlCR}{000D}
\pdfglyphtounicode{controlDC1}{0011}
\pdfglyphtounicode{controlDC2}{0012}
\pdfglyphtounicode{controlDC3}{0013}
\pdfglyphtounicode{controlDC4}{0014}
\pdfglyphtounicode{controlDEL}{007F}
\pdfglyphtounicode{controlDLE}{0010}
\pdfglyphtounicode{controlEM}{0019}
\pdfglyphtounicode{controlENQ}{0005}
\pdfglyphtounicode{controlEOT}{0004}
\pdfglyphtounicode{controlESC}{001B}
\pdfglyphtounicode{controlETB}{0017}
\pdfglyphtounicode{controlETX}{0003}
\pdfglyphtounicode{controlFF}{000C}
\pdfglyphtounicode{controlFS}{001C}
\pdfglyphtounicode{controlGS}{001D}
\pdfglyphtounicode{controlHT}{0009}
\pdfglyphtounicode{controlLF}{000A}
\pdfglyphtounicode{controlNAK}{0015}
\pdfglyphtounicode{controlRS}{001E}
\pdfglyphtounicode{controlSI}{000F}
\pdfglyphtounicode{controlSO}{000E}
\pdfglyphtounicode{controlSOT}{0002}
\pdfglyphtounicode{controlSTX}{0001}
\pdfglyphtounicode{controlSUB}{001A}
\pdfglyphtounicode{controlSYN}{0016}
\pdfglyphtounicode{controlUS}{001F}
\pdfglyphtounicode{controlVT}{000B}
\pdfglyphtounicode{coproduct}{2A3F}
\pdfglyphtounicode{copyright}{00A9}
\pdfglyphtounicode{copyrightsans}{00A9}
\pdfglyphtounicode{copyrightserif}{00A9}
\pdfglyphtounicode{cornerbracketleft}{300C}
\pdfglyphtounicode{cornerbracketlefthalfwidth}{FF62}
\pdfglyphtounicode{cornerbracketleftvertical}{FE41}
\pdfglyphtounicode{cornerbracketright}{300D}
\pdfglyphtounicode{cornerbracketrighthalfwidth}{FF63}
\pdfglyphtounicode{cornerbracketrightvertical}{FE42}
\pdfglyphtounicode{corporationsquare}{337F}
\pdfglyphtounicode{cosquare}{33C7}
\pdfglyphtounicode{coverkgsquare}{33C6}
\pdfglyphtounicode{cparen}{249E}
\pdfglyphtounicode{cruzeiro}{20A2}
\pdfglyphtounicode{cstretched}{0297}
\pdfglyphtounicode{ct}{0063 0074}
\pdfglyphtounicode{curlyand}{22CF}
\pdfglyphtounicode{curlyleft}{21AB}
\pdfglyphtounicode{curlyor}{22CE}
\pdfglyphtounicode{curlyright}{21AC}
\pdfglyphtounicode{currency}{00A4}
\pdfglyphtounicode{cwm}{200C}
\pdfglyphtounicode{cyrBreve}{02D8}
\pdfglyphtounicode{cyrFlex}{00A0 0311}
\pdfglyphtounicode{cyrbreve}{02D8}
\pdfglyphtounicode{cyrflex}{00A0 0311}
\pdfglyphtounicode{d}{0064}
\pdfglyphtounicode{daarmenian}{0564}
\pdfglyphtounicode{dabengali}{09A6}
\pdfglyphtounicode{dadarabic}{0636}
\pdfglyphtounicode{dadeva}{0926}
\pdfglyphtounicode{dadfinalarabic}{FEBE}
\pdfglyphtounicode{dadinitialarabic}{FEBF}
\pdfglyphtounicode{dadmedialarabic}{FEC0}
\pdfglyphtounicode{dagesh}{05BC}
\pdfglyphtounicode{dageshhebrew}{05BC}
\pdfglyphtounicode{dagger}{2020}
\pdfglyphtounicode{daggerdbl}{2021}
\pdfglyphtounicode{dagujarati}{0AA6}
\pdfglyphtounicode{dagurmukhi}{0A26}
\pdfglyphtounicode{dahiragana}{3060}
\pdfglyphtounicode{dakatakana}{30C0}
\pdfglyphtounicode{dalarabic}{062F}
\pdfglyphtounicode{dalet}{05D3}
\pdfglyphtounicode{daletdagesh}{FB33}
\pdfglyphtounicode{daletdageshhebrew}{FB33}
\pdfglyphtounicode{daleth}{2138}
\pdfglyphtounicode{dalethatafpatah}{05D3 05B2}
\pdfglyphtounicode{dalethatafpatahhebrew}{05D3 05B2}
\pdfglyphtounicode{dalethatafsegol}{05D3 05B1}
\pdfglyphtounicode{dalethatafsegolhebrew}{05D3 05B1}
\pdfglyphtounicode{dalethebrew}{05D3}
\pdfglyphtounicode{dalethiriq}{05D3 05B4}
\pdfglyphtounicode{dalethiriqhebrew}{05D3 05B4}
\pdfglyphtounicode{daletholam}{05D3 05B9}
\pdfglyphtounicode{daletholamhebrew}{05D3 05B9}
\pdfglyphtounicode{daletpatah}{05D3 05B7}
\pdfglyphtounicode{daletpatahhebrew}{05D3 05B7}
\pdfglyphtounicode{daletqamats}{05D3 05B8}
\pdfglyphtounicode{daletqamatshebrew}{05D3 05B8}
\pdfglyphtounicode{daletqubuts}{05D3 05BB}
\pdfglyphtounicode{daletqubutshebrew}{05D3 05BB}
\pdfglyphtounicode{daletsegol}{05D3 05B6}
\pdfglyphtounicode{daletsegolhebrew}{05D3 05B6}
\pdfglyphtounicode{daletsheva}{05D3 05B0}
\pdfglyphtounicode{daletshevahebrew}{05D3 05B0}
\pdfglyphtounicode{dalettsere}{05D3 05B5}
\pdfglyphtounicode{dalettserehebrew}{05D3 05B5}
\pdfglyphtounicode{dalfinalarabic}{FEAA}
\pdfglyphtounicode{dammaarabic}{064F}
\pdfglyphtounicode{dammalowarabic}{064F}
\pdfglyphtounicode{dammatanaltonearabic}{064C}
\pdfglyphtounicode{dammatanarabic}{064C}
\pdfglyphtounicode{danda}{0964}
\pdfglyphtounicode{dargahebrew}{05A7}
\pdfglyphtounicode{dargalefthebrew}{05A7}
\pdfglyphtounicode{dasiapneumatacyrilliccmb}{0485}
\pdfglyphtounicode{dbar}{0111}
\pdfglyphtounicode{dblGrave}{00A0 030F}
\pdfglyphtounicode{dblanglebracketleft}{300A}
\pdfglyphtounicode{dblanglebracketleftvertical}{FE3D}
\pdfglyphtounicode{dblanglebracketright}{300B}
\pdfglyphtounicode{dblanglebracketrightvertical}{FE3E}
\pdfglyphtounicode{dblarchinvertedbelowcmb}{032B}
\pdfglyphtounicode{dblarrowdwn}{21CA}
\pdfglyphtounicode{dblarrowheadleft}{219E}
\pdfglyphtounicode{dblarrowheadright}{21A0}
\pdfglyphtounicode{dblarrowleft}{21D4}
\pdfglyphtounicode{dblarrowright}{21D2}
\pdfglyphtounicode{dblarrowup}{21C8}
\pdfglyphtounicode{dblbracketleft}{27E6}
\pdfglyphtounicode{dblbracketright}{27E7}
\pdfglyphtounicode{dbldanda}{0965}
\pdfglyphtounicode{dblgrave}{00A0 030F}
\pdfglyphtounicode{dblgravecmb}{030F}
\pdfglyphtounicode{dblintegral}{222C}
\pdfglyphtounicode{dbllowline}{2017}
\pdfglyphtounicode{dbllowlinecmb}{0333}
\pdfglyphtounicode{dbloverlinecmb}{033F}
\pdfglyphtounicode{dblprimemod}{02BA}
\pdfglyphtounicode{dblverticalbar}{2016}
\pdfglyphtounicode{dblverticallineabovecmb}{030E}
\pdfglyphtounicode{dbopomofo}{3109}
\pdfglyphtounicode{dbsquare}{33C8}
\pdfglyphtounicode{dcaron}{010F}
\pdfglyphtounicode{dcedilla}{1E11}
\pdfglyphtounicode{dcircle}{24D3}
\pdfglyphtounicode{dcircumflexbelow}{1E13}
\pdfglyphtounicode{dcroat}{0111}
\pdfglyphtounicode{ddabengali}{09A1}
\pdfglyphtounicode{ddadeva}{0921}
\pdfglyphtounicode{ddagujarati}{0AA1}
\pdfglyphtounicode{ddagurmukhi}{0A21}
\pdfglyphtounicode{ddalarabic}{0688}
\pdfglyphtounicode{ddalfinalarabic}{FB89}
\pdfglyphtounicode{dddhadeva}{095C}
\pdfglyphtounicode{ddhabengali}{09A2}
\pdfglyphtounicode{ddhadeva}{0922}
\pdfglyphtounicode{ddhagujarati}{0AA2}
\pdfglyphtounicode{ddhagurmukhi}{0A22}
\pdfglyphtounicode{ddotaccent}{1E0B}
\pdfglyphtounicode{ddotbelow}{1E0D}
\pdfglyphtounicode{decimalseparatorarabic}{066B}
\pdfglyphtounicode{decimalseparatorpersian}{066B}
\pdfglyphtounicode{decyrillic}{0434}
\pdfglyphtounicode{defines}{225C}
\pdfglyphtounicode{degree}{00B0}
\pdfglyphtounicode{dehihebrew}{05AD}
\pdfglyphtounicode{dehiragana}{3067}
\pdfglyphtounicode{deicoptic}{03EF}
\pdfglyphtounicode{dekatakana}{30C7}
\pdfglyphtounicode{deleteleft}{232B}
\pdfglyphtounicode{deleteright}{2326}
\pdfglyphtounicode{delta}{03B4}
\pdfglyphtounicode{deltaturned}{018D}
\pdfglyphtounicode{denominatorminusonenumeratorbengali}{09F8}
\pdfglyphtounicode{dezh}{02A4}
\pdfglyphtounicode{dhabengali}{09A7}
\pdfglyphtounicode{dhadeva}{0927}
\pdfglyphtounicode{dhagujarati}{0AA7}
\pdfglyphtounicode{dhagurmukhi}{0A27}
\pdfglyphtounicode{dhook}{0257}
\pdfglyphtounicode{dialytikatonos}{0385}
\pdfglyphtounicode{dialytikatonoscmb}{0344}
\pdfglyphtounicode{diamond}{2666}
\pdfglyphtounicode{diamondmath}{22C4}
\pdfglyphtounicode{diamondsolid}{2666}
\pdfglyphtounicode{diamondsuitwhite}{2662}
\pdfglyphtounicode{dieresis}{00A8}
\pdfglyphtounicode{dieresisacute}{00A0 0308 0301}
\pdfglyphtounicode{dieresisbelowcmb}{0324}
\pdfglyphtounicode{dieresiscmb}{0308}
\pdfglyphtounicode{dieresisgrave}{00A0 0308 0300}
\pdfglyphtounicode{dieresistonos}{0385}
\pdfglyphtounicode{difference}{224F}
\pdfglyphtounicode{dihiragana}{3062}
\pdfglyphtounicode{dikatakana}{30C2}
\pdfglyphtounicode{dittomark}{3003}
\pdfglyphtounicode{divide}{00F7}
\pdfglyphtounicode{dividemultiply}{22C7}
\pdfglyphtounicode{divides}{2223}
\pdfglyphtounicode{divisionslash}{2215}
\pdfglyphtounicode{djecyrillic}{0452}
\pdfglyphtounicode{dkshade}{2593}
\pdfglyphtounicode{dlinebelow}{1E0F}
\pdfglyphtounicode{dlsquare}{3397}
\pdfglyphtounicode{dmacron}{0111}
\pdfglyphtounicode{dmonospace}{FF44}
\pdfglyphtounicode{dnblock}{2584}
\pdfglyphtounicode{dochadathai}{0E0E}
\pdfglyphtounicode{dodekthai}{0E14}
\pdfglyphtounicode{dohiragana}{3069}
\pdfglyphtounicode{dokatakana}{30C9}
\pdfglyphtounicode{dollar}{0024}
\pdfglyphtounicode{dollarinferior}{0024}
\pdfglyphtounicode{dollarmonospace}{FF04}
\pdfglyphtounicode{dollaroldstyle}{0024}
\pdfglyphtounicode{dollarsmall}{FE69}
\pdfglyphtounicode{dollarsuperior}{0024}
\pdfglyphtounicode{dong}{20AB}
\pdfglyphtounicode{dorusquare}{3326}
\pdfglyphtounicode{dotaccent}{02D9}
\pdfglyphtounicode{dotaccentcmb}{0307}
\pdfglyphtounicode{dotbelowcmb}{0323}
\pdfglyphtounicode{dotbelowcomb}{0323}
\pdfglyphtounicode{dotkatakana}{30FB}
\pdfglyphtounicode{dotlessi}{0131}
\pdfglyphtounicode{dotlessj}{0237}
\pdfglyphtounicode{dotlessjstrokehook}{0284}
\pdfglyphtounicode{dotmath}{22C5}
\pdfglyphtounicode{dotplus}{2214}
\pdfglyphtounicode{dottedcircle}{25CC}
\pdfglyphtounicode{doubleyodpatah}{FB1F}
\pdfglyphtounicode{doubleyodpatahhebrew}{FB1F}
\pdfglyphtounicode{downfall}{22CE}
\pdfglyphtounicode{downslope}{29F9}
\pdfglyphtounicode{downtackbelowcmb}{031E}
\pdfglyphtounicode{downtackmod}{02D5}
\pdfglyphtounicode{dparen}{249F}
\pdfglyphtounicode{dsuperior}{0064}
\pdfglyphtounicode{dtail}{0256}
\pdfglyphtounicode{dtopbar}{018C}
\pdfglyphtounicode{duhiragana}{3065}
\pdfglyphtounicode{dukatakana}{30C5}
\pdfglyphtounicode{dz}{01F3}
\pdfglyphtounicode{dzaltone}{02A3}
\pdfglyphtounicode{dzcaron}{01C6}
\pdfglyphtounicode{dzcurl}{02A5}
\pdfglyphtounicode{dzeabkhasiancyrillic}{04E1}
\pdfglyphtounicode{dzecyrillic}{0455}
\pdfglyphtounicode{dzhecyrillic}{045F}
\pdfglyphtounicode{e}{0065}
\pdfglyphtounicode{eacute}{00E9}
\pdfglyphtounicode{earth}{2641}
\pdfglyphtounicode{ebengali}{098F}
\pdfglyphtounicode{ebopomofo}{311C}
\pdfglyphtounicode{ebreve}{0115}
\pdfglyphtounicode{ecandradeva}{090D}
\pdfglyphtounicode{ecandragujarati}{0A8D}
\pdfglyphtounicode{ecandravowelsigndeva}{0945}
\pdfglyphtounicode{ecandravowelsigngujarati}{0AC5}
\pdfglyphtounicode{ecaron}{011B}
\pdfglyphtounicode{ecedillabreve}{1E1D}
\pdfglyphtounicode{echarmenian}{0565}
\pdfglyphtounicode{echyiwnarmenian}{0587}
\pdfglyphtounicode{ecircle}{24D4}
\pdfglyphtounicode{ecircumflex}{00EA}
\pdfglyphtounicode{ecircumflexacute}{1EBF}
\pdfglyphtounicode{ecircumflexbelow}{1E19}
\pdfglyphtounicode{ecircumflexdotbelow}{1EC7}
\pdfglyphtounicode{ecircumflexgrave}{1EC1}
\pdfglyphtounicode{ecircumflexhookabove}{1EC3}
\pdfglyphtounicode{ecircumflextilde}{1EC5}
\pdfglyphtounicode{ecyrillic}{0454}
\pdfglyphtounicode{edblgrave}{0205}
\pdfglyphtounicode{edeva}{090F}
\pdfglyphtounicode{edieresis}{00EB}
\pdfglyphtounicode{edot}{0117}
\pdfglyphtounicode{edotaccent}{0117}
\pdfglyphtounicode{edotbelow}{1EB9}
\pdfglyphtounicode{eegurmukhi}{0A0F}
\pdfglyphtounicode{eematragurmukhi}{0A47}
\pdfglyphtounicode{efcyrillic}{0444}
\pdfglyphtounicode{egrave}{00E8}
\pdfglyphtounicode{egujarati}{0A8F}
\pdfglyphtounicode{eharmenian}{0567}
\pdfglyphtounicode{ehbopomofo}{311D}
\pdfglyphtounicode{ehiragana}{3048}
\pdfglyphtounicode{ehookabove}{1EBB}
\pdfglyphtounicode{eibopomofo}{311F}
\pdfglyphtounicode{eight}{0038}
\pdfglyphtounicode{eightarabic}{0668}
\pdfglyphtounicode{eightbengali}{09EE}
\pdfglyphtounicode{eightcircle}{2467}
\pdfglyphtounicode{eightcircleinversesansserif}{2791}
\pdfglyphtounicode{eightdeva}{096E}
\pdfglyphtounicode{eighteencircle}{2471}
\pdfglyphtounicode{eighteenparen}{2485}
\pdfglyphtounicode{eighteenperiod}{2499}
\pdfglyphtounicode{eightgujarati}{0AEE}
\pdfglyphtounicode{eightgurmukhi}{0A6E}
\pdfglyphtounicode{eighthackarabic}{0668}
\pdfglyphtounicode{eighthangzhou}{3028}
\pdfglyphtounicode{eighthnotebeamed}{266B}
\pdfglyphtounicode{eightideographicparen}{3227}
\pdfglyphtounicode{eightinferior}{2088}
\pdfglyphtounicode{eightmonospace}{FF18}
\pdfglyphtounicode{eightoldstyle}{0038}
\pdfglyphtounicode{eightparen}{247B}
\pdfglyphtounicode{eightperiod}{248F}
\pdfglyphtounicode{eightpersian}{06F8}
\pdfglyphtounicode{eightroman}{2177}
\pdfglyphtounicode{eightsuperior}{2078}
\pdfglyphtounicode{eightthai}{0E58}
\pdfglyphtounicode{einvertedbreve}{0207}
\pdfglyphtounicode{eiotifiedcyrillic}{0465}
\pdfglyphtounicode{ekatakana}{30A8}
\pdfglyphtounicode{ekatakanahalfwidth}{FF74}
\pdfglyphtounicode{ekonkargurmukhi}{0A74}
\pdfglyphtounicode{ekorean}{3154}
\pdfglyphtounicode{elcyrillic}{043B}
\pdfglyphtounicode{element}{2208}
\pdfglyphtounicode{elevencircle}{246A}
\pdfglyphtounicode{elevenparen}{247E}
\pdfglyphtounicode{elevenperiod}{2492}
\pdfglyphtounicode{elevenroman}{217A}
\pdfglyphtounicode{ellipsis}{2026}
\pdfglyphtounicode{ellipsisvertical}{22EE}
\pdfglyphtounicode{emacron}{0113}
\pdfglyphtounicode{emacronacute}{1E17}
\pdfglyphtounicode{emacrongrave}{1E15}
\pdfglyphtounicode{emcyrillic}{043C}
\pdfglyphtounicode{emdash}{2014}
\pdfglyphtounicode{emdashvertical}{FE31}
\pdfglyphtounicode{emonospace}{FF45}
\pdfglyphtounicode{emphasismarkarmenian}{055B}
\pdfglyphtounicode{emptyset}{2205}
\pdfglyphtounicode{enbopomofo}{3123}
\pdfglyphtounicode{encyrillic}{043D}
\pdfglyphtounicode{endash}{2013}
\pdfglyphtounicode{endashvertical}{FE32}
\pdfglyphtounicode{endescendercyrillic}{04A3}
\pdfglyphtounicode{eng}{014B}
\pdfglyphtounicode{engbopomofo}{3125}
\pdfglyphtounicode{enghecyrillic}{04A5}
\pdfglyphtounicode{enhookcyrillic}{04C8}
\pdfglyphtounicode{enspace}{2002}
\pdfglyphtounicode{eogonek}{0119}
\pdfglyphtounicode{eokorean}{3153}
\pdfglyphtounicode{eopen}{025B}
\pdfglyphtounicode{eopenclosed}{029A}
\pdfglyphtounicode{eopenreversed}{025C}
\pdfglyphtounicode{eopenreversedclosed}{025E}
\pdfglyphtounicode{eopenreversedhook}{025D}
\pdfglyphtounicode{eparen}{24A0}
\pdfglyphtounicode{epsilon}{03B5}
\pdfglyphtounicode{epsilon1}{03F5}
\pdfglyphtounicode{epsiloninv}{03F6}
\pdfglyphtounicode{epsilontonos}{03AD}
\pdfglyphtounicode{equal}{003D}
\pdfglyphtounicode{equaldotleftright}{2252}
\pdfglyphtounicode{equaldotrightleft}{2253}
\pdfglyphtounicode{equalmonospace}{FF1D}
\pdfglyphtounicode{equalorfollows}{22DF}
\pdfglyphtounicode{equalorgreater}{2A96}
\pdfglyphtounicode{equalorless}{2A95}
\pdfglyphtounicode{equalorprecedes}{22DE}
\pdfglyphtounicode{equalorsimilar}{2242}
\pdfglyphtounicode{equalsdots}{2251}
\pdfglyphtounicode{equalsmall}{FE66}
\pdfglyphtounicode{equalsuperior}{207C}
\pdfglyphtounicode{equivalence}{2261}
\pdfglyphtounicode{equivasymptotic}{224D}
\pdfglyphtounicode{erbopomofo}{3126}
\pdfglyphtounicode{ercyrillic}{0440}
\pdfglyphtounicode{ereversed}{0258}
\pdfglyphtounicode{ereversedcyrillic}{044D}
\pdfglyphtounicode{escyrillic}{0441}
\pdfglyphtounicode{esdescendercyrillic}{04AB}
\pdfglyphtounicode{esh}{0283}
\pdfglyphtounicode{eshcurl}{0286}
\pdfglyphtounicode{eshortdeva}{090E}
\pdfglyphtounicode{eshortvowelsigndeva}{0946}
\pdfglyphtounicode{eshreversedloop}{01AA}
\pdfglyphtounicode{eshsquatreversed}{0285}
\pdfglyphtounicode{esmallhiragana}{3047}
\pdfglyphtounicode{esmallkatakana}{30A7}
\pdfglyphtounicode{esmallkatakanahalfwidth}{FF6A}
\pdfglyphtounicode{estimated}{212E}
\pdfglyphtounicode{esuperior}{0065}
\pdfglyphtounicode{eta}{03B7}
\pdfglyphtounicode{etarmenian}{0568}
\pdfglyphtounicode{etatonos}{03AE}
\pdfglyphtounicode{eth}{00F0}
\pdfglyphtounicode{etilde}{1EBD}
\pdfglyphtounicode{etildebelow}{1E1B}
\pdfglyphtounicode{etnahtafoukhhebrew}{0591}
\pdfglyphtounicode{etnahtafoukhlefthebrew}{0591}
\pdfglyphtounicode{etnahtahebrew}{0591}
\pdfglyphtounicode{etnahtalefthebrew}{0591}
\pdfglyphtounicode{eturned}{01DD}
\pdfglyphtounicode{eukorean}{3161}
\pdfglyphtounicode{euro}{20AC}
\pdfglyphtounicode{evowelsignbengali}{09C7}
\pdfglyphtounicode{evowelsigndeva}{0947}
\pdfglyphtounicode{evowelsigngujarati}{0AC7}
\pdfglyphtounicode{exclam}{0021}
\pdfglyphtounicode{exclamarmenian}{055C}
\pdfglyphtounicode{exclamdbl}{203C}
\pdfglyphtounicode{exclamdown}{00A1}
\pdfglyphtounicode{exclamdownsmall}{00A1}
\pdfglyphtounicode{exclammonospace}{FF01}
\pdfglyphtounicode{exclamsmall}{0021}
\pdfglyphtounicode{existential}{2203}
\pdfglyphtounicode{ezh}{0292}
\pdfglyphtounicode{ezhcaron}{01EF}
\pdfglyphtounicode{ezhcurl}{0293}
\pdfglyphtounicode{ezhreversed}{01B9}
\pdfglyphtounicode{ezhtail}{01BA}
\pdfglyphtounicode{f}{0066}
\pdfglyphtounicode{fadeva}{095E}
\pdfglyphtounicode{fagurmukhi}{0A5E}
\pdfglyphtounicode{fahrenheit}{2109}
\pdfglyphtounicode{fathaarabic}{064E}
\pdfglyphtounicode{fathalowarabic}{064E}
\pdfglyphtounicode{fathatanarabic}{064B}
\pdfglyphtounicode{fbopomofo}{3108}
\pdfglyphtounicode{fcircle}{24D5}
\pdfglyphtounicode{fdotaccent}{1E1F}
\pdfglyphtounicode{feharabic}{0641}
\pdfglyphtounicode{feharmenian}{0586}
\pdfglyphtounicode{fehfinalarabic}{FED2}
\pdfglyphtounicode{fehinitialarabic}{FED3}
\pdfglyphtounicode{fehmedialarabic}{FED4}
\pdfglyphtounicode{feicoptic}{03E5}
\pdfglyphtounicode{female}{2640}
\pdfglyphtounicode{ff}{0066 0066}
\pdfglyphtounicode{ffi}{0066 0066 0069}
\pdfglyphtounicode{ffl}{0066 0066 006C}
\pdfglyphtounicode{fi}{0066 0069}
\pdfglyphtounicode{fifteencircle}{246E}
\pdfglyphtounicode{fifteenparen}{2482}
\pdfglyphtounicode{fifteenperiod}{2496}
\pdfglyphtounicode{figuredash}{2012}
\pdfglyphtounicode{filledbox}{25A0}
\pdfglyphtounicode{filledrect}{25AC}
\pdfglyphtounicode{finalkaf}{05DA}
\pdfglyphtounicode{finalkafdagesh}{FB3A}
\pdfglyphtounicode{finalkafdageshhebrew}{FB3A}
\pdfglyphtounicode{finalkafhebrew}{05DA}
\pdfglyphtounicode{finalkafqamats}{05DA 05B8}
\pdfglyphtounicode{finalkafqamatshebrew}{05DA 05B8}
\pdfglyphtounicode{finalkafsheva}{05DA 05B0}
\pdfglyphtounicode{finalkafshevahebrew}{05DA 05B0}
\pdfglyphtounicode{finalmem}{05DD}
\pdfglyphtounicode{finalmemhebrew}{05DD}
\pdfglyphtounicode{finalnun}{05DF}
\pdfglyphtounicode{finalnunhebrew}{05DF}
\pdfglyphtounicode{finalpe}{05E3}
\pdfglyphtounicode{finalpehebrew}{05E3}
\pdfglyphtounicode{finaltsadi}{05E5}
\pdfglyphtounicode{finaltsadihebrew}{05E5}
\pdfglyphtounicode{firsttonechinese}{02C9}
\pdfglyphtounicode{fisheye}{25C9}
\pdfglyphtounicode{fitacyrillic}{0473}
\pdfglyphtounicode{five}{0035}
\pdfglyphtounicode{fivearabic}{0665}
\pdfglyphtounicode{fivebengali}{09EB}
\pdfglyphtounicode{fivecircle}{2464}
\pdfglyphtounicode{fivecircleinversesansserif}{278E}
\pdfglyphtounicode{fivedeva}{096B}
\pdfglyphtounicode{fiveeighths}{215D}
\pdfglyphtounicode{fivegujarati}{0AEB}
\pdfglyphtounicode{fivegurmukhi}{0A6B}
\pdfglyphtounicode{fivehackarabic}{0665}
\pdfglyphtounicode{fivehangzhou}{3025}
\pdfglyphtounicode{fiveideographicparen}{3224}
\pdfglyphtounicode{fiveinferior}{2085}
\pdfglyphtounicode{fivemonospace}{FF15}
\pdfglyphtounicode{fiveoldstyle}{0035}
\pdfglyphtounicode{fiveparen}{2478}
\pdfglyphtounicode{fiveperiod}{248C}
\pdfglyphtounicode{fivepersian}{06F5}
\pdfglyphtounicode{fiveroman}{2174}
\pdfglyphtounicode{fivesuperior}{2075}
\pdfglyphtounicode{fivethai}{0E55}
\pdfglyphtounicode{fl}{0066 006C}
\pdfglyphtounicode{flat}{266D}
\pdfglyphtounicode{floorleft}{230A}
\pdfglyphtounicode{floorright}{230B}
\pdfglyphtounicode{florin}{0192}
\pdfglyphtounicode{fmonospace}{FF46}
\pdfglyphtounicode{fmsquare}{3399}
\pdfglyphtounicode{fofanthai}{0E1F}
\pdfglyphtounicode{fofathai}{0E1D}
\pdfglyphtounicode{follownotdbleqv}{2ABA}
\pdfglyphtounicode{follownotslnteql}{2AB6}
\pdfglyphtounicode{followornoteqvlnt}{22E9}
\pdfglyphtounicode{follows}{227B}
\pdfglyphtounicode{followsequal}{2AB0}
\pdfglyphtounicode{followsorcurly}{227D}
\pdfglyphtounicode{followsorequal}{227F}
\pdfglyphtounicode{fongmanthai}{0E4F}
\pdfglyphtounicode{forall}{2200}
\pdfglyphtounicode{forces}{22A9}
\pdfglyphtounicode{forcesbar}{22AA}
\pdfglyphtounicode{fork}{22D4}
\pdfglyphtounicode{four}{0034}
\pdfglyphtounicode{fourarabic}{0664}
\pdfglyphtounicode{fourbengali}{09EA}
\pdfglyphtounicode{fourcircle}{2463}
\pdfglyphtounicode{fourcircleinversesansserif}{278D}
\pdfglyphtounicode{fourdeva}{096A}
\pdfglyphtounicode{fourgujarati}{0AEA}
\pdfglyphtounicode{fourgurmukhi}{0A6A}
\pdfglyphtounicode{fourhackarabic}{0664}
\pdfglyphtounicode{fourhangzhou}{3024}
\pdfglyphtounicode{fourideographicparen}{3223}
\pdfglyphtounicode{fourinferior}{2084}
\pdfglyphtounicode{fourmonospace}{FF14}
\pdfglyphtounicode{fournumeratorbengali}{09F7}
\pdfglyphtounicode{fouroldstyle}{0034}
\pdfglyphtounicode{fourparen}{2477}
\pdfglyphtounicode{fourperiod}{248B}
\pdfglyphtounicode{fourpersian}{06F4}
\pdfglyphtounicode{fourroman}{2173}
\pdfglyphtounicode{foursuperior}{2074}
\pdfglyphtounicode{fourteencircle}{246D}
\pdfglyphtounicode{fourteenparen}{2481}
\pdfglyphtounicode{fourteenperiod}{2495}
\pdfglyphtounicode{fourthai}{0E54}
\pdfglyphtounicode{fourthtonechinese}{02CB}
\pdfglyphtounicode{fparen}{24A1}
\pdfglyphtounicode{fraction}{2044}
\pdfglyphtounicode{franc}{20A3}
\pdfglyphtounicode{frown}{2322}
\pdfglyphtounicode{g}{0067}
\pdfglyphtounicode{gabengali}{0997}
\pdfglyphtounicode{gacute}{01F5}
\pdfglyphtounicode{gadeva}{0917}
\pdfglyphtounicode{gafarabic}{06AF}
\pdfglyphtounicode{gaffinalarabic}{FB93}
\pdfglyphtounicode{gafinitialarabic}{FB94}
\pdfglyphtounicode{gafmedialarabic}{FB95}
\pdfglyphtounicode{gagujarati}{0A97}
\pdfglyphtounicode{gagurmukhi}{0A17}
\pdfglyphtounicode{gahiragana}{304C}
\pdfglyphtounicode{gakatakana}{30AC}
\pdfglyphtounicode{gamma}{03B3}
\pdfglyphtounicode{gammalatinsmall}{0263}
\pdfglyphtounicode{gammasuperior}{02E0}
\pdfglyphtounicode{gangiacoptic}{03EB}
\pdfglyphtounicode{gbopomofo}{310D}
\pdfglyphtounicode{gbreve}{011F}
\pdfglyphtounicode{gcaron}{01E7}
\pdfglyphtounicode{gcedilla}{0123}
\pdfglyphtounicode{gcircle}{24D6}
\pdfglyphtounicode{gcircumflex}{011D}
\pdfglyphtounicode{gcommaaccent}{0123}
\pdfglyphtounicode{gdot}{0121}
\pdfglyphtounicode{gdotaccent}{0121}
\pdfglyphtounicode{gecyrillic}{0433}
\pdfglyphtounicode{gehiragana}{3052}
\pdfglyphtounicode{gekatakana}{30B2}
\pdfglyphtounicode{geomequivalent}{224E}
\pdfglyphtounicode{geometricallyequal}{2251}
\pdfglyphtounicode{gereshaccenthebrew}{059C}
\pdfglyphtounicode{gereshhebrew}{05F3}
\pdfglyphtounicode{gereshmuqdamhebrew}{059D}
\pdfglyphtounicode{germandbls}{00DF}
\pdfglyphtounicode{gershayimaccenthebrew}{059E}
\pdfglyphtounicode{gershayimhebrew}{05F4}
\pdfglyphtounicode{getamark}{3013}
\pdfglyphtounicode{ghabengali}{0998}
\pdfglyphtounicode{ghadarmenian}{0572}
\pdfglyphtounicode{ghadeva}{0918}
\pdfglyphtounicode{ghagujarati}{0A98}
\pdfglyphtounicode{ghagurmukhi}{0A18}
\pdfglyphtounicode{ghainarabic}{063A}
\pdfglyphtounicode{ghainfinalarabic}{FECE}
\pdfglyphtounicode{ghaininitialarabic}{FECF}
\pdfglyphtounicode{ghainmedialarabic}{FED0}
\pdfglyphtounicode{ghemiddlehookcyrillic}{0495}
\pdfglyphtounicode{ghestrokecyrillic}{0493}
\pdfglyphtounicode{gheupturncyrillic}{0491}
\pdfglyphtounicode{ghhadeva}{095A}
\pdfglyphtounicode{ghhagurmukhi}{0A5A}
\pdfglyphtounicode{ghook}{0260}
\pdfglyphtounicode{ghzsquare}{3393}
\pdfglyphtounicode{gihiragana}{304E}
\pdfglyphtounicode{gikatakana}{30AE}
\pdfglyphtounicode{gimarmenian}{0563}
\pdfglyphtounicode{gimel}{05D2}
\pdfglyphtounicode{gimeldagesh}{FB32}
\pdfglyphtounicode{gimeldageshhebrew}{FB32}
\pdfglyphtounicode{gimelhebrew}{05D2}
\pdfglyphtounicode{gjecyrillic}{0453}
\pdfglyphtounicode{glottalinvertedstroke}{01BE}
\pdfglyphtounicode{glottalstop}{0294}
\pdfglyphtounicode{glottalstopinverted}{0296}
\pdfglyphtounicode{glottalstopmod}{02C0}
\pdfglyphtounicode{glottalstopreversed}{0295}
\pdfglyphtounicode{glottalstopreversedmod}{02C1}
\pdfglyphtounicode{glottalstopreversedsuperior}{02E4}
\pdfglyphtounicode{glottalstopstroke}{02A1}
\pdfglyphtounicode{glottalstopstrokereversed}{02A2}
\pdfglyphtounicode{gmacron}{1E21}
\pdfglyphtounicode{gmonospace}{FF47}
\pdfglyphtounicode{gohiragana}{3054}
\pdfglyphtounicode{gokatakana}{30B4}
\pdfglyphtounicode{gparen}{24A2}
\pdfglyphtounicode{gpasquare}{33AC}
\pdfglyphtounicode{gradient}{2207}
\pdfglyphtounicode{grave}{0060}
\pdfglyphtounicode{gravebelowcmb}{0316}
\pdfglyphtounicode{gravecmb}{0300}
\pdfglyphtounicode{gravecomb}{0300}
\pdfglyphtounicode{gravedeva}{0953}
\pdfglyphtounicode{gravelowmod}{02CE}
\pdfglyphtounicode{gravemonospace}{FF40}
\pdfglyphtounicode{gravetonecmb}{0340}
\pdfglyphtounicode{greater}{003E}
\pdfglyphtounicode{greaterdbleqlless}{2A8C}
\pdfglyphtounicode{greaterdblequal}{2267}
\pdfglyphtounicode{greaterdot}{22D7}
\pdfglyphtounicode{greaterequal}{2265}
\pdfglyphtounicode{greaterequalorless}{22DB}
\pdfglyphtounicode{greaterlessequal}{22DB}
\pdfglyphtounicode{greatermonospace}{FF1E}
\pdfglyphtounicode{greatermuch}{226B}
\pdfglyphtounicode{greaternotdblequal}{2A8A}
\pdfglyphtounicode{greaternotequal}{2A88}
\pdfglyphtounicode{greaterorapproxeql}{2A86}
\pdfglyphtounicode{greaterorequalslant}{2A7E}
\pdfglyphtounicode{greaterorequivalent}{2273}
\pdfglyphtounicode{greaterorless}{2277}
\pdfglyphtounicode{greaterornotdbleql}{2269}
\pdfglyphtounicode{greaterornotequal}{2269}
\pdfglyphtounicode{greaterorsimilar}{2273}
\pdfglyphtounicode{greateroverequal}{2267}
\pdfglyphtounicode{greatersmall}{FE65}
\pdfglyphtounicode{gscript}{0261}
\pdfglyphtounicode{gstroke}{01E5}
\pdfglyphtounicode{guhiragana}{3050}
\pdfglyphtounicode{guillemotleft}{00AB}
\pdfglyphtounicode{guillemotright}{00BB}
\pdfglyphtounicode{guilsinglleft}{2039}
\pdfglyphtounicode{guilsinglright}{203A}
\pdfglyphtounicode{gukatakana}{30B0}
\pdfglyphtounicode{guramusquare}{3318}
\pdfglyphtounicode{gysquare}{33C9}
\pdfglyphtounicode{h}{0068}
\pdfglyphtounicode{haabkhasiancyrillic}{04A9}
\pdfglyphtounicode{haaltonearabic}{06C1}
\pdfglyphtounicode{habengali}{09B9}
\pdfglyphtounicode{hadescendercyrillic}{04B3}
\pdfglyphtounicode{hadeva}{0939}
\pdfglyphtounicode{hagujarati}{0AB9}
\pdfglyphtounicode{hagurmukhi}{0A39}
\pdfglyphtounicode{haharabic}{062D}
\pdfglyphtounicode{hahfinalarabic}{FEA2}
\pdfglyphtounicode{hahinitialarabic}{FEA3}
\pdfglyphtounicode{hahiragana}{306F}
\pdfglyphtounicode{hahmedialarabic}{FEA4}
\pdfglyphtounicode{haitusquare}{332A}
\pdfglyphtounicode{hakatakana}{30CF}
\pdfglyphtounicode{hakatakanahalfwidth}{FF8A}
\pdfglyphtounicode{halantgurmukhi}{0A4D}
\pdfglyphtounicode{hamzaarabic}{0621}
\pdfglyphtounicode{hamzadammaarabic}{0621 064F}
\pdfglyphtounicode{hamzadammatanarabic}{0621 064C}
\pdfglyphtounicode{hamzafathaarabic}{0621 064E}
\pdfglyphtounicode{hamzafathatanarabic}{0621 064B}
\pdfglyphtounicode{hamzalowarabic}{0621}
\pdfglyphtounicode{hamzalowkasraarabic}{0621 0650}
\pdfglyphtounicode{hamzalowkasratanarabic}{0621 064D}
\pdfglyphtounicode{hamzasukunarabic}{0621 0652}
\pdfglyphtounicode{hangulfiller}{3164}
\pdfglyphtounicode{hardsigncyrillic}{044A}
\pdfglyphtounicode{harpoondownleft}{21C3}
\pdfglyphtounicode{harpoondownright}{21C2}
\pdfglyphtounicode{harpoonleftbarbup}{21BC}
\pdfglyphtounicode{harpoonleftright}{21CC}
\pdfglyphtounicode{harpoonrightbarbup}{21C0}
\pdfglyphtounicode{harpoonrightleft}{21CB}
\pdfglyphtounicode{harpoonupleft}{21BF}
\pdfglyphtounicode{harpoonupright}{21BE}
\pdfglyphtounicode{hasquare}{33CA}
\pdfglyphtounicode{hatafpatah}{05B2}
\pdfglyphtounicode{hatafpatah16}{05B2}
\pdfglyphtounicode{hatafpatah23}{05B2}
\pdfglyphtounicode{hatafpatah2f}{05B2}
\pdfglyphtounicode{hatafpatahhebrew}{05B2}
\pdfglyphtounicode{hatafpatahnarrowhebrew}{05B2}
\pdfglyphtounicode{hatafpatahquarterhebrew}{05B2}
\pdfglyphtounicode{hatafpatahwidehebrew}{05B2}
\pdfglyphtounicode{hatafqamats}{05B3}
\pdfglyphtounicode{hatafqamats1b}{05B3}
\pdfglyphtounicode{hatafqamats28}{05B3}
\pdfglyphtounicode{hatafqamats34}{05B3}
\pdfglyphtounicode{hatafqamatshebrew}{05B3}
\pdfglyphtounicode{hatafqamatsnarrowhebrew}{05B3}
\pdfglyphtounicode{hatafqamatsquarterhebrew}{05B3}
\pdfglyphtounicode{hatafqamatswidehebrew}{05B3}
\pdfglyphtounicode{hatafsegol}{05B1}
\pdfglyphtounicode{hatafsegol17}{05B1}
\pdfglyphtounicode{hatafsegol24}{05B1}
\pdfglyphtounicode{hatafsegol30}{05B1}
\pdfglyphtounicode{hatafsegolhebrew}{05B1}
\pdfglyphtounicode{hatafsegolnarrowhebrew}{05B1}
\pdfglyphtounicode{hatafsegolquarterhebrew}{05B1}
\pdfglyphtounicode{hatafsegolwidehebrew}{05B1}
\pdfglyphtounicode{hbar}{0127}
\pdfglyphtounicode{hbopomofo}{310F}
\pdfglyphtounicode{hbrevebelow}{1E2B}
\pdfglyphtounicode{hcedilla}{1E29}
\pdfglyphtounicode{hcircle}{24D7}
\pdfglyphtounicode{hcircumflex}{0125}
\pdfglyphtounicode{hdieresis}{1E27}
\pdfglyphtounicode{hdotaccent}{1E23}
\pdfglyphtounicode{hdotbelow}{1E25}
\pdfglyphtounicode{he}{05D4}
\pdfglyphtounicode{heart}{2665}
\pdfglyphtounicode{heartsuitblack}{2665}
\pdfglyphtounicode{heartsuitwhite}{2661}
\pdfglyphtounicode{hedagesh}{FB34}
\pdfglyphtounicode{hedageshhebrew}{FB34}
\pdfglyphtounicode{hehaltonearabic}{06C1}
\pdfglyphtounicode{heharabic}{0647}
\pdfglyphtounicode{hehebrew}{05D4}
\pdfglyphtounicode{hehfinalaltonearabic}{FBA7}
\pdfglyphtounicode{hehfinalalttwoarabic}{FEEA}
\pdfglyphtounicode{hehfinalarabic}{FEEA}
\pdfglyphtounicode{hehhamzaabovefinalarabic}{FBA5}
\pdfglyphtounicode{hehhamzaaboveisolatedarabic}{FBA4}
\pdfglyphtounicode{hehinitialaltonearabic}{FBA8}
\pdfglyphtounicode{hehinitialarabic}{FEEB}
\pdfglyphtounicode{hehiragana}{3078}
\pdfglyphtounicode{hehmedialaltonearabic}{FBA9}
\pdfglyphtounicode{hehmedialarabic}{FEEC}
\pdfglyphtounicode{heiseierasquare}{337B}
\pdfglyphtounicode{hekatakana}{30D8}
\pdfglyphtounicode{hekatakanahalfwidth}{FF8D}
\pdfglyphtounicode{hekutaarusquare}{3336}
\pdfglyphtounicode{henghook}{0267}
\pdfglyphtounicode{herutusquare}{3339}
\pdfglyphtounicode{het}{05D7}
\pdfglyphtounicode{hethebrew}{05D7}
\pdfglyphtounicode{hhook}{0266}
\pdfglyphtounicode{hhooksuperior}{02B1}
\pdfglyphtounicode{hieuhacirclekorean}{327B}
\pdfglyphtounicode{hieuhaparenkorean}{321B}
\pdfglyphtounicode{hieuhcirclekorean}{326D}
\pdfglyphtounicode{hieuhkorean}{314E}
\pdfglyphtounicode{hieuhparenkorean}{320D}
\pdfglyphtounicode{hihiragana}{3072}
\pdfglyphtounicode{hikatakana}{30D2}
\pdfglyphtounicode{hikatakanahalfwidth}{FF8B}
\pdfglyphtounicode{hiriq}{05B4}
\pdfglyphtounicode{hiriq14}{05B4}
\pdfglyphtounicode{hiriq21}{05B4}
\pdfglyphtounicode{hiriq2d}{05B4}
\pdfglyphtounicode{hiriqhebrew}{05B4}
\pdfglyphtounicode{hiriqnarrowhebrew}{05B4}
\pdfglyphtounicode{hiriqquarterhebrew}{05B4}
\pdfglyphtounicode{hiriqwidehebrew}{05B4}
\pdfglyphtounicode{hlinebelow}{1E96}
\pdfglyphtounicode{hmonospace}{FF48}
\pdfglyphtounicode{hoarmenian}{0570}
\pdfglyphtounicode{hohipthai}{0E2B}
\pdfglyphtounicode{hohiragana}{307B}
\pdfglyphtounicode{hokatakana}{30DB}
\pdfglyphtounicode{hokatakanahalfwidth}{FF8E}
\pdfglyphtounicode{holam}{05B9}
\pdfglyphtounicode{holam19}{05B9}
\pdfglyphtounicode{holam26}{05B9}
\pdfglyphtounicode{holam32}{05B9}
\pdfglyphtounicode{holamhebrew}{05B9}
\pdfglyphtounicode{holamnarrowhebrew}{05B9}
\pdfglyphtounicode{holamquarterhebrew}{05B9}
\pdfglyphtounicode{holamwidehebrew}{05B9}
\pdfglyphtounicode{honokhukthai}{0E2E}
\pdfglyphtounicode{hookabovecomb}{0309}
\pdfglyphtounicode{hookcmb}{0309}
\pdfglyphtounicode{hookpalatalizedbelowcmb}{0321}
\pdfglyphtounicode{hookretroflexbelowcmb}{0322}
\pdfglyphtounicode{hoonsquare}{3342}
\pdfglyphtounicode{horicoptic}{03E9}
\pdfglyphtounicode{horizontalbar}{2015}
\pdfglyphtounicode{horncmb}{031B}
\pdfglyphtounicode{hotsprings}{2668}
\pdfglyphtounicode{house}{2302}
\pdfglyphtounicode{hparen}{24A3}
\pdfglyphtounicode{hsuperior}{02B0}
\pdfglyphtounicode{hturned}{0265}
\pdfglyphtounicode{huhiragana}{3075}
\pdfglyphtounicode{huiitosquare}{3333}
\pdfglyphtounicode{hukatakana}{30D5}
\pdfglyphtounicode{hukatakanahalfwidth}{FF8C}
\pdfglyphtounicode{hungarumlaut}{02DD}
\pdfglyphtounicode{hungarumlautcmb}{030B}
\pdfglyphtounicode{hv}{0195}
\pdfglyphtounicode{hyphen}{002D}
\pdfglyphtounicode{hyphenchar}{002D}
\pdfglyphtounicode{hypheninferior}{002D}
\pdfglyphtounicode{hyphenmonospace}{FF0D}
\pdfglyphtounicode{hyphensmall}{FE63}
\pdfglyphtounicode{hyphensuperior}{002D}
\pdfglyphtounicode{hyphentwo}{2010}
\pdfglyphtounicode{i}{0069}
\pdfglyphtounicode{iacute}{00ED}
\pdfglyphtounicode{iacyrillic}{044F}
\pdfglyphtounicode{ibengali}{0987}
\pdfglyphtounicode{ibopomofo}{3127}
\pdfglyphtounicode{ibreve}{012D}
\pdfglyphtounicode{icaron}{01D0}
\pdfglyphtounicode{icircle}{24D8}
\pdfglyphtounicode{icircumflex}{00EE}
\pdfglyphtounicode{icyrillic}{0456}
\pdfglyphtounicode{idblgrave}{0209}
\pdfglyphtounicode{ideographearthcircle}{328F}
\pdfglyphtounicode{ideographfirecircle}{328B}
\pdfglyphtounicode{ideographicallianceparen}{323F}
\pdfglyphtounicode{ideographiccallparen}{323A}
\pdfglyphtounicode{ideographiccentrecircle}{32A5}
\pdfglyphtounicode{ideographicclose}{3006}
\pdfglyphtounicode{ideographiccomma}{3001}
\pdfglyphtounicode{ideographiccommaleft}{FF64}
\pdfglyphtounicode{ideographiccongratulationparen}{3237}
\pdfglyphtounicode{ideographiccorrectcircle}{32A3}
\pdfglyphtounicode{ideographicearthparen}{322F}
\pdfglyphtounicode{ideographicenterpriseparen}{323D}
\pdfglyphtounicode{ideographicexcellentcircle}{329D}
\pdfglyphtounicode{ideographicfestivalparen}{3240}
\pdfglyphtounicode{ideographicfinancialcircle}{3296}
\pdfglyphtounicode{ideographicfinancialparen}{3236}
\pdfglyphtounicode{ideographicfireparen}{322B}
\pdfglyphtounicode{ideographichaveparen}{3232}
\pdfglyphtounicode{ideographichighcircle}{32A4}
\pdfglyphtounicode{ideographiciterationmark}{3005}
\pdfglyphtounicode{ideographiclaborcircle}{3298}
\pdfglyphtounicode{ideographiclaborparen}{3238}
\pdfglyphtounicode{ideographicleftcircle}{32A7}
\pdfglyphtounicode{ideographiclowcircle}{32A6}
\pdfglyphtounicode{ideographicmedicinecircle}{32A9}
\pdfglyphtounicode{ideographicmetalparen}{322E}
\pdfglyphtounicode{ideographicmoonparen}{322A}
\pdfglyphtounicode{ideographicnameparen}{3234}
\pdfglyphtounicode{ideographicperiod}{3002}
\pdfglyphtounicode{ideographicprintcircle}{329E}
\pdfglyphtounicode{ideographicreachparen}{3243}
\pdfglyphtounicode{ideographicrepresentparen}{3239}
\pdfglyphtounicode{ideographicresourceparen}{323E}
\pdfglyphtounicode{ideographicrightcircle}{32A8}
\pdfglyphtounicode{ideographicsecretcircle}{3299}
\pdfglyphtounicode{ideographicselfparen}{3242}
\pdfglyphtounicode{ideographicsocietyparen}{3233}
\pdfglyphtounicode{ideographicspace}{3000}
\pdfglyphtounicode{ideographicspecialparen}{3235}
\pdfglyphtounicode{ideographicstockparen}{3231}
\pdfglyphtounicode{ideographicstudyparen}{323B}
\pdfglyphtounicode{ideographicsunparen}{3230}
\pdfglyphtounicode{ideographicsuperviseparen}{323C}
\pdfglyphtounicode{ideographicwaterparen}{322C}
\pdfglyphtounicode{ideographicwoodparen}{322D}
\pdfglyphtounicode{ideographiczero}{3007}
\pdfglyphtounicode{ideographmetalcircle}{328E}
\pdfglyphtounicode{ideographmooncircle}{328A}
\pdfglyphtounicode{ideographnamecircle}{3294}
\pdfglyphtounicode{ideographsuncircle}{3290}
\pdfglyphtounicode{ideographwatercircle}{328C}
\pdfglyphtounicode{ideographwoodcircle}{328D}
\pdfglyphtounicode{ideva}{0907}
\pdfglyphtounicode{idieresis}{00EF}
\pdfglyphtounicode{idieresisacute}{1E2F}
\pdfglyphtounicode{idieresiscyrillic}{04E5}
\pdfglyphtounicode{idotbelow}{1ECB}
\pdfglyphtounicode{iebrevecyrillic}{04D7}
\pdfglyphtounicode{iecyrillic}{0435}
\pdfglyphtounicode{ieungacirclekorean}{3275}
\pdfglyphtounicode{ieungaparenkorean}{3215}
\pdfglyphtounicode{ieungcirclekorean}{3267}
\pdfglyphtounicode{ieungkorean}{3147}
\pdfglyphtounicode{ieungparenkorean}{3207}
\pdfglyphtounicode{igrave}{00EC}
\pdfglyphtounicode{igujarati}{0A87}
\pdfglyphtounicode{igurmukhi}{0A07}
\pdfglyphtounicode{ihiragana}{3044}
\pdfglyphtounicode{ihookabove}{1EC9}
\pdfglyphtounicode{iibengali}{0988}
\pdfglyphtounicode{iicyrillic}{0438}
\pdfglyphtounicode{iideva}{0908}
\pdfglyphtounicode{iigujarati}{0A88}
\pdfglyphtounicode{iigurmukhi}{0A08}
\pdfglyphtounicode{iimatragurmukhi}{0A40}
\pdfglyphtounicode{iinvertedbreve}{020B}
\pdfglyphtounicode{iishortcyrillic}{0439}
\pdfglyphtounicode{iivowelsignbengali}{09C0}
\pdfglyphtounicode{iivowelsigndeva}{0940}
\pdfglyphtounicode{iivowelsigngujarati}{0AC0}
\pdfglyphtounicode{ij}{0133}
\pdfglyphtounicode{ikatakana}{30A4}
\pdfglyphtounicode{ikatakanahalfwidth}{FF72}
\pdfglyphtounicode{ikorean}{3163}
\pdfglyphtounicode{ilde}{02DC}
\pdfglyphtounicode{iluyhebrew}{05AC}
\pdfglyphtounicode{imacron}{012B}
\pdfglyphtounicode{imacroncyrillic}{04E3}
\pdfglyphtounicode{imageorapproximatelyequal}{2253}
\pdfglyphtounicode{imatragurmukhi}{0A3F}
\pdfglyphtounicode{imonospace}{FF49}
\pdfglyphtounicode{increment}{2206}
\pdfglyphtounicode{infinity}{221E}
\pdfglyphtounicode{iniarmenian}{056B}
\pdfglyphtounicode{integerdivide}{2216}
\pdfglyphtounicode{integral}{222B}
\pdfglyphtounicode{integralbottom}{2321}
\pdfglyphtounicode{integralbt}{2321}
\pdfglyphtounicode{integralex}{F8F5}
\pdfglyphtounicode{integraltop}{2320}
\pdfglyphtounicode{integraltp}{2320}
\pdfglyphtounicode{intercal}{22BA}
\pdfglyphtounicode{interrobang}{203D}
\pdfglyphtounicode{interrobangdown}{2E18}
\pdfglyphtounicode{intersection}{2229}
\pdfglyphtounicode{intersectiondbl}{22D2}
\pdfglyphtounicode{intersectionsq}{2293}
\pdfglyphtounicode{intisquare}{3305}
\pdfglyphtounicode{invbullet}{25D8}
\pdfglyphtounicode{invcircle}{25D9}
\pdfglyphtounicode{invsmileface}{263B}
\pdfglyphtounicode{iocyrillic}{0451}
\pdfglyphtounicode{iogonek}{012F}
\pdfglyphtounicode{iota}{03B9}
\pdfglyphtounicode{iotadieresis}{03CA}
\pdfglyphtounicode{iotadieresistonos}{0390}
\pdfglyphtounicode{iotalatin}{0269}
\pdfglyphtounicode{iotatonos}{03AF}
\pdfglyphtounicode{iparen}{24A4}
\pdfglyphtounicode{irigurmukhi}{0A72}
\pdfglyphtounicode{ismallhiragana}{3043}
\pdfglyphtounicode{ismallkatakana}{30A3}
\pdfglyphtounicode{ismallkatakanahalfwidth}{FF68}
\pdfglyphtounicode{issharbengali}{09FA}
\pdfglyphtounicode{istroke}{0268}
\pdfglyphtounicode{isuperior}{0069}
\pdfglyphtounicode{iterationhiragana}{309D}
\pdfglyphtounicode{iterationkatakana}{30FD}
\pdfglyphtounicode{itilde}{0129}
\pdfglyphtounicode{itildebelow}{1E2D}
\pdfglyphtounicode{iubopomofo}{3129}
\pdfglyphtounicode{iucyrillic}{044E}
\pdfglyphtounicode{ivowelsignbengali}{09BF}
\pdfglyphtounicode{ivowelsigndeva}{093F}
\pdfglyphtounicode{ivowelsigngujarati}{0ABF}
\pdfglyphtounicode{izhitsacyrillic}{0475}
\pdfglyphtounicode{izhitsadblgravecyrillic}{0477}
\pdfglyphtounicode{j}{006A}
\pdfglyphtounicode{jaarmenian}{0571}
\pdfglyphtounicode{jabengali}{099C}
\pdfglyphtounicode{jadeva}{091C}
\pdfglyphtounicode{jagujarati}{0A9C}
\pdfglyphtounicode{jagurmukhi}{0A1C}
\pdfglyphtounicode{jbopomofo}{3110}
\pdfglyphtounicode{jcaron}{01F0}
\pdfglyphtounicode{jcircle}{24D9}
\pdfglyphtounicode{jcircumflex}{0135}
\pdfglyphtounicode{jcrossedtail}{029D}
\pdfglyphtounicode{jdotlessstroke}{025F}
\pdfglyphtounicode{jecyrillic}{0458}
\pdfglyphtounicode{jeemarabic}{062C}
\pdfglyphtounicode{jeemfinalarabic}{FE9E}
\pdfglyphtounicode{jeeminitialarabic}{FE9F}
\pdfglyphtounicode{jeemmedialarabic}{FEA0}
\pdfglyphtounicode{jeharabic}{0698}
\pdfglyphtounicode{jehfinalarabic}{FB8B}
\pdfglyphtounicode{jhabengali}{099D}
\pdfglyphtounicode{jhadeva}{091D}
\pdfglyphtounicode{jhagujarati}{0A9D}
\pdfglyphtounicode{jhagurmukhi}{0A1D}
\pdfglyphtounicode{jheharmenian}{057B}
\pdfglyphtounicode{jis}{3004}
\pdfglyphtounicode{jmonospace}{FF4A}
\pdfglyphtounicode{jparen}{24A5}
\pdfglyphtounicode{jsuperior}{02B2}
\pdfglyphtounicode{k}{006B}
\pdfglyphtounicode{kabashkircyrillic}{04A1}
\pdfglyphtounicode{kabengali}{0995}
\pdfglyphtounicode{kacute}{1E31}
\pdfglyphtounicode{kacyrillic}{043A}
\pdfglyphtounicode{kadescendercyrillic}{049B}
\pdfglyphtounicode{kadeva}{0915}
\pdfglyphtounicode{kaf}{05DB}
\pdfglyphtounicode{kafarabic}{0643}
\pdfglyphtounicode{kafdagesh}{FB3B}
\pdfglyphtounicode{kafdageshhebrew}{FB3B}
\pdfglyphtounicode{kaffinalarabic}{FEDA}
\pdfglyphtounicode{kafhebrew}{05DB}
\pdfglyphtounicode{kafinitialarabic}{FEDB}
\pdfglyphtounicode{kafmedialarabic}{FEDC}
\pdfglyphtounicode{kafrafehebrew}{FB4D}
\pdfglyphtounicode{kagujarati}{0A95}
\pdfglyphtounicode{kagurmukhi}{0A15}
\pdfglyphtounicode{kahiragana}{304B}
\pdfglyphtounicode{kahookcyrillic}{04C4}
\pdfglyphtounicode{kakatakana}{30AB}
\pdfglyphtounicode{kakatakanahalfwidth}{FF76}
\pdfglyphtounicode{kappa}{03BA}
\pdfglyphtounicode{kappasymbolgreek}{03F0}
\pdfglyphtounicode{kapyeounmieumkorean}{3171}
\pdfglyphtounicode{kapyeounphieuphkorean}{3184}
\pdfglyphtounicode{kapyeounpieupkorean}{3178}
\pdfglyphtounicode{kapyeounssangpieupkorean}{3179}
\pdfglyphtounicode{karoriisquare}{330D}
\pdfglyphtounicode{kashidaautoarabic}{0640}
\pdfglyphtounicode{kashidaautonosidebearingarabic}{0640}
\pdfglyphtounicode{kasmallkatakana}{30F5}
\pdfglyphtounicode{kasquare}{3384}
\pdfglyphtounicode{kasraarabic}{0650}
\pdfglyphtounicode{kasratanarabic}{064D}
\pdfglyphtounicode{kastrokecyrillic}{049F}
\pdfglyphtounicode{katahiraprolongmarkhalfwidth}{FF70}
\pdfglyphtounicode{kaverticalstrokecyrillic}{049D}
\pdfglyphtounicode{kbopomofo}{310E}
\pdfglyphtounicode{kcalsquare}{3389}
\pdfglyphtounicode{kcaron}{01E9}
\pdfglyphtounicode{kcedilla}{0137}
\pdfglyphtounicode{kcircle}{24DA}
\pdfglyphtounicode{kcommaaccent}{0137}
\pdfglyphtounicode{kdotbelow}{1E33}
\pdfglyphtounicode{keharmenian}{0584}
\pdfglyphtounicode{kehiragana}{3051}
\pdfglyphtounicode{kekatakana}{30B1}
\pdfglyphtounicode{kekatakanahalfwidth}{FF79}
\pdfglyphtounicode{kenarmenian}{056F}
\pdfglyphtounicode{kesmallkatakana}{30F6}
\pdfglyphtounicode{kgreenlandic}{0138}
\pdfglyphtounicode{khabengali}{0996}
\pdfglyphtounicode{khacyrillic}{0445}
\pdfglyphtounicode{khadeva}{0916}
\pdfglyphtounicode{khagujarati}{0A96}
\pdfglyphtounicode{khagurmukhi}{0A16}
\pdfglyphtounicode{khaharabic}{062E}
\pdfglyphtounicode{khahfinalarabic}{FEA6}
\pdfglyphtounicode{khahinitialarabic}{FEA7}
\pdfglyphtounicode{khahmedialarabic}{FEA8}
\pdfglyphtounicode{kheicoptic}{03E7}
\pdfglyphtounicode{khhadeva}{0959}
\pdfglyphtounicode{khhagurmukhi}{0A59}
\pdfglyphtounicode{khieukhacirclekorean}{3278}
\pdfglyphtounicode{khieukhaparenkorean}{3218}
\pdfglyphtounicode{khieukhcirclekorean}{326A}
\pdfglyphtounicode{khieukhkorean}{314B}
\pdfglyphtounicode{khieukhparenkorean}{320A}
\pdfglyphtounicode{khokhaithai}{0E02}
\pdfglyphtounicode{khokhonthai}{0E05}
\pdfglyphtounicode{khokhuatthai}{0E03}
\pdfglyphtounicode{khokhwaithai}{0E04}
\pdfglyphtounicode{khomutthai}{0E5B}
\pdfglyphtounicode{khook}{0199}
\pdfglyphtounicode{khorakhangthai}{0E06}
\pdfglyphtounicode{khzsquare}{3391}
\pdfglyphtounicode{kihiragana}{304D}
\pdfglyphtounicode{kikatakana}{30AD}
\pdfglyphtounicode{kikatakanahalfwidth}{FF77}
\pdfglyphtounicode{kiroguramusquare}{3315}
\pdfglyphtounicode{kiromeetorusquare}{3316}
\pdfglyphtounicode{kirosquare}{3314}
\pdfglyphtounicode{kiyeokacirclekorean}{326E}
\pdfglyphtounicode{kiyeokaparenkorean}{320E}
\pdfglyphtounicode{kiyeokcirclekorean}{3260}
\pdfglyphtounicode{kiyeokkorean}{3131}
\pdfglyphtounicode{kiyeokparenkorean}{3200}
\pdfglyphtounicode{kiyeoksioskorean}{3133}
\pdfglyphtounicode{kjecyrillic}{045C}
\pdfglyphtounicode{klinebelow}{1E35}
\pdfglyphtounicode{klsquare}{3398}
\pdfglyphtounicode{kmcubedsquare}{33A6}
\pdfglyphtounicode{kmonospace}{FF4B}
\pdfglyphtounicode{kmsquaredsquare}{33A2}
\pdfglyphtounicode{kohiragana}{3053}
\pdfglyphtounicode{kohmsquare}{33C0}
\pdfglyphtounicode{kokaithai}{0E01}
\pdfglyphtounicode{kokatakana}{30B3}
\pdfglyphtounicode{kokatakanahalfwidth}{FF7A}
\pdfglyphtounicode{kooposquare}{331E}
\pdfglyphtounicode{koppacyrillic}{0481}
\pdfglyphtounicode{koreanstandardsymbol}{327F}
\pdfglyphtounicode{koroniscmb}{0343}
\pdfglyphtounicode{kparen}{24A6}
\pdfglyphtounicode{kpasquare}{33AA}
\pdfglyphtounicode{ksicyrillic}{046F}
\pdfglyphtounicode{ktsquare}{33CF}
\pdfglyphtounicode{kturned}{029E}
\pdfglyphtounicode{kuhiragana}{304F}
\pdfglyphtounicode{kukatakana}{30AF}
\pdfglyphtounicode{kukatakanahalfwidth}{FF78}
\pdfglyphtounicode{kvsquare}{33B8}
\pdfglyphtounicode{kwsquare}{33BE}
\pdfglyphtounicode{l}{006C}
\pdfglyphtounicode{labengali}{09B2}
\pdfglyphtounicode{lacute}{013A}
\pdfglyphtounicode{ladeva}{0932}
\pdfglyphtounicode{lagujarati}{0AB2}
\pdfglyphtounicode{lagurmukhi}{0A32}
\pdfglyphtounicode{lakkhangyaothai}{0E45}
\pdfglyphtounicode{lamaleffinalarabic}{FEFC}
\pdfglyphtounicode{lamalefhamzaabovefinalarabic}{FEF8}
\pdfglyphtounicode{lamalefhamzaaboveisolatedarabic}{FEF7}
\pdfglyphtounicode{lamalefhamzabelowfinalarabic}{FEFA}
\pdfglyphtounicode{lamalefhamzabelowisolatedarabic}{FEF9}
\pdfglyphtounicode{lamalefisolatedarabic}{FEFB}
\pdfglyphtounicode{lamalefmaddaabovefinalarabic}{FEF6}
\pdfglyphtounicode{lamalefmaddaaboveisolatedarabic}{FEF5}
\pdfglyphtounicode{lamarabic}{0644}
\pdfglyphtounicode{lambda}{03BB}
\pdfglyphtounicode{lambdastroke}{019B}
\pdfglyphtounicode{lamed}{05DC}
\pdfglyphtounicode{lameddagesh}{FB3C}
\pdfglyphtounicode{lameddageshhebrew}{FB3C}
\pdfglyphtounicode{lamedhebrew}{05DC}
\pdfglyphtounicode{lamedholam}{05DC 05B9}
\pdfglyphtounicode{lamedholamdagesh}{05DC 05B9 05BC}
\pdfglyphtounicode{lamedholamdageshhebrew}{05DC 05B9 05BC}
\pdfglyphtounicode{lamedholamhebrew}{05DC 05B9}
\pdfglyphtounicode{lamfinalarabic}{FEDE}
\pdfglyphtounicode{lamhahinitialarabic}{FCCA}
\pdfglyphtounicode{laminitialarabic}{FEDF}
\pdfglyphtounicode{lamjeeminitialarabic}{FCC9}
\pdfglyphtounicode{lamkhahinitialarabic}{FCCB}
\pdfglyphtounicode{lamlamhehisolatedarabic}{FDF2}
\pdfglyphtounicode{lammedialarabic}{FEE0}
\pdfglyphtounicode{lammeemhahinitialarabic}{FD88}
\pdfglyphtounicode{lammeeminitialarabic}{FCCC}
\pdfglyphtounicode{lammeemjeeminitialarabic}{FEDF FEE4 FEA0}
\pdfglyphtounicode{lammeemkhahinitialarabic}{FEDF FEE4 FEA8}
\pdfglyphtounicode{largecircle}{25EF}
\pdfglyphtounicode{latticetop}{22A4}
\pdfglyphtounicode{lbar}{019A}
\pdfglyphtounicode{lbelt}{026C}
\pdfglyphtounicode{lbopomofo}{310C}
\pdfglyphtounicode{lcaron}{013E}
\pdfglyphtounicode{lcedilla}{013C}
\pdfglyphtounicode{lcircle}{24DB}
\pdfglyphtounicode{lcircumflexbelow}{1E3D}
\pdfglyphtounicode{lcommaaccent}{013C}
\pdfglyphtounicode{ldot}{0140}
\pdfglyphtounicode{ldotaccent}{0140}
\pdfglyphtounicode{ldotbelow}{1E37}
\pdfglyphtounicode{ldotbelowmacron}{1E39}
\pdfglyphtounicode{leftangleabovecmb}{031A}
\pdfglyphtounicode{lefttackbelowcmb}{0318}
\pdfglyphtounicode{less}{003C}
\pdfglyphtounicode{lessdbleqlgreater}{2A8B}
\pdfglyphtounicode{lessdblequal}{2266}
\pdfglyphtounicode{lessdot}{22D6}
\pdfglyphtounicode{lessequal}{2264}
\pdfglyphtounicode{lessequalgreater}{22DA}
\pdfglyphtounicode{lessequalorgreater}{22DA}
\pdfglyphtounicode{lessmonospace}{FF1C}
\pdfglyphtounicode{lessmuch}{226A}
\pdfglyphtounicode{lessnotdblequal}{2A89}
\pdfglyphtounicode{lessnotequal}{2A87}
\pdfglyphtounicode{lessorapproxeql}{2A85}
\pdfglyphtounicode{lessorequalslant}{2A7D}
\pdfglyphtounicode{lessorequivalent}{2272}
\pdfglyphtounicode{lessorgreater}{2276}
\pdfglyphtounicode{lessornotdbleql}{2268}
\pdfglyphtounicode{lessornotequal}{2268}
\pdfglyphtounicode{lessorsimilar}{2272}
\pdfglyphtounicode{lessoverequal}{2266}
\pdfglyphtounicode{lesssmall}{FE64}
\pdfglyphtounicode{lezh}{026E}
\pdfglyphtounicode{lfblock}{258C}
\pdfglyphtounicode{lhookretroflex}{026D}
\pdfglyphtounicode{lira}{20A4}
\pdfglyphtounicode{liwnarmenian}{056C}
\pdfglyphtounicode{lj}{01C9}
\pdfglyphtounicode{ljecyrillic}{0459}
\pdfglyphtounicode{ll}{006C 006C}
\pdfglyphtounicode{lladeva}{0933}
\pdfglyphtounicode{llagujarati}{0AB3}
\pdfglyphtounicode{llinebelow}{1E3B}
\pdfglyphtounicode{llladeva}{0934}
\pdfglyphtounicode{llvocalicbengali}{09E1}
\pdfglyphtounicode{llvocalicdeva}{0961}
\pdfglyphtounicode{llvocalicvowelsignbengali}{09E3}
\pdfglyphtounicode{llvocalicvowelsigndeva}{0963}
\pdfglyphtounicode{lmiddletilde}{026B}
\pdfglyphtounicode{lmonospace}{FF4C}
\pdfglyphtounicode{lmsquare}{33D0}
\pdfglyphtounicode{lochulathai}{0E2C}
\pdfglyphtounicode{logicaland}{2227}
\pdfglyphtounicode{logicalnot}{00AC}
\pdfglyphtounicode{logicalnotreversed}{2310}
\pdfglyphtounicode{logicalor}{2228}
\pdfglyphtounicode{lolingthai}{0E25}
\pdfglyphtounicode{longdbls}{017F 017F}
\pdfglyphtounicode{longs}{017F}
\pdfglyphtounicode{longsh}{017F 0068}
\pdfglyphtounicode{longsi}{017F 0069}
\pdfglyphtounicode{longsl}{017F 006C}
\pdfglyphtounicode{longst}{017F 0074}
\pdfglyphtounicode{lowlinecenterline}{FE4E}
\pdfglyphtounicode{lowlinecmb}{0332}
\pdfglyphtounicode{lowlinedashed}{FE4D}
\pdfglyphtounicode{lozenge}{25CA}
\pdfglyphtounicode{lparen}{24A7}
\pdfglyphtounicode{lscript}{2113}
\pdfglyphtounicode{lslash}{0142}
\pdfglyphtounicode{lsquare}{2113}
\pdfglyphtounicode{lsuperior}{006C}
\pdfglyphtounicode{ltshade}{2591}
\pdfglyphtounicode{luthai}{0E26}
\pdfglyphtounicode{lvocalicbengali}{098C}
\pdfglyphtounicode{lvocalicdeva}{090C}
\pdfglyphtounicode{lvocalicvowelsignbengali}{09E2}
\pdfglyphtounicode{lvocalicvowelsigndeva}{0962}
\pdfglyphtounicode{lxsquare}{33D3}
\pdfglyphtounicode{m}{006D}
\pdfglyphtounicode{mabengali}{09AE}
\pdfglyphtounicode{macron}{00AF}
\pdfglyphtounicode{macronbelowcmb}{0331}
\pdfglyphtounicode{macroncmb}{0304}
\pdfglyphtounicode{macronlowmod}{02CD}
\pdfglyphtounicode{macronmonospace}{FFE3}
\pdfglyphtounicode{macute}{1E3F}
\pdfglyphtounicode{madeva}{092E}
\pdfglyphtounicode{magujarati}{0AAE}
\pdfglyphtounicode{magurmukhi}{0A2E}
\pdfglyphtounicode{mahapakhhebrew}{05A4}
\pdfglyphtounicode{mahapakhlefthebrew}{05A4}
\pdfglyphtounicode{mahiragana}{307E}
\pdfglyphtounicode{maichattawalowleftthai}{F895}
\pdfglyphtounicode{maichattawalowrightthai}{F894}
\pdfglyphtounicode{maichattawathai}{0E4B}
\pdfglyphtounicode{maichattawaupperleftthai}{F893}
\pdfglyphtounicode{maieklowleftthai}{F88C}
\pdfglyphtounicode{maieklowrightthai}{F88B}
\pdfglyphtounicode{maiekthai}{0E48}
\pdfglyphtounicode{maiekupperleftthai}{F88A}
\pdfglyphtounicode{maihanakatleftthai}{F884}
\pdfglyphtounicode{maihanakatthai}{0E31}
\pdfglyphtounicode{maitaikhuleftthai}{F889}
\pdfglyphtounicode{maitaikhuthai}{0E47}
\pdfglyphtounicode{maitholowleftthai}{F88F}
\pdfglyphtounicode{maitholowrightthai}{F88E}
\pdfglyphtounicode{maithothai}{0E49}
\pdfglyphtounicode{maithoupperleftthai}{F88D}
\pdfglyphtounicode{maitrilowleftthai}{F892}
\pdfglyphtounicode{maitrilowrightthai}{F891}
\pdfglyphtounicode{maitrithai}{0E4A}
\pdfglyphtounicode{maitriupperleftthai}{F890}
\pdfglyphtounicode{maiyamokthai}{0E46}
\pdfglyphtounicode{makatakana}{30DE}
\pdfglyphtounicode{makatakanahalfwidth}{FF8F}
\pdfglyphtounicode{male}{2642}
\pdfglyphtounicode{maltesecross}{2720}
\pdfglyphtounicode{mansyonsquare}{3347}
\pdfglyphtounicode{maqafhebrew}{05BE}
\pdfglyphtounicode{mars}{2642}
\pdfglyphtounicode{masoracirclehebrew}{05AF}
\pdfglyphtounicode{masquare}{3383}
\pdfglyphtounicode{mbopomofo}{3107}
\pdfglyphtounicode{mbsquare}{33D4}
\pdfglyphtounicode{mcircle}{24DC}
\pdfglyphtounicode{mcubedsquare}{33A5}
\pdfglyphtounicode{mdotaccent}{1E41}
\pdfglyphtounicode{mdotbelow}{1E43}
\pdfglyphtounicode{measuredangle}{2221}
\pdfglyphtounicode{meemarabic}{0645}
\pdfglyphtounicode{meemfinalarabic}{FEE2}
\pdfglyphtounicode{meeminitialarabic}{FEE3}
\pdfglyphtounicode{meemmedialarabic}{FEE4}
\pdfglyphtounicode{meemmeeminitialarabic}{FCD1}
\pdfglyphtounicode{meemmeemisolatedarabic}{FC48}
\pdfglyphtounicode{meetorusquare}{334D}
\pdfglyphtounicode{mehiragana}{3081}
\pdfglyphtounicode{meizierasquare}{337E}
\pdfglyphtounicode{mekatakana}{30E1}
\pdfglyphtounicode{mekatakanahalfwidth}{FF92}
\pdfglyphtounicode{mem}{05DE}
\pdfglyphtounicode{memdagesh}{FB3E}
\pdfglyphtounicode{memdageshhebrew}{FB3E}
\pdfglyphtounicode{memhebrew}{05DE}
\pdfglyphtounicode{menarmenian}{0574}
\pdfglyphtounicode{merkhahebrew}{05A5}
\pdfglyphtounicode{merkhakefulahebrew}{05A6}
\pdfglyphtounicode{merkhakefulalefthebrew}{05A6}
\pdfglyphtounicode{merkhalefthebrew}{05A5}
\pdfglyphtounicode{mhook}{0271}
\pdfglyphtounicode{mhzsquare}{3392}
\pdfglyphtounicode{middledotkatakanahalfwidth}{FF65}
\pdfglyphtounicode{middot}{00B7}
\pdfglyphtounicode{mieumacirclekorean}{3272}
\pdfglyphtounicode{mieumaparenkorean}{3212}
\pdfglyphtounicode{mieumcirclekorean}{3264}
\pdfglyphtounicode{mieumkorean}{3141}
\pdfglyphtounicode{mieumpansioskorean}{3170}
\pdfglyphtounicode{mieumparenkorean}{3204}
\pdfglyphtounicode{mieumpieupkorean}{316E}
\pdfglyphtounicode{mieumsioskorean}{316F}
\pdfglyphtounicode{mihiragana}{307F}
\pdfglyphtounicode{mikatakana}{30DF}
\pdfglyphtounicode{mikatakanahalfwidth}{FF90}
\pdfglyphtounicode{minus}{2212}
\pdfglyphtounicode{minusbelowcmb}{0320}
\pdfglyphtounicode{minuscircle}{2296}
\pdfglyphtounicode{minusmod}{02D7}
\pdfglyphtounicode{minusplus}{2213}
\pdfglyphtounicode{minute}{2032}
\pdfglyphtounicode{miribaarusquare}{334A}
\pdfglyphtounicode{mirisquare}{3349}
\pdfglyphtounicode{mlonglegturned}{0270}
\pdfglyphtounicode{mlsquare}{3396}
\pdfglyphtounicode{mmcubedsquare}{33A3}
\pdfglyphtounicode{mmonospace}{FF4D}
\pdfglyphtounicode{mmsquaredsquare}{339F}
\pdfglyphtounicode{mohiragana}{3082}
\pdfglyphtounicode{mohmsquare}{33C1}
\pdfglyphtounicode{mokatakana}{30E2}
\pdfglyphtounicode{mokatakanahalfwidth}{FF93}
\pdfglyphtounicode{molsquare}{33D6}
\pdfglyphtounicode{momathai}{0E21}
\pdfglyphtounicode{moverssquare}{33A7}
\pdfglyphtounicode{moverssquaredsquare}{33A8}
\pdfglyphtounicode{mparen}{24A8}
\pdfglyphtounicode{mpasquare}{33AB}
\pdfglyphtounicode{mssquare}{33B3}
\pdfglyphtounicode{msuperior}{006D}
\pdfglyphtounicode{mturned}{026F}
\pdfglyphtounicode{mu}{00B5}
\pdfglyphtounicode{mu1}{00B5}
\pdfglyphtounicode{muasquare}{3382}
\pdfglyphtounicode{muchgreater}{226B}
\pdfglyphtounicode{muchless}{226A}
\pdfglyphtounicode{mufsquare}{338C}
\pdfglyphtounicode{mugreek}{03BC}
\pdfglyphtounicode{mugsquare}{338D}
\pdfglyphtounicode{muhiragana}{3080}
\pdfglyphtounicode{mukatakana}{30E0}
\pdfglyphtounicode{mukatakanahalfwidth}{FF91}
\pdfglyphtounicode{mulsquare}{3395}
\pdfglyphtounicode{multicloseleft}{22C9}
\pdfglyphtounicode{multicloseright}{22CA}
\pdfglyphtounicode{multimap}{22B8}
\pdfglyphtounicode{multiopenleft}{22CB}
\pdfglyphtounicode{multiopenright}{22CC}
\pdfglyphtounicode{multiply}{00D7}
\pdfglyphtounicode{mumsquare}{339B}
\pdfglyphtounicode{munahhebrew}{05A3}
\pdfglyphtounicode{munahlefthebrew}{05A3}
\pdfglyphtounicode{musicalnote}{266A}
\pdfglyphtounicode{musicalnotedbl}{266B}
\pdfglyphtounicode{musicflatsign}{266D}
\pdfglyphtounicode{musicsharpsign}{266F}
\pdfglyphtounicode{mussquare}{33B2}
\pdfglyphtounicode{muvsquare}{33B6}
\pdfglyphtounicode{muwsquare}{33BC}
\pdfglyphtounicode{mvmegasquare}{33B9}
\pdfglyphtounicode{mvsquare}{33B7}
\pdfglyphtounicode{mwmegasquare}{33BF}
\pdfglyphtounicode{mwsquare}{33BD}
\pdfglyphtounicode{n}{006E}
\pdfglyphtounicode{nabengali}{09A8}
\pdfglyphtounicode{nabla}{2207}
\pdfglyphtounicode{nacute}{0144}
\pdfglyphtounicode{nadeva}{0928}
\pdfglyphtounicode{nagujarati}{0AA8}
\pdfglyphtounicode{nagurmukhi}{0A28}
\pdfglyphtounicode{nahiragana}{306A}
\pdfglyphtounicode{nakatakana}{30CA}
\pdfglyphtounicode{nakatakanahalfwidth}{FF85}
\pdfglyphtounicode{nand}{22BC}
\pdfglyphtounicode{napostrophe}{0149}
\pdfglyphtounicode{nasquare}{3381}
\pdfglyphtounicode{natural}{266E}
\pdfglyphtounicode{nbopomofo}{310B}
\pdfglyphtounicode{nbspace}{00A0}
\pdfglyphtounicode{ncaron}{0148}
\pdfglyphtounicode{ncedilla}{0146}
\pdfglyphtounicode{ncircle}{24DD}
\pdfglyphtounicode{ncircumflexbelow}{1E4B}
\pdfglyphtounicode{ncommaaccent}{0146}
\pdfglyphtounicode{ndotaccent}{1E45}
\pdfglyphtounicode{ndotbelow}{1E47}
\pdfglyphtounicode{negationslash}{0338}
\pdfglyphtounicode{nehiragana}{306D}
\pdfglyphtounicode{nekatakana}{30CD}
\pdfglyphtounicode{nekatakanahalfwidth}{FF88}
\pdfglyphtounicode{newsheqelsign}{20AA}
\pdfglyphtounicode{nfsquare}{338B}
\pdfglyphtounicode{ng}{014B}
\pdfglyphtounicode{ngabengali}{0999}
\pdfglyphtounicode{ngadeva}{0919}
\pdfglyphtounicode{ngagujarati}{0A99}
\pdfglyphtounicode{ngagurmukhi}{0A19}
\pdfglyphtounicode{ngonguthai}{0E07}
\pdfglyphtounicode{nhiragana}{3093}
\pdfglyphtounicode{nhookleft}{0272}
\pdfglyphtounicode{nhookretroflex}{0273}
\pdfglyphtounicode{nieunacirclekorean}{326F}
\pdfglyphtounicode{nieunaparenkorean}{320F}
\pdfglyphtounicode{nieuncieuckorean}{3135}
\pdfglyphtounicode{nieuncirclekorean}{3261}
\pdfglyphtounicode{nieunhieuhkorean}{3136}
\pdfglyphtounicode{nieunkorean}{3134}
\pdfglyphtounicode{nieunpansioskorean}{3168}
\pdfglyphtounicode{nieunparenkorean}{3201}
\pdfglyphtounicode{nieunsioskorean}{3167}
\pdfglyphtounicode{nieuntikeutkorean}{3166}
\pdfglyphtounicode{nihiragana}{306B}
\pdfglyphtounicode{nikatakana}{30CB}
\pdfglyphtounicode{nikatakanahalfwidth}{FF86}
\pdfglyphtounicode{nikhahitleftthai}{F899}
\pdfglyphtounicode{nikhahitthai}{0E4D}
\pdfglyphtounicode{nine}{0039}
\pdfglyphtounicode{ninearabic}{0669}
\pdfglyphtounicode{ninebengali}{09EF}
\pdfglyphtounicode{ninecircle}{2468}
\pdfglyphtounicode{ninecircleinversesansserif}{2792}
\pdfglyphtounicode{ninedeva}{096F}
\pdfglyphtounicode{ninegujarati}{0AEF}
\pdfglyphtounicode{ninegurmukhi}{0A6F}
\pdfglyphtounicode{ninehackarabic}{0669}
\pdfglyphtounicode{ninehangzhou}{3029}
\pdfglyphtounicode{nineideographicparen}{3228}
\pdfglyphtounicode{nineinferior}{2089}
\pdfglyphtounicode{ninemonospace}{FF19}
\pdfglyphtounicode{nineoldstyle}{0039}
\pdfglyphtounicode{nineparen}{247C}
\pdfglyphtounicode{nineperiod}{2490}
\pdfglyphtounicode{ninepersian}{06F9}
\pdfglyphtounicode{nineroman}{2178}
\pdfglyphtounicode{ninesuperior}{2079}
\pdfglyphtounicode{nineteencircle}{2472}
\pdfglyphtounicode{nineteenparen}{2486}
\pdfglyphtounicode{nineteenperiod}{249A}
\pdfglyphtounicode{ninethai}{0E59}
\pdfglyphtounicode{nj}{01CC}
\pdfglyphtounicode{njecyrillic}{045A}
\pdfglyphtounicode{nkatakana}{30F3}
\pdfglyphtounicode{nkatakanahalfwidth}{FF9D}
\pdfglyphtounicode{nlegrightlong}{019E}
\pdfglyphtounicode{nlinebelow}{1E49}
\pdfglyphtounicode{nmonospace}{FF4E}
\pdfglyphtounicode{nmsquare}{339A}
\pdfglyphtounicode{nnabengali}{09A3}
\pdfglyphtounicode{nnadeva}{0923}
\pdfglyphtounicode{nnagujarati}{0AA3}
\pdfglyphtounicode{nnagurmukhi}{0A23}
\pdfglyphtounicode{nnnadeva}{0929}
\pdfglyphtounicode{nohiragana}{306E}
\pdfglyphtounicode{nokatakana}{30CE}
\pdfglyphtounicode{nokatakanahalfwidth}{FF89}
\pdfglyphtounicode{nonbreakingspace}{00A0}
\pdfglyphtounicode{nonenthai}{0E13}
\pdfglyphtounicode{nonuthai}{0E19}
\pdfglyphtounicode{noonarabic}{0646}
\pdfglyphtounicode{noonfinalarabic}{FEE6}
\pdfglyphtounicode{noonghunnaarabic}{06BA}
\pdfglyphtounicode{noonghunnafinalarabic}{FB9F}
\pdfglyphtounicode{noonhehinitialarabic}{FEE7 FEEC}
\pdfglyphtounicode{nooninitialarabic}{FEE7}
\pdfglyphtounicode{noonjeeminitialarabic}{FCD2}
\pdfglyphtounicode{noonjeemisolatedarabic}{FC4B}
\pdfglyphtounicode{noonmedialarabic}{FEE8}
\pdfglyphtounicode{noonmeeminitialarabic}{FCD5}
\pdfglyphtounicode{noonmeemisolatedarabic}{FC4E}
\pdfglyphtounicode{noonnoonfinalarabic}{FC8D}
\pdfglyphtounicode{notapproxequal}{2247}
\pdfglyphtounicode{notarrowboth}{21AE}
\pdfglyphtounicode{notarrowleft}{219A}
\pdfglyphtounicode{notarrowright}{219B}
\pdfglyphtounicode{notbar}{2224}
\pdfglyphtounicode{notcontains}{220C}
\pdfglyphtounicode{notdblarrowboth}{21CE}
\pdfglyphtounicode{notdblarrowleft}{21CD}
\pdfglyphtounicode{notdblarrowright}{21CF}
\pdfglyphtounicode{notelement}{2209}
\pdfglyphtounicode{notelementof}{2209}
\pdfglyphtounicode{notequal}{2260}
\pdfglyphtounicode{notexistential}{2204}
\pdfglyphtounicode{notfollows}{2281}
\pdfglyphtounicode{notfollowsoreql}{2AB0 0338}
\pdfglyphtounicode{notforces}{22AE}
\pdfglyphtounicode{notforcesextra}{22AF}
\pdfglyphtounicode{notgreater}{226F}
\pdfglyphtounicode{notgreaterdblequal}{2267 0338}
\pdfglyphtounicode{notgreaterequal}{2271}
\pdfglyphtounicode{notgreaternorequal}{2271}
\pdfglyphtounicode{notgreaternorless}{2279}
\pdfglyphtounicode{notgreaterorslnteql}{2A7E 0338}
\pdfglyphtounicode{notidentical}{2262}
\pdfglyphtounicode{notless}{226E}
\pdfglyphtounicode{notlessdblequal}{2266 0338}
\pdfglyphtounicode{notlessequal}{2270}
\pdfglyphtounicode{notlessnorequal}{2270}
\pdfglyphtounicode{notlessorslnteql}{2A7D 0338}
\pdfglyphtounicode{notparallel}{2226}
\pdfglyphtounicode{notprecedes}{2280}
\pdfglyphtounicode{notprecedesoreql}{2AAF 0338}
\pdfglyphtounicode{notsatisfies}{22AD}
\pdfglyphtounicode{notsimilar}{2241}
\pdfglyphtounicode{notsubset}{2284}
\pdfglyphtounicode{notsubseteql}{2288}
\pdfglyphtounicode{notsubsetordbleql}{2AC5 0338}
\pdfglyphtounicode{notsubsetoreql}{228A}
\pdfglyphtounicode{notsucceeds}{2281}
\pdfglyphtounicode{notsuperset}{2285}
\pdfglyphtounicode{notsuperseteql}{2289}
\pdfglyphtounicode{notsupersetordbleql}{2AC6 0338}
\pdfglyphtounicode{notsupersetoreql}{228B}
\pdfglyphtounicode{nottriangeqlleft}{22EC}
\pdfglyphtounicode{nottriangeqlright}{22ED}
\pdfglyphtounicode{nottriangleleft}{22EA}
\pdfglyphtounicode{nottriangleright}{22EB}
\pdfglyphtounicode{notturnstile}{22AC}
\pdfglyphtounicode{nowarmenian}{0576}
\pdfglyphtounicode{nparen}{24A9}
\pdfglyphtounicode{nssquare}{33B1}
\pdfglyphtounicode{nsuperior}{207F}
\pdfglyphtounicode{ntilde}{00F1}
\pdfglyphtounicode{nu}{03BD}
\pdfglyphtounicode{nuhiragana}{306C}
\pdfglyphtounicode{nukatakana}{30CC}
\pdfglyphtounicode{nukatakanahalfwidth}{FF87}
\pdfglyphtounicode{nuktabengali}{09BC}
\pdfglyphtounicode{nuktadeva}{093C}
\pdfglyphtounicode{nuktagujarati}{0ABC}
\pdfglyphtounicode{nuktagurmukhi}{0A3C}
\pdfglyphtounicode{numbersign}{0023}
\pdfglyphtounicode{numbersignmonospace}{FF03}
\pdfglyphtounicode{numbersignsmall}{FE5F}
\pdfglyphtounicode{numeralsigngreek}{0374}
\pdfglyphtounicode{numeralsignlowergreek}{0375}
\pdfglyphtounicode{numero}{2116}
\pdfglyphtounicode{nun}{05E0}
\pdfglyphtounicode{nundagesh}{FB40}
\pdfglyphtounicode{nundageshhebrew}{FB40}
\pdfglyphtounicode{nunhebrew}{05E0}
\pdfglyphtounicode{nvsquare}{33B5}
\pdfglyphtounicode{nwsquare}{33BB}
\pdfglyphtounicode{nyabengali}{099E}
\pdfglyphtounicode{nyadeva}{091E}
\pdfglyphtounicode{nyagujarati}{0A9E}
\pdfglyphtounicode{nyagurmukhi}{0A1E}
\pdfglyphtounicode{o}{006F}
\pdfglyphtounicode{oacute}{00F3}
\pdfglyphtounicode{oangthai}{0E2D}
\pdfglyphtounicode{obarred}{0275}
\pdfglyphtounicode{obarredcyrillic}{04E9}
\pdfglyphtounicode{obarreddieresiscyrillic}{04EB}
\pdfglyphtounicode{obengali}{0993}
\pdfglyphtounicode{obopomofo}{311B}
\pdfglyphtounicode{obreve}{014F}
\pdfglyphtounicode{ocandradeva}{0911}
\pdfglyphtounicode{ocandragujarati}{0A91}
\pdfglyphtounicode{ocandravowelsigndeva}{0949}
\pdfglyphtounicode{ocandravowelsigngujarati}{0AC9}
\pdfglyphtounicode{ocaron}{01D2}
\pdfglyphtounicode{ocircle}{24DE}
\pdfglyphtounicode{ocircumflex}{00F4}
\pdfglyphtounicode{ocircumflexacute}{1ED1}
\pdfglyphtounicode{ocircumflexdotbelow}{1ED9}
\pdfglyphtounicode{ocircumflexgrave}{1ED3}
\pdfglyphtounicode{ocircumflexhookabove}{1ED5}
\pdfglyphtounicode{ocircumflextilde}{1ED7}
\pdfglyphtounicode{ocyrillic}{043E}
\pdfglyphtounicode{odblacute}{0151}
\pdfglyphtounicode{odblgrave}{020D}
\pdfglyphtounicode{odeva}{0913}
\pdfglyphtounicode{odieresis}{00F6}
\pdfglyphtounicode{odieresiscyrillic}{04E7}
\pdfglyphtounicode{odotbelow}{1ECD}
\pdfglyphtounicode{oe}{0153}
\pdfglyphtounicode{oekorean}{315A}
\pdfglyphtounicode{ogonek}{02DB}
\pdfglyphtounicode{ogonekcmb}{0328}
\pdfglyphtounicode{ograve}{00F2}
\pdfglyphtounicode{ogujarati}{0A93}
\pdfglyphtounicode{oharmenian}{0585}
\pdfglyphtounicode{ohiragana}{304A}
\pdfglyphtounicode{ohookabove}{1ECF}
\pdfglyphtounicode{ohorn}{01A1}
\pdfglyphtounicode{ohornacute}{1EDB}
\pdfglyphtounicode{ohorndotbelow}{1EE3}
\pdfglyphtounicode{ohorngrave}{1EDD}
\pdfglyphtounicode{ohornhookabove}{1EDF}
\pdfglyphtounicode{ohorntilde}{1EE1}
\pdfglyphtounicode{ohungarumlaut}{0151}
\pdfglyphtounicode{oi}{01A3}
\pdfglyphtounicode{oinvertedbreve}{020F}
\pdfglyphtounicode{okatakana}{30AA}
\pdfglyphtounicode{okatakanahalfwidth}{FF75}
\pdfglyphtounicode{okorean}{3157}
\pdfglyphtounicode{olehebrew}{05AB}
\pdfglyphtounicode{omacron}{014D}
\pdfglyphtounicode{omacronacute}{1E53}
\pdfglyphtounicode{omacrongrave}{1E51}
\pdfglyphtounicode{omdeva}{0950}
\pdfglyphtounicode{omega}{03C9}
\pdfglyphtounicode{omega1}{03D6}
\pdfglyphtounicode{omegacyrillic}{0461}
\pdfglyphtounicode{omegalatinclosed}{0277}
\pdfglyphtounicode{omegaroundcyrillic}{047B}
\pdfglyphtounicode{omegatitlocyrillic}{047D}
\pdfglyphtounicode{omegatonos}{03CE}
\pdfglyphtounicode{omgujarati}{0AD0}
\pdfglyphtounicode{omicron}{03BF}
\pdfglyphtounicode{omicrontonos}{03CC}
\pdfglyphtounicode{omonospace}{FF4F}
\pdfglyphtounicode{one}{0031}
\pdfglyphtounicode{onearabic}{0661}
\pdfglyphtounicode{onebengali}{09E7}
\pdfglyphtounicode{onecircle}{2460}
\pdfglyphtounicode{onecircleinversesansserif}{278A}
\pdfglyphtounicode{onedeva}{0967}
\pdfglyphtounicode{onedotenleader}{2024}
\pdfglyphtounicode{oneeighth}{215B}
\pdfglyphtounicode{onefitted}{0031}
\pdfglyphtounicode{onegujarati}{0AE7}
\pdfglyphtounicode{onegurmukhi}{0A67}
\pdfglyphtounicode{onehackarabic}{0661}
\pdfglyphtounicode{onehalf}{00BD}
\pdfglyphtounicode{onehangzhou}{3021}
\pdfglyphtounicode{oneideographicparen}{3220}
\pdfglyphtounicode{oneinferior}{2081}
\pdfglyphtounicode{onemonospace}{FF11}
\pdfglyphtounicode{onenumeratorbengali}{09F4}
\pdfglyphtounicode{oneoldstyle}{0031}
\pdfglyphtounicode{oneparen}{2474}
\pdfglyphtounicode{oneperiod}{2488}
\pdfglyphtounicode{onepersian}{06F1}
\pdfglyphtounicode{onequarter}{00BC}
\pdfglyphtounicode{oneroman}{2170}
\pdfglyphtounicode{onesuperior}{00B9}
\pdfglyphtounicode{onethai}{0E51}
\pdfglyphtounicode{onethird}{2153}
\pdfglyphtounicode{oogonek}{01EB}
\pdfglyphtounicode{oogonekmacron}{01ED}
\pdfglyphtounicode{oogurmukhi}{0A13}
\pdfglyphtounicode{oomatragurmukhi}{0A4B}
\pdfglyphtounicode{oopen}{0254}
\pdfglyphtounicode{oparen}{24AA}
\pdfglyphtounicode{openbullet}{25E6}
\pdfglyphtounicode{option}{2325}
\pdfglyphtounicode{ordfeminine}{00AA}
\pdfglyphtounicode{ordmasculine}{00BA}
\pdfglyphtounicode{orthogonal}{221F}
\pdfglyphtounicode{orunderscore}{22BB}
\pdfglyphtounicode{oshortdeva}{0912}
\pdfglyphtounicode{oshortvowelsigndeva}{094A}
\pdfglyphtounicode{oslash}{00F8}
\pdfglyphtounicode{oslashacute}{01FF}
\pdfglyphtounicode{osmallhiragana}{3049}
\pdfglyphtounicode{osmallkatakana}{30A9}
\pdfglyphtounicode{osmallkatakanahalfwidth}{FF6B}
\pdfglyphtounicode{ostrokeacute}{01FF}
\pdfglyphtounicode{osuperior}{006F}
\pdfglyphtounicode{otcyrillic}{047F}
\pdfglyphtounicode{otilde}{00F5}
\pdfglyphtounicode{otildeacute}{1E4D}
\pdfglyphtounicode{otildedieresis}{1E4F}
\pdfglyphtounicode{oubopomofo}{3121}
\pdfglyphtounicode{overline}{203E}
\pdfglyphtounicode{overlinecenterline}{FE4A}
\pdfglyphtounicode{overlinecmb}{0305}
\pdfglyphtounicode{overlinedashed}{FE49}
\pdfglyphtounicode{overlinedblwavy}{FE4C}
\pdfglyphtounicode{overlinewavy}{FE4B}
\pdfglyphtounicode{overscore}{00AF}
\pdfglyphtounicode{ovowelsignbengali}{09CB}
\pdfglyphtounicode{ovowelsigndeva}{094B}
\pdfglyphtounicode{ovowelsigngujarati}{0ACB}
\pdfglyphtounicode{owner}{220B}
\pdfglyphtounicode{p}{0070}
\pdfglyphtounicode{paampssquare}{3380}
\pdfglyphtounicode{paasentosquare}{332B}
\pdfglyphtounicode{pabengali}{09AA}
\pdfglyphtounicode{pacute}{1E55}
\pdfglyphtounicode{padeva}{092A}
\pdfglyphtounicode{pagedown}{21DF}
\pdfglyphtounicode{pageup}{21DE}
\pdfglyphtounicode{pagujarati}{0AAA}
\pdfglyphtounicode{pagurmukhi}{0A2A}
\pdfglyphtounicode{pahiragana}{3071}
\pdfglyphtounicode{paiyannoithai}{0E2F}
\pdfglyphtounicode{pakatakana}{30D1}
\pdfglyphtounicode{palatalizationcyrilliccmb}{0484}
\pdfglyphtounicode{palochkacyrillic}{04C0}
\pdfglyphtounicode{pansioskorean}{317F}
\pdfglyphtounicode{paragraph}{00B6}
\pdfglyphtounicode{parallel}{2225}
\pdfglyphtounicode{parenleft}{0028}
\pdfglyphtounicode{parenleftaltonearabic}{FD3E}
\pdfglyphtounicode{parenleftbt}{F8ED}
\pdfglyphtounicode{parenleftex}{F8EC}
\pdfglyphtounicode{parenleftinferior}{208D}
\pdfglyphtounicode{parenleftmonospace}{FF08}
\pdfglyphtounicode{parenleftsmall}{FE59}
\pdfglyphtounicode{parenleftsuperior}{207D}
\pdfglyphtounicode{parenlefttp}{F8EB}
\pdfglyphtounicode{parenleftvertical}{FE35}
\pdfglyphtounicode{parenright}{0029}
\pdfglyphtounicode{parenrightaltonearabic}{FD3F}
\pdfglyphtounicode{parenrightbt}{F8F8}
\pdfglyphtounicode{parenrightex}{F8F7}
\pdfglyphtounicode{parenrightinferior}{208E}
\pdfglyphtounicode{parenrightmonospace}{FF09}
\pdfglyphtounicode{parenrightsmall}{FE5A}
\pdfglyphtounicode{parenrightsuperior}{207E}
\pdfglyphtounicode{parenrighttp}{F8F6}
\pdfglyphtounicode{parenrightvertical}{FE36}
\pdfglyphtounicode{partialdiff}{2202}
\pdfglyphtounicode{paseqhebrew}{05C0}
\pdfglyphtounicode{pashtahebrew}{0599}
\pdfglyphtounicode{pasquare}{33A9}
\pdfglyphtounicode{patah}{05B7}
\pdfglyphtounicode{patah11}{05B7}
\pdfglyphtounicode{patah1d}{05B7}
\pdfglyphtounicode{patah2a}{05B7}
\pdfglyphtounicode{patahhebrew}{05B7}
\pdfglyphtounicode{patahnarrowhebrew}{05B7}
\pdfglyphtounicode{patahquarterhebrew}{05B7}
\pdfglyphtounicode{patahwidehebrew}{05B7}
\pdfglyphtounicode{pazerhebrew}{05A1}
\pdfglyphtounicode{pbopomofo}{3106}
\pdfglyphtounicode{pcircle}{24DF}
\pdfglyphtounicode{pdotaccent}{1E57}
\pdfglyphtounicode{pe}{05E4}
\pdfglyphtounicode{pecyrillic}{043F}
\pdfglyphtounicode{pedagesh}{FB44}
\pdfglyphtounicode{pedageshhebrew}{FB44}
\pdfglyphtounicode{peezisquare}{333B}
\pdfglyphtounicode{pefinaldageshhebrew}{FB43}
\pdfglyphtounicode{peharabic}{067E}
\pdfglyphtounicode{peharmenian}{057A}
\pdfglyphtounicode{pehebrew}{05E4}
\pdfglyphtounicode{pehfinalarabic}{FB57}
\pdfglyphtounicode{pehinitialarabic}{FB58}
\pdfglyphtounicode{pehiragana}{307A}
\pdfglyphtounicode{pehmedialarabic}{FB59}
\pdfglyphtounicode{pekatakana}{30DA}
\pdfglyphtounicode{pemiddlehookcyrillic}{04A7}
\pdfglyphtounicode{perafehebrew}{FB4E}
\pdfglyphtounicode{percent}{0025}
\pdfglyphtounicode{percentarabic}{066A}
\pdfglyphtounicode{percentmonospace}{FF05}
\pdfglyphtounicode{percentsmall}{FE6A}
\pdfglyphtounicode{period}{002E}
\pdfglyphtounicode{periodarmenian}{0589}
\pdfglyphtounicode{periodcentered}{00B7}
\pdfglyphtounicode{periodhalfwidth}{FF61}
\pdfglyphtounicode{periodinferior}{002E}
\pdfglyphtounicode{periodmonospace}{FF0E}
\pdfglyphtounicode{periodsmall}{FE52}
\pdfglyphtounicode{periodsuperior}{002E}
\pdfglyphtounicode{perispomenigreekcmb}{0342}
\pdfglyphtounicode{perpcorrespond}{2A5E}
\pdfglyphtounicode{perpendicular}{22A5}
\pdfglyphtounicode{pertenthousand}{2031}
\pdfglyphtounicode{perthousand}{2030}
\pdfglyphtounicode{peseta}{20A7}
\pdfglyphtounicode{pfsquare}{338A}
\pdfglyphtounicode{phabengali}{09AB}
\pdfglyphtounicode{phadeva}{092B}
\pdfglyphtounicode{phagujarati}{0AAB}
\pdfglyphtounicode{phagurmukhi}{0A2B}
\pdfglyphtounicode{phi}{03C6}
\pdfglyphtounicode{phi1}{03D5}
\pdfglyphtounicode{phieuphacirclekorean}{327A}
\pdfglyphtounicode{phieuphaparenkorean}{321A}
\pdfglyphtounicode{phieuphcirclekorean}{326C}
\pdfglyphtounicode{phieuphkorean}{314D}
\pdfglyphtounicode{phieuphparenkorean}{320C}
\pdfglyphtounicode{philatin}{0278}
\pdfglyphtounicode{phinthuthai}{0E3A}
\pdfglyphtounicode{phisymbolgreek}{03D5}
\pdfglyphtounicode{phook}{01A5}
\pdfglyphtounicode{phophanthai}{0E1E}
\pdfglyphtounicode{phophungthai}{0E1C}
\pdfglyphtounicode{phosamphaothai}{0E20}
\pdfglyphtounicode{pi}{03C0}
\pdfglyphtounicode{pi1}{03D6}
\pdfglyphtounicode{pieupacirclekorean}{3273}
\pdfglyphtounicode{pieupaparenkorean}{3213}
\pdfglyphtounicode{pieupcieuckorean}{3176}
\pdfglyphtounicode{pieupcirclekorean}{3265}
\pdfglyphtounicode{pieupkiyeokkorean}{3172}
\pdfglyphtounicode{pieupkorean}{3142}
\pdfglyphtounicode{pieupparenkorean}{3205}
\pdfglyphtounicode{pieupsioskiyeokkorean}{3174}
\pdfglyphtounicode{pieupsioskorean}{3144}
\pdfglyphtounicode{pieupsiostikeutkorean}{3175}
\pdfglyphtounicode{pieupthieuthkorean}{3177}
\pdfglyphtounicode{pieuptikeutkorean}{3173}
\pdfglyphtounicode{pihiragana}{3074}
\pdfglyphtounicode{pikatakana}{30D4}
\pdfglyphtounicode{pisymbolgreek}{03D6}
\pdfglyphtounicode{piwrarmenian}{0583}
\pdfglyphtounicode{planckover2pi}{210F}
\pdfglyphtounicode{planckover2pi1}{210F}
\pdfglyphtounicode{plus}{002B}
\pdfglyphtounicode{plusbelowcmb}{031F}
\pdfglyphtounicode{pluscircle}{2295}
\pdfglyphtounicode{plusminus}{00B1}
\pdfglyphtounicode{plusmod}{02D6}
\pdfglyphtounicode{plusmonospace}{FF0B}
\pdfglyphtounicode{plussmall}{FE62}
\pdfglyphtounicode{plussuperior}{207A}
\pdfglyphtounicode{pmonospace}{FF50}
\pdfglyphtounicode{pmsquare}{33D8}
\pdfglyphtounicode{pohiragana}{307D}
\pdfglyphtounicode{pointingindexdownwhite}{261F}
\pdfglyphtounicode{pointingindexleftwhite}{261C}
\pdfglyphtounicode{pointingindexrightwhite}{261E}
\pdfglyphtounicode{pointingindexupwhite}{261D}
\pdfglyphtounicode{pokatakana}{30DD}
\pdfglyphtounicode{poplathai}{0E1B}
\pdfglyphtounicode{postalmark}{3012}
\pdfglyphtounicode{postalmarkface}{3020}
\pdfglyphtounicode{pparen}{24AB}
\pdfglyphtounicode{precedenotdbleqv}{2AB9}
\pdfglyphtounicode{precedenotslnteql}{2AB5}
\pdfglyphtounicode{precedeornoteqvlnt}{22E8}
\pdfglyphtounicode{precedes}{227A}
\pdfglyphtounicode{precedesequal}{2AAF}
\pdfglyphtounicode{precedesorcurly}{227C}
\pdfglyphtounicode{precedesorequal}{227E}
\pdfglyphtounicode{prescription}{211E}
\pdfglyphtounicode{prime}{2032}
\pdfglyphtounicode{primemod}{02B9}
\pdfglyphtounicode{primereverse}{2035}
\pdfglyphtounicode{primereversed}{2035}
\pdfglyphtounicode{product}{220F}
\pdfglyphtounicode{projective}{2305}
\pdfglyphtounicode{prolongedkana}{30FC}
\pdfglyphtounicode{propellor}{2318}
\pdfglyphtounicode{propersubset}{2282}
\pdfglyphtounicode{propersuperset}{2283}
\pdfglyphtounicode{proportion}{2237}
\pdfglyphtounicode{proportional}{221D}
\pdfglyphtounicode{psi}{03C8}
\pdfglyphtounicode{psicyrillic}{0471}
\pdfglyphtounicode{psilipneumatacyrilliccmb}{0486}
\pdfglyphtounicode{pssquare}{33B0}
\pdfglyphtounicode{puhiragana}{3077}
\pdfglyphtounicode{pukatakana}{30D7}
\pdfglyphtounicode{punctdash}{2014}
\pdfglyphtounicode{pvsquare}{33B4}
\pdfglyphtounicode{pwsquare}{33BA}
\pdfglyphtounicode{q}{0071}
\pdfglyphtounicode{qadeva}{0958}
\pdfglyphtounicode{qadmahebrew}{05A8}
\pdfglyphtounicode{qafarabic}{0642}
\pdfglyphtounicode{qaffinalarabic}{FED6}
\pdfglyphtounicode{qafinitialarabic}{FED7}
\pdfglyphtounicode{qafmedialarabic}{FED8}
\pdfglyphtounicode{qamats}{05B8}
\pdfglyphtounicode{qamats10}{05B8}
\pdfglyphtounicode{qamats1a}{05B8}
\pdfglyphtounicode{qamats1c}{05B8}
\pdfglyphtounicode{qamats27}{05B8}
\pdfglyphtounicode{qamats29}{05B8}
\pdfglyphtounicode{qamats33}{05B8}
\pdfglyphtounicode{qamatsde}{05B8}
\pdfglyphtounicode{qamatshebrew}{05B8}
\pdfglyphtounicode{qamatsnarrowhebrew}{05B8}
\pdfglyphtounicode{qamatsqatanhebrew}{05B8}
\pdfglyphtounicode{qamatsqatannarrowhebrew}{05B8}
\pdfglyphtounicode{qamatsqatanquarterhebrew}{05B8}
\pdfglyphtounicode{qamatsqatanwidehebrew}{05B8}
\pdfglyphtounicode{qamatsquarterhebrew}{05B8}
\pdfglyphtounicode{qamatswidehebrew}{05B8}
\pdfglyphtounicode{qarneyparahebrew}{059F}
\pdfglyphtounicode{qbopomofo}{3111}
\pdfglyphtounicode{qcircle}{24E0}
\pdfglyphtounicode{qhook}{02A0}
\pdfglyphtounicode{qmonospace}{FF51}
\pdfglyphtounicode{qof}{05E7}
\pdfglyphtounicode{qofdagesh}{FB47}
\pdfglyphtounicode{qofdageshhebrew}{FB47}
\pdfglyphtounicode{qofhatafpatah}{05E7 05B2}
\pdfglyphtounicode{qofhatafpatahhebrew}{05E7 05B2}
\pdfglyphtounicode{qofhatafsegol}{05E7 05B1}
\pdfglyphtounicode{qofhatafsegolhebrew}{05E7 05B1}
\pdfglyphtounicode{qofhebrew}{05E7}
\pdfglyphtounicode{qofhiriq}{05E7 05B4}
\pdfglyphtounicode{qofhiriqhebrew}{05E7 05B4}
\pdfglyphtounicode{qofholam}{05E7 05B9}
\pdfglyphtounicode{qofholamhebrew}{05E7 05B9}
\pdfglyphtounicode{qofpatah}{05E7 05B7}
\pdfglyphtounicode{qofpatahhebrew}{05E7 05B7}
\pdfglyphtounicode{qofqamats}{05E7 05B8}
\pdfglyphtounicode{qofqamatshebrew}{05E7 05B8}
\pdfglyphtounicode{qofqubuts}{05E7 05BB}
\pdfglyphtounicode{qofqubutshebrew}{05E7 05BB}
\pdfglyphtounicode{qofsegol}{05E7 05B6}
\pdfglyphtounicode{qofsegolhebrew}{05E7 05B6}
\pdfglyphtounicode{qofsheva}{05E7 05B0}
\pdfglyphtounicode{qofshevahebrew}{05E7 05B0}
\pdfglyphtounicode{qoftsere}{05E7 05B5}
\pdfglyphtounicode{qoftserehebrew}{05E7 05B5}
\pdfglyphtounicode{qparen}{24AC}
\pdfglyphtounicode{quarternote}{2669}
\pdfglyphtounicode{qubuts}{05BB}
\pdfglyphtounicode{qubuts18}{05BB}
\pdfglyphtounicode{qubuts25}{05BB}
\pdfglyphtounicode{qubuts31}{05BB}
\pdfglyphtounicode{qubutshebrew}{05BB}
\pdfglyphtounicode{qubutsnarrowhebrew}{05BB}
\pdfglyphtounicode{qubutsquarterhebrew}{05BB}
\pdfglyphtounicode{qubutswidehebrew}{05BB}
\pdfglyphtounicode{question}{003F}
\pdfglyphtounicode{questionarabic}{061F}
\pdfglyphtounicode{questionarmenian}{055E}
\pdfglyphtounicode{questiondown}{00BF}
\pdfglyphtounicode{questiondownsmall}{00BF}
\pdfglyphtounicode{questiongreek}{037E}
\pdfglyphtounicode{questionmonospace}{FF1F}
\pdfglyphtounicode{questionsmall}{003F}
\pdfglyphtounicode{quotedbl}{0022}
\pdfglyphtounicode{quotedblbase}{201E}
\pdfglyphtounicode{quotedblleft}{201C}
\pdfglyphtounicode{quotedblmonospace}{FF02}
\pdfglyphtounicode{quotedblprime}{301E}
\pdfglyphtounicode{quotedblprimereversed}{301D}
\pdfglyphtounicode{quotedblright}{201D}
\pdfglyphtounicode{quoteleft}{2018}
\pdfglyphtounicode{quoteleftreversed}{201B}
\pdfglyphtounicode{quotereversed}{201B}
\pdfglyphtounicode{quoteright}{2019}
\pdfglyphtounicode{quoterightn}{0149}
\pdfglyphtounicode{quotesinglbase}{201A}
\pdfglyphtounicode{quotesingle}{0027}
\pdfglyphtounicode{quotesinglemonospace}{FF07}
\pdfglyphtounicode{r}{0072}
\pdfglyphtounicode{raarmenian}{057C}
\pdfglyphtounicode{rabengali}{09B0}
\pdfglyphtounicode{racute}{0155}
\pdfglyphtounicode{radeva}{0930}
\pdfglyphtounicode{radical}{221A}
\pdfglyphtounicode{radicalex}{F8E5}
\pdfglyphtounicode{radoverssquare}{33AE}
\pdfglyphtounicode{radoverssquaredsquare}{33AF}
\pdfglyphtounicode{radsquare}{33AD}
\pdfglyphtounicode{rafe}{05BF}
\pdfglyphtounicode{rafehebrew}{05BF}
\pdfglyphtounicode{ragujarati}{0AB0}
\pdfglyphtounicode{ragurmukhi}{0A30}
\pdfglyphtounicode{rahiragana}{3089}
\pdfglyphtounicode{rakatakana}{30E9}
\pdfglyphtounicode{rakatakanahalfwidth}{FF97}
\pdfglyphtounicode{ralowerdiagonalbengali}{09F1}
\pdfglyphtounicode{ramiddlediagonalbengali}{09F0}
\pdfglyphtounicode{ramshorn}{0264}
\pdfglyphtounicode{rangedash}{2013}
\pdfglyphtounicode{ratio}{2236}
\pdfglyphtounicode{rbopomofo}{3116}
\pdfglyphtounicode{rcaron}{0159}
\pdfglyphtounicode{rcedilla}{0157}
\pdfglyphtounicode{rcircle}{24E1}
\pdfglyphtounicode{rcommaaccent}{0157}
\pdfglyphtounicode{rdblgrave}{0211}
\pdfglyphtounicode{rdotaccent}{1E59}
\pdfglyphtounicode{rdotbelow}{1E5B}
\pdfglyphtounicode{rdotbelowmacron}{1E5D}
\pdfglyphtounicode{referencemark}{203B}
\pdfglyphtounicode{reflexsubset}{2286}
\pdfglyphtounicode{reflexsuperset}{2287}
\pdfglyphtounicode{registered}{00AE}
\pdfglyphtounicode{registersans}{00AE}
\pdfglyphtounicode{registerserif}{00AE}
\pdfglyphtounicode{reharabic}{0631}
\pdfglyphtounicode{reharmenian}{0580}
\pdfglyphtounicode{rehfinalarabic}{FEAE}
\pdfglyphtounicode{rehiragana}{308C}
\pdfglyphtounicode{rehyehaleflamarabic}{0631 FEF3 FE8E 0644}
\pdfglyphtounicode{rekatakana}{30EC}
\pdfglyphtounicode{rekatakanahalfwidth}{FF9A}
\pdfglyphtounicode{resh}{05E8}
\pdfglyphtounicode{reshdageshhebrew}{FB48}
\pdfglyphtounicode{reshhatafpatah}{05E8 05B2}
\pdfglyphtounicode{reshhatafpatahhebrew}{05E8 05B2}
\pdfglyphtounicode{reshhatafsegol}{05E8 05B1}
\pdfglyphtounicode{reshhatafsegolhebrew}{05E8 05B1}
\pdfglyphtounicode{reshhebrew}{05E8}
\pdfglyphtounicode{reshhiriq}{05E8 05B4}
\pdfglyphtounicode{reshhiriqhebrew}{05E8 05B4}
\pdfglyphtounicode{reshholam}{05E8 05B9}
\pdfglyphtounicode{reshholamhebrew}{05E8 05B9}
\pdfglyphtounicode{reshpatah}{05E8 05B7}
\pdfglyphtounicode{reshpatahhebrew}{05E8 05B7}
\pdfglyphtounicode{reshqamats}{05E8 05B8}
\pdfglyphtounicode{reshqamatshebrew}{05E8 05B8}
\pdfglyphtounicode{reshqubuts}{05E8 05BB}
\pdfglyphtounicode{reshqubutshebrew}{05E8 05BB}
\pdfglyphtounicode{reshsegol}{05E8 05B6}
\pdfglyphtounicode{reshsegolhebrew}{05E8 05B6}
\pdfglyphtounicode{reshsheva}{05E8 05B0}
\pdfglyphtounicode{reshshevahebrew}{05E8 05B0}
\pdfglyphtounicode{reshtsere}{05E8 05B5}
\pdfglyphtounicode{reshtserehebrew}{05E8 05B5}
\pdfglyphtounicode{revasymptequal}{22CD}
\pdfglyphtounicode{reversedtilde}{223D}
\pdfglyphtounicode{reviahebrew}{0597}
\pdfglyphtounicode{reviamugrashhebrew}{0597}
\pdfglyphtounicode{revlogicalnot}{2310}
\pdfglyphtounicode{revsimilar}{223D}
\pdfglyphtounicode{rfishhook}{027E}
\pdfglyphtounicode{rfishhookreversed}{027F}
\pdfglyphtounicode{rhabengali}{09DD}
\pdfglyphtounicode{rhadeva}{095D}
\pdfglyphtounicode{rho}{03C1}
\pdfglyphtounicode{rho1}{03F1}
\pdfglyphtounicode{rhook}{027D}
\pdfglyphtounicode{rhookturned}{027B}
\pdfglyphtounicode{rhookturnedsuperior}{02B5}
\pdfglyphtounicode{rhosymbolgreek}{03F1}
\pdfglyphtounicode{rhotichookmod}{02DE}
\pdfglyphtounicode{rieulacirclekorean}{3271}
\pdfglyphtounicode{rieulaparenkorean}{3211}
\pdfglyphtounicode{rieulcirclekorean}{3263}
\pdfglyphtounicode{rieulhieuhkorean}{3140}
\pdfglyphtounicode{rieulkiyeokkorean}{313A}
\pdfglyphtounicode{rieulkiyeoksioskorean}{3169}
\pdfglyphtounicode{rieulkorean}{3139}
\pdfglyphtounicode{rieulmieumkorean}{313B}
\pdfglyphtounicode{rieulpansioskorean}{316C}
\pdfglyphtounicode{rieulparenkorean}{3203}
\pdfglyphtounicode{rieulphieuphkorean}{313F}
\pdfglyphtounicode{rieulpieupkorean}{313C}
\pdfglyphtounicode{rieulpieupsioskorean}{316B}
\pdfglyphtounicode{rieulsioskorean}{313D}
\pdfglyphtounicode{rieulthieuthkorean}{313E}
\pdfglyphtounicode{rieultikeutkorean}{316A}
\pdfglyphtounicode{rieulyeorinhieuhkorean}{316D}
\pdfglyphtounicode{rightangle}{221F}
\pdfglyphtounicode{rightanglene}{231D}
\pdfglyphtounicode{rightanglenw}{231C}
\pdfglyphtounicode{rightanglese}{231F}
\pdfglyphtounicode{rightanglesw}{231E}
\pdfglyphtounicode{righttackbelowcmb}{0319}
\pdfglyphtounicode{righttriangle}{22BF}
\pdfglyphtounicode{rihiragana}{308A}
\pdfglyphtounicode{rikatakana}{30EA}
\pdfglyphtounicode{rikatakanahalfwidth}{FF98}
\pdfglyphtounicode{ring}{02DA}
\pdfglyphtounicode{ringbelowcmb}{0325}
\pdfglyphtounicode{ringcmb}{030A}
\pdfglyphtounicode{ringhalfleft}{02BF}
\pdfglyphtounicode{ringhalfleftarmenian}{0559}
\pdfglyphtounicode{ringhalfleftbelowcmb}{031C}
\pdfglyphtounicode{ringhalfleftcentered}{02D3}
\pdfglyphtounicode{ringhalfright}{02BE}
\pdfglyphtounicode{ringhalfrightbelowcmb}{0339}
\pdfglyphtounicode{ringhalfrightcentered}{02D2}
\pdfglyphtounicode{ringinequal}{2256}
\pdfglyphtounicode{rinvertedbreve}{0213}
\pdfglyphtounicode{rittorusquare}{3351}
\pdfglyphtounicode{rlinebelow}{1E5F}
\pdfglyphtounicode{rlongleg}{027C}
\pdfglyphtounicode{rlonglegturned}{027A}
\pdfglyphtounicode{rmonospace}{FF52}
\pdfglyphtounicode{rohiragana}{308D}
\pdfglyphtounicode{rokatakana}{30ED}
\pdfglyphtounicode{rokatakanahalfwidth}{FF9B}
\pdfglyphtounicode{roruathai}{0E23}
\pdfglyphtounicode{rparen}{24AD}
\pdfglyphtounicode{rrabengali}{09DC}
\pdfglyphtounicode{rradeva}{0931}
\pdfglyphtounicode{rragurmukhi}{0A5C}
\pdfglyphtounicode{rreharabic}{0691}
\pdfglyphtounicode{rrehfinalarabic}{FB8D}
\pdfglyphtounicode{rrvocalicbengali}{09E0}
\pdfglyphtounicode{rrvocalicdeva}{0960}
\pdfglyphtounicode{rrvocalicgujarati}{0AE0}
\pdfglyphtounicode{rrvocalicvowelsignbengali}{09C4}
\pdfglyphtounicode{rrvocalicvowelsigndeva}{0944}
\pdfglyphtounicode{rrvocalicvowelsigngujarati}{0AC4}
\pdfglyphtounicode{rsuperior}{0072}
\pdfglyphtounicode{rtblock}{2590}
\pdfglyphtounicode{rturned}{0279}
\pdfglyphtounicode{rturnedsuperior}{02B4}
\pdfglyphtounicode{ruhiragana}{308B}
\pdfglyphtounicode{rukatakana}{30EB}
\pdfglyphtounicode{rukatakanahalfwidth}{FF99}
\pdfglyphtounicode{rupeemarkbengali}{09F2}
\pdfglyphtounicode{rupeesignbengali}{09F3}
\pdfglyphtounicode{rupiah}{20A8}
\pdfglyphtounicode{ruthai}{0E24}
\pdfglyphtounicode{rvocalicbengali}{098B}
\pdfglyphtounicode{rvocalicdeva}{090B}
\pdfglyphtounicode{rvocalicgujarati}{0A8B}
\pdfglyphtounicode{rvocalicvowelsignbengali}{09C3}
\pdfglyphtounicode{rvocalicvowelsigndeva}{0943}
\pdfglyphtounicode{rvocalicvowelsigngujarati}{0AC3}
\pdfglyphtounicode{s}{0073}
\pdfglyphtounicode{sabengali}{09B8}
\pdfglyphtounicode{sacute}{015B}
\pdfglyphtounicode{sacutedotaccent}{1E65}
\pdfglyphtounicode{sadarabic}{0635}
\pdfglyphtounicode{sadeva}{0938}
\pdfglyphtounicode{sadfinalarabic}{FEBA}
\pdfglyphtounicode{sadinitialarabic}{FEBB}
\pdfglyphtounicode{sadmedialarabic}{FEBC}
\pdfglyphtounicode{sagujarati}{0AB8}
\pdfglyphtounicode{sagurmukhi}{0A38}
\pdfglyphtounicode{sahiragana}{3055}
\pdfglyphtounicode{sakatakana}{30B5}
\pdfglyphtounicode{sakatakanahalfwidth}{FF7B}
\pdfglyphtounicode{sallallahoualayhewasallamarabic}{FDFA}
\pdfglyphtounicode{samekh}{05E1}
\pdfglyphtounicode{samekhdagesh}{FB41}
\pdfglyphtounicode{samekhdageshhebrew}{FB41}
\pdfglyphtounicode{samekhhebrew}{05E1}
\pdfglyphtounicode{saraaathai}{0E32}
\pdfglyphtounicode{saraaethai}{0E41}
\pdfglyphtounicode{saraaimaimalaithai}{0E44}
\pdfglyphtounicode{saraaimaimuanthai}{0E43}
\pdfglyphtounicode{saraamthai}{0E33}
\pdfglyphtounicode{saraathai}{0E30}
\pdfglyphtounicode{saraethai}{0E40}
\pdfglyphtounicode{saraiileftthai}{F886}
\pdfglyphtounicode{saraiithai}{0E35}
\pdfglyphtounicode{saraileftthai}{F885}
\pdfglyphtounicode{saraithai}{0E34}
\pdfglyphtounicode{saraothai}{0E42}
\pdfglyphtounicode{saraueeleftthai}{F888}
\pdfglyphtounicode{saraueethai}{0E37}
\pdfglyphtounicode{saraueleftthai}{F887}
\pdfglyphtounicode{sarauethai}{0E36}
\pdfglyphtounicode{sarauthai}{0E38}
\pdfglyphtounicode{sarauuthai}{0E39}
\pdfglyphtounicode{satisfies}{22A8}
\pdfglyphtounicode{sbopomofo}{3119}
\pdfglyphtounicode{scaron}{0161}
\pdfglyphtounicode{scarondotaccent}{1E67}
\pdfglyphtounicode{scedilla}{015F}
\pdfglyphtounicode{schwa}{0259}
\pdfglyphtounicode{schwacyrillic}{04D9}
\pdfglyphtounicode{schwadieresiscyrillic}{04DB}
\pdfglyphtounicode{schwahook}{025A}
\pdfglyphtounicode{scircle}{24E2}
\pdfglyphtounicode{scircumflex}{015D}
\pdfglyphtounicode{scommaaccent}{0219}
\pdfglyphtounicode{sdotaccent}{1E61}
\pdfglyphtounicode{sdotbelow}{1E63}
\pdfglyphtounicode{sdotbelowdotaccent}{1E69}
\pdfglyphtounicode{seagullbelowcmb}{033C}
\pdfglyphtounicode{second}{2033}
\pdfglyphtounicode{secondtonechinese}{02CA}
\pdfglyphtounicode{section}{00A7}
\pdfglyphtounicode{seenarabic}{0633}
\pdfglyphtounicode{seenfinalarabic}{FEB2}
\pdfglyphtounicode{seeninitialarabic}{FEB3}
\pdfglyphtounicode{seenmedialarabic}{FEB4}
\pdfglyphtounicode{segol}{05B6}
\pdfglyphtounicode{segol13}{05B6}
\pdfglyphtounicode{segol1f}{05B6}
\pdfglyphtounicode{segol2c}{05B6}
\pdfglyphtounicode{segolhebrew}{05B6}
\pdfglyphtounicode{segolnarrowhebrew}{05B6}
\pdfglyphtounicode{segolquarterhebrew}{05B6}
\pdfglyphtounicode{segoltahebrew}{0592}
\pdfglyphtounicode{segolwidehebrew}{05B6}
\pdfglyphtounicode{seharmenian}{057D}
\pdfglyphtounicode{sehiragana}{305B}
\pdfglyphtounicode{sekatakana}{30BB}
\pdfglyphtounicode{sekatakanahalfwidth}{FF7E}
\pdfglyphtounicode{semicolon}{003B}
\pdfglyphtounicode{semicolonarabic}{061B}
\pdfglyphtounicode{semicolonmonospace}{FF1B}
\pdfglyphtounicode{semicolonsmall}{FE54}
\pdfglyphtounicode{semivoicedmarkkana}{309C}
\pdfglyphtounicode{semivoicedmarkkanahalfwidth}{FF9F}
\pdfglyphtounicode{sentisquare}{3322}
\pdfglyphtounicode{sentosquare}{3323}
\pdfglyphtounicode{seven}{0037}
\pdfglyphtounicode{sevenarabic}{0667}
\pdfglyphtounicode{sevenbengali}{09ED}
\pdfglyphtounicode{sevencircle}{2466}
\pdfglyphtounicode{sevencircleinversesansserif}{2790}
\pdfglyphtounicode{sevendeva}{096D}
\pdfglyphtounicode{seveneighths}{215E}
\pdfglyphtounicode{sevengujarati}{0AED}
\pdfglyphtounicode{sevengurmukhi}{0A6D}
\pdfglyphtounicode{sevenhackarabic}{0667}
\pdfglyphtounicode{sevenhangzhou}{3027}
\pdfglyphtounicode{sevenideographicparen}{3226}
\pdfglyphtounicode{seveninferior}{2087}
\pdfglyphtounicode{sevenmonospace}{FF17}
\pdfglyphtounicode{sevenoldstyle}{0037}
\pdfglyphtounicode{sevenparen}{247A}
\pdfglyphtounicode{sevenperiod}{248E}
\pdfglyphtounicode{sevenpersian}{06F7}
\pdfglyphtounicode{sevenroman}{2176}
\pdfglyphtounicode{sevensuperior}{2077}
\pdfglyphtounicode{seventeencircle}{2470}
\pdfglyphtounicode{seventeenparen}{2484}
\pdfglyphtounicode{seventeenperiod}{2498}
\pdfglyphtounicode{seventhai}{0E57}
\pdfglyphtounicode{sfthyphen}{00AD}
\pdfglyphtounicode{shaarmenian}{0577}
\pdfglyphtounicode{shabengali}{09B6}
\pdfglyphtounicode{shacyrillic}{0448}
\pdfglyphtounicode{shaddaarabic}{0651}
\pdfglyphtounicode{shaddadammaarabic}{FC61}
\pdfglyphtounicode{shaddadammatanarabic}{FC5E}
\pdfglyphtounicode{shaddafathaarabic}{FC60}
\pdfglyphtounicode{shaddafathatanarabic}{0651 064B}
\pdfglyphtounicode{shaddakasraarabic}{FC62}
\pdfglyphtounicode{shaddakasratanarabic}{FC5F}
\pdfglyphtounicode{shade}{2592}
\pdfglyphtounicode{shadedark}{2593}
\pdfglyphtounicode{shadelight}{2591}
\pdfglyphtounicode{shademedium}{2592}
\pdfglyphtounicode{shadeva}{0936}
\pdfglyphtounicode{shagujarati}{0AB6}
\pdfglyphtounicode{shagurmukhi}{0A36}
\pdfglyphtounicode{shalshelethebrew}{0593}
\pdfglyphtounicode{sharp}{266F}
\pdfglyphtounicode{shbopomofo}{3115}
\pdfglyphtounicode{shchacyrillic}{0449}
\pdfglyphtounicode{sheenarabic}{0634}
\pdfglyphtounicode{sheenfinalarabic}{FEB6}
\pdfglyphtounicode{sheeninitialarabic}{FEB7}
\pdfglyphtounicode{sheenmedialarabic}{FEB8}
\pdfglyphtounicode{sheicoptic}{03E3}
\pdfglyphtounicode{sheqel}{20AA}
\pdfglyphtounicode{sheqelhebrew}{20AA}
\pdfglyphtounicode{sheva}{05B0}
\pdfglyphtounicode{sheva115}{05B0}
\pdfglyphtounicode{sheva15}{05B0}
\pdfglyphtounicode{sheva22}{05B0}
\pdfglyphtounicode{sheva2e}{05B0}
\pdfglyphtounicode{shevahebrew}{05B0}
\pdfglyphtounicode{shevanarrowhebrew}{05B0}
\pdfglyphtounicode{shevaquarterhebrew}{05B0}
\pdfglyphtounicode{shevawidehebrew}{05B0}
\pdfglyphtounicode{shhacyrillic}{04BB}
\pdfglyphtounicode{shiftleft}{21B0}
\pdfglyphtounicode{shiftright}{21B1}
\pdfglyphtounicode{shimacoptic}{03ED}
\pdfglyphtounicode{shin}{05E9}
\pdfglyphtounicode{shindagesh}{FB49}
\pdfglyphtounicode{shindageshhebrew}{FB49}
\pdfglyphtounicode{shindageshshindot}{FB2C}
\pdfglyphtounicode{shindageshshindothebrew}{FB2C}
\pdfglyphtounicode{shindageshsindot}{FB2D}
\pdfglyphtounicode{shindageshsindothebrew}{FB2D}
\pdfglyphtounicode{shindothebrew}{05C1}
\pdfglyphtounicode{shinhebrew}{05E9}
\pdfglyphtounicode{shinshindot}{FB2A}
\pdfglyphtounicode{shinshindothebrew}{FB2A}
\pdfglyphtounicode{shinsindot}{FB2B}
\pdfglyphtounicode{shinsindothebrew}{FB2B}
\pdfglyphtounicode{shook}{0282}
\pdfglyphtounicode{sigma}{03C3}
\pdfglyphtounicode{sigma1}{03C2}
\pdfglyphtounicode{sigmafinal}{03C2}
\pdfglyphtounicode{sigmalunatesymbolgreek}{03F2}
\pdfglyphtounicode{sihiragana}{3057}
\pdfglyphtounicode{sikatakana}{30B7}
\pdfglyphtounicode{sikatakanahalfwidth}{FF7C}
\pdfglyphtounicode{siluqhebrew}{05BD}
\pdfglyphtounicode{siluqlefthebrew}{05BD}
\pdfglyphtounicode{similar}{223C}
\pdfglyphtounicode{similarequal}{2243}
\pdfglyphtounicode{sindothebrew}{05C2}
\pdfglyphtounicode{siosacirclekorean}{3274}
\pdfglyphtounicode{siosaparenkorean}{3214}
\pdfglyphtounicode{sioscieuckorean}{317E}
\pdfglyphtounicode{sioscirclekorean}{3266}
\pdfglyphtounicode{sioskiyeokkorean}{317A}
\pdfglyphtounicode{sioskorean}{3145}
\pdfglyphtounicode{siosnieunkorean}{317B}
\pdfglyphtounicode{siosparenkorean}{3206}
\pdfglyphtounicode{siospieupkorean}{317D}
\pdfglyphtounicode{siostikeutkorean}{317C}
\pdfglyphtounicode{six}{0036}
\pdfglyphtounicode{sixarabic}{0666}
\pdfglyphtounicode{sixbengali}{09EC}
\pdfglyphtounicode{sixcircle}{2465}
\pdfglyphtounicode{sixcircleinversesansserif}{278F}
\pdfglyphtounicode{sixdeva}{096C}
\pdfglyphtounicode{sixgujarati}{0AEC}
\pdfglyphtounicode{sixgurmukhi}{0A6C}
\pdfglyphtounicode{sixhackarabic}{0666}
\pdfglyphtounicode{sixhangzhou}{3026}
\pdfglyphtounicode{sixideographicparen}{3225}
\pdfglyphtounicode{sixinferior}{2086}
\pdfglyphtounicode{sixmonospace}{FF16}
\pdfglyphtounicode{sixoldstyle}{0036}
\pdfglyphtounicode{sixparen}{2479}
\pdfglyphtounicode{sixperiod}{248D}
\pdfglyphtounicode{sixpersian}{06F6}
\pdfglyphtounicode{sixroman}{2175}
\pdfglyphtounicode{sixsuperior}{2076}
\pdfglyphtounicode{sixteencircle}{246F}
\pdfglyphtounicode{sixteencurrencydenominatorbengali}{09F9}
\pdfglyphtounicode{sixteenparen}{2483}
\pdfglyphtounicode{sixteenperiod}{2497}
\pdfglyphtounicode{sixthai}{0E56}
\pdfglyphtounicode{slash}{002F}
\pdfglyphtounicode{slashmonospace}{FF0F}
\pdfglyphtounicode{slong}{017F}
\pdfglyphtounicode{slongdotaccent}{1E9B}
\pdfglyphtounicode{slurabove}{2322}
\pdfglyphtounicode{slurbelow}{2323}
\pdfglyphtounicode{smile}{2323}
\pdfglyphtounicode{smileface}{263A}
\pdfglyphtounicode{smonospace}{FF53}
\pdfglyphtounicode{sofpasuqhebrew}{05C3}
\pdfglyphtounicode{softhyphen}{00AD}
\pdfglyphtounicode{softsigncyrillic}{044C}
\pdfglyphtounicode{sohiragana}{305D}
\pdfglyphtounicode{sokatakana}{30BD}
\pdfglyphtounicode{sokatakanahalfwidth}{FF7F}
\pdfglyphtounicode{soliduslongoverlaycmb}{0338}
\pdfglyphtounicode{solidusshortoverlaycmb}{0337}
\pdfglyphtounicode{sorusithai}{0E29}
\pdfglyphtounicode{sosalathai}{0E28}
\pdfglyphtounicode{sosothai}{0E0B}
\pdfglyphtounicode{sosuathai}{0E2A}
\pdfglyphtounicode{space}{0020}
\pdfglyphtounicode{spacehackarabic}{0020}
\pdfglyphtounicode{spade}{2660}
\pdfglyphtounicode{spadesuitblack}{2660}
\pdfglyphtounicode{spadesuitwhite}{2664}
\pdfglyphtounicode{sparen}{24AE}
\pdfglyphtounicode{sphericalangle}{2222}
\pdfglyphtounicode{square}{25A1}
\pdfglyphtounicode{squarebelowcmb}{033B}
\pdfglyphtounicode{squarecc}{33C4}
\pdfglyphtounicode{squarecm}{339D}
\pdfglyphtounicode{squarediagonalcrosshatchfill}{25A9}
\pdfglyphtounicode{squaredot}{22A1}
\pdfglyphtounicode{squarehorizontalfill}{25A4}
\pdfglyphtounicode{squareimage}{228F}
\pdfglyphtounicode{squarekg}{338F}
\pdfglyphtounicode{squarekm}{339E}
\pdfglyphtounicode{squarekmcapital}{33CE}
\pdfglyphtounicode{squareln}{33D1}
\pdfglyphtounicode{squarelog}{33D2}
\pdfglyphtounicode{squaremg}{338E}
\pdfglyphtounicode{squaremil}{33D5}
\pdfglyphtounicode{squareminus}{229F}
\pdfglyphtounicode{squaremm}{339C}
\pdfglyphtounicode{squaremsquared}{33A1}
\pdfglyphtounicode{squaremultiply}{22A0}
\pdfglyphtounicode{squareoriginal}{2290}
\pdfglyphtounicode{squareorthogonalcrosshatchfill}{25A6}
\pdfglyphtounicode{squareplus}{229E}
\pdfglyphtounicode{squaresolid}{25A0}
\pdfglyphtounicode{squareupperlefttolowerrightfill}{25A7}
\pdfglyphtounicode{squareupperrighttolowerleftfill}{25A8}
\pdfglyphtounicode{squareverticalfill}{25A5}
\pdfglyphtounicode{squarewhitewithsmallblack}{25A3}
\pdfglyphtounicode{squiggleleftright}{21AD}
\pdfglyphtounicode{squiggleright}{21DD}
\pdfglyphtounicode{srsquare}{33DB}
\pdfglyphtounicode{ssabengali}{09B7}
\pdfglyphtounicode{ssadeva}{0937}
\pdfglyphtounicode{ssagujarati}{0AB7}
\pdfglyphtounicode{ssangcieuckorean}{3149}
\pdfglyphtounicode{ssanghieuhkorean}{3185}
\pdfglyphtounicode{ssangieungkorean}{3180}
\pdfglyphtounicode{ssangkiyeokkorean}{3132}
\pdfglyphtounicode{ssangnieunkorean}{3165}
\pdfglyphtounicode{ssangpieupkorean}{3143}
\pdfglyphtounicode{ssangsioskorean}{3146}
\pdfglyphtounicode{ssangtikeutkorean}{3138}
\pdfglyphtounicode{ssuperior}{0073}
\pdfglyphtounicode{st}{0073 0074}
\pdfglyphtounicode{star}{22C6}
\pdfglyphtounicode{sterling}{00A3}
\pdfglyphtounicode{sterlingmonospace}{FFE1}
\pdfglyphtounicode{strokelongoverlaycmb}{0336}
\pdfglyphtounicode{strokeshortoverlaycmb}{0335}
\pdfglyphtounicode{subset}{2282}
\pdfglyphtounicode{subsetdbl}{22D0}
\pdfglyphtounicode{subsetdblequal}{2AC5}
\pdfglyphtounicode{subsetnoteql}{228A}
\pdfglyphtounicode{subsetnotequal}{228A}
\pdfglyphtounicode{subsetorequal}{2286}
\pdfglyphtounicode{subsetornotdbleql}{2ACB}
\pdfglyphtounicode{subsetsqequal}{2291}
\pdfglyphtounicode{succeeds}{227B}
\pdfglyphtounicode{suchthat}{220B}
\pdfglyphtounicode{suhiragana}{3059}
\pdfglyphtounicode{sukatakana}{30B9}
\pdfglyphtounicode{sukatakanahalfwidth}{FF7D}
\pdfglyphtounicode{sukunarabic}{0652}
\pdfglyphtounicode{summation}{2211}
\pdfglyphtounicode{sun}{263C}
\pdfglyphtounicode{superset}{2283}
\pdfglyphtounicode{supersetdbl}{22D1}
\pdfglyphtounicode{supersetdblequal}{2AC6}
\pdfglyphtounicode{supersetnoteql}{228B}
\pdfglyphtounicode{supersetnotequal}{228B}
\pdfglyphtounicode{supersetorequal}{2287}
\pdfglyphtounicode{supersetornotdbleql}{2ACC}
\pdfglyphtounicode{supersetsqequal}{2292}
\pdfglyphtounicode{svsquare}{33DC}
\pdfglyphtounicode{syouwaerasquare}{337C}
\pdfglyphtounicode{t}{0074}
\pdfglyphtounicode{tabengali}{09A4}
\pdfglyphtounicode{tackdown}{22A4}
\pdfglyphtounicode{tackleft}{22A3}
\pdfglyphtounicode{tadeva}{0924}
\pdfglyphtounicode{tagujarati}{0AA4}
\pdfglyphtounicode{tagurmukhi}{0A24}
\pdfglyphtounicode{taharabic}{0637}
\pdfglyphtounicode{tahfinalarabic}{FEC2}
\pdfglyphtounicode{tahinitialarabic}{FEC3}
\pdfglyphtounicode{tahiragana}{305F}
\pdfglyphtounicode{tahmedialarabic}{FEC4}
\pdfglyphtounicode{taisyouerasquare}{337D}
\pdfglyphtounicode{takatakana}{30BF}
\pdfglyphtounicode{takatakanahalfwidth}{FF80}
\pdfglyphtounicode{tatweelarabic}{0640}
\pdfglyphtounicode{tau}{03C4}
\pdfglyphtounicode{tav}{05EA}
\pdfglyphtounicode{tavdages}{FB4A}
\pdfglyphtounicode{tavdagesh}{FB4A}
\pdfglyphtounicode{tavdageshhebrew}{FB4A}
\pdfglyphtounicode{tavhebrew}{05EA}
\pdfglyphtounicode{tbar}{0167}
\pdfglyphtounicode{tbopomofo}{310A}
\pdfglyphtounicode{tcaron}{0165}
\pdfglyphtounicode{tccurl}{02A8}
\pdfglyphtounicode{tcedilla}{0163}
\pdfglyphtounicode{tcheharabic}{0686}
\pdfglyphtounicode{tchehfinalarabic}{FB7B}
\pdfglyphtounicode{tchehinitialarabic}{FB7C}
\pdfglyphtounicode{tchehmedialarabic}{FB7D}
\pdfglyphtounicode{tchehmeeminitialarabic}{FB7C FEE4}
\pdfglyphtounicode{tcircle}{24E3}
\pdfglyphtounicode{tcircumflexbelow}{1E71}
\pdfglyphtounicode{tcommaaccent}{0163}
\pdfglyphtounicode{tdieresis}{1E97}
\pdfglyphtounicode{tdotaccent}{1E6B}
\pdfglyphtounicode{tdotbelow}{1E6D}
\pdfglyphtounicode{tecyrillic}{0442}
\pdfglyphtounicode{tedescendercyrillic}{04AD}
\pdfglyphtounicode{teharabic}{062A}
\pdfglyphtounicode{tehfinalarabic}{FE96}
\pdfglyphtounicode{tehhahinitialarabic}{FCA2}
\pdfglyphtounicode{tehhahisolatedarabic}{FC0C}
\pdfglyphtounicode{tehinitialarabic}{FE97}
\pdfglyphtounicode{tehiragana}{3066}
\pdfglyphtounicode{tehjeeminitialarabic}{FCA1}
\pdfglyphtounicode{tehjeemisolatedarabic}{FC0B}
\pdfglyphtounicode{tehmarbutaarabic}{0629}
\pdfglyphtounicode{tehmarbutafinalarabic}{FE94}
\pdfglyphtounicode{tehmedialarabic}{FE98}
\pdfglyphtounicode{tehmeeminitialarabic}{FCA4}
\pdfglyphtounicode{tehmeemisolatedarabic}{FC0E}
\pdfglyphtounicode{tehnoonfinalarabic}{FC73}
\pdfglyphtounicode{tekatakana}{30C6}
\pdfglyphtounicode{tekatakanahalfwidth}{FF83}
\pdfglyphtounicode{telephone}{2121}
\pdfglyphtounicode{telephoneblack}{260E}
\pdfglyphtounicode{telishagedolahebrew}{05A0}
\pdfglyphtounicode{telishaqetanahebrew}{05A9}
\pdfglyphtounicode{tencircle}{2469}
\pdfglyphtounicode{tenideographicparen}{3229}
\pdfglyphtounicode{tenparen}{247D}
\pdfglyphtounicode{tenperiod}{2491}
\pdfglyphtounicode{tenroman}{2179}
\pdfglyphtounicode{tesh}{02A7}
\pdfglyphtounicode{tet}{05D8}
\pdfglyphtounicode{tetdagesh}{FB38}
\pdfglyphtounicode{tetdageshhebrew}{FB38}
\pdfglyphtounicode{tethebrew}{05D8}
\pdfglyphtounicode{tetsecyrillic}{04B5}
\pdfglyphtounicode{tevirhebrew}{059B}
\pdfglyphtounicode{tevirlefthebrew}{059B}
\pdfglyphtounicode{tfm:cmbsy10/diamond}{2662}
\pdfglyphtounicode{tfm:cmbsy10/heart}{2661}
\pdfglyphtounicode{tfm:cmbsy5/diamond}{2662}
\pdfglyphtounicode{tfm:cmbsy5/heart}{2661}
\pdfglyphtounicode{tfm:cmbsy6/diamond}{2662}
\pdfglyphtounicode{tfm:cmbsy6/heart}{2661}
\pdfglyphtounicode{tfm:cmbsy7/diamond}{2662}
\pdfglyphtounicode{tfm:cmbsy7/heart}{2661}
\pdfglyphtounicode{tfm:cmbsy8/diamond}{2662}
\pdfglyphtounicode{tfm:cmbsy8/heart}{2661}
\pdfglyphtounicode{tfm:cmbsy9/diamond}{2662}
\pdfglyphtounicode{tfm:cmbsy9/heart}{2661}
\pdfglyphtounicode{tfm:cmmi10/phi}{03D5}
\pdfglyphtounicode{tfm:cmmi10/phi1}{03C6}
\pdfglyphtounicode{tfm:cmmi12/phi}{03D5}
\pdfglyphtounicode{tfm:cmmi12/phi1}{03C6}
\pdfglyphtounicode{tfm:cmmi5/phi}{03D5}
\pdfglyphtounicode{tfm:cmmi5/phi1}{03C6}
\pdfglyphtounicode{tfm:cmmi6/phi}{03D5}
\pdfglyphtounicode{tfm:cmmi6/phi1}{03C6}
\pdfglyphtounicode{tfm:cmmi7/phi}{03D5}
\pdfglyphtounicode{tfm:cmmi7/phi1}{03C6}
\pdfglyphtounicode{tfm:cmmi8/phi}{03D5}
\pdfglyphtounicode{tfm:cmmi8/phi1}{03C6}
\pdfglyphtounicode{tfm:cmmi9/phi}{03D5}
\pdfglyphtounicode{tfm:cmmi9/phi1}{03C6}
\pdfglyphtounicode{tfm:cmmib10/phi}{03D5}
\pdfglyphtounicode{tfm:cmmib10/phi1}{03C6}
\pdfglyphtounicode{tfm:cmmib5/phi}{03D5}
\pdfglyphtounicode{tfm:cmmib5/phi1}{03C6}
\pdfglyphtounicode{tfm:cmmib6/phi}{03D5}
\pdfglyphtounicode{tfm:cmmib6/phi1}{03C6}
\pdfglyphtounicode{tfm:cmmib7/phi}{03D5}
\pdfglyphtounicode{tfm:cmmib7/phi1}{03C6}
\pdfglyphtounicode{tfm:cmmib8/phi}{03D5}
\pdfglyphtounicode{tfm:cmmib8/phi1}{03C6}
\pdfglyphtounicode{tfm:cmmib9/phi}{03D5}
\pdfglyphtounicode{tfm:cmmib9/phi1}{03C6}
\pdfglyphtounicode{tfm:cmsy10/diamond}{2662}
\pdfglyphtounicode{tfm:cmsy10/heart}{2661}
\pdfglyphtounicode{tfm:cmsy5/heart}{2661}
\pdfglyphtounicode{tfm:cmsy6/diamond}{2662}
\pdfglyphtounicode{tfm:cmsy6/heart}{2661}
\pdfglyphtounicode{tfm:cmsy7/diamond}{2662}
\pdfglyphtounicode{tfm:cmsy7/heart}{2661}
\pdfglyphtounicode{tfm:cmsy8/diamond}{2662}
\pdfglyphtounicode{tfm:cmsy8/heart}{2661}
\pdfglyphtounicode{tfm:cmsy9/diamond}{2662}
\pdfglyphtounicode{tfm:cmsy9/heart}{2661}
\pdfglyphtounicode{tfm:eurb10/phi}{03D5}
\pdfglyphtounicode{tfm:eurb10/phi1}{03C6}
\pdfglyphtounicode{tfm:eurb5/phi}{03D5}
\pdfglyphtounicode{tfm:eurb5/phi1}{03C6}
\pdfglyphtounicode{tfm:eurb6/phi}{03D5}
\pdfglyphtounicode{tfm:eurb6/phi1}{03C6}
\pdfglyphtounicode{tfm:eurb7/phi}{03D5}
\pdfglyphtounicode{tfm:eurb7/phi1}{03C6}
\pdfglyphtounicode{tfm:eurb8/phi}{03D5}
\pdfglyphtounicode{tfm:eurb8/phi1}{03C6}
\pdfglyphtounicode{tfm:eurb9/phi}{03D5}
\pdfglyphtounicode{tfm:eurb9/phi1}{03C6}
\pdfglyphtounicode{tfm:eurm10/phi}{03D5}
\pdfglyphtounicode{tfm:eurm10/phi1}{03C6}
\pdfglyphtounicode{tfm:eurm5/phi}{03D5}
\pdfglyphtounicode{tfm:eurm5/phi1}{03C6}
\pdfglyphtounicode{tfm:eurm6/phi}{03D5}
\pdfglyphtounicode{tfm:eurm6/phi1}{03C6}
\pdfglyphtounicode{tfm:eurm7/phi}{03D5}
\pdfglyphtounicode{tfm:eurm7/phi1}{03C6}
\pdfglyphtounicode{tfm:eurm8/phi}{03D5}
\pdfglyphtounicode{tfm:eurm8/phi1}{03C6}
\pdfglyphtounicode{tfm:eurm9/phi}{03D5}
\pdfglyphtounicode{tfm:eurm9/phi1}{03C6}
\pdfglyphtounicode{tfm:fplmbi/phi}{03D5}
\pdfglyphtounicode{tfm:fplmbi/phi1}{03C6}
\pdfglyphtounicode{tfm:fplmri/phi}{03D5}
\pdfglyphtounicode{tfm:fplmri/phi1}{03C6}
\pdfglyphtounicode{tfm:lmbsy10/diamond}{2662}
\pdfglyphtounicode{tfm:lmbsy10/heart}{2661}
\pdfglyphtounicode{tfm:lmbsy5/diamond}{2662}
\pdfglyphtounicode{tfm:lmbsy5/heart}{2661}
\pdfglyphtounicode{tfm:lmbsy7/diamond}{2662}
\pdfglyphtounicode{tfm:lmbsy7/heart}{2661}
\pdfglyphtounicode{tfm:lmmi10/phi}{03D5}
\pdfglyphtounicode{tfm:lmmi10/phi1}{03C6}
\pdfglyphtounicode{tfm:lmmi12/phi}{03D5}
\pdfglyphtounicode{tfm:lmmi12/phi1}{03C6}
\pdfglyphtounicode{tfm:lmmi5/phi}{03D5}
\pdfglyphtounicode{tfm:lmmi5/phi1}{03C6}
\pdfglyphtounicode{tfm:lmmi6/phi}{03D5}
\pdfglyphtounicode{tfm:lmmi6/phi1}{03C6}
\pdfglyphtounicode{tfm:lmmi7/phi}{03D5}
\pdfglyphtounicode{tfm:lmmi7/phi1}{03C6}
\pdfglyphtounicode{tfm:lmmi8/phi}{03D5}
\pdfglyphtounicode{tfm:lmmi8/phi1}{03C6}
\pdfglyphtounicode{tfm:lmmi9/phi}{03D5}
\pdfglyphtounicode{tfm:lmmi9/phi1}{03C6}
\pdfglyphtounicode{tfm:lmmib10/phi}{03D5}
\pdfglyphtounicode{tfm:lmmib10/phi1}{03C6}
\pdfglyphtounicode{tfm:lmmib5/phi}{03D5}
\pdfglyphtounicode{tfm:lmmib5/phi1}{03C6}
\pdfglyphtounicode{tfm:lmmib7/phi}{03D5}
\pdfglyphtounicode{tfm:lmmib7/phi1}{03C6}
\pdfglyphtounicode{tfm:lmsy10/diamond}{2662}
\pdfglyphtounicode{tfm:lmsy10/heart}{2661}
\pdfglyphtounicode{tfm:lmsy5/diamond}{2662}
\pdfglyphtounicode{tfm:lmsy5/heart}{2661}
\pdfglyphtounicode{tfm:lmsy6/diamond}{2662}
\pdfglyphtounicode{tfm:lmsy6/heart}{2661}
\pdfglyphtounicode{tfm:lmsy7/diamond}{2662}
\pdfglyphtounicode{tfm:lmsy7/heart}{2661}
\pdfglyphtounicode{tfm:lmsy8/diamond}{2662}
\pdfglyphtounicode{tfm:lmsy8/heart}{2661}
\pdfglyphtounicode{tfm:lmsy9/diamond}{2662}
\pdfglyphtounicode{tfm:lmsy9/heart}{2661}
\pdfglyphtounicode{tfm:msam10/diamond}{2662}
\pdfglyphtounicode{tfm:msam5/diamond}{2662}
\pdfglyphtounicode{tfm:msam6/diamond}{2662}
\pdfglyphtounicode{tfm:msam7/diamond}{2662}
\pdfglyphtounicode{tfm:msam8/diamond}{2662}
\pdfglyphtounicode{tfm:msam9/diamond}{2662}
\pdfglyphtounicode{tfm:pxbmia/phi}{03D5}
\pdfglyphtounicode{tfm:pxbmia/phi1}{03C6}
\pdfglyphtounicode{tfm:pxbsy/diamond}{2662}
\pdfglyphtounicode{tfm:pxbsy/heart}{2661}
\pdfglyphtounicode{tfm:pxbsya/diamond}{2662}
\pdfglyphtounicode{tfm:pxmia/phi}{03D5}
\pdfglyphtounicode{tfm:pxmia/phi1}{03C6}
\pdfglyphtounicode{tfm:pxsy/diamond}{2662}
\pdfglyphtounicode{tfm:pxsy/heart}{2661}
\pdfglyphtounicode{tfm:pxsya/diamond}{2662}
\pdfglyphtounicode{tfm:pzdr/a1}{2701}
\pdfglyphtounicode{tfm:pzdr/a10}{2721}
\pdfglyphtounicode{tfm:pzdr/a100}{275E}
\pdfglyphtounicode{tfm:pzdr/a101}{2761}
\pdfglyphtounicode{tfm:pzdr/a102}{2762}
\pdfglyphtounicode{tfm:pzdr/a103}{2763}
\pdfglyphtounicode{tfm:pzdr/a104}{2764}
\pdfglyphtounicode{tfm:pzdr/a105}{2710}
\pdfglyphtounicode{tfm:pzdr/a106}{2765}
\pdfglyphtounicode{tfm:pzdr/a107}{2766}
\pdfglyphtounicode{tfm:pzdr/a108}{2767}
\pdfglyphtounicode{tfm:pzdr/a109}{2660}
\pdfglyphtounicode{tfm:pzdr/a11}{261B}
\pdfglyphtounicode{tfm:pzdr/a110}{2665}
\pdfglyphtounicode{tfm:pzdr/a111}{2666}
\pdfglyphtounicode{tfm:pzdr/a112}{2663}
\pdfglyphtounicode{tfm:pzdr/a117}{2709}
\pdfglyphtounicode{tfm:pzdr/a118}{2708}
\pdfglyphtounicode{tfm:pzdr/a119}{2707}
\pdfglyphtounicode{tfm:pzdr/a12}{261E}
\pdfglyphtounicode{tfm:pzdr/a120}{2460}
\pdfglyphtounicode{tfm:pzdr/a121}{2461}
\pdfglyphtounicode{tfm:pzdr/a122}{2462}
\pdfglyphtounicode{tfm:pzdr/a123}{2463}
\pdfglyphtounicode{tfm:pzdr/a124}{2464}
\pdfglyphtounicode{tfm:pzdr/a125}{2465}
\pdfglyphtounicode{tfm:pzdr/a126}{2466}
\pdfglyphtounicode{tfm:pzdr/a127}{2467}
\pdfglyphtounicode{tfm:pzdr/a128}{2468}
\pdfglyphtounicode{tfm:pzdr/a129}{2469}
\pdfglyphtounicode{tfm:pzdr/a13}{270C}
\pdfglyphtounicode{tfm:pzdr/a130}{2776}
\pdfglyphtounicode{tfm:pzdr/a131}{2777}
\pdfglyphtounicode{tfm:pzdr/a132}{2778}
\pdfglyphtounicode{tfm:pzdr/a133}{2779}
\pdfglyphtounicode{tfm:pzdr/a134}{277A}
\pdfglyphtounicode{tfm:pzdr/a135}{277B}
\pdfglyphtounicode{tfm:pzdr/a136}{277C}
\pdfglyphtounicode{tfm:pzdr/a137}{277D}
\pdfglyphtounicode{tfm:pzdr/a138}{277E}
\pdfglyphtounicode{tfm:pzdr/a139}{277F}
\pdfglyphtounicode{tfm:pzdr/a14}{270D}
\pdfglyphtounicode{tfm:pzdr/a140}{2780}
\pdfglyphtounicode{tfm:pzdr/a141}{2781}
\pdfglyphtounicode{tfm:pzdr/a142}{2782}
\pdfglyphtounicode{tfm:pzdr/a143}{2783}
\pdfglyphtounicode{tfm:pzdr/a144}{2784}
\pdfglyphtounicode{tfm:pzdr/a145}{2785}
\pdfglyphtounicode{tfm:pzdr/a146}{2786}
\pdfglyphtounicode{tfm:pzdr/a147}{2787}
\pdfglyphtounicode{tfm:pzdr/a148}{2788}
\pdfglyphtounicode{tfm:pzdr/a149}{2789}
\pdfglyphtounicode{tfm:pzdr/a15}{270E}
\pdfglyphtounicode{tfm:pzdr/a150}{278A}
\pdfglyphtounicode{tfm:pzdr/a151}{278B}
\pdfglyphtounicode{tfm:pzdr/a152}{278C}
\pdfglyphtounicode{tfm:pzdr/a153}{278D}
\pdfglyphtounicode{tfm:pzdr/a154}{278E}
\pdfglyphtounicode{tfm:pzdr/a155}{278F}
\pdfglyphtounicode{tfm:pzdr/a156}{2790}
\pdfglyphtounicode{tfm:pzdr/a157}{2791}
\pdfglyphtounicode{tfm:pzdr/a158}{2792}
\pdfglyphtounicode{tfm:pzdr/a159}{2793}
\pdfglyphtounicode{tfm:pzdr/a16}{270F}
\pdfglyphtounicode{tfm:pzdr/a160}{2794}
\pdfglyphtounicode{tfm:pzdr/a161}{2192}
\pdfglyphtounicode{tfm:pzdr/a162}{27A3}
\pdfglyphtounicode{tfm:pzdr/a163}{2194}
\pdfglyphtounicode{tfm:pzdr/a164}{2195}
\pdfglyphtounicode{tfm:pzdr/a165}{2799}
\pdfglyphtounicode{tfm:pzdr/a166}{279B}
\pdfglyphtounicode{tfm:pzdr/a167}{279C}
\pdfglyphtounicode{tfm:pzdr/a168}{279D}
\pdfglyphtounicode{tfm:pzdr/a169}{279E}
\pdfglyphtounicode{tfm:pzdr/a17}{2711}
\pdfglyphtounicode{tfm:pzdr/a170}{279F}
\pdfglyphtounicode{tfm:pzdr/a171}{27A0}
\pdfglyphtounicode{tfm:pzdr/a172}{27A1}
\pdfglyphtounicode{tfm:pzdr/a173}{27A2}
\pdfglyphtounicode{tfm:pzdr/a174}{27A4}
\pdfglyphtounicode{tfm:pzdr/a175}{27A5}
\pdfglyphtounicode{tfm:pzdr/a176}{27A6}
\pdfglyphtounicode{tfm:pzdr/a177}{27A7}
\pdfglyphtounicode{tfm:pzdr/a178}{27A8}
\pdfglyphtounicode{tfm:pzdr/a179}{27A9}
\pdfglyphtounicode{tfm:pzdr/a18}{2712}
\pdfglyphtounicode{tfm:pzdr/a180}{27AB}
\pdfglyphtounicode{tfm:pzdr/a181}{27AD}
\pdfglyphtounicode{tfm:pzdr/a182}{27AF}
\pdfglyphtounicode{tfm:pzdr/a183}{27B2}
\pdfglyphtounicode{tfm:pzdr/a184}{27B3}
\pdfglyphtounicode{tfm:pzdr/a185}{27B5}
\pdfglyphtounicode{tfm:pzdr/a186}{27B8}
\pdfglyphtounicode{tfm:pzdr/a187}{27BA}
\pdfglyphtounicode{tfm:pzdr/a188}{27BB}
\pdfglyphtounicode{tfm:pzdr/a189}{27BC}
\pdfglyphtounicode{tfm:pzdr/a19}{2713}
\pdfglyphtounicode{tfm:pzdr/a190}{27BD}
\pdfglyphtounicode{tfm:pzdr/a191}{27BE}
\pdfglyphtounicode{tfm:pzdr/a192}{279A}
\pdfglyphtounicode{tfm:pzdr/a193}{27AA}
\pdfglyphtounicode{tfm:pzdr/a194}{27B6}
\pdfglyphtounicode{tfm:pzdr/a195}{27B9}
\pdfglyphtounicode{tfm:pzdr/a196}{2798}
\pdfglyphtounicode{tfm:pzdr/a197}{27B4}
\pdfglyphtounicode{tfm:pzdr/a198}{27B7}
\pdfglyphtounicode{tfm:pzdr/a199}{27AC}
\pdfglyphtounicode{tfm:pzdr/a2}{2702}
\pdfglyphtounicode{tfm:pzdr/a20}{2714}
\pdfglyphtounicode{tfm:pzdr/a200}{27AE}
\pdfglyphtounicode{tfm:pzdr/a201}{27B1}
\pdfglyphtounicode{tfm:pzdr/a202}{2703}
\pdfglyphtounicode{tfm:pzdr/a203}{2750}
\pdfglyphtounicode{tfm:pzdr/a204}{2752}
\pdfglyphtounicode{tfm:pzdr/a205}{276E}
\pdfglyphtounicode{tfm:pzdr/a206}{2770}
\pdfglyphtounicode{tfm:pzdr/a21}{2715}
\pdfglyphtounicode{tfm:pzdr/a22}{2716}
\pdfglyphtounicode{tfm:pzdr/a23}{2717}
\pdfglyphtounicode{tfm:pzdr/a24}{2718}
\pdfglyphtounicode{tfm:pzdr/a25}{2719}
\pdfglyphtounicode{tfm:pzdr/a26}{271A}
\pdfglyphtounicode{tfm:pzdr/a27}{271B}
\pdfglyphtounicode{tfm:pzdr/a28}{271C}
\pdfglyphtounicode{tfm:pzdr/a29}{2722}
\pdfglyphtounicode{tfm:pzdr/a3}{2704}
\pdfglyphtounicode{tfm:pzdr/a30}{2723}
\pdfglyphtounicode{tfm:pzdr/a31}{2724}
\pdfglyphtounicode{tfm:pzdr/a32}{2725}
\pdfglyphtounicode{tfm:pzdr/a33}{2726}
\pdfglyphtounicode{tfm:pzdr/a34}{2727}
\pdfglyphtounicode{tfm:pzdr/a35}{2605}
\pdfglyphtounicode{tfm:pzdr/a36}{2729}
\pdfglyphtounicode{tfm:pzdr/a37}{272A}
\pdfglyphtounicode{tfm:pzdr/a38}{272B}
\pdfglyphtounicode{tfm:pzdr/a39}{272C}
\pdfglyphtounicode{tfm:pzdr/a4}{260E}
\pdfglyphtounicode{tfm:pzdr/a40}{272D}
\pdfglyphtounicode{tfm:pzdr/a41}{272E}
\pdfglyphtounicode{tfm:pzdr/a42}{272F}
\pdfglyphtounicode{tfm:pzdr/a43}{2730}
\pdfglyphtounicode{tfm:pzdr/a44}{2731}
\pdfglyphtounicode{tfm:pzdr/a45}{2732}
\pdfglyphtounicode{tfm:pzdr/a46}{2733}
\pdfglyphtounicode{tfm:pzdr/a47}{2734}
\pdfglyphtounicode{tfm:pzdr/a48}{2735}
\pdfglyphtounicode{tfm:pzdr/a49}{2736}
\pdfglyphtounicode{tfm:pzdr/a5}{2706}
\pdfglyphtounicode{tfm:pzdr/a50}{2737}
\pdfglyphtounicode{tfm:pzdr/a51}{2738}
\pdfglyphtounicode{tfm:pzdr/a52}{2739}
\pdfglyphtounicode{tfm:pzdr/a53}{273A}
\pdfglyphtounicode{tfm:pzdr/a54}{273B}
\pdfglyphtounicode{tfm:pzdr/a55}{273C}
\pdfglyphtounicode{tfm:pzdr/a56}{273D}
\pdfglyphtounicode{tfm:pzdr/a57}{273E}
\pdfglyphtounicode{tfm:pzdr/a58}{273F}
\pdfglyphtounicode{tfm:pzdr/a59}{2740}
\pdfglyphtounicode{tfm:pzdr/a6}{271D}
\pdfglyphtounicode{tfm:pzdr/a60}{2741}
\pdfglyphtounicode{tfm:pzdr/a61}{2742}
\pdfglyphtounicode{tfm:pzdr/a62}{2743}
\pdfglyphtounicode{tfm:pzdr/a63}{2744}
\pdfglyphtounicode{tfm:pzdr/a64}{2745}
\pdfglyphtounicode{tfm:pzdr/a65}{2746}
\pdfglyphtounicode{tfm:pzdr/a66}{2747}
\pdfglyphtounicode{tfm:pzdr/a67}{2748}
\pdfglyphtounicode{tfm:pzdr/a68}{2749}
\pdfglyphtounicode{tfm:pzdr/a69}{274A}
\pdfglyphtounicode{tfm:pzdr/a7}{271E}
\pdfglyphtounicode{tfm:pzdr/a70}{274B}
\pdfglyphtounicode{tfm:pzdr/a71}{25CF}
\pdfglyphtounicode{tfm:pzdr/a72}{274D}
\pdfglyphtounicode{tfm:pzdr/a73}{25A0}
\pdfglyphtounicode{tfm:pzdr/a74}{274F}
\pdfglyphtounicode{tfm:pzdr/a75}{2751}
\pdfglyphtounicode{tfm:pzdr/a76}{25B2}
\pdfglyphtounicode{tfm:pzdr/a77}{25BC}
\pdfglyphtounicode{tfm:pzdr/a78}{25C6}
\pdfglyphtounicode{tfm:pzdr/a79}{2756}
\pdfglyphtounicode{tfm:pzdr/a8}{271F}
\pdfglyphtounicode{tfm:pzdr/a81}{25D7}
\pdfglyphtounicode{tfm:pzdr/a82}{2758}
\pdfglyphtounicode{tfm:pzdr/a83}{2759}
\pdfglyphtounicode{tfm:pzdr/a84}{275A}
\pdfglyphtounicode{tfm:pzdr/a85}{276F}
\pdfglyphtounicode{tfm:pzdr/a86}{2771}
\pdfglyphtounicode{tfm:pzdr/a87}{2772}
\pdfglyphtounicode{tfm:pzdr/a88}{2773}
\pdfglyphtounicode{tfm:pzdr/a89}{2768}
\pdfglyphtounicode{tfm:pzdr/a9}{2720}
\pdfglyphtounicode{tfm:pzdr/a90}{2769}
\pdfglyphtounicode{tfm:pzdr/a91}{276C}
\pdfglyphtounicode{tfm:pzdr/a92}{276D}
\pdfglyphtounicode{tfm:pzdr/a93}{276A}
\pdfglyphtounicode{tfm:pzdr/a94}{276B}
\pdfglyphtounicode{tfm:pzdr/a95}{2774}
\pdfglyphtounicode{tfm:pzdr/a96}{2775}
\pdfglyphtounicode{tfm:pzdr/a97}{275B}
\pdfglyphtounicode{tfm:pzdr/a98}{275C}
\pdfglyphtounicode{tfm:pzdr/a99}{275D}
\pdfglyphtounicode{tfm:rpxbmi/phi}{03D5}
\pdfglyphtounicode{tfm:rpxbmi/phi1}{03C6}
\pdfglyphtounicode{tfm:rpxmi/phi}{03D5}
\pdfglyphtounicode{tfm:rpxmi/phi1}{03C6}
\pdfglyphtounicode{tfm:rpzdr/a1}{2701}
\pdfglyphtounicode{tfm:rpzdr/a10}{2721}
\pdfglyphtounicode{tfm:rpzdr/a100}{275E}
\pdfglyphtounicode{tfm:rpzdr/a101}{2761}
\pdfglyphtounicode{tfm:rpzdr/a102}{2762}
\pdfglyphtounicode{tfm:rpzdr/a103}{2763}
\pdfglyphtounicode{tfm:rpzdr/a104}{2764}
\pdfglyphtounicode{tfm:rpzdr/a105}{2710}
\pdfglyphtounicode{tfm:rpzdr/a106}{2765}
\pdfglyphtounicode{tfm:rpzdr/a107}{2766}
\pdfglyphtounicode{tfm:rpzdr/a108}{2767}
\pdfglyphtounicode{tfm:rpzdr/a109}{2660}
\pdfglyphtounicode{tfm:rpzdr/a11}{261B}
\pdfglyphtounicode{tfm:rpzdr/a110}{2665}
\pdfglyphtounicode{tfm:rpzdr/a111}{2666}
\pdfglyphtounicode{tfm:rpzdr/a112}{2663}
\pdfglyphtounicode{tfm:rpzdr/a117}{2709}
\pdfglyphtounicode{tfm:rpzdr/a118}{2708}
\pdfglyphtounicode{tfm:rpzdr/a119}{2707}
\pdfglyphtounicode{tfm:rpzdr/a12}{261E}
\pdfglyphtounicode{tfm:rpzdr/a120}{2460}
\pdfglyphtounicode{tfm:rpzdr/a121}{2461}
\pdfglyphtounicode{tfm:rpzdr/a122}{2462}
\pdfglyphtounicode{tfm:rpzdr/a123}{2463}
\pdfglyphtounicode{tfm:rpzdr/a124}{2464}
\pdfglyphtounicode{tfm:rpzdr/a125}{2465}
\pdfglyphtounicode{tfm:rpzdr/a126}{2466}
\pdfglyphtounicode{tfm:rpzdr/a127}{2467}
\pdfglyphtounicode{tfm:rpzdr/a128}{2468}
\pdfglyphtounicode{tfm:rpzdr/a129}{2469}
\pdfglyphtounicode{tfm:rpzdr/a13}{270C}
\pdfglyphtounicode{tfm:rpzdr/a130}{2776}
\pdfglyphtounicode{tfm:rpzdr/a131}{2777}
\pdfglyphtounicode{tfm:rpzdr/a132}{2778}
\pdfglyphtounicode{tfm:rpzdr/a133}{2779}
\pdfglyphtounicode{tfm:rpzdr/a134}{277A}
\pdfglyphtounicode{tfm:rpzdr/a135}{277B}
\pdfglyphtounicode{tfm:rpzdr/a136}{277C}
\pdfglyphtounicode{tfm:rpzdr/a137}{277D}
\pdfglyphtounicode{tfm:rpzdr/a138}{277E}
\pdfglyphtounicode{tfm:rpzdr/a139}{277F}
\pdfglyphtounicode{tfm:rpzdr/a14}{270D}
\pdfglyphtounicode{tfm:rpzdr/a140}{2780}
\pdfglyphtounicode{tfm:rpzdr/a141}{2781}
\pdfglyphtounicode{tfm:rpzdr/a142}{2782}
\pdfglyphtounicode{tfm:rpzdr/a143}{2783}
\pdfglyphtounicode{tfm:rpzdr/a144}{2784}
\pdfglyphtounicode{tfm:rpzdr/a145}{2785}
\pdfglyphtounicode{tfm:rpzdr/a146}{2786}
\pdfglyphtounicode{tfm:rpzdr/a147}{2787}
\pdfglyphtounicode{tfm:rpzdr/a148}{2788}
\pdfglyphtounicode{tfm:rpzdr/a149}{2789}
\pdfglyphtounicode{tfm:rpzdr/a15}{270E}
\pdfglyphtounicode{tfm:rpzdr/a150}{278A}
\pdfglyphtounicode{tfm:rpzdr/a151}{278B}
\pdfglyphtounicode{tfm:rpzdr/a152}{278C}
\pdfglyphtounicode{tfm:rpzdr/a153}{278D}
\pdfglyphtounicode{tfm:rpzdr/a154}{278E}
\pdfglyphtounicode{tfm:rpzdr/a155}{278F}
\pdfglyphtounicode{tfm:rpzdr/a156}{2790}
\pdfglyphtounicode{tfm:rpzdr/a157}{2791}
\pdfglyphtounicode{tfm:rpzdr/a158}{2792}
\pdfglyphtounicode{tfm:rpzdr/a159}{2793}
\pdfglyphtounicode{tfm:rpzdr/a16}{270F}
\pdfglyphtounicode{tfm:rpzdr/a160}{2794}
\pdfglyphtounicode{tfm:rpzdr/a161}{2192}
\pdfglyphtounicode{tfm:rpzdr/a162}{27A3}
\pdfglyphtounicode{tfm:rpzdr/a163}{2194}
\pdfglyphtounicode{tfm:rpzdr/a164}{2195}
\pdfglyphtounicode{tfm:rpzdr/a165}{2799}
\pdfglyphtounicode{tfm:rpzdr/a166}{279B}
\pdfglyphtounicode{tfm:rpzdr/a167}{279C}
\pdfglyphtounicode{tfm:rpzdr/a168}{279D}
\pdfglyphtounicode{tfm:rpzdr/a169}{279E}
\pdfglyphtounicode{tfm:rpzdr/a17}{2711}
\pdfglyphtounicode{tfm:rpzdr/a170}{279F}
\pdfglyphtounicode{tfm:rpzdr/a171}{27A0}
\pdfglyphtounicode{tfm:rpzdr/a172}{27A1}
\pdfglyphtounicode{tfm:rpzdr/a173}{27A2}
\pdfglyphtounicode{tfm:rpzdr/a174}{27A4}
\pdfglyphtounicode{tfm:rpzdr/a175}{27A5}
\pdfglyphtounicode{tfm:rpzdr/a176}{27A6}
\pdfglyphtounicode{tfm:rpzdr/a177}{27A7}
\pdfglyphtounicode{tfm:rpzdr/a178}{27A8}
\pdfglyphtounicode{tfm:rpzdr/a179}{27A9}
\pdfglyphtounicode{tfm:rpzdr/a18}{2712}
\pdfglyphtounicode{tfm:rpzdr/a180}{27AB}
\pdfglyphtounicode{tfm:rpzdr/a181}{27AD}
\pdfglyphtounicode{tfm:rpzdr/a182}{27AF}
\pdfglyphtounicode{tfm:rpzdr/a183}{27B2}
\pdfglyphtounicode{tfm:rpzdr/a184}{27B3}
\pdfglyphtounicode{tfm:rpzdr/a185}{27B5}
\pdfglyphtounicode{tfm:rpzdr/a186}{27B8}
\pdfglyphtounicode{tfm:rpzdr/a187}{27BA}
\pdfglyphtounicode{tfm:rpzdr/a188}{27BB}
\pdfglyphtounicode{tfm:rpzdr/a189}{27BC}
\pdfglyphtounicode{tfm:rpzdr/a19}{2713}
\pdfglyphtounicode{tfm:rpzdr/a190}{27BD}
\pdfglyphtounicode{tfm:rpzdr/a191}{27BE}
\pdfglyphtounicode{tfm:rpzdr/a192}{279A}
\pdfglyphtounicode{tfm:rpzdr/a193}{27AA}
\pdfglyphtounicode{tfm:rpzdr/a194}{27B6}
\pdfglyphtounicode{tfm:rpzdr/a195}{27B9}
\pdfglyphtounicode{tfm:rpzdr/a196}{2798}
\pdfglyphtounicode{tfm:rpzdr/a197}{27B4}
\pdfglyphtounicode{tfm:rpzdr/a198}{27B7}
\pdfglyphtounicode{tfm:rpzdr/a199}{27AC}
\pdfglyphtounicode{tfm:rpzdr/a2}{2702}
\pdfglyphtounicode{tfm:rpzdr/a20}{2714}
\pdfglyphtounicode{tfm:rpzdr/a200}{27AE}
\pdfglyphtounicode{tfm:rpzdr/a201}{27B1}
\pdfglyphtounicode{tfm:rpzdr/a202}{2703}
\pdfglyphtounicode{tfm:rpzdr/a203}{2750}
\pdfglyphtounicode{tfm:rpzdr/a204}{2752}
\pdfglyphtounicode{tfm:rpzdr/a205}{276E}
\pdfglyphtounicode{tfm:rpzdr/a206}{2770}
\pdfglyphtounicode{tfm:rpzdr/a21}{2715}
\pdfglyphtounicode{tfm:rpzdr/a22}{2716}
\pdfglyphtounicode{tfm:rpzdr/a23}{2717}
\pdfglyphtounicode{tfm:rpzdr/a24}{2718}
\pdfglyphtounicode{tfm:rpzdr/a25}{2719}
\pdfglyphtounicode{tfm:rpzdr/a26}{271A}
\pdfglyphtounicode{tfm:rpzdr/a27}{271B}
\pdfglyphtounicode{tfm:rpzdr/a28}{271C}
\pdfglyphtounicode{tfm:rpzdr/a29}{2722}
\pdfglyphtounicode{tfm:rpzdr/a3}{2704}
\pdfglyphtounicode{tfm:rpzdr/a30}{2723}
\pdfglyphtounicode{tfm:rpzdr/a31}{2724}
\pdfglyphtounicode{tfm:rpzdr/a32}{2725}
\pdfglyphtounicode{tfm:rpzdr/a33}{2726}
\pdfglyphtounicode{tfm:rpzdr/a34}{2727}
\pdfglyphtounicode{tfm:rpzdr/a35}{2605}
\pdfglyphtounicode{tfm:rpzdr/a36}{2729}
\pdfglyphtounicode{tfm:rpzdr/a37}{272A}
\pdfglyphtounicode{tfm:rpzdr/a38}{272B}
\pdfglyphtounicode{tfm:rpzdr/a39}{272C}
\pdfglyphtounicode{tfm:rpzdr/a4}{260E}
\pdfglyphtounicode{tfm:rpzdr/a40}{272D}
\pdfglyphtounicode{tfm:rpzdr/a41}{272E}
\pdfglyphtounicode{tfm:rpzdr/a42}{272F}
\pdfglyphtounicode{tfm:rpzdr/a43}{2730}
\pdfglyphtounicode{tfm:rpzdr/a44}{2731}
\pdfglyphtounicode{tfm:rpzdr/a45}{2732}
\pdfglyphtounicode{tfm:rpzdr/a46}{2733}
\pdfglyphtounicode{tfm:rpzdr/a47}{2734}
\pdfglyphtounicode{tfm:rpzdr/a48}{2735}
\pdfglyphtounicode{tfm:rpzdr/a49}{2736}
\pdfglyphtounicode{tfm:rpzdr/a5}{2706}
\pdfglyphtounicode{tfm:rpzdr/a50}{2737}
\pdfglyphtounicode{tfm:rpzdr/a51}{2738}
\pdfglyphtounicode{tfm:rpzdr/a52}{2739}
\pdfglyphtounicode{tfm:rpzdr/a53}{273A}
\pdfglyphtounicode{tfm:rpzdr/a54}{273B}
\pdfglyphtounicode{tfm:rpzdr/a55}{273C}
\pdfglyphtounicode{tfm:rpzdr/a56}{273D}
\pdfglyphtounicode{tfm:rpzdr/a57}{273E}
\pdfglyphtounicode{tfm:rpzdr/a58}{273F}
\pdfglyphtounicode{tfm:rpzdr/a59}{2740}
\pdfglyphtounicode{tfm:rpzdr/a6}{271D}
\pdfglyphtounicode{tfm:rpzdr/a60}{2741}
\pdfglyphtounicode{tfm:rpzdr/a61}{2742}
\pdfglyphtounicode{tfm:rpzdr/a62}{2743}
\pdfglyphtounicode{tfm:rpzdr/a63}{2744}
\pdfglyphtounicode{tfm:rpzdr/a64}{2745}
\pdfglyphtounicode{tfm:rpzdr/a65}{2746}
\pdfglyphtounicode{tfm:rpzdr/a66}{2747}
\pdfglyphtounicode{tfm:rpzdr/a67}{2748}
\pdfglyphtounicode{tfm:rpzdr/a68}{2749}
\pdfglyphtounicode{tfm:rpzdr/a69}{274A}
\pdfglyphtounicode{tfm:rpzdr/a7}{271E}
\pdfglyphtounicode{tfm:rpzdr/a70}{274B}
\pdfglyphtounicode{tfm:rpzdr/a71}{25CF}
\pdfglyphtounicode{tfm:rpzdr/a72}{274D}
\pdfglyphtounicode{tfm:rpzdr/a73}{25A0}
\pdfglyphtounicode{tfm:rpzdr/a74}{274F}
\pdfglyphtounicode{tfm:rpzdr/a75}{2751}
\pdfglyphtounicode{tfm:rpzdr/a76}{25B2}
\pdfglyphtounicode{tfm:rpzdr/a77}{25BC}
\pdfglyphtounicode{tfm:rpzdr/a78}{25C6}
\pdfglyphtounicode{tfm:rpzdr/a79}{2756}
\pdfglyphtounicode{tfm:rpzdr/a8}{271F}
\pdfglyphtounicode{tfm:rpzdr/a81}{25D7}
\pdfglyphtounicode{tfm:rpzdr/a82}{2758}
\pdfglyphtounicode{tfm:rpzdr/a83}{2759}
\pdfglyphtounicode{tfm:rpzdr/a84}{275A}
\pdfglyphtounicode{tfm:rpzdr/a85}{276F}
\pdfglyphtounicode{tfm:rpzdr/a86}{2771}
\pdfglyphtounicode{tfm:rpzdr/a87}{2772}
\pdfglyphtounicode{tfm:rpzdr/a88}{2773}
\pdfglyphtounicode{tfm:rpzdr/a89}{2768}
\pdfglyphtounicode{tfm:rpzdr/a9}{2720}
\pdfglyphtounicode{tfm:rpzdr/a90}{2769}
\pdfglyphtounicode{tfm:rpzdr/a91}{276C}
\pdfglyphtounicode{tfm:rpzdr/a92}{276D}
\pdfglyphtounicode{tfm:rpzdr/a93}{276A}
\pdfglyphtounicode{tfm:rpzdr/a94}{276B}
\pdfglyphtounicode{tfm:rpzdr/a95}{2774}
\pdfglyphtounicode{tfm:rpzdr/a96}{2775}
\pdfglyphtounicode{tfm:rpzdr/a97}{275B}
\pdfglyphtounicode{tfm:rpzdr/a98}{275C}
\pdfglyphtounicode{tfm:rpzdr/a99}{275D}
\pdfglyphtounicode{tfm:rtxbmi/phi}{03D5}
\pdfglyphtounicode{tfm:rtxbmi/phi1}{03C6}
\pdfglyphtounicode{tfm:rtxmi/phi}{03D5}
\pdfglyphtounicode{tfm:rtxmi/phi1}{03C6}
\pdfglyphtounicode{tfm:txbmia/phi}{03D5}
\pdfglyphtounicode{tfm:txbmia/phi1}{03C6}
\pdfglyphtounicode{tfm:txbsy/diamond}{2662}
\pdfglyphtounicode{tfm:txbsy/heart}{2661}
\pdfglyphtounicode{tfm:txbsya/diamond}{2662}
\pdfglyphtounicode{tfm:txmia/phi}{03D5}
\pdfglyphtounicode{tfm:txmia/phi1}{03C6}
\pdfglyphtounicode{tfm:txsy/diamond}{2662}
\pdfglyphtounicode{tfm:txsy/heart}{2661}
\pdfglyphtounicode{tfm:txsya/diamond}{2662}
\pdfglyphtounicode{tfm:zd/a1}{2701}
\pdfglyphtounicode{tfm:zd/a10}{2721}
\pdfglyphtounicode{tfm:zd/a100}{275E}
\pdfglyphtounicode{tfm:zd/a101}{2761}
\pdfglyphtounicode{tfm:zd/a102}{2762}
\pdfglyphtounicode{tfm:zd/a103}{2763}
\pdfglyphtounicode{tfm:zd/a104}{2764}
\pdfglyphtounicode{tfm:zd/a105}{2710}
\pdfglyphtounicode{tfm:zd/a106}{2765}
\pdfglyphtounicode{tfm:zd/a107}{2766}
\pdfglyphtounicode{tfm:zd/a108}{2767}
\pdfglyphtounicode{tfm:zd/a109}{2660}
\pdfglyphtounicode{tfm:zd/a11}{261B}
\pdfglyphtounicode{tfm:zd/a110}{2665}
\pdfglyphtounicode{tfm:zd/a111}{2666}
\pdfglyphtounicode{tfm:zd/a112}{2663}
\pdfglyphtounicode{tfm:zd/a117}{2709}
\pdfglyphtounicode{tfm:zd/a118}{2708}
\pdfglyphtounicode{tfm:zd/a119}{2707}
\pdfglyphtounicode{tfm:zd/a12}{261E}
\pdfglyphtounicode{tfm:zd/a120}{2460}
\pdfglyphtounicode{tfm:zd/a121}{2461}
\pdfglyphtounicode{tfm:zd/a122}{2462}
\pdfglyphtounicode{tfm:zd/a123}{2463}
\pdfglyphtounicode{tfm:zd/a124}{2464}
\pdfglyphtounicode{tfm:zd/a125}{2465}
\pdfglyphtounicode{tfm:zd/a126}{2466}
\pdfglyphtounicode{tfm:zd/a127}{2467}
\pdfglyphtounicode{tfm:zd/a128}{2468}
\pdfglyphtounicode{tfm:zd/a129}{2469}
\pdfglyphtounicode{tfm:zd/a13}{270C}
\pdfglyphtounicode{tfm:zd/a130}{2776}
\pdfglyphtounicode{tfm:zd/a131}{2777}
\pdfglyphtounicode{tfm:zd/a132}{2778}
\pdfglyphtounicode{tfm:zd/a133}{2779}
\pdfglyphtounicode{tfm:zd/a134}{277A}
\pdfglyphtounicode{tfm:zd/a135}{277B}
\pdfglyphtounicode{tfm:zd/a136}{277C}
\pdfglyphtounicode{tfm:zd/a137}{277D}
\pdfglyphtounicode{tfm:zd/a138}{277E}
\pdfglyphtounicode{tfm:zd/a139}{277F}
\pdfglyphtounicode{tfm:zd/a14}{270D}
\pdfglyphtounicode{tfm:zd/a140}{2780}
\pdfglyphtounicode{tfm:zd/a141}{2781}
\pdfglyphtounicode{tfm:zd/a142}{2782}
\pdfglyphtounicode{tfm:zd/a143}{2783}
\pdfglyphtounicode{tfm:zd/a144}{2784}
\pdfglyphtounicode{tfm:zd/a145}{2785}
\pdfglyphtounicode{tfm:zd/a146}{2786}
\pdfglyphtounicode{tfm:zd/a147}{2787}
\pdfglyphtounicode{tfm:zd/a148}{2788}
\pdfglyphtounicode{tfm:zd/a149}{2789}
\pdfglyphtounicode{tfm:zd/a15}{270E}
\pdfglyphtounicode{tfm:zd/a150}{278A}
\pdfglyphtounicode{tfm:zd/a151}{278B}
\pdfglyphtounicode{tfm:zd/a152}{278C}
\pdfglyphtounicode{tfm:zd/a153}{278D}
\pdfglyphtounicode{tfm:zd/a154}{278E}
\pdfglyphtounicode{tfm:zd/a155}{278F}
\pdfglyphtounicode{tfm:zd/a156}{2790}
\pdfglyphtounicode{tfm:zd/a157}{2791}
\pdfglyphtounicode{tfm:zd/a158}{2792}
\pdfglyphtounicode{tfm:zd/a159}{2793}
\pdfglyphtounicode{tfm:zd/a16}{270F}
\pdfglyphtounicode{tfm:zd/a160}{2794}
\pdfglyphtounicode{tfm:zd/a161}{2192}
\pdfglyphtounicode{tfm:zd/a162}{27A3}
\pdfglyphtounicode{tfm:zd/a163}{2194}
\pdfglyphtounicode{tfm:zd/a164}{2195}
\pdfglyphtounicode{tfm:zd/a165}{2799}
\pdfglyphtounicode{tfm:zd/a166}{279B}
\pdfglyphtounicode{tfm:zd/a167}{279C}
\pdfglyphtounicode{tfm:zd/a168}{279D}
\pdfglyphtounicode{tfm:zd/a169}{279E}
\pdfglyphtounicode{tfm:zd/a17}{2711}
\pdfglyphtounicode{tfm:zd/a170}{279F}
\pdfglyphtounicode{tfm:zd/a171}{27A0}
\pdfglyphtounicode{tfm:zd/a172}{27A1}
\pdfglyphtounicode{tfm:zd/a173}{27A2}
\pdfglyphtounicode{tfm:zd/a174}{27A4}
\pdfglyphtounicode{tfm:zd/a175}{27A5}
\pdfglyphtounicode{tfm:zd/a176}{27A6}
\pdfglyphtounicode{tfm:zd/a177}{27A7}
\pdfglyphtounicode{tfm:zd/a178}{27A8}
\pdfglyphtounicode{tfm:zd/a179}{27A9}
\pdfglyphtounicode{tfm:zd/a18}{2712}
\pdfglyphtounicode{tfm:zd/a180}{27AB}
\pdfglyphtounicode{tfm:zd/a181}{27AD}
\pdfglyphtounicode{tfm:zd/a182}{27AF}
\pdfglyphtounicode{tfm:zd/a183}{27B2}
\pdfglyphtounicode{tfm:zd/a184}{27B3}
\pdfglyphtounicode{tfm:zd/a185}{27B5}
\pdfglyphtounicode{tfm:zd/a186}{27B8}
\pdfglyphtounicode{tfm:zd/a187}{27BA}
\pdfglyphtounicode{tfm:zd/a188}{27BB}
\pdfglyphtounicode{tfm:zd/a189}{27BC}
\pdfglyphtounicode{tfm:zd/a19}{2713}
\pdfglyphtounicode{tfm:zd/a190}{27BD}
\pdfglyphtounicode{tfm:zd/a191}{27BE}
\pdfglyphtounicode{tfm:zd/a192}{279A}
\pdfglyphtounicode{tfm:zd/a193}{27AA}
\pdfglyphtounicode{tfm:zd/a194}{27B6}
\pdfglyphtounicode{tfm:zd/a195}{27B9}
\pdfglyphtounicode{tfm:zd/a196}{2798}
\pdfglyphtounicode{tfm:zd/a197}{27B4}
\pdfglyphtounicode{tfm:zd/a198}{27B7}
\pdfglyphtounicode{tfm:zd/a199}{27AC}
\pdfglyphtounicode{tfm:zd/a2}{2702}
\pdfglyphtounicode{tfm:zd/a20}{2714}
\pdfglyphtounicode{tfm:zd/a200}{27AE}
\pdfglyphtounicode{tfm:zd/a201}{27B1}
\pdfglyphtounicode{tfm:zd/a202}{2703}
\pdfglyphtounicode{tfm:zd/a203}{2750}
\pdfglyphtounicode{tfm:zd/a204}{2752}
\pdfglyphtounicode{tfm:zd/a205}{276E}
\pdfglyphtounicode{tfm:zd/a206}{2770}
\pdfglyphtounicode{tfm:zd/a21}{2715}
\pdfglyphtounicode{tfm:zd/a22}{2716}
\pdfglyphtounicode{tfm:zd/a23}{2717}
\pdfglyphtounicode{tfm:zd/a24}{2718}
\pdfglyphtounicode{tfm:zd/a25}{2719}
\pdfglyphtounicode{tfm:zd/a26}{271A}
\pdfglyphtounicode{tfm:zd/a27}{271B}
\pdfglyphtounicode{tfm:zd/a28}{271C}
\pdfglyphtounicode{tfm:zd/a29}{2722}
\pdfglyphtounicode{tfm:zd/a3}{2704}
\pdfglyphtounicode{tfm:zd/a30}{2723}
\pdfglyphtounicode{tfm:zd/a31}{2724}
\pdfglyphtounicode{tfm:zd/a32}{2725}
\pdfglyphtounicode{tfm:zd/a33}{2726}
\pdfglyphtounicode{tfm:zd/a34}{2727}
\pdfglyphtounicode{tfm:zd/a35}{2605}
\pdfglyphtounicode{tfm:zd/a36}{2729}
\pdfglyphtounicode{tfm:zd/a37}{272A}
\pdfglyphtounicode{tfm:zd/a38}{272B}
\pdfglyphtounicode{tfm:zd/a39}{272C}
\pdfglyphtounicode{tfm:zd/a4}{260E}
\pdfglyphtounicode{tfm:zd/a40}{272D}
\pdfglyphtounicode{tfm:zd/a41}{272E}
\pdfglyphtounicode{tfm:zd/a42}{272F}
\pdfglyphtounicode{tfm:zd/a43}{2730}
\pdfglyphtounicode{tfm:zd/a44}{2731}
\pdfglyphtounicode{tfm:zd/a45}{2732}
\pdfglyphtounicode{tfm:zd/a46}{2733}
\pdfglyphtounicode{tfm:zd/a47}{2734}
\pdfglyphtounicode{tfm:zd/a48}{2735}
\pdfglyphtounicode{tfm:zd/a49}{2736}
\pdfglyphtounicode{tfm:zd/a5}{2706}
\pdfglyphtounicode{tfm:zd/a50}{2737}
\pdfglyphtounicode{tfm:zd/a51}{2738}
\pdfglyphtounicode{tfm:zd/a52}{2739}
\pdfglyphtounicode{tfm:zd/a53}{273A}
\pdfglyphtounicode{tfm:zd/a54}{273B}
\pdfglyphtounicode{tfm:zd/a55}{273C}
\pdfglyphtounicode{tfm:zd/a56}{273D}
\pdfglyphtounicode{tfm:zd/a57}{273E}
\pdfglyphtounicode{tfm:zd/a58}{273F}
\pdfglyphtounicode{tfm:zd/a59}{2740}
\pdfglyphtounicode{tfm:zd/a6}{271D}
\pdfglyphtounicode{tfm:zd/a60}{2741}
\pdfglyphtounicode{tfm:zd/a61}{2742}
\pdfglyphtounicode{tfm:zd/a62}{2743}
\pdfglyphtounicode{tfm:zd/a63}{2744}
\pdfglyphtounicode{tfm:zd/a64}{2745}
\pdfglyphtounicode{tfm:zd/a65}{2746}
\pdfglyphtounicode{tfm:zd/a66}{2747}
\pdfglyphtounicode{tfm:zd/a67}{2748}
\pdfglyphtounicode{tfm:zd/a68}{2749}
\pdfglyphtounicode{tfm:zd/a69}{274A}
\pdfglyphtounicode{tfm:zd/a7}{271E}
\pdfglyphtounicode{tfm:zd/a70}{274B}
\pdfglyphtounicode{tfm:zd/a71}{25CF}
\pdfglyphtounicode{tfm:zd/a72}{274D}
\pdfglyphtounicode{tfm:zd/a73}{25A0}
\pdfglyphtounicode{tfm:zd/a74}{274F}
\pdfglyphtounicode{tfm:zd/a75}{2751}
\pdfglyphtounicode{tfm:zd/a76}{25B2}
\pdfglyphtounicode{tfm:zd/a77}{25BC}
\pdfglyphtounicode{tfm:zd/a78}{25C6}
\pdfglyphtounicode{tfm:zd/a79}{2756}
\pdfglyphtounicode{tfm:zd/a8}{271F}
\pdfglyphtounicode{tfm:zd/a81}{25D7}
\pdfglyphtounicode{tfm:zd/a82}{2758}
\pdfglyphtounicode{tfm:zd/a83}{2759}
\pdfglyphtounicode{tfm:zd/a84}{275A}
\pdfglyphtounicode{tfm:zd/a85}{276F}
\pdfglyphtounicode{tfm:zd/a86}{2771}
\pdfglyphtounicode{tfm:zd/a87}{2772}
\pdfglyphtounicode{tfm:zd/a88}{2773}
\pdfglyphtounicode{tfm:zd/a89}{2768}
\pdfglyphtounicode{tfm:zd/a9}{2720}
\pdfglyphtounicode{tfm:zd/a90}{2769}
\pdfglyphtounicode{tfm:zd/a91}{276C}
\pdfglyphtounicode{tfm:zd/a92}{276D}
\pdfglyphtounicode{tfm:zd/a93}{276A}
\pdfglyphtounicode{tfm:zd/a94}{276B}
\pdfglyphtounicode{tfm:zd/a95}{2774}
\pdfglyphtounicode{tfm:zd/a96}{2775}
\pdfglyphtounicode{tfm:zd/a97}{275B}
\pdfglyphtounicode{tfm:zd/a98}{275C}
\pdfglyphtounicode{tfm:zd/a99}{275D}
\pdfglyphtounicode{tfm:zpzdr-reversed/a1}{2701}
\pdfglyphtounicode{tfm:zpzdr-reversed/a10}{2721}
\pdfglyphtounicode{tfm:zpzdr-reversed/a100}{275E}
\pdfglyphtounicode{tfm:zpzdr-reversed/a101}{2761}
\pdfglyphtounicode{tfm:zpzdr-reversed/a102}{2762}
\pdfglyphtounicode{tfm:zpzdr-reversed/a103}{2763}
\pdfglyphtounicode{tfm:zpzdr-reversed/a104}{2764}
\pdfglyphtounicode{tfm:zpzdr-reversed/a105}{2710}
\pdfglyphtounicode{tfm:zpzdr-reversed/a106}{2765}
\pdfglyphtounicode{tfm:zpzdr-reversed/a107}{2766}
\pdfglyphtounicode{tfm:zpzdr-reversed/a108}{2767}
\pdfglyphtounicode{tfm:zpzdr-reversed/a109}{2660}
\pdfglyphtounicode{tfm:zpzdr-reversed/a11}{261B}
\pdfglyphtounicode{tfm:zpzdr-reversed/a110}{2665}
\pdfglyphtounicode{tfm:zpzdr-reversed/a111}{2666}
\pdfglyphtounicode{tfm:zpzdr-reversed/a112}{2663}
\pdfglyphtounicode{tfm:zpzdr-reversed/a117}{2709}
\pdfglyphtounicode{tfm:zpzdr-reversed/a118}{2708}
\pdfglyphtounicode{tfm:zpzdr-reversed/a119}{2707}
\pdfglyphtounicode{tfm:zpzdr-reversed/a12}{261E}
\pdfglyphtounicode{tfm:zpzdr-reversed/a120}{2460}
\pdfglyphtounicode{tfm:zpzdr-reversed/a121}{2461}
\pdfglyphtounicode{tfm:zpzdr-reversed/a122}{2462}
\pdfglyphtounicode{tfm:zpzdr-reversed/a123}{2463}
\pdfglyphtounicode{tfm:zpzdr-reversed/a124}{2464}
\pdfglyphtounicode{tfm:zpzdr-reversed/a125}{2465}
\pdfglyphtounicode{tfm:zpzdr-reversed/a126}{2466}
\pdfglyphtounicode{tfm:zpzdr-reversed/a127}{2467}
\pdfglyphtounicode{tfm:zpzdr-reversed/a128}{2468}
\pdfglyphtounicode{tfm:zpzdr-reversed/a129}{2469}
\pdfglyphtounicode{tfm:zpzdr-reversed/a13}{270C}
\pdfglyphtounicode{tfm:zpzdr-reversed/a130}{2776}
\pdfglyphtounicode{tfm:zpzdr-reversed/a131}{2777}
\pdfglyphtounicode{tfm:zpzdr-reversed/a132}{2778}
\pdfglyphtounicode{tfm:zpzdr-reversed/a133}{2779}
\pdfglyphtounicode{tfm:zpzdr-reversed/a134}{277A}
\pdfglyphtounicode{tfm:zpzdr-reversed/a135}{277B}
\pdfglyphtounicode{tfm:zpzdr-reversed/a136}{277C}
\pdfglyphtounicode{tfm:zpzdr-reversed/a137}{277D}
\pdfglyphtounicode{tfm:zpzdr-reversed/a138}{277E}
\pdfglyphtounicode{tfm:zpzdr-reversed/a139}{277F}
\pdfglyphtounicode{tfm:zpzdr-reversed/a14}{270D}
\pdfglyphtounicode{tfm:zpzdr-reversed/a140}{2780}
\pdfglyphtounicode{tfm:zpzdr-reversed/a141}{2781}
\pdfglyphtounicode{tfm:zpzdr-reversed/a142}{2782}
\pdfglyphtounicode{tfm:zpzdr-reversed/a143}{2783}
\pdfglyphtounicode{tfm:zpzdr-reversed/a144}{2784}
\pdfglyphtounicode{tfm:zpzdr-reversed/a145}{2785}
\pdfglyphtounicode{tfm:zpzdr-reversed/a146}{2786}
\pdfglyphtounicode{tfm:zpzdr-reversed/a147}{2787}
\pdfglyphtounicode{tfm:zpzdr-reversed/a148}{2788}
\pdfglyphtounicode{tfm:zpzdr-reversed/a149}{2789}
\pdfglyphtounicode{tfm:zpzdr-reversed/a15}{270E}
\pdfglyphtounicode{tfm:zpzdr-reversed/a150}{278A}
\pdfglyphtounicode{tfm:zpzdr-reversed/a151}{278B}
\pdfglyphtounicode{tfm:zpzdr-reversed/a152}{278C}
\pdfglyphtounicode{tfm:zpzdr-reversed/a153}{278D}
\pdfglyphtounicode{tfm:zpzdr-reversed/a154}{278E}
\pdfglyphtounicode{tfm:zpzdr-reversed/a155}{278F}
\pdfglyphtounicode{tfm:zpzdr-reversed/a156}{2790}
\pdfglyphtounicode{tfm:zpzdr-reversed/a157}{2791}
\pdfglyphtounicode{tfm:zpzdr-reversed/a158}{2792}
\pdfglyphtounicode{tfm:zpzdr-reversed/a159}{2793}
\pdfglyphtounicode{tfm:zpzdr-reversed/a16}{270F}
\pdfglyphtounicode{tfm:zpzdr-reversed/a160}{2794}
\pdfglyphtounicode{tfm:zpzdr-reversed/a161}{2192}
\pdfglyphtounicode{tfm:zpzdr-reversed/a162}{27A3}
\pdfglyphtounicode{tfm:zpzdr-reversed/a163}{2194}
\pdfglyphtounicode{tfm:zpzdr-reversed/a164}{2195}
\pdfglyphtounicode{tfm:zpzdr-reversed/a165}{2799}
\pdfglyphtounicode{tfm:zpzdr-reversed/a166}{279B}
\pdfglyphtounicode{tfm:zpzdr-reversed/a167}{279C}
\pdfglyphtounicode{tfm:zpzdr-reversed/a168}{279D}
\pdfglyphtounicode{tfm:zpzdr-reversed/a169}{279E}
\pdfglyphtounicode{tfm:zpzdr-reversed/a17}{2711}
\pdfglyphtounicode{tfm:zpzdr-reversed/a170}{279F}
\pdfglyphtounicode{tfm:zpzdr-reversed/a171}{27A0}
\pdfglyphtounicode{tfm:zpzdr-reversed/a172}{27A1}
\pdfglyphtounicode{tfm:zpzdr-reversed/a173}{27A2}
\pdfglyphtounicode{tfm:zpzdr-reversed/a174}{27A4}
\pdfglyphtounicode{tfm:zpzdr-reversed/a175}{27A5}
\pdfglyphtounicode{tfm:zpzdr-reversed/a176}{27A6}
\pdfglyphtounicode{tfm:zpzdr-reversed/a177}{27A7}
\pdfglyphtounicode{tfm:zpzdr-reversed/a178}{27A8}
\pdfglyphtounicode{tfm:zpzdr-reversed/a179}{27A9}
\pdfglyphtounicode{tfm:zpzdr-reversed/a18}{2712}
\pdfglyphtounicode{tfm:zpzdr-reversed/a180}{27AB}
\pdfglyphtounicode{tfm:zpzdr-reversed/a181}{27AD}
\pdfglyphtounicode{tfm:zpzdr-reversed/a182}{27AF}
\pdfglyphtounicode{tfm:zpzdr-reversed/a183}{27B2}
\pdfglyphtounicode{tfm:zpzdr-reversed/a184}{27B3}
\pdfglyphtounicode{tfm:zpzdr-reversed/a185}{27B5}
\pdfglyphtounicode{tfm:zpzdr-reversed/a186}{27B8}
\pdfglyphtounicode{tfm:zpzdr-reversed/a187}{27BA}
\pdfglyphtounicode{tfm:zpzdr-reversed/a188}{27BB}
\pdfglyphtounicode{tfm:zpzdr-reversed/a189}{27BC}
\pdfglyphtounicode{tfm:zpzdr-reversed/a19}{2713}
\pdfglyphtounicode{tfm:zpzdr-reversed/a190}{27BD}
\pdfglyphtounicode{tfm:zpzdr-reversed/a191}{27BE}
\pdfglyphtounicode{tfm:zpzdr-reversed/a192}{279A}
\pdfglyphtounicode{tfm:zpzdr-reversed/a193}{27AA}
\pdfglyphtounicode{tfm:zpzdr-reversed/a194}{27B6}
\pdfglyphtounicode{tfm:zpzdr-reversed/a195}{27B9}
\pdfglyphtounicode{tfm:zpzdr-reversed/a196}{2798}
\pdfglyphtounicode{tfm:zpzdr-reversed/a197}{27B4}
\pdfglyphtounicode{tfm:zpzdr-reversed/a198}{27B7}
\pdfglyphtounicode{tfm:zpzdr-reversed/a199}{27AC}
\pdfglyphtounicode{tfm:zpzdr-reversed/a2}{2702}
\pdfglyphtounicode{tfm:zpzdr-reversed/a20}{2714}
\pdfglyphtounicode{tfm:zpzdr-reversed/a200}{27AE}
\pdfglyphtounicode{tfm:zpzdr-reversed/a201}{27B1}
\pdfglyphtounicode{tfm:zpzdr-reversed/a202}{2703}
\pdfglyphtounicode{tfm:zpzdr-reversed/a203}{2750}
\pdfglyphtounicode{tfm:zpzdr-reversed/a204}{2752}
\pdfglyphtounicode{tfm:zpzdr-reversed/a205}{276E}
\pdfglyphtounicode{tfm:zpzdr-reversed/a206}{2770}
\pdfglyphtounicode{tfm:zpzdr-reversed/a21}{2715}
\pdfglyphtounicode{tfm:zpzdr-reversed/a22}{2716}
\pdfglyphtounicode{tfm:zpzdr-reversed/a23}{2717}
\pdfglyphtounicode{tfm:zpzdr-reversed/a24}{2718}
\pdfglyphtounicode{tfm:zpzdr-reversed/a25}{2719}
\pdfglyphtounicode{tfm:zpzdr-reversed/a26}{271A}
\pdfglyphtounicode{tfm:zpzdr-reversed/a27}{271B}
\pdfglyphtounicode{tfm:zpzdr-reversed/a28}{271C}
\pdfglyphtounicode{tfm:zpzdr-reversed/a29}{2722}
\pdfglyphtounicode{tfm:zpzdr-reversed/a3}{2704}
\pdfglyphtounicode{tfm:zpzdr-reversed/a30}{2723}
\pdfglyphtounicode{tfm:zpzdr-reversed/a31}{2724}
\pdfglyphtounicode{tfm:zpzdr-reversed/a32}{2725}
\pdfglyphtounicode{tfm:zpzdr-reversed/a33}{2726}
\pdfglyphtounicode{tfm:zpzdr-reversed/a34}{2727}
\pdfglyphtounicode{tfm:zpzdr-reversed/a35}{2605}
\pdfglyphtounicode{tfm:zpzdr-reversed/a36}{2729}
\pdfglyphtounicode{tfm:zpzdr-reversed/a37}{272A}
\pdfglyphtounicode{tfm:zpzdr-reversed/a38}{272B}
\pdfglyphtounicode{tfm:zpzdr-reversed/a39}{272C}
\pdfglyphtounicode{tfm:zpzdr-reversed/a4}{260E}
\pdfglyphtounicode{tfm:zpzdr-reversed/a40}{272D}
\pdfglyphtounicode{tfm:zpzdr-reversed/a41}{272E}
\pdfglyphtounicode{tfm:zpzdr-reversed/a42}{272F}
\pdfglyphtounicode{tfm:zpzdr-reversed/a43}{2730}
\pdfglyphtounicode{tfm:zpzdr-reversed/a44}{2731}
\pdfglyphtounicode{tfm:zpzdr-reversed/a45}{2732}
\pdfglyphtounicode{tfm:zpzdr-reversed/a46}{2733}
\pdfglyphtounicode{tfm:zpzdr-reversed/a47}{2734}
\pdfglyphtounicode{tfm:zpzdr-reversed/a48}{2735}
\pdfglyphtounicode{tfm:zpzdr-reversed/a49}{2736}
\pdfglyphtounicode{tfm:zpzdr-reversed/a5}{2706}
\pdfglyphtounicode{tfm:zpzdr-reversed/a50}{2737}
\pdfglyphtounicode{tfm:zpzdr-reversed/a51}{2738}
\pdfglyphtounicode{tfm:zpzdr-reversed/a52}{2739}
\pdfglyphtounicode{tfm:zpzdr-reversed/a53}{273A}
\pdfglyphtounicode{tfm:zpzdr-reversed/a54}{273B}
\pdfglyphtounicode{tfm:zpzdr-reversed/a55}{273C}
\pdfglyphtounicode{tfm:zpzdr-reversed/a56}{273D}
\pdfglyphtounicode{tfm:zpzdr-reversed/a57}{273E}
\pdfglyphtounicode{tfm:zpzdr-reversed/a58}{273F}
\pdfglyphtounicode{tfm:zpzdr-reversed/a59}{2740}
\pdfglyphtounicode{tfm:zpzdr-reversed/a6}{271D}
\pdfglyphtounicode{tfm:zpzdr-reversed/a60}{2741}
\pdfglyphtounicode{tfm:zpzdr-reversed/a61}{2742}
\pdfglyphtounicode{tfm:zpzdr-reversed/a62}{2743}
\pdfglyphtounicode{tfm:zpzdr-reversed/a63}{2744}
\pdfglyphtounicode{tfm:zpzdr-reversed/a64}{2745}
\pdfglyphtounicode{tfm:zpzdr-reversed/a65}{2746}
\pdfglyphtounicode{tfm:zpzdr-reversed/a66}{2747}
\pdfglyphtounicode{tfm:zpzdr-reversed/a67}{2748}
\pdfglyphtounicode{tfm:zpzdr-reversed/a68}{2749}
\pdfglyphtounicode{tfm:zpzdr-reversed/a69}{274A}
\pdfglyphtounicode{tfm:zpzdr-reversed/a7}{271E}
\pdfglyphtounicode{tfm:zpzdr-reversed/a70}{274B}
\pdfglyphtounicode{tfm:zpzdr-reversed/a71}{25CF}
\pdfglyphtounicode{tfm:zpzdr-reversed/a72}{274D}
\pdfglyphtounicode{tfm:zpzdr-reversed/a73}{25A0}
\pdfglyphtounicode{tfm:zpzdr-reversed/a74}{274F}
\pdfglyphtounicode{tfm:zpzdr-reversed/a75}{2751}
\pdfglyphtounicode{tfm:zpzdr-reversed/a76}{25B2}
\pdfglyphtounicode{tfm:zpzdr-reversed/a77}{25BC}
\pdfglyphtounicode{tfm:zpzdr-reversed/a78}{25C6}
\pdfglyphtounicode{tfm:zpzdr-reversed/a79}{2756}
\pdfglyphtounicode{tfm:zpzdr-reversed/a8}{271F}
\pdfglyphtounicode{tfm:zpzdr-reversed/a81}{25D7}
\pdfglyphtounicode{tfm:zpzdr-reversed/a82}{2758}
\pdfglyphtounicode{tfm:zpzdr-reversed/a83}{2759}
\pdfglyphtounicode{tfm:zpzdr-reversed/a84}{275A}
\pdfglyphtounicode{tfm:zpzdr-reversed/a85}{276F}
\pdfglyphtounicode{tfm:zpzdr-reversed/a86}{2771}
\pdfglyphtounicode{tfm:zpzdr-reversed/a87}{2772}
\pdfglyphtounicode{tfm:zpzdr-reversed/a88}{2773}
\pdfglyphtounicode{tfm:zpzdr-reversed/a89}{2768}
\pdfglyphtounicode{tfm:zpzdr-reversed/a9}{2720}
\pdfglyphtounicode{tfm:zpzdr-reversed/a90}{2769}
\pdfglyphtounicode{tfm:zpzdr-reversed/a91}{276C}
\pdfglyphtounicode{tfm:zpzdr-reversed/a92}{276D}
\pdfglyphtounicode{tfm:zpzdr-reversed/a93}{276A}
\pdfglyphtounicode{tfm:zpzdr-reversed/a94}{276B}
\pdfglyphtounicode{tfm:zpzdr-reversed/a95}{2774}
\pdfglyphtounicode{tfm:zpzdr-reversed/a96}{2775}
\pdfglyphtounicode{tfm:zpzdr-reversed/a97}{275B}
\pdfglyphtounicode{tfm:zpzdr-reversed/a98}{275C}
\pdfglyphtounicode{tfm:zpzdr-reversed/a99}{275D}
\pdfglyphtounicode{thabengali}{09A5}
\pdfglyphtounicode{thadeva}{0925}
\pdfglyphtounicode{thagujarati}{0AA5}
\pdfglyphtounicode{thagurmukhi}{0A25}
\pdfglyphtounicode{thalarabic}{0630}
\pdfglyphtounicode{thalfinalarabic}{FEAC}
\pdfglyphtounicode{thanthakhatlowleftthai}{F898}
\pdfglyphtounicode{thanthakhatlowrightthai}{F897}
\pdfglyphtounicode{thanthakhatthai}{0E4C}
\pdfglyphtounicode{thanthakhatupperleftthai}{F896}
\pdfglyphtounicode{theharabic}{062B}
\pdfglyphtounicode{thehfinalarabic}{FE9A}
\pdfglyphtounicode{thehinitialarabic}{FE9B}
\pdfglyphtounicode{thehmedialarabic}{FE9C}
\pdfglyphtounicode{thereexists}{2203}
\pdfglyphtounicode{therefore}{2234}
\pdfglyphtounicode{theta}{03B8}
\pdfglyphtounicode{theta1}{03D1}
\pdfglyphtounicode{thetasymbolgreek}{03D1}
\pdfglyphtounicode{thieuthacirclekorean}{3279}
\pdfglyphtounicode{thieuthaparenkorean}{3219}
\pdfglyphtounicode{thieuthcirclekorean}{326B}
\pdfglyphtounicode{thieuthkorean}{314C}
\pdfglyphtounicode{thieuthparenkorean}{320B}
\pdfglyphtounicode{thirteencircle}{246C}
\pdfglyphtounicode{thirteenparen}{2480}
\pdfglyphtounicode{thirteenperiod}{2494}
\pdfglyphtounicode{thonangmonthothai}{0E11}
\pdfglyphtounicode{thook}{01AD}
\pdfglyphtounicode{thophuthaothai}{0E12}
\pdfglyphtounicode{thorn}{00FE}
\pdfglyphtounicode{thothahanthai}{0E17}
\pdfglyphtounicode{thothanthai}{0E10}
\pdfglyphtounicode{thothongthai}{0E18}
\pdfglyphtounicode{thothungthai}{0E16}
\pdfglyphtounicode{thousandcyrillic}{0482}
\pdfglyphtounicode{thousandsseparatorarabic}{066C}
\pdfglyphtounicode{thousandsseparatorpersian}{066C}
\pdfglyphtounicode{three}{0033}
\pdfglyphtounicode{threearabic}{0663}
\pdfglyphtounicode{threebengali}{09E9}
\pdfglyphtounicode{threecircle}{2462}
\pdfglyphtounicode{threecircleinversesansserif}{278C}
\pdfglyphtounicode{threedeva}{0969}
\pdfglyphtounicode{threeeighths}{215C}
\pdfglyphtounicode{threegujarati}{0AE9}
\pdfglyphtounicode{threegurmukhi}{0A69}
\pdfglyphtounicode{threehackarabic}{0663}
\pdfglyphtounicode{threehangzhou}{3023}
\pdfglyphtounicode{threeideographicparen}{3222}
\pdfglyphtounicode{threeinferior}{2083}
\pdfglyphtounicode{threemonospace}{FF13}
\pdfglyphtounicode{threenumeratorbengali}{09F6}
\pdfglyphtounicode{threeoldstyle}{0033}
\pdfglyphtounicode{threeparen}{2476}
\pdfglyphtounicode{threeperiod}{248A}
\pdfglyphtounicode{threepersian}{06F3}
\pdfglyphtounicode{threequarters}{00BE}
\pdfglyphtounicode{threequartersemdash}{F6DE}
\pdfglyphtounicode{threeroman}{2172}
\pdfglyphtounicode{threesuperior}{00B3}
\pdfglyphtounicode{threethai}{0E53}
\pdfglyphtounicode{thzsquare}{3394}
\pdfglyphtounicode{tihiragana}{3061}
\pdfglyphtounicode{tikatakana}{30C1}
\pdfglyphtounicode{tikatakanahalfwidth}{FF81}
\pdfglyphtounicode{tikeutacirclekorean}{3270}
\pdfglyphtounicode{tikeutaparenkorean}{3210}
\pdfglyphtounicode{tikeutcirclekorean}{3262}
\pdfglyphtounicode{tikeutkorean}{3137}
\pdfglyphtounicode{tikeutparenkorean}{3202}
\pdfglyphtounicode{tilde}{02DC}
\pdfglyphtounicode{tildebelowcmb}{0330}
\pdfglyphtounicode{tildecmb}{0303}
\pdfglyphtounicode{tildecomb}{0303}
\pdfglyphtounicode{tildedoublecmb}{0360}
\pdfglyphtounicode{tildeoperator}{223C}
\pdfglyphtounicode{tildeoverlaycmb}{0334}
\pdfglyphtounicode{tildeverticalcmb}{033E}
\pdfglyphtounicode{timescircle}{2297}
\pdfglyphtounicode{tipehahebrew}{0596}
\pdfglyphtounicode{tipehalefthebrew}{0596}
\pdfglyphtounicode{tippigurmukhi}{0A70}
\pdfglyphtounicode{titlocyrilliccmb}{0483}
\pdfglyphtounicode{tiwnarmenian}{057F}
\pdfglyphtounicode{tlinebelow}{1E6F}
\pdfglyphtounicode{tmonospace}{FF54}
\pdfglyphtounicode{toarmenian}{0569}
\pdfglyphtounicode{tohiragana}{3068}
\pdfglyphtounicode{tokatakana}{30C8}
\pdfglyphtounicode{tokatakanahalfwidth}{FF84}
\pdfglyphtounicode{tonebarextrahighmod}{02E5}
\pdfglyphtounicode{tonebarextralowmod}{02E9}
\pdfglyphtounicode{tonebarhighmod}{02E6}
\pdfglyphtounicode{tonebarlowmod}{02E8}
\pdfglyphtounicode{tonebarmidmod}{02E7}
\pdfglyphtounicode{tonefive}{01BD}
\pdfglyphtounicode{tonesix}{0185}
\pdfglyphtounicode{tonetwo}{01A8}
\pdfglyphtounicode{tonos}{0384}
\pdfglyphtounicode{tonsquare}{3327}
\pdfglyphtounicode{topatakthai}{0E0F}
\pdfglyphtounicode{tortoiseshellbracketleft}{3014}
\pdfglyphtounicode{tortoiseshellbracketleftsmall}{FE5D}
\pdfglyphtounicode{tortoiseshellbracketleftvertical}{FE39}
\pdfglyphtounicode{tortoiseshellbracketright}{3015}
\pdfglyphtounicode{tortoiseshellbracketrightsmall}{FE5E}
\pdfglyphtounicode{tortoiseshellbracketrightvertical}{FE3A}
\pdfglyphtounicode{totaothai}{0E15}
\pdfglyphtounicode{tpalatalhook}{01AB}
\pdfglyphtounicode{tparen}{24AF}
\pdfglyphtounicode{trademark}{2122}
\pdfglyphtounicode{trademarksans}{2122}
\pdfglyphtounicode{trademarkserif}{2122}
\pdfglyphtounicode{tretroflexhook}{0288}
\pdfglyphtounicode{triagdn}{25BC}
\pdfglyphtounicode{triaglf}{25C4}
\pdfglyphtounicode{triagrt}{25BA}
\pdfglyphtounicode{triagup}{25B2}
\pdfglyphtounicode{triangle}{25B3}
\pdfglyphtounicode{triangledownsld}{25BC}
\pdfglyphtounicode{triangleinv}{25BD}
\pdfglyphtounicode{triangleleft}{25C1}
\pdfglyphtounicode{triangleleftequal}{22B4}
\pdfglyphtounicode{triangleleftsld}{25C0}
\pdfglyphtounicode{triangleright}{25B7}
\pdfglyphtounicode{trianglerightequal}{22B5}
\pdfglyphtounicode{trianglerightsld}{25B6}
\pdfglyphtounicode{trianglesolid}{25B2}
\pdfglyphtounicode{ts}{02A6}
\pdfglyphtounicode{tsadi}{05E6}
\pdfglyphtounicode{tsadidagesh}{FB46}
\pdfglyphtounicode{tsadidageshhebrew}{FB46}
\pdfglyphtounicode{tsadihebrew}{05E6}
\pdfglyphtounicode{tsecyrillic}{0446}
\pdfglyphtounicode{tsere}{05B5}
\pdfglyphtounicode{tsere12}{05B5}
\pdfglyphtounicode{tsere1e}{05B5}
\pdfglyphtounicode{tsere2b}{05B5}
\pdfglyphtounicode{tserehebrew}{05B5}
\pdfglyphtounicode{tserenarrowhebrew}{05B5}
\pdfglyphtounicode{tserequarterhebrew}{05B5}
\pdfglyphtounicode{tserewidehebrew}{05B5}
\pdfglyphtounicode{tshecyrillic}{045B}
\pdfglyphtounicode{tsuperior}{0074}
\pdfglyphtounicode{ttabengali}{099F}
\pdfglyphtounicode{ttadeva}{091F}
\pdfglyphtounicode{ttagujarati}{0A9F}
\pdfglyphtounicode{ttagurmukhi}{0A1F}
\pdfglyphtounicode{tteharabic}{0679}
\pdfglyphtounicode{ttehfinalarabic}{FB67}
\pdfglyphtounicode{ttehinitialarabic}{FB68}
\pdfglyphtounicode{ttehmedialarabic}{FB69}
\pdfglyphtounicode{tthabengali}{09A0}
\pdfglyphtounicode{tthadeva}{0920}
\pdfglyphtounicode{tthagujarati}{0AA0}
\pdfglyphtounicode{tthagurmukhi}{0A20}
\pdfglyphtounicode{tturned}{0287}
\pdfglyphtounicode{tuhiragana}{3064}
\pdfglyphtounicode{tukatakana}{30C4}
\pdfglyphtounicode{tukatakanahalfwidth}{FF82}
\pdfglyphtounicode{turnstileleft}{22A2}
\pdfglyphtounicode{turnstileright}{22A3}
\pdfglyphtounicode{tusmallhiragana}{3063}
\pdfglyphtounicode{tusmallkatakana}{30C3}
\pdfglyphtounicode{tusmallkatakanahalfwidth}{FF6F}
\pdfglyphtounicode{twelvecircle}{246B}
\pdfglyphtounicode{twelveparen}{247F}
\pdfglyphtounicode{twelveperiod}{2493}
\pdfglyphtounicode{twelveroman}{217B}
\pdfglyphtounicode{twentycircle}{2473}
\pdfglyphtounicode{twentyhangzhou}{5344}
\pdfglyphtounicode{twentyparen}{2487}
\pdfglyphtounicode{twentyperiod}{249B}
\pdfglyphtounicode{two}{0032}
\pdfglyphtounicode{twoarabic}{0662}
\pdfglyphtounicode{twobengali}{09E8}
\pdfglyphtounicode{twocircle}{2461}
\pdfglyphtounicode{twocircleinversesansserif}{278B}
\pdfglyphtounicode{twodeva}{0968}
\pdfglyphtounicode{twodotenleader}{2025}
\pdfglyphtounicode{twodotleader}{2025}
\pdfglyphtounicode{twodotleadervertical}{FE30}
\pdfglyphtounicode{twogujarati}{0AE8}
\pdfglyphtounicode{twogurmukhi}{0A68}
\pdfglyphtounicode{twohackarabic}{0662}
\pdfglyphtounicode{twohangzhou}{3022}
\pdfglyphtounicode{twoideographicparen}{3221}
\pdfglyphtounicode{twoinferior}{2082}
\pdfglyphtounicode{twomonospace}{FF12}
\pdfglyphtounicode{twonumeratorbengali}{09F5}
\pdfglyphtounicode{twooldstyle}{0032}
\pdfglyphtounicode{twoparen}{2475}
\pdfglyphtounicode{twoperiod}{2489}
\pdfglyphtounicode{twopersian}{06F2}
\pdfglyphtounicode{tworoman}{2171}
\pdfglyphtounicode{twostroke}{01BB}
\pdfglyphtounicode{twosuperior}{00B2}
\pdfglyphtounicode{twothai}{0E52}
\pdfglyphtounicode{twothirds}{2154}
\pdfglyphtounicode{u}{0075}
\pdfglyphtounicode{uacute}{00FA}
\pdfglyphtounicode{ubar}{0289}
\pdfglyphtounicode{ubengali}{0989}
\pdfglyphtounicode{ubopomofo}{3128}
\pdfglyphtounicode{ubreve}{016D}
\pdfglyphtounicode{ucaron}{01D4}
\pdfglyphtounicode{ucircle}{24E4}
\pdfglyphtounicode{ucircumflex}{00FB}
\pdfglyphtounicode{ucircumflexbelow}{1E77}
\pdfglyphtounicode{ucyrillic}{0443}
\pdfglyphtounicode{udattadeva}{0951}
\pdfglyphtounicode{udblacute}{0171}
\pdfglyphtounicode{udblgrave}{0215}
\pdfglyphtounicode{udeva}{0909}
\pdfglyphtounicode{udieresis}{00FC}
\pdfglyphtounicode{udieresisacute}{01D8}
\pdfglyphtounicode{udieresisbelow}{1E73}
\pdfglyphtounicode{udieresiscaron}{01DA}
\pdfglyphtounicode{udieresiscyrillic}{04F1}
\pdfglyphtounicode{udieresisgrave}{01DC}
\pdfglyphtounicode{udieresismacron}{01D6}
\pdfglyphtounicode{udotbelow}{1EE5}
\pdfglyphtounicode{ugrave}{00F9}
\pdfglyphtounicode{ugujarati}{0A89}
\pdfglyphtounicode{ugurmukhi}{0A09}
\pdfglyphtounicode{uhiragana}{3046}
\pdfglyphtounicode{uhookabove}{1EE7}
\pdfglyphtounicode{uhorn}{01B0}
\pdfglyphtounicode{uhornacute}{1EE9}
\pdfglyphtounicode{uhorndotbelow}{1EF1}
\pdfglyphtounicode{uhorngrave}{1EEB}
\pdfglyphtounicode{uhornhookabove}{1EED}
\pdfglyphtounicode{uhorntilde}{1EEF}
\pdfglyphtounicode{uhungarumlaut}{0171}
\pdfglyphtounicode{uhungarumlautcyrillic}{04F3}
\pdfglyphtounicode{uinvertedbreve}{0217}
\pdfglyphtounicode{ukatakana}{30A6}
\pdfglyphtounicode{ukatakanahalfwidth}{FF73}
\pdfglyphtounicode{ukcyrillic}{0479}
\pdfglyphtounicode{ukorean}{315C}
\pdfglyphtounicode{umacron}{016B}
\pdfglyphtounicode{umacroncyrillic}{04EF}
\pdfglyphtounicode{umacrondieresis}{1E7B}
\pdfglyphtounicode{umatragurmukhi}{0A41}
\pdfglyphtounicode{umonospace}{FF55}
\pdfglyphtounicode{underscore}{005F}
\pdfglyphtounicode{underscoredbl}{2017}
\pdfglyphtounicode{underscoremonospace}{FF3F}
\pdfglyphtounicode{underscorevertical}{FE33}
\pdfglyphtounicode{underscorewavy}{FE4F}
\pdfglyphtounicode{union}{222A}
\pdfglyphtounicode{uniondbl}{22D3}
\pdfglyphtounicode{unionmulti}{228E}
\pdfglyphtounicode{unionsq}{2294}
\pdfglyphtounicode{universal}{2200}
\pdfglyphtounicode{uogonek}{0173}
\pdfglyphtounicode{uparen}{24B0}
\pdfglyphtounicode{upblock}{2580}
\pdfglyphtounicode{upperdothebrew}{05C4}
\pdfglyphtounicode{uprise}{22CF}
\pdfglyphtounicode{upsilon}{03C5}
\pdfglyphtounicode{upsilondieresis}{03CB}
\pdfglyphtounicode{upsilondieresistonos}{03B0}
\pdfglyphtounicode{upsilonlatin}{028A}
\pdfglyphtounicode{upsilontonos}{03CD}
\pdfglyphtounicode{upslope}{29F8}
\pdfglyphtounicode{uptackbelowcmb}{031D}
\pdfglyphtounicode{uptackmod}{02D4}
\pdfglyphtounicode{uragurmukhi}{0A73}
\pdfglyphtounicode{uring}{016F}
\pdfglyphtounicode{ushortcyrillic}{045E}
\pdfglyphtounicode{usmallhiragana}{3045}
\pdfglyphtounicode{usmallkatakana}{30A5}
\pdfglyphtounicode{usmallkatakanahalfwidth}{FF69}
\pdfglyphtounicode{ustraightcyrillic}{04AF}
\pdfglyphtounicode{ustraightstrokecyrillic}{04B1}
\pdfglyphtounicode{utilde}{0169}
\pdfglyphtounicode{utildeacute}{1E79}
\pdfglyphtounicode{utildebelow}{1E75}
\pdfglyphtounicode{uubengali}{098A}
\pdfglyphtounicode{uudeva}{090A}
\pdfglyphtounicode{uugujarati}{0A8A}
\pdfglyphtounicode{uugurmukhi}{0A0A}
\pdfglyphtounicode{uumatragurmukhi}{0A42}
\pdfglyphtounicode{uuvowelsignbengali}{09C2}
\pdfglyphtounicode{uuvowelsigndeva}{0942}
\pdfglyphtounicode{uuvowelsigngujarati}{0AC2}
\pdfglyphtounicode{uvowelsignbengali}{09C1}
\pdfglyphtounicode{uvowelsigndeva}{0941}
\pdfglyphtounicode{uvowelsigngujarati}{0AC1}
\pdfglyphtounicode{v}{0076}
\pdfglyphtounicode{vadeva}{0935}
\pdfglyphtounicode{vagujarati}{0AB5}
\pdfglyphtounicode{vagurmukhi}{0A35}
\pdfglyphtounicode{vakatakana}{30F7}
\pdfglyphtounicode{vav}{05D5}
\pdfglyphtounicode{vavdagesh}{FB35}
\pdfglyphtounicode{vavdagesh65}{FB35}
\pdfglyphtounicode{vavdageshhebrew}{FB35}
\pdfglyphtounicode{vavhebrew}{05D5}
\pdfglyphtounicode{vavholam}{FB4B}
\pdfglyphtounicode{vavholamhebrew}{FB4B}
\pdfglyphtounicode{vavvavhebrew}{05F0}
\pdfglyphtounicode{vavyodhebrew}{05F1}
\pdfglyphtounicode{vcircle}{24E5}
\pdfglyphtounicode{vdotbelow}{1E7F}
\pdfglyphtounicode{vector}{20D7}
\pdfglyphtounicode{vecyrillic}{0432}
\pdfglyphtounicode{veharabic}{06A4}
\pdfglyphtounicode{vehfinalarabic}{FB6B}
\pdfglyphtounicode{vehinitialarabic}{FB6C}
\pdfglyphtounicode{vehmedialarabic}{FB6D}
\pdfglyphtounicode{vekatakana}{30F9}
\pdfglyphtounicode{venus}{2640}
\pdfglyphtounicode{verticalbar}{007C}
\pdfglyphtounicode{verticallineabovecmb}{030D}
\pdfglyphtounicode{verticallinebelowcmb}{0329}
\pdfglyphtounicode{verticallinelowmod}{02CC}
\pdfglyphtounicode{verticallinemod}{02C8}
\pdfglyphtounicode{vewarmenian}{057E}
\pdfglyphtounicode{vhook}{028B}
\pdfglyphtounicode{vikatakana}{30F8}
\pdfglyphtounicode{viramabengali}{09CD}
\pdfglyphtounicode{viramadeva}{094D}
\pdfglyphtounicode{viramagujarati}{0ACD}
\pdfglyphtounicode{visargabengali}{0983}
\pdfglyphtounicode{visargadeva}{0903}
\pdfglyphtounicode{visargagujarati}{0A83}
\pdfglyphtounicode{visiblespace}{2423}
\pdfglyphtounicode{visualspace}{2423}
\pdfglyphtounicode{vmonospace}{FF56}
\pdfglyphtounicode{voarmenian}{0578}
\pdfglyphtounicode{voicediterationhiragana}{309E}
\pdfglyphtounicode{voicediterationkatakana}{30FE}
\pdfglyphtounicode{voicedmarkkana}{309B}
\pdfglyphtounicode{voicedmarkkanahalfwidth}{FF9E}
\pdfglyphtounicode{vokatakana}{30FA}
\pdfglyphtounicode{vparen}{24B1}
\pdfglyphtounicode{vtilde}{1E7D}
\pdfglyphtounicode{vturned}{028C}
\pdfglyphtounicode{vuhiragana}{3094}
\pdfglyphtounicode{vukatakana}{30F4}
\pdfglyphtounicode{w}{0077}
\pdfglyphtounicode{wacute}{1E83}
\pdfglyphtounicode{waekorean}{3159}
\pdfglyphtounicode{wahiragana}{308F}
\pdfglyphtounicode{wakatakana}{30EF}
\pdfglyphtounicode{wakatakanahalfwidth}{FF9C}
\pdfglyphtounicode{wakorean}{3158}
\pdfglyphtounicode{wasmallhiragana}{308E}
\pdfglyphtounicode{wasmallkatakana}{30EE}
\pdfglyphtounicode{wattosquare}{3357}
\pdfglyphtounicode{wavedash}{301C}
\pdfglyphtounicode{wavyunderscorevertical}{FE34}
\pdfglyphtounicode{wawarabic}{0648}
\pdfglyphtounicode{wawfinalarabic}{FEEE}
\pdfglyphtounicode{wawhamzaabovearabic}{0624}
\pdfglyphtounicode{wawhamzaabovefinalarabic}{FE86}
\pdfglyphtounicode{wbsquare}{33DD}
\pdfglyphtounicode{wcircle}{24E6}
\pdfglyphtounicode{wcircumflex}{0175}
\pdfglyphtounicode{wdieresis}{1E85}
\pdfglyphtounicode{wdotaccent}{1E87}
\pdfglyphtounicode{wdotbelow}{1E89}
\pdfglyphtounicode{wehiragana}{3091}
\pdfglyphtounicode{weierstrass}{2118}
\pdfglyphtounicode{wekatakana}{30F1}
\pdfglyphtounicode{wekorean}{315E}
\pdfglyphtounicode{weokorean}{315D}
\pdfglyphtounicode{wgrave}{1E81}
\pdfglyphtounicode{whitebullet}{25E6}
\pdfglyphtounicode{whitecircle}{25CB}
\pdfglyphtounicode{whitecircleinverse}{25D9}
\pdfglyphtounicode{whitecornerbracketleft}{300E}
\pdfglyphtounicode{whitecornerbracketleftvertical}{FE43}
\pdfglyphtounicode{whitecornerbracketright}{300F}
\pdfglyphtounicode{whitecornerbracketrightvertical}{FE44}
\pdfglyphtounicode{whitediamond}{25C7}
\pdfglyphtounicode{whitediamondcontainingblacksmalldiamond}{25C8}
\pdfglyphtounicode{whitedownpointingsmalltriangle}{25BF}
\pdfglyphtounicode{whitedownpointingtriangle}{25BD}
\pdfglyphtounicode{whiteleftpointingsmalltriangle}{25C3}
\pdfglyphtounicode{whiteleftpointingtriangle}{25C1}
\pdfglyphtounicode{whitelenticularbracketleft}{3016}
\pdfglyphtounicode{whitelenticularbracketright}{3017}
\pdfglyphtounicode{whiterightpointingsmalltriangle}{25B9}
\pdfglyphtounicode{whiterightpointingtriangle}{25B7}
\pdfglyphtounicode{whitesmallsquare}{25AB}
\pdfglyphtounicode{whitesmilingface}{263A}
\pdfglyphtounicode{whitesquare}{25A1}
\pdfglyphtounicode{whitestar}{2606}
\pdfglyphtounicode{whitetelephone}{260F}
\pdfglyphtounicode{whitetortoiseshellbracketleft}{3018}
\pdfglyphtounicode{whitetortoiseshellbracketright}{3019}
\pdfglyphtounicode{whiteuppointingsmalltriangle}{25B5}
\pdfglyphtounicode{whiteuppointingtriangle}{25B3}
\pdfglyphtounicode{wihiragana}{3090}
\pdfglyphtounicode{wikatakana}{30F0}
\pdfglyphtounicode{wikorean}{315F}
\pdfglyphtounicode{wmonospace}{FF57}
\pdfglyphtounicode{wohiragana}{3092}
\pdfglyphtounicode{wokatakana}{30F2}
\pdfglyphtounicode{wokatakanahalfwidth}{FF66}
\pdfglyphtounicode{won}{20A9}
\pdfglyphtounicode{wonmonospace}{FFE6}
\pdfglyphtounicode{wowaenthai}{0E27}
\pdfglyphtounicode{wparen}{24B2}
\pdfglyphtounicode{wreathproduct}{2240}
\pdfglyphtounicode{wring}{1E98}
\pdfglyphtounicode{wsuperior}{02B7}
\pdfglyphtounicode{wturned}{028D}
\pdfglyphtounicode{wynn}{01BF}
\pdfglyphtounicode{x}{0078}
\pdfglyphtounicode{xabovecmb}{033D}
\pdfglyphtounicode{xbopomofo}{3112}
\pdfglyphtounicode{xcircle}{24E7}
\pdfglyphtounicode{xdieresis}{1E8D}
\pdfglyphtounicode{xdotaccent}{1E8B}
\pdfglyphtounicode{xeharmenian}{056D}
\pdfglyphtounicode{xi}{03BE}
\pdfglyphtounicode{xmonospace}{FF58}
\pdfglyphtounicode{xparen}{24B3}
\pdfglyphtounicode{xsuperior}{02E3}
\pdfglyphtounicode{y}{0079}
\pdfglyphtounicode{yaadosquare}{334E}
\pdfglyphtounicode{yabengali}{09AF}
\pdfglyphtounicode{yacute}{00FD}
\pdfglyphtounicode{yadeva}{092F}
\pdfglyphtounicode{yaekorean}{3152}
\pdfglyphtounicode{yagujarati}{0AAF}
\pdfglyphtounicode{yagurmukhi}{0A2F}
\pdfglyphtounicode{yahiragana}{3084}
\pdfglyphtounicode{yakatakana}{30E4}
\pdfglyphtounicode{yakatakanahalfwidth}{FF94}
\pdfglyphtounicode{yakorean}{3151}
\pdfglyphtounicode{yamakkanthai}{0E4E}
\pdfglyphtounicode{yasmallhiragana}{3083}
\pdfglyphtounicode{yasmallkatakana}{30E3}
\pdfglyphtounicode{yasmallkatakanahalfwidth}{FF6C}
\pdfglyphtounicode{yatcyrillic}{0463}
\pdfglyphtounicode{ycircle}{24E8}
\pdfglyphtounicode{ycircumflex}{0177}
\pdfglyphtounicode{ydieresis}{00FF}
\pdfglyphtounicode{ydotaccent}{1E8F}
\pdfglyphtounicode{ydotbelow}{1EF5}
\pdfglyphtounicode{yeharabic}{064A}
\pdfglyphtounicode{yehbarreearabic}{06D2}
\pdfglyphtounicode{yehbarreefinalarabic}{FBAF}
\pdfglyphtounicode{yehfinalarabic}{FEF2}
\pdfglyphtounicode{yehhamzaabovearabic}{0626}
\pdfglyphtounicode{yehhamzaabovefinalarabic}{FE8A}
\pdfglyphtounicode{yehhamzaaboveinitialarabic}{FE8B}
\pdfglyphtounicode{yehhamzaabovemedialarabic}{FE8C}
\pdfglyphtounicode{yehinitialarabic}{FEF3}
\pdfglyphtounicode{yehmedialarabic}{FEF4}
\pdfglyphtounicode{yehmeeminitialarabic}{FCDD}
\pdfglyphtounicode{yehmeemisolatedarabic}{FC58}
\pdfglyphtounicode{yehnoonfinalarabic}{FC94}
\pdfglyphtounicode{yehthreedotsbelowarabic}{06D1}
\pdfglyphtounicode{yekorean}{3156}
\pdfglyphtounicode{yen}{00A5}
\pdfglyphtounicode{yenmonospace}{FFE5}
\pdfglyphtounicode{yeokorean}{3155}
\pdfglyphtounicode{yeorinhieuhkorean}{3186}
\pdfglyphtounicode{yerahbenyomohebrew}{05AA}
\pdfglyphtounicode{yerahbenyomolefthebrew}{05AA}
\pdfglyphtounicode{yericyrillic}{044B}
\pdfglyphtounicode{yerudieresiscyrillic}{04F9}
\pdfglyphtounicode{yesieungkorean}{3181}
\pdfglyphtounicode{yesieungpansioskorean}{3183}
\pdfglyphtounicode{yesieungsioskorean}{3182}
\pdfglyphtounicode{yetivhebrew}{059A}
\pdfglyphtounicode{ygrave}{1EF3}
\pdfglyphtounicode{yhook}{01B4}
\pdfglyphtounicode{yhookabove}{1EF7}
\pdfglyphtounicode{yiarmenian}{0575}
\pdfglyphtounicode{yicyrillic}{0457}
\pdfglyphtounicode{yikorean}{3162}
\pdfglyphtounicode{yinyang}{262F}
\pdfglyphtounicode{yiwnarmenian}{0582}
\pdfglyphtounicode{ymonospace}{FF59}
\pdfglyphtounicode{yod}{05D9}
\pdfglyphtounicode{yoddagesh}{FB39}
\pdfglyphtounicode{yoddageshhebrew}{FB39}
\pdfglyphtounicode{yodhebrew}{05D9}
\pdfglyphtounicode{yodyodhebrew}{05F2}
\pdfglyphtounicode{yodyodpatahhebrew}{FB1F}
\pdfglyphtounicode{yohiragana}{3088}
\pdfglyphtounicode{yoikorean}{3189}
\pdfglyphtounicode{yokatakana}{30E8}
\pdfglyphtounicode{yokatakanahalfwidth}{FF96}
\pdfglyphtounicode{yokorean}{315B}
\pdfglyphtounicode{yosmallhiragana}{3087}
\pdfglyphtounicode{yosmallkatakana}{30E7}
\pdfglyphtounicode{yosmallkatakanahalfwidth}{FF6E}
\pdfglyphtounicode{yotgreek}{03F3}
\pdfglyphtounicode{yoyaekorean}{3188}
\pdfglyphtounicode{yoyakorean}{3187}
\pdfglyphtounicode{yoyakthai}{0E22}
\pdfglyphtounicode{yoyingthai}{0E0D}
\pdfglyphtounicode{yparen}{24B4}
\pdfglyphtounicode{ypogegrammeni}{037A}
\pdfglyphtounicode{ypogegrammenigreekcmb}{0345}
\pdfglyphtounicode{yr}{01A6}
\pdfglyphtounicode{yring}{1E99}
\pdfglyphtounicode{ysuperior}{02B8}
\pdfglyphtounicode{ytilde}{1EF9}
\pdfglyphtounicode{yturned}{028E}
\pdfglyphtounicode{yuhiragana}{3086}
\pdfglyphtounicode{yuikorean}{318C}
\pdfglyphtounicode{yukatakana}{30E6}
\pdfglyphtounicode{yukatakanahalfwidth}{FF95}
\pdfglyphtounicode{yukorean}{3160}
\pdfglyphtounicode{yusbigcyrillic}{046B}
\pdfglyphtounicode{yusbigiotifiedcyrillic}{046D}
\pdfglyphtounicode{yuslittlecyrillic}{0467}
\pdfglyphtounicode{yuslittleiotifiedcyrillic}{0469}
\pdfglyphtounicode{yusmallhiragana}{3085}
\pdfglyphtounicode{yusmallkatakana}{30E5}
\pdfglyphtounicode{yusmallkatakanahalfwidth}{FF6D}
\pdfglyphtounicode{yuyekorean}{318B}
\pdfglyphtounicode{yuyeokorean}{318A}
\pdfglyphtounicode{yyabengali}{09DF}
\pdfglyphtounicode{yyadeva}{095F}
\pdfglyphtounicode{z}{007A}
\pdfglyphtounicode{zaarmenian}{0566}
\pdfglyphtounicode{zacute}{017A}
\pdfglyphtounicode{zadeva}{095B}
\pdfglyphtounicode{zagurmukhi}{0A5B}
\pdfglyphtounicode{zaharabic}{0638}
\pdfglyphtounicode{zahfinalarabic}{FEC6}
\pdfglyphtounicode{zahinitialarabic}{FEC7}
\pdfglyphtounicode{zahiragana}{3056}
\pdfglyphtounicode{zahmedialarabic}{FEC8}
\pdfglyphtounicode{zainarabic}{0632}
\pdfglyphtounicode{zainfinalarabic}{FEB0}
\pdfglyphtounicode{zakatakana}{30B6}
\pdfglyphtounicode{zaqefgadolhebrew}{0595}
\pdfglyphtounicode{zaqefqatanhebrew}{0594}
\pdfglyphtounicode{zarqahebrew}{0598}
\pdfglyphtounicode{zayin}{05D6}
\pdfglyphtounicode{zayindagesh}{FB36}
\pdfglyphtounicode{zayindageshhebrew}{FB36}
\pdfglyphtounicode{zayinhebrew}{05D6}
\pdfglyphtounicode{zbopomofo}{3117}
\pdfglyphtounicode{zcaron}{017E}
\pdfglyphtounicode{zcircle}{24E9}
\pdfglyphtounicode{zcircumflex}{1E91}
\pdfglyphtounicode{zcurl}{0291}
\pdfglyphtounicode{zdot}{017C}
\pdfglyphtounicode{zdotaccent}{017C}
\pdfglyphtounicode{zdotbelow}{1E93}
\pdfglyphtounicode{zecyrillic}{0437}
\pdfglyphtounicode{zedescendercyrillic}{0499}
\pdfglyphtounicode{zedieresiscyrillic}{04DF}
\pdfglyphtounicode{zehiragana}{305C}
\pdfglyphtounicode{zekatakana}{30BC}
\pdfglyphtounicode{zero}{0030}
\pdfglyphtounicode{zeroarabic}{0660}
\pdfglyphtounicode{zerobengali}{09E6}
\pdfglyphtounicode{zerodeva}{0966}
\pdfglyphtounicode{zerogujarati}{0AE6}
\pdfglyphtounicode{zerogurmukhi}{0A66}
\pdfglyphtounicode{zerohackarabic}{0660}
\pdfglyphtounicode{zeroinferior}{2080}
\pdfglyphtounicode{zeromonospace}{FF10}
\pdfglyphtounicode{zerooldstyle}{0030}
\pdfglyphtounicode{zeropersian}{06F0}
\pdfglyphtounicode{zerosuperior}{2070}
\pdfglyphtounicode{zerothai}{0E50}
\pdfglyphtounicode{zerowidthjoiner}{FEFF}
\pdfglyphtounicode{zerowidthnonjoiner}{200C}
\pdfglyphtounicode{zerowidthspace}{200B}
\pdfglyphtounicode{zeta}{03B6}
\pdfglyphtounicode{zhbopomofo}{3113}
\pdfglyphtounicode{zhearmenian}{056A}
\pdfglyphtounicode{zhebrevecyrillic}{04C2}
\pdfglyphtounicode{zhecyrillic}{0436}
\pdfglyphtounicode{zhedescendercyrillic}{0497}
\pdfglyphtounicode{zhedieresiscyrillic}{04DD}
\pdfglyphtounicode{zihiragana}{3058}
\pdfglyphtounicode{zikatakana}{30B8}
\pdfglyphtounicode{zinorhebrew}{05AE}
\pdfglyphtounicode{zlinebelow}{1E95}
\pdfglyphtounicode{zmonospace}{FF5A}
\pdfglyphtounicode{zohiragana}{305E}
\pdfglyphtounicode{zokatakana}{30BE}
\pdfglyphtounicode{zparen}{24B5}
\pdfglyphtounicode{zretroflexhook}{0290}
\pdfglyphtounicode{zstroke}{01B6}
\pdfglyphtounicode{zuhiragana}{305A}
\pdfglyphtounicode{zukatakana}{30BA}
\fi

\usepackage{graphicx}
\graphicspath{{figures/}}
\usepackage{adjustbox}
\usepackage{booktabs}
\usepackage{multirow}
\usepackage{array}
\usepackage{tabularx}
\usepackage{threeparttable}

\usepackage{amsmath}
\usepackage{amssymb}
\usepackage{textcomp}

\usepackage[ruled,vlined,linesnumbered]{algorithm2e}

\usepackage{tikz}
\usetikzlibrary{arrows.meta,positioning,fit,backgrounds,calc,shapes.geometric,decorations.pathreplacing}

\usepackage{url}
\usepackage{xcolor}
\usepackage{comment}
\usepackage{cite}                  %
\usepackage[hidelinks]{hyperref}   %

\newif\ifblind  \blindfalse   %
\newif\ifmarkup \markupfalse

\ifmarkup

  \DeclareRobustCommand{\omesso}[1]{\textcolor{red}{\relax#1}}
  
  \newenvironment{omessoblock}{\par\color{red}}{\par}
\else

  \ifblind
    \DeclareRobustCommand{\omesso}[1]{#1}
    
  \else
    \DeclareRobustCommand{\omesso}[1]{}
    \excludecomment{omessoblock}
  \fi
\fi
\DeclareRobustCommand{\dacompletare}[1]{\textcolor{orange}{\textbf{[TBD: #1]}}}

\newcommand{\fwname}{S1Mal}   %
\newcommand{\nuovo}[1]{#1}
\newcolumntype{L}[1]{>{\raggedright\arraybackslash}p{#1}}
\newcolumntype{C}[1]{>{\centering\arraybackslash}p{#1}}
\providecommand{\keywords}[1]{\par\medskip\noindent\textbf{Keywords:} #1}

\makeatletter
\newenvironment{widefigure}[1][t]%
  {\if@twocolumn\def\wide@env{figure*}\else\def\wide@env{figure}\fi
   \begin{\wide@env}[#1]}%
  {\end{\wide@env}}
\newenvironment{widetable}[1][t]%
  {\if@twocolumn\def\wide@env{table*}\else\def\wide@env{table}\fi
   \begin{\wide@env}[#1]}%
  {\end{\wide@env}}
\makeatother

\definecolor{neuralfill}{HTML}{E8F0FE}  \definecolor{neuralline}{HTML}{1A56A8}
\definecolor{symbfill}{HTML}{FDF0E3}    \definecolor{symbline}{HTML}{A85A1A}
\definecolor{iofill}{HTML}{F0F0F2}      \definecolor{ioline}{HTML}{55555C}
\definecolor{outfill}{HTML}{E7F4EC}     \definecolor{outline}{HTML}{1E7A45}
\definecolor{expertfill}{HTML}{F5E8FD}  \definecolor{expertline}{HTML}{7A1AA8}
\tikzset{
  blk/.style    = {draw, rounded corners=2pt, align=center, font=\footnotesize,
                   minimum height=9mm, minimum width=15mm, inner sep=3pt,
                   line width=0.6pt},
  neural/.style = {blk, fill=neuralfill, draw=neuralline},
  symb/.style   = {blk, fill=symbfill,   draw=symbline},
  io/.style     = {blk, fill=iofill,     draw=ioline},
  result/.style = {blk, fill=outfill,    draw=outline},
  expert/.style = {blk, fill=expertfill, draw=expertline},
  fusion/.style = {blk, left color=neuralfill, right color=symbfill, draw=ioline},
  key/.style    = {minimum width=3.5mm, minimum height=2.5mm, inner sep=0pt},
  flow/.style   = {-{Stealth[length=2mm]}, line width=0.6pt, draw=ioline},
  lbl/.style    = {font=\scriptsize\itshape, text=ioline},
}

\usepackage{orcidlink}
\hypersetup{pdftitle={A First Look at System One Decision Models for Malware Detection}}
\ifblind\else
  \hypersetup{pdfauthor={Pier Mattia Basciano, Giovanni Farina, Salvatore Monteleone}}
\fi

\begin{document}
\makeatletter
\providecommand{\resset}[2]{\expandafter\def\csname res@#1\endcsname{#2}}
\providecommand{\res}[1]{\ifcsname res@#1\endcsname\csname res@#1\endcsname\else\dacompletare{\detokenize{#1}}\fi}
\providecommand{\ifres}[3]{\ifcsname res@#1\endcsname\expandafter\@firstoftwo\else\expandafter\@secondoftwo\fi{#2}{#3}}
\makeatother
\resset{a3.dev.auroc_a2}{0.9965}
\resset{a3.dev.auroc_a3}{0.9966}
\resset{a3.dev.k}{128}
\resset{a3.dev.loss_a2}{0.0786}
\resset{a3.dev.loss_pca}{0.1198}
\resset{e1.A3-C1.abs}{0.0004}
\resset{e1.A3-C1.ci}{[-0.0004, +0.0012]}
\resset{e1.A3-C1.d}{+0.0004}
\resset{e1.A3-C1.fpr_ci}{[+0.02, +0.70]}
\resset{e1.A3-C1.fpr_d}{+0.34}
\resset{e1.A3-C1.fpr_p}{0.006}
\resset{e1.A3-C1.p}{0.370}
\resset{e1.A3-C1.tpr_ci}{[+0.88, +4.97]}
\resset{e1.A3-C1.tpr_d}{+2.57}
\resset{e1.A3-C1.tpr_p}{$<0.001$}
\resset{e1.A3-C2.abs}{0.0184}
\resset{e1.A3-C2.ci}{[+0.0120, +0.0262]}
\resset{e1.A3-C2.d}{+0.0184}
\resset{e1.A3-C2.fpr_ci}{[-0.17, +0.80]}
\resset{e1.A3-C2.fpr_d}{+0.26}
\resset{e1.A3-C2.fpr_p}{0.121}
\resset{e1.A3-C2.p}{0.002}
\resset{e1.A3-C2.tpr_ci}{[+23.64, +49.55]}
\resset{e1.A3-C2.tpr_d}{+35.57}
\resset{e1.A3-C2.tpr_p}{$<0.001$}
\resset{e1.E1-C1.abs}{0.0013}
\resset{e1.E1-C1.ci}{[-0.0023, -0.0005]}
\resset{e1.E1-C1.d}{-0.0013}
\resset{e1.E1-C1.p}{0.007}
\resset{e1.E1-C2.abs}{0.2749}
\resset{e1.E1-C2.ci}{[+0.1989, +0.3511]}
\resset{e1.E1-C2.d}{+0.2749}
\resset{e1.E1-C2.p}{0.007}
\resset{e1.E1-C3.abs}{0.2569}
\resset{e1.E1-C3.ci}{[+0.1790, +0.3365]}
\resset{e1.E1-C3.d}{+0.2569}
\resset{e1.E1-C3.p}{0.007}
\resset{e1.E1-C4.abs}{0.0041}
\resset{e1.E1-C4.ci}{[+0.0012, +0.0078]}
\resset{e1.E1-C4.d}{+0.0041}
\resset{e1.E1-C4.p}{0.010}
\resset{e1.E1-C6.abs}{0.0108}
\resset{e1.E1-C6.ci}{[+0.0063, +0.0177]}
\resset{e1.E1-C6.d}{+0.0108}
\resset{e1.E1-C6.p}{0.007}
\resset{e1.E1-C7.abs}{0.0449}
\resset{e1.E1-C7.ci}{[+0.0314, +0.0638]}
\resset{e1.E1-C7.d}{+0.0449}
\resset{e1.E1-C7.p}{0.007}
\resset{e1.E1-S-d13-a2.abs}{0.0012}
\resset{e1.E1-S-d13-a2.ci}{[+0.0002, +0.0025]}
\resset{e1.E1-S-d13-a2.d}{+0.0012}
\resset{e1.E1-S-d13-a2.p}{0.045}
\resset{e1.E1-S-d13-det.abs}{0.0006}
\resset{e1.E1-S-d13-det.ci}{[-0.0008, +0.0020]}
\resset{e1.E1-S-d13-det.d}{+0.0006}
\resset{e1.E1-S-d13-det.p}{0.756}
\resset{e1.E1-S-open-d1-3b.abs}{0.3890}
\resset{e1.E1-S-open-d1-3b.ci}{[+0.2904, +0.4818]}
\resset{e1.E1-S-open-d1-3b.d}{+0.3890}
\resset{e1.E1-S-open-d1-3b.p}{0.018}
\resset{e1.E1-S-open-d1-omni-600m.abs}{0.5412}
\resset{e1.E1-S-open-d1-omni-600m.ci}{[+0.4232, +0.6507]}
\resset{e1.E1-S-open-d1-omni-600m.d}{+0.5412}
\resset{e1.E1-S-open-d1-omni-600m.p}{0.018}
\resset{e1.E1-S-open-decider-0.8b.abs}{0.4099}
\resset{e1.E1-S-open-decider-0.8b.ci}{[+0.3221, +0.4933]}
\resset{e1.E1-S-open-decider-0.8b.d}{+0.4099}
\resset{e1.E1-S-open-decider-0.8b.p}{0.018}
\resset{e1.E1-S-open-decider-2b.abs}{0.3141}
\resset{e1.E1-S-open-decider-2b.ci}{[+0.2253, +0.4089]}
\resset{e1.E1-S-open-decider-2b.d}{+0.3141}
\resset{e1.E1-S-open-decider-2b.p}{0.018}
\resset{e1.E1-S-open-decider-4b-gguf.abs}{0.3167}
\resset{e1.E1-S-open-decider-4b-gguf.ci}{[+0.2433, +0.3882]}
\resset{e1.E1-S-open-decider-4b-gguf.d}{+0.3167}
\resset{e1.E1-S-open-decider-4b-gguf.p}{0.018}
\resset{e1.E1-S-open-kev-0.8b.abs}{0.2742}
\resset{e1.E1-S-open-kev-0.8b.ci}{[+0.2181, +0.3338]}
\resset{e1.E1-S-open-kev-0.8b.d}{+0.2742}
\resset{e1.E1-S-open-kev-0.8b.p}{0.018}
\resset{e1.E1-S-open-laya-multilingual.abs}{0.2606}
\resset{e1.E1-S-open-laya-multilingual.ci}{[+0.1712, +0.3551]}
\resset{e1.E1-S-open-laya-multilingual.d}{+0.2606}
\resset{e1.E1-S-open-laya-multilingual.p}{0.018}
\resset{e1.E1-S-open-laya-typed-decisions.abs}{0.4842}
\resset{e1.E1-S-open-laya-typed-decisions.ci}{[+0.3656, +0.6062]}
\resset{e1.E1-S-open-laya-typed-decisions.d}{+0.4842}
\resset{e1.E1-S-open-laya-typed-decisions.p}{0.018}
\resset{e1.E1-S-open-strands-decider-2b-v1-2610.abs}{0.2899}
\resset{e1.E1-S-open-strands-decider-2b-v1-2610.ci}{[+0.2113, +0.3722]}
\resset{e1.E1-S-open-strands-decider-2b-v1-2610.d}{+0.2899}
\resset{e1.E1-S-open-strands-decider-2b-v1-2610.p}{0.018}
\resset{e1.E1-S-open-von.abs}{0.2645}
\resset{e1.E1-S-open-von.ci}{[+0.1712, +0.3835]}
\resset{e1.E1-S-open-von.d}{+0.2645}
\resset{e1.E1-S-open-von.p}{0.018}
\resset{e1.E1-S5.abs}{0.0014}
\resset{e1.E1-S5.ci}{[-0.0035, +0.0005]}
\resset{e1.E1-S5.d}{-0.0014}
\resset{e1.E1-S5.p}{0.620}
\resset{e1.E1-S6.abs}{0.0209}
\resset{e1.E1-S6.ci}{[+0.0138, +0.0307]}
\resset{e1.E1-S6.d}{+0.0209}
\resset{e1.E1-S6.p}{0.018}
\resset{e1.a0.max}{0.8590}
\resset{e1.a0.min}{0.4025}
\resset{e1.a1.max}{0.9834}
\resset{e1.a1.min}{0.9126}
\resset{e1.best_open}{decider-4b (GGUF Q4\_K\_M)}
\resset{e1.clm_a0.abstained}{100.00\%}
\resset{e1.clm_a0.aurc_margin}{0.2679}
\resset{e1.clm_a0.auroc}{0.7194}
\resset{e1.clm_a0.balanced_accuracy}{0.5383}
\resset{e1.clm_a0.brier}{0.2139}
\resset{e1.clm_a0.ci}{[0.643, 0.796]}
\resset{e1.clm_a0.coverage_at_risk_1pct_margin}{0.01\%}
\resset{e1.clm_a0.ece}{0.1024}
\resset{e1.clm_a0.fpr}{1.16\%}
\resset{e1.clm_a0.pauc}{0.5258}
\resset{e1.clm_a0.tpr}{8.82\%}
\resset{e1.clm_a1.abstained}{36.10\%}
\resset{e1.clm_a1.aurc_margin}{0.0783}
\resset{e1.clm_a1.auroc}{0.9763}
\resset{e1.clm_a1.balanced_accuracy}{0.7835}
\resset{e1.clm_a1.brier}{0.0526}
\resset{e1.clm_a1.ci}{[0.965, 0.985]}
\resset{e1.clm_a1.coverage_at_risk_1pct_margin}{39.87\%}
\resset{e1.clm_a1.ece}{0.0161}
\resset{e1.clm_a1.fpr}{1.19\%}
\resset{e1.clm_a1.pauc}{0.6713}
\resset{e1.clm_a1.tpr}{57.90\%}
\resset{e1.clm_a2.abstained}{7.64\%}
\resset{e1.clm_a2.aurc_margin}{0.0166}
\resset{e1.clm_a2.auroc}{0.9943}
\resset{e1.clm_a2.balanced_accuracy}{0.9489}
\resset{e1.clm_a2.brier}{0.0262}
\resset{e1.clm_a2.ci}{[0.991, 0.997]}
\resset{e1.clm_a2.coverage_at_risk_1pct_margin}{52.90\%}
\resset{e1.clm_a2.ece}{0.0060}
\resset{e1.clm_a2.fpr}{1.11\%}
\resset{e1.clm_a2.pauc}{0.8760}
\resset{e1.clm_a2.tpr}{90.90\%}
\resset{e1.clm_a2.tpr01}{46.59\%}
\resset{e1.clm_a2.tpr_at1}{90.09\%}
\resset{e1.clm_a2.tpr_at_a3fpr}{92.15\%}
\resset{e1.clm_a2_ctrl_lr.abstained}{5.62\%}
\resset{e1.clm_a2_ctrl_lr.aurc_margin}{0.0134}
\resset{e1.clm_a2_ctrl_lr.auroc}{0.9955}
\resset{e1.clm_a2_ctrl_lr.balanced_accuracy}{0.9592}
\resset{e1.clm_a2_ctrl_lr.brier}{0.0229}
\resset{e1.clm_a2_ctrl_lr.ci}{[0.993, 0.997]}
\resset{e1.clm_a2_ctrl_lr.coverage_at_risk_1pct_margin}{53.77\%}
\resset{e1.clm_a2_ctrl_lr.ece}{0.0104}
\resset{e1.clm_a2_ctrl_lr.fpr}{1.11\%}
\resset{e1.clm_a2_ctrl_lr.pauc}{0.8974}
\resset{e1.clm_a2_ctrl_lr.tpr}{92.94\%}
\resset{e1.clm_a3.abstained}{6.02\%}
\resset{e1.clm_a3.aurc_margin}{0.0124}
\resset{e1.clm_a3.auroc}{0.9946}
\resset{e1.clm_a3.balanced_accuracy}{0.9601}
\resset{e1.clm_a3.brier}{0.0250}
\resset{e1.clm_a3.ci}{[0.991, 0.997]}
\resset{e1.clm_a3.coverage_at_risk_1pct_margin}{52.33\%}
\resset{e1.clm_a3.ece}{0.0078}
\resset{e1.clm_a3.fpr}{1.45\%}
\resset{e1.clm_a3.pauc}{0.8851}
\resset{e1.clm_a3.tpr}{93.47\%}
\resset{e1.clm_a3.tpr01}{59.15\%}
\resset{e1.clm_a3.tpr_at1}{89.94\%}
\resset{e1.det.abstained}{25.19\%}
\resset{e1.det.aurc_margin}{0.0215}
\resset{e1.det.auroc}{0.9901}
\resset{e1.det.balanced_accuracy}{0.9317}
\resset{e1.det.brier}{0.0310}
\resset{e1.det.ci}{[0.985, 0.994]}
\resset{e1.det.coverage_at_risk_1pct_margin}{51.94\%}
\resset{e1.det.ece}{0.0139}
\resset{e1.det.fpr}{1.91\%}
\resset{e1.det.pauc}{0.8284}
\resset{e1.det.tpr}{88.24\%}
\resset{e1.det.tpr01}{44.99\%}
\resset{e1.fpr.max}{1.91\%}
\resset{e1.fpr.min}{0.95\%}
\resset{e1.lift.max}{+0.5412}
\resset{e1.lift.min}{+0.1143}
\resset{e1.lift.n_sig}{11}
\resset{e1.lift.n_tested}{11}
\resset{e1.lift.pmax}{0.018}
\resset{e1.lift_rel.clm}{35.71\%}
\resset{e1.lift_rel.max}{134.44\%}
\resset{e1.lift_rel.mean}{54.61\%}
\resset{e1.lift_rel.min}{13.30\%}
\resset{e1.llm_fsec_a1.abstained}{35.04\%}
\resset{e1.llm_fsec_a1.aurc_margin}{0.0715}
\resset{e1.llm_fsec_a1.auroc}{0.9733}
\resset{e1.llm_fsec_a1.balanced_accuracy}{0.8130}
\resset{e1.llm_fsec_a1.brier}{0.0537}
\resset{e1.llm_fsec_a1.ci}{[0.962, 0.981]}
\resset{e1.llm_fsec_a1.coverage_at_risk_1pct_margin}{33.37\%}
\resset{e1.llm_fsec_a1.ece}{0.0184}
\resset{e1.llm_fsec_a1.fpr}{1.58\%}
\resset{e1.llm_fsec_a1.pauc}{0.6595}
\resset{e1.llm_fsec_a1.tpr}{64.19\%}
\resset{e1.llm_qwen3_a1.abstained}{75.61\%}
\resset{e1.llm_qwen3_a1.aurc_margin}{0.1184}
\resset{e1.llm_qwen3_a1.auroc}{0.9493}
\resset{e1.llm_qwen3_a1.balanced_accuracy}{0.6767}
\resset{e1.llm_qwen3_a1.brier}{0.0809}
\resset{e1.llm_qwen3_a1.ci}{[0.929, 0.964]}
\resset{e1.llm_qwen3_a1.coverage_at_risk_1pct_margin}{10.87\%}
\resset{e1.llm_qwen3_a1.ece}{0.0235}
\resset{e1.llm_qwen3_a1.fpr}{0.95\%}
\resset{e1.llm_qwen3_a1.pauc}{0.6013}
\resset{e1.llm_qwen3_a1.tpr}{36.30\%}
\resset{e1.llm_qwen3_a1.tpr01}{3.38\%}
\resset{e1.n_benign}{6,126}
\resset{e1.n_malware}{4,703}
\resset{e1.n_models}{13}
\resset{e1.n_no_score}{1,001}
\resset{e1.no_state}{1,001}
\resset{e1.no_state.benign}{971}
\resset{e1.no_state.capa}{968}
\resset{e1.no_state.ghidra}{948}
\resset{e1.no_state.malware}{30}
\resset{e1.no_state.strings}{122}
\resset{e1.open_decider-4b-gguf_a1.abstained}{28.71\%}
\resset{e1.open_decider-4b-gguf_a1.aurc_margin}{0.0472}
\resset{e1.open_decider-4b-gguf_a1.auroc}{0.9834}
\resset{e1.open_decider-4b-gguf_a1.balanced_accuracy}{0.8708}
\resset{e1.open_decider-4b-gguf_a1.brier}{0.0470}
\resset{e1.open_decider-4b-gguf_a1.ci}{[0.975, 0.989]}
\resset{e1.open_decider-4b-gguf_a1.coverage_at_risk_1pct_margin}{44.56\%}
\resset{e1.open_decider-4b-gguf_a1.ece}{0.0167}
\resset{e1.open_decider-4b-gguf_a1.fpr}{1.29\%}
\resset{e1.open_decider-4b-gguf_a1.pauc}{0.7530}
\resset{e1.open_decider-4b-gguf_a1.tpr}{75.44\%}
\resset{e1.open_decider-4b-gguf_a1.tpr01}{24.75\%}
\resset{e1.qwen3_emb_probe.abstained}{4.50\%}
\resset{e1.qwen3_emb_probe.aurc_margin}{0.0107}
\resset{e1.qwen3_emb_probe.auroc}{0.9956}
\resset{e1.qwen3_emb_probe.balanced_accuracy}{0.9610}
\resset{e1.qwen3_emb_probe.brier}{0.0202}
\resset{e1.qwen3_emb_probe.ci}{[0.993, 0.997]}
\resset{e1.qwen3_emb_probe.coverage_at_risk_1pct_margin}{54.68\%}
\resset{e1.qwen3_emb_probe.ece}{0.0066}
\resset{e1.qwen3_emb_probe.fpr}{1.42\%}
\resset{e1.qwen3_emb_probe.pauc}{0.8678}
\resset{e1.qwen3_emb_probe.tpr}{93.62\%}
\resset{e1.tpr01.others_max}{46.59\%}
\resset{e10.clm_a0.auroc}{0.5992}
\resset{e10.clm_a0.auroc_ci}{[0.588, 0.611]}
\resset{e10.clm_a0.fpr}{1.16\%}
\resset{e10.clm_a0.n}{4,725}
\resset{e10.clm_a0.tpr}{2.98\%}
\resset{e10.clm_a0.tpr_ci}{[2.52\%, 3.51\%]}
\resset{e10.clm_a1.auroc}{0.8997}
\resset{e10.clm_a1.auroc_ci}{[0.894, 0.905]}
\resset{e10.clm_a1.fpr}{1.19\%}
\resset{e10.clm_a1.n}{4,725}
\resset{e10.clm_a1.tpr}{22.20\%}
\resset{e10.clm_a1.tpr_ci}{[21.02\%, 23.41\%]}
\resset{e10.clm_a2.auroc}{0.9472}
\resset{e10.clm_a2.auroc_ci}{[0.943, 0.951]}
\resset{e10.clm_a2.fpr}{1.11\%}
\resset{e10.clm_a2.n}{4,725}
\resset{e10.clm_a2.tpr}{51.87\%}
\resset{e10.clm_a2.tpr_ci}{[50.44\%, 53.31\%]}
\resset{e10.clm_a2.tpr_packed}{53.60\%}
\resset{e10.clm_a2.tpr_unpacked}{50.04\%}
\resset{e10.clm_a3.auroc}{0.9453}
\resset{e10.clm_a3.auroc_ci}{[0.941, 0.949]}
\resset{e10.clm_a3.fpr}{1.45\%}
\resset{e10.clm_a3.n}{4,725}
\resset{e10.clm_a3.tpr}{54.75\%}
\resset{e10.clm_a3.tpr_ci}{[53.32\%, 56.18\%]}
\resset{e10.clm_a3.tpr_packed}{57.58\%}
\resset{e10.clm_a3.tpr_unpacked}{51.75\%}
\resset{e10.contam.clm_a0.clean.fpr}{0.23\%}
\resset{e10.contam.clm_a0.clean.tpr}{4.45\%}
\resset{e10.contam.clm_a0.noid.d}{-0.40}
\resset{e10.contam.clm_a0.noid.fpr}{0.94\%}
\resset{e10.contam.clm_a0.noid.p}{0.754}
\resset{e10.contam.clm_a0.noid.tpr}{4.05\%}
\resset{e10.contam.clm_a1.clean.fpr}{1.17\%}
\resset{e10.contam.clm_a1.clean.tpr}{21.66\%}
\resset{e10.contam.clm_a1.noid.d}{+1.21}
\resset{e10.contam.clm_a1.noid.fpr}{1.87\%}
\resset{e10.contam.clm_a1.noid.p}{0.405}
\resset{e10.contam.clm_a1.noid.tpr}{22.87\%}
\resset{e10.contam.clm_a2.clean.fpr}{1.17\%}
\resset{e10.contam.clm_a2.clean.tpr}{52.63\%}
\resset{e10.contam.clm_a2.noid.d}{+8.70}
\resset{e10.contam.clm_a2.noid.fpr}{2.58\%}
\resset{e10.contam.clm_a2.noid.p}{$<0.001$}
\resset{e10.contam.clm_a2.noid.tpr}{61.34\%}
\resset{e10.contam.clm_a3.clean.fpr}{1.17\%}
\resset{e10.contam.clm_a3.clean.tpr}{55.47\%}
\resset{e10.contam.clm_a3.noid.d}{+6.07}
\resset{e10.contam.clm_a3.noid.fpr}{2.34\%}
\resset{e10.contam.clm_a3.noid.p}{$<0.001$}
\resset{e10.contam.clm_a3.noid.tpr}{61.54\%}
\resset{e10.contam.n_benign}{427}
\resset{e10.contam.n_malware}{494}
\resset{e10.det.auroc}{0.9501}
\resset{e10.det.auroc_ci}{[0.946, 0.954]}
\resset{e10.det.fpr}{1.91\%}
\resset{e10.det.n}{4,725}
\resset{e10.det.tpr}{66.46\%}
\resset{e10.det.tpr_ci}{[65.09\%, 67.80\%]}
\resset{e10.idonly.clm_a0.clean.auroc}{0.6047}
\resset{e10.idonly.clm_a0.clean.ci}{[0.570, 0.639]}
\resset{e10.idonly.clm_a0.idonly.auroc}{0.5398}
\resset{e10.idonly.clm_a0.idonly.ci}{[0.504, 0.577]}
\resset{e10.idonly.clm_a0.noid.auroc}{0.5792}
\resset{e10.idonly.clm_a0.noid.ci}{[0.542, 0.614]}
\resset{e10.idonly.clm_a1.clean.auroc}{0.9059}
\resset{e10.idonly.clm_a1.clean.ci}{[0.886, 0.924]}
\resset{e10.idonly.clm_a1.idonly.auroc}{0.6611}
\resset{e10.idonly.clm_a1.idonly.ci}{[0.626, 0.696]}
\resset{e10.idonly.clm_a1.noid.auroc}{0.9027}
\resset{e10.idonly.clm_a1.noid.ci}{[0.882, 0.921]}
\resset{e10.idonly.clm_a2.clean.auroc}{0.9474}
\resset{e10.idonly.clm_a2.clean.ci}{[0.933, 0.960]}
\resset{e10.idonly.clm_a2.idonly.auroc}{0.6548}
\resset{e10.idonly.clm_a2.idonly.ci}{[0.620, 0.690]}
\resset{e10.idonly.clm_a2.noid.auroc}{0.9408}
\resset{e10.idonly.clm_a2.noid.ci}{[0.926, 0.954]}
\resset{e10.idonly.clm_a3.clean.auroc}{0.9484}
\resset{e10.idonly.clm_a3.clean.ci}{[0.934, 0.961]}
\resset{e10.idonly.clm_a3.idonly.auroc}{0.6408}
\resset{e10.idonly.clm_a3.idonly.ci}{[0.606, 0.676]}
\resset{e10.idonly.clm_a3.noid.auroc}{0.9375}
\resset{e10.idonly.clm_a3.noid.ci}{[0.922, 0.952]}
\resset{e10.idonly.flag_lo}{0.60}
\resset{e1b.E1b-P1-d1-3b.abs}{0.0116}
\resset{e1b.E1b-P1-d1-3b.ci}{[+0.0071, +0.0182]}
\resset{e1b.E1b-P1-d1-3b.d}{+0.0116}
\resset{e1b.E1b-P1-d1-3b.p}{0.009}
\resset{e1b.E1b-P1-decider-0.8b.abs}{0.0143}
\resset{e1b.E1b-P1-decider-0.8b.ci}{[+0.0097, +0.0199]}
\resset{e1b.E1b-P1-decider-0.8b.d}{+0.0143}
\resset{e1b.E1b-P1-decider-0.8b.p}{0.009}
\resset{e1b.E1b-P1-decider-2b.abs}{0.0114}
\resset{e1b.E1b-P1-decider-2b.ci}{[+0.0072, +0.0176]}
\resset{e1b.E1b-P1-decider-2b.d}{+0.0114}
\resset{e1b.E1b-P1-decider-2b.p}{0.009}
\resset{e1b.E1b-P1-decider-4b-gguf.abs}{0.0102}
\resset{e1b.E1b-P1-decider-4b-gguf.ci}{[+0.0060, +0.0165]}
\resset{e1b.E1b-P1-decider-4b-gguf.d}{+0.0102}
\resset{e1b.E1b-P1-decider-4b-gguf.p}{0.009}
\resset{e1b.E1b-P1-kev-0.8b.abs}{0.0262}
\resset{e1b.E1b-P1-kev-0.8b.ci}{[+0.0179, +0.0367]}
\resset{e1b.E1b-P1-kev-0.8b.d}{+0.0262}
\resset{e1b.E1b-P1-kev-0.8b.p}{0.009}
\resset{e1b.E1b-P1-laya-multilingual.abs}{0.0724}
\resset{e1b.E1b-P1-laya-multilingual.ci}{[+0.0535, +0.0970]}
\resset{e1b.E1b-P1-laya-multilingual.d}{+0.0724}
\resset{e1b.E1b-P1-laya-multilingual.p}{0.009}
\resset{e1b.E1b-P1-laya-typed-decisions.abs}{0.0517}
\resset{e1b.E1b-P1-laya-typed-decisions.ci}{[+0.0336, +0.0765]}
\resset{e1b.E1b-P1-laya-typed-decisions.d}{+0.0517}
\resset{e1b.E1b-P1-laya-typed-decisions.p}{0.009}
\resset{e1b.E1b-P1-strands-decider-2b-v1-2610.abs}{0.0116}
\resset{e1b.E1b-P1-strands-decider-2b-v1-2610.ci}{[+0.0077, +0.0171]}
\resset{e1b.E1b-P1-strands-decider-2b-v1-2610.d}{+0.0116}
\resset{e1b.E1b-P1-strands-decider-2b-v1-2610.p}{0.009}
\resset{e1b.E1b-P1-von.abs}{0.0293}
\resset{e1b.E1b-P1-von.ci}{[+0.0198, +0.0424]}
\resset{e1b.E1b-P1-von.d}{+0.0293}
\resset{e1b.E1b-P1-von.p}{0.009}
\resset{e1b.E1b-S1-d1-3b.abs}{0.4006}
\resset{e1b.E1b-S1-d1-3b.ci}{[+0.3029, +0.4921]}
\resset{e1b.E1b-S1-d1-3b.d}{+0.4006}
\resset{e1b.E1b-S1-d1-3b.p}{0.009}
\resset{e1b.E1b-S1-decider-0.8b.abs}{0.4242}
\resset{e1b.E1b-S1-decider-0.8b.ci}{[+0.3383, +0.5062]}
\resset{e1b.E1b-S1-decider-0.8b.d}{+0.4242}
\resset{e1b.E1b-S1-decider-0.8b.p}{0.009}
\resset{e1b.E1b-S1-decider-2b.abs}{0.3255}
\resset{e1b.E1b-S1-decider-2b.ci}{[+0.2382, +0.4192]}
\resset{e1b.E1b-S1-decider-2b.d}{+0.3255}
\resset{e1b.E1b-S1-decider-2b.p}{0.009}
\resset{e1b.E1b-S1-decider-4b-gguf.abs}{0.3269}
\resset{e1b.E1b-S1-decider-4b-gguf.ci}{[+0.2531, +0.3993]}
\resset{e1b.E1b-S1-decider-4b-gguf.d}{+0.3269}
\resset{e1b.E1b-S1-decider-4b-gguf.p}{0.009}
\resset{e1b.E1b-S1-kev-0.8b.abs}{0.3004}
\resset{e1b.E1b-S1-kev-0.8b.ci}{[+0.2459, +0.3586]}
\resset{e1b.E1b-S1-kev-0.8b.d}{+0.3004}
\resset{e1b.E1b-S1-kev-0.8b.p}{0.009}
\resset{e1b.E1b-S1-laya-multilingual.abs}{0.3329}
\resset{e1b.E1b-S1-laya-multilingual.ci}{[+0.2468, +0.4255]}
\resset{e1b.E1b-S1-laya-multilingual.d}{+0.3329}
\resset{e1b.E1b-S1-laya-multilingual.p}{0.009}
\resset{e1b.E1b-S1-laya-typed-decisions.abs}{0.5353}
\resset{e1b.E1b-S1-laya-typed-decisions.ci}{[+0.4106, +0.6669]}
\resset{e1b.E1b-S1-laya-typed-decisions.d}{+0.5353}
\resset{e1b.E1b-S1-laya-typed-decisions.p}{0.009}
\resset{e1b.E1b-S1-strands-decider-2b-v1-2610.abs}{0.3015}
\resset{e1b.E1b-S1-strands-decider-2b-v1-2610.ci}{[+0.2223, +0.3839]}
\resset{e1b.E1b-S1-strands-decider-2b-v1-2610.d}{+0.3015}
\resset{e1b.E1b-S1-strands-decider-2b-v1-2610.p}{0.009}
\resset{e1b.E1b-S1-von.abs}{0.2939}
\resset{e1b.E1b-S1-von.ci}{[+0.1955, +0.4163]}
\resset{e1b.E1b-S1-von.d}{+0.2939}
\resset{e1b.E1b-S1-von.p}{0.009}
\resset{e1b.E1b-S2-d1-3b.abs}{0.0035}
\resset{e1b.E1b-S2-d1-3b.ci}{[-0.0067, -0.0010]}
\resset{e1b.E1b-S2-d1-3b.d}{-0.0035}
\resset{e1b.E1b-S2-d1-3b.p}{0.035}
\resset{e1b.E1b-S2-decider-0.8b.abs}{0.0026}
\resset{e1b.E1b-S2-decider-0.8b.ci}{[-0.0053, -0.0007]}
\resset{e1b.E1b-S2-decider-0.8b.d}{-0.0026}
\resset{e1b.E1b-S2-decider-0.8b.p}{0.055}
\resset{e1b.E1b-S2-decider-2b.abs}{0.0005}
\resset{e1b.E1b-S2-decider-2b.ci}{[-0.0027, +0.0015]}
\resset{e1b.E1b-S2-decider-2b.d}{-0.0005}
\resset{e1b.E1b-S2-decider-2b.p}{1.000}
\resset{e1b.E1b-S2-decider-4b-gguf.abs}{0.0006}
\resset{e1b.E1b-S2-decider-4b-gguf.ci}{[-0.0027, +0.0012]}
\resset{e1b.E1b-S2-decider-4b-gguf.d}{-0.0006}
\resset{e1b.E1b-S2-decider-4b-gguf.p}{1.000}
\resset{e1b.E1b-S2-kev-0.8b.abs}{0.0008}
\resset{e1b.E1b-S2-kev-0.8b.ci}{[-0.0027, +0.0011]}
\resset{e1b.E1b-S2-kev-0.8b.d}{-0.0008}
\resset{e1b.E1b-S2-kev-0.8b.p}{1.000}
\resset{e1b.E1b-S2-laya-multilingual.abs}{0.0096}
\resset{e1b.E1b-S2-laya-multilingual.ci}{[-0.0154, -0.0048]}
\resset{e1b.E1b-S2-laya-multilingual.d}{-0.0096}
\resset{e1b.E1b-S2-laya-multilingual.p}{0.009}
\resset{e1b.E1b-S2-laya-typed-decisions.abs}{0.0034}
\resset{e1b.E1b-S2-laya-typed-decisions.ci}{[-0.0067, -0.0008]}
\resset{e1b.E1b-S2-laya-typed-decisions.d}{-0.0034}
\resset{e1b.E1b-S2-laya-typed-decisions.p}{0.054}
\resset{e1b.E1b-S2-strands-decider-2b-v1-2610.abs}{0.0006}
\resset{e1b.E1b-S2-strands-decider-2b-v1-2610.ci}{[-0.0018, +0.0027]}
\resset{e1b.E1b-S2-strands-decider-2b-v1-2610.d}{+0.0006}
\resset{e1b.E1b-S2-strands-decider-2b-v1-2610.p}{1.000}
\resset{e1b.E1b-S2-von.abs}{0.0047}
\resset{e1b.E1b-S2-von.ci}{[-0.0086, -0.0019]}
\resset{e1b.E1b-S2-von.d}{-0.0047}
\resset{e1b.E1b-S2-von.p}{0.009}
\resset{e1b.E1b-S3-d1-3b.abs}{0.0002}
\resset{e1b.E1b-S3-d1-3b.ci}{[-0.0013, +0.0016]}
\resset{e1b.E1b-S3-d1-3b.d}{+0.0002}
\resset{e1b.E1b-S3-d1-3b.p}{1.000}
\resset{e1b.E1b-S3-decider-0.8b.abs}{0.0012}
\resset{e1b.E1b-S3-decider-0.8b.ci}{[-0.0010, +0.0036]}
\resset{e1b.E1b-S3-decider-0.8b.d}{+0.0012}
\resset{e1b.E1b-S3-decider-0.8b.p}{1.000}
\resset{e1b.E1b-S3-decider-2b.abs}{0.0008}
\resset{e1b.E1b-S3-decider-2b.ci}{[-0.0038, +0.0013]}
\resset{e1b.E1b-S3-decider-2b.d}{-0.0008}
\resset{e1b.E1b-S3-decider-2b.p}{1.000}
\resset{e1b.E1b-S3-decider-4b-gguf.abs}{0.0011}
\resset{e1b.E1b-S3-decider-4b-gguf.ci}{[-0.0026, +0.0003]}
\resset{e1b.E1b-S3-decider-4b-gguf.d}{-0.0011}
\resset{e1b.E1b-S3-decider-4b-gguf.p}{0.864}
\resset{e1b.E1b-S3-kev-0.8b.abs}{0.0001}
\resset{e1b.E1b-S3-kev-0.8b.ci}{[-0.0015, +0.0015]}
\resset{e1b.E1b-S3-kev-0.8b.d}{-0.0001}
\resset{e1b.E1b-S3-kev-0.8b.p}{1.000}
\resset{e1b.E1b-S3-laya-multilingual.abs}{0.0019}
\resset{e1b.E1b-S3-laya-multilingual.ci}{[-0.0010, +0.0051]}
\resset{e1b.E1b-S3-laya-multilingual.d}{+0.0019}
\resset{e1b.E1b-S3-laya-multilingual.p}{1.000}
\resset{e1b.E1b-S3-laya-typed-decisions.abs}{0.0025}
\resset{e1b.E1b-S3-laya-typed-decisions.ci}{[-0.0000, +0.0052]}
\resset{e1b.E1b-S3-laya-typed-decisions.d}{+0.0025}
\resset{e1b.E1b-S3-laya-typed-decisions.p}{0.531}
\resset{e1b.E1b-S3-strands-decider-2b-v1-2610.abs}{0.0010}
\resset{e1b.E1b-S3-strands-decider-2b-v1-2610.ci}{[-0.0010, +0.0034]}
\resset{e1b.E1b-S3-strands-decider-2b-v1-2610.d}{+0.0010}
\resset{e1b.E1b-S3-strands-decider-2b-v1-2610.p}{1.000}
\resset{e1b.E1b-S3-von.abs}{0.0004}
\resset{e1b.E1b-S3-von.ci}{[-0.0014, +0.0020]}
\resset{e1b.E1b-S3-von.d}{+0.0004}
\resset{e1b.E1b-S3-von.p}{1.000}
\resset{e1b.a2.best}{Strands Decider 2B}
\resset{e1b.a2.max}{0.9948}
\resset{e1b.a2.min}{0.9853}
\resset{e1b.a2.tpr_max}{95.73\%}
\resset{e1b.a2.tpr_min}{83.66\%}
\resset{e1b.a2.worst}{Laya multilingual 322M}
\resset{e1b.n_models}{9}
\resset{e1b.open_d1-3b_a2.auroc}{0.9908}
\resset{e1b.open_d1-3b_a2.ci}{[0.986, 0.994]}
\resset{e1b.open_d1-3b_a2.fpr}{1.37\%}
\resset{e1b.open_d1-3b_a2.tpr}{88.26\%}
\resset{e1b.open_decider-0.8b_a2.auroc}{0.9916}
\resset{e1b.open_decider-0.8b_a2.ci}{[0.987, 0.995]}
\resset{e1b.open_decider-0.8b_a2.fpr}{0.88\%}
\resset{e1b.open_decider-0.8b_a2.tpr}{85.05\%}
\resset{e1b.open_decider-2b_a2.auroc}{0.9938}
\resset{e1b.open_decider-2b_a2.ci}{[0.989, 0.997]}
\resset{e1b.open_decider-2b_a2.fpr}{1.37\%}
\resset{e1b.open_decider-2b_a2.tpr}{95.36\%}
\resset{e1b.open_decider-4b-gguf_a2.auroc}{0.9936}
\resset{e1b.open_decider-4b-gguf_a2.ci}{[0.990, 0.996]}
\resset{e1b.open_decider-4b-gguf_a2.fpr}{1.42\%}
\resset{e1b.open_decider-4b-gguf_a2.tpr}{92.86\%}
\resset{e1b.open_kev-0.8b_a2.auroc}{0.9934}
\resset{e1b.open_kev-0.8b_a2.ci}{[0.990, 0.996]}
\resset{e1b.open_kev-0.8b_a2.fpr}{1.52\%}
\resset{e1b.open_kev-0.8b_a2.tpr}{91.69\%}
\resset{e1b.open_laya-multilingual_a2.auroc}{0.9853}
\resset{e1b.open_laya-multilingual_a2.ci}{[0.978, 0.991]}
\resset{e1b.open_laya-multilingual_a2.fpr}{1.49\%}
\resset{e1b.open_laya-multilingual_a2.tpr}{83.66\%}
\resset{e1b.open_laya-typed-decisions_a2.auroc}{0.9915}
\resset{e1b.open_laya-typed-decisions_a2.ci}{[0.987, 0.994]}
\resset{e1b.open_laya-typed-decisions_a2.fpr}{1.18\%}
\resset{e1b.open_laya-typed-decisions_a2.tpr}{85.83\%}
\resset{e1b.open_strands-decider-2b-v1-2610_a2.auroc}{0.9948}
\resset{e1b.open_strands-decider-2b-v1-2610_a2.ci}{[0.992, 0.997]}
\resset{e1b.open_strands-decider-2b-v1-2610_a2.fpr}{1.50\%}
\resset{e1b.open_strands-decider-2b-v1-2610_a2.tpr}{95.73\%}
\resset{e1b.open_von_a2.auroc}{0.9896}
\resset{e1b.open_von_a2.ci}{[0.983, 0.994]}
\resset{e1b.open_von_a2.fpr}{1.78\%}
\resset{e1b.open_von_a2.tpr}{93.26\%}
\resset{e1b.p1.max}{0.0724}
\resset{e1b.p1.min}{0.0102}
\resset{e1b.p1.n_sig}{9}
\resset{e1b.p1.pmax}{0.009}
\resset{e1b.s2.max}{+0.0006}
\resset{e1b.s2.min}{-0.0096}
\resset{e1b.s2.n_lower}{3}
\resset{e1b.s2.n_ns}{6}
\resset{e1b.s3.n_sig}{0}
\resset{e1b.s3.pmin}{0.531}
\resset{e1b.stage}{test}
\resset{e2.A.max}{+0.0136}
\resset{e2.A.min}{-0.0001}
\resset{e2.A.n}{4}
\resset{e2.A.n_sig}{0}
\resset{e2.A.n_sig_neg}{0}
\resset{e2.A.n_sig_pos}{0}
\resset{e2.A1.clm_a0.a}{0.6379}
\resset{e2.A1.clm_a0.b}{0.6379}
\resset{e2.A1.clm_a0.ci}{[+0.0000, +0.0000]}
\resset{e2.A1.clm_a0.d}{+0.0000}
\resset{e2.A1.clm_a0.p}{1.000}
\resset{e2.A1.clm_a1.a}{0.8898}
\resset{e2.A1.clm_a1.b}{0.8762}
\resset{e2.A1.clm_a1.ci}{[-0.0018, +0.0341]}
\resset{e2.A1.clm_a1.d}{+0.0136}
\resset{e2.A1.clm_a1.p}{0.110}
\resset{e2.A1.clm_a2.a}{0.9523}
\resset{e2.A1.clm_a2.b}{0.9524}
\resset{e2.A1.clm_a2.ci}{[-0.0019, +0.0016]}
\resset{e2.A1.clm_a2.d}{-0.0001}
\resset{e2.A1.clm_a2.p}{0.972}
\resset{e2.A1.kev.a}{0.8839}
\resset{e2.A1.kev.b}{0.8833}
\resset{e2.A1.kev.ci}{[-0.0051, +0.0060]}
\resset{e2.A1.kev.d}{+0.0006}
\resset{e2.A1.kev.p}{0.855}
\resset{e2.B.max}{+0.0502}
\resset{e2.B.min}{-0.0487}
\resset{e2.B.n}{12}
\resset{e2.B.n_sig}{2}
\resset{e2.B.n_sig_neg}{0}
\resset{e2.B.n_sig_pos}{2}
\resset{e2.B1.clm_a0.a}{0.7216}
\resset{e2.B1.clm_a0.b}{0.7498}
\resset{e2.B1.clm_a0.ci}{[-0.0702, +0.0176]}
\resset{e2.B1.clm_a0.d}{-0.0282}
\resset{e2.B1.clm_a0.p}{0.231}
\resset{e2.B1.clm_a1.a}{0.9120}
\resset{e2.B1.clm_a1.b}{0.8909}
\resset{e2.B1.clm_a1.ci}{[-0.0140, +0.0669]}
\resset{e2.B1.clm_a1.d}{+0.0211}
\resset{e2.B1.clm_a1.p}{0.265}
\resset{e2.B1.clm_a2.a}{0.9642}
\resset{e2.B1.clm_a2.b}{0.9606}
\resset{e2.B1.clm_a2.ci}{[-0.0052, +0.0152]}
\resset{e2.B1.clm_a2.d}{+0.0035}
\resset{e2.B1.clm_a2.p}{1.000}
\resset{e2.B1.kev.a}{0.9171}
\resset{e2.B1.kev.b}{0.8669}
\resset{e2.B1.kev.ci}{[+0.0072, +0.1009]}
\resset{e2.B1.kev.d}{+0.0502}
\resset{e2.B1.kev.p}{0.078}
\resset{e2.B2.clm_a0.a}{0.6772}
\resset{e2.B2.clm_a0.b}{0.7052}
\resset{e2.B2.clm_a0.ci}{[-0.0586, +0.0019]}
\resset{e2.B2.clm_a0.d}{-0.0280}
\resset{e2.B2.clm_a0.p}{0.204}
\resset{e2.B2.clm_a1.a}{0.9334}
\resset{e2.B2.clm_a1.b}{0.9062}
\resset{e2.B2.clm_a1.ci}{[+0.0081, +0.0523]}
\resset{e2.B2.clm_a1.d}{+0.0271}
\resset{e2.B2.clm_a1.p}{0.009}
\resset{e2.B2.clm_a2.a}{0.9645}
\resset{e2.B2.clm_a2.b}{0.9638}
\resset{e2.B2.clm_a2.ci}{[-0.0049, +0.0086]}
\resset{e2.B2.clm_a2.d}{+0.0007}
\resset{e2.B2.clm_a2.p}{1.000}
\resset{e2.B2.kev.a}{0.9149}
\resset{e2.B2.kev.b}{0.9090}
\resset{e2.B2.kev.ci}{[-0.0426, +0.0569]}
\resset{e2.B2.kev.d}{+0.0059}
\resset{e2.B2.kev.p}{0.896}
\resset{e2.B3.clm_a0.a}{0.7012}
\resset{e2.B3.clm_a0.b}{0.7498}
\resset{e2.B3.clm_a0.ci}{[-0.1035, +0.0082]}
\resset{e2.B3.clm_a0.d}{-0.0487}
\resset{e2.B3.clm_a0.p}{0.204}
\resset{e2.B3.clm_a1.a}{0.9370}
\resset{e2.B3.clm_a1.b}{0.8909}
\resset{e2.B3.clm_a1.ci}{[+0.0114, +0.0898]}
\resset{e2.B3.clm_a1.d}{+0.0461}
\resset{e2.B3.clm_a1.p}{0.012}
\resset{e2.B3.clm_a2.a}{0.9645}
\resset{e2.B3.clm_a2.b}{0.9606}
\resset{e2.B3.clm_a2.ci}{[-0.0073, +0.0216]}
\resset{e2.B3.clm_a2.d}{+0.0039}
\resset{e2.B3.clm_a2.p}{1.000}
\resset{e2.B3.kev.a}{0.9282}
\resset{e2.B3.kev.b}{0.8856}
\resset{e2.B3.kev.ci}{[-0.0057, +0.0969]}
\resset{e2.B3.kev.d}{+0.0427}
\resset{e2.B3.kev.p}{0.196}
\resset{e2.C.max}{+0.0361}
\resset{e2.C.min}{-0.0430}
\resset{e2.C.n}{20}
\resset{e2.C.n_sig}{1}
\resset{e2.C.n_sig_neg}{0}
\resset{e2.C.n_sig_pos}{1}
\resset{e2.C1.clm_a0.a}{0.6709}
\resset{e2.C1.clm_a0.b}{0.6379}
\resset{e2.C1.clm_a0.ci}{[-0.0078, +0.0828]}
\resset{e2.C1.clm_a0.d}{+0.0330}
\resset{e2.C1.clm_a0.p}{0.348}
\resset{e2.C1.clm_a1.a}{0.8798}
\resset{e2.C1.clm_a1.b}{0.8762}
\resset{e2.C1.clm_a1.ci}{[-0.0222, +0.0279]}
\resset{e2.C1.clm_a1.d}{+0.0036}
\resset{e2.C1.clm_a1.p}{0.966}
\resset{e2.C1.clm_a2.a}{0.9600}
\resset{e2.C1.clm_a2.b}{0.9524}
\resset{e2.C1.clm_a2.ci}{[-0.0010, +0.0199]}
\resset{e2.C1.clm_a2.d}{+0.0076}
\resset{e2.C1.clm_a2.p}{0.368}
\resset{e2.C1.kev.a}{0.9133}
\resset{e2.C1.kev.b}{0.8833}
\resset{e2.C1.kev.ci}{[-0.0272, +0.0890]}
\resset{e2.C1.kev.d}{+0.0300}
\resset{e2.C1.kev.p}{1.000}
\resset{e2.C4.clm_a0.a}{0.6621}
\resset{e2.C4.clm_a0.b}{0.6379}
\resset{e2.C4.clm_a0.ci}{[-0.0013, +0.0547]}
\resset{e2.C4.clm_a0.d}{+0.0243}
\resset{e2.C4.clm_a0.p}{0.252}
\resset{e2.C4.clm_a1.a}{0.9015}
\resset{e2.C4.clm_a1.b}{0.8762}
\resset{e2.C4.clm_a1.ci}{[+0.0099, +0.0457]}
\resset{e2.C4.clm_a1.d}{+0.0253}
\resset{e2.C4.clm_a1.p}{0.005}
\resset{e2.C4.clm_a2.a}{0.9464}
\resset{e2.C4.clm_a2.b}{0.9524}
\resset{e2.C4.clm_a2.ci}{[-0.0131, -0.0007]}
\resset{e2.C4.clm_a2.d}{-0.0060}
\resset{e2.C4.clm_a2.p}{0.110}
\resset{e2.C4.kev.a}{0.9003}
\resset{e2.C4.kev.b}{0.8833}
\resset{e2.C4.kev.ci}{[-0.0103, +0.0481]}
\resset{e2.C4.kev.d}{+0.0170}
\resset{e2.C4.kev.p}{1.000}
\resset{e2.D.max}{-0.0252}
\resset{e2.D.min}{-0.3397}
\resset{e2.D.n}{4}
\resset{e2.D.n_sig}{4}
\resset{e2.D.n_sig_neg}{4}
\resset{e2.D.n_sig_pos}{0}
\resset{e2.D1.clm_a0.a}{0.6379}
\resset{e2.D1.clm_a0.b}{0.9776}
\resset{e2.D1.clm_a0.ci}{[-0.4356, -0.2547]}
\resset{e2.D1.clm_a0.d}{-0.3397}
\resset{e2.D1.clm_a0.p}{$<0.001$}
\resset{e2.D1.clm_a1.a}{0.8762}
\resset{e2.D1.clm_a1.b}{0.9776}
\resset{e2.D1.clm_a1.ci}{[-0.1509, -0.0586]}
\resset{e2.D1.clm_a1.d}{-0.1014}
\resset{e2.D1.clm_a1.p}{$<0.001$}
\resset{e2.D1.clm_a2.a}{0.9524}
\resset{e2.D1.clm_a2.b}{0.9776}
\resset{e2.D1.clm_a2.ci}{[-0.0592, -0.0012]}
\resset{e2.D1.clm_a2.d}{-0.0252}
\resset{e2.D1.clm_a2.p}{0.041}
\resset{e2.D1.kev.a}{0.8833}
\resset{e2.D1.kev.b}{0.9776}
\resset{e2.D1.kev.ci}{[-0.1698, -0.0370]}
\resset{e2.D1.kev.d}{-0.0943}
\resset{e2.D1.kev.p}{0.002}
\resset{e2.K.max}{+0.0278}
\resset{e2.K.min}{-0.0121}
\resset{e2.K.n}{12}
\resset{e2.K.n_sig}{0}
\resset{e2.K.n_sig_neg}{0}
\resset{e2.K.n_sig_pos}{0}
\resset{e2.K1.clm_a0.a}{0.7052}
\resset{e2.K1.clm_a0.b}{0.6848}
\resset{e2.K1.clm_a0.ci}{[+0.0013, +0.0401]}
\resset{e2.K1.clm_a0.d}{+0.0204}
\resset{e2.K1.clm_a0.p}{0.090}
\resset{e2.K1.clm_a1.a}{0.9062}
\resset{e2.K1.clm_a1.b}{0.9183}
\resset{e2.K1.clm_a1.ci}{[-0.0274, +0.0047]}
\resset{e2.K1.clm_a1.d}{-0.0121}
\resset{e2.K1.clm_a1.p}{0.666}
\resset{e2.K1.clm_a2.a}{0.9638}
\resset{e2.K1.clm_a2.b}{0.9630}
\resset{e2.K1.clm_a2.ci}{[-0.0032, +0.0055]}
\resset{e2.K1.clm_a2.d}{+0.0008}
\resset{e2.K1.clm_a2.p}{1.000}
\resset{e2.K1.kev.a}{0.8991}
\resset{e2.K1.kev.b}{0.8984}
\resset{e2.K1.kev.ci}{[-0.0141, +0.0185]}
\resset{e2.K1.kev.d}{+0.0007}
\resset{e2.K1.kev.p}{1.000}
\resset{e2.KBP.clm_a0.n}{11}
\resset{e2.KBP.clm_a0.n_sig}{2}
\resset{e2.KBP.clm_a1.n}{11}
\resset{e2.KBP.clm_a1.n_sig}{2}
\resset{e2.KBP.clm_a2.n}{11}
\resset{e2.KBP.clm_a2.n_sig}{0}
\resset{e2.KBP.kev.n}{11}
\resset{e2.KBP.kev.n_sig}{0}
\resset{e2.L.clm_a0.n}{7}
\resset{e2.L.clm_a0.n_sig}{1}
\resset{e2.L.clm_a0.rest_blocks}{File, Sections, Packing, Imports, Capabilities and Code}
\resset{e2.L.clm_a0.rest_max}{+0.0487}
\resset{e2.L.clm_a0.sig_blocks}{Strings}
\resset{e2.L.clm_a0.sig_max}{+0.1373}
\resset{e2.L.clm_a0.sig_min}{+0.1373}
\resset{e2.L.clm_a1.n}{7}
\resset{e2.L.clm_a1.n_sig}{0}
\resset{e2.L.clm_a1.rest_blocks}{File, Sections, Packing, Imports, Capabilities, Strings and Code}
\resset{e2.L.clm_a1.rest_max}{+0.0048}
\resset{e2.L.clm_a1.sig_blocks}{none}
\resset{e2.L.clm_a2.n}{7}
\resset{e2.L.clm_a2.n_sig}{0}
\resset{e2.L.clm_a2.rest_blocks}{File, Sections, Packing, Imports, Capabilities, Strings and Code}
\resset{e2.L.clm_a2.rest_max}{+0.0095}
\resset{e2.L.clm_a2.sig_blocks}{none}
\resset{e2.L.kev.n}{7}
\resset{e2.L.kev.n_sig}{0}
\resset{e2.L.kev.rest_blocks}{File, Sections, Packing, Imports, Capabilities, Strings and Code}
\resset{e2.L.kev.rest_max}{+0.0445}
\resset{e2.L.kev.sig_blocks}{none}
\resset{e2.L.max}{+0.1373}
\resset{e2.L.min}{-0.0623}
\resset{e2.L.n}{28}
\resset{e2.L.n_sig}{3}
\resset{e2.L.n_sig_neg}{2}
\resset{e2.L.n_sig_pos}{1}
\resset{e2.M.max}{+0.0334}
\resset{e2.M.min}{-0.0292}
\resset{e2.M.n}{2}
\resset{e2.M.n_sig}{0}
\resset{e2.M.n_sig_neg}{0}
\resset{e2.M.n_sig_pos}{0}
\resset{e2.M1.clm_a1.a}{0.9350}
\resset{e2.M1.clm_a1.b}{0.9015}
\resset{e2.M1.clm_a1.ci}{[-0.0057, +0.0766]}
\resset{e2.M1.clm_a1.d}{+0.0334}
\resset{e2.M1.clm_a1.p}{0.098}
\resset{e2.M1.kev.a}{0.8999}
\resset{e2.M1.kev.b}{0.9291}
\resset{e2.M1.kev.ci}{[-0.0823, +0.0182]}
\resset{e2.M1.kev.d}{-0.0292}
\resset{e2.M1.kev.p}{0.235}
\resset{e2.O.clm_a0.n}{7}
\resset{e2.O.clm_a0.n_sig}{4}
\resset{e2.O.clm_a0.rest_blocks}{Sections, Packing and Imports}
\resset{e2.O.clm_a0.rest_max}{+0.0555}
\resset{e2.O.clm_a0.sig_blocks}{File, Capabilities, Strings and Code}
\resset{e2.O.clm_a0.sig_max}{+0.2880}
\resset{e2.O.clm_a0.sig_min}{+0.1519}
\resset{e2.O.clm_a1.n}{7}
\resset{e2.O.clm_a1.n_sig}{2}
\resset{e2.O.clm_a1.rest_blocks}{File, Sections, Imports, Capabilities and Strings}
\resset{e2.O.clm_a1.rest_max}{+0.0417}
\resset{e2.O.clm_a1.sig_blocks}{Packing and Code}
\resset{e2.O.clm_a1.sig_max}{+0.1662}
\resset{e2.O.clm_a1.sig_min}{+0.1632}
\resset{e2.O.clm_a2.n}{7}
\resset{e2.O.clm_a2.n_sig}{3}
\resset{e2.O.clm_a2.rest_blocks}{Sections, Imports, Capabilities and Strings}
\resset{e2.O.clm_a2.rest_max}{+0.0167}
\resset{e2.O.clm_a2.sig_blocks}{File, Packing and Code}
\resset{e2.O.clm_a2.sig_max}{+0.2433}
\resset{e2.O.clm_a2.sig_min}{+0.1175}
\resset{e2.O.kev.n}{7}
\resset{e2.O.kev.n_sig}{1}
\resset{e2.O.kev.rest_blocks}{File, Sections, Packing, Imports, Capabilities and Strings}
\resset{e2.O.kev.rest_max}{+0.1743}
\resset{e2.O.kev.sig_blocks}{Code}
\resset{e2.O.kev.sig_max}{+0.1533}
\resset{e2.O.kev.sig_min}{+0.1533}
\resset{e2.O.max}{+0.2880}
\resset{e2.O.min}{-0.0674}
\resset{e2.O.n}{28}
\resset{e2.O.n_sig}{10}
\resset{e2.O.n_sig_neg}{0}
\resset{e2.O.n_sig_pos}{10}
\resset{e2.P.max}{+0.1060}
\resset{e2.P.min}{-0.0747}
\resset{e2.P.n}{20}
\resset{e2.P.n_sig}{2}
\resset{e2.P.n_sig_neg}{1}
\resset{e2.P.n_sig_pos}{1}
\resset{e2.P1.clm_a0.a}{0.6848}
\resset{e2.P1.clm_a0.b}{0.7595}
\resset{e2.P1.clm_a0.ci}{[-0.1365, -0.0181]}
\resset{e2.P1.clm_a0.d}{-0.0747}
\resset{e2.P1.clm_a0.p}{0.040}
\resset{e2.P1.clm_a1.a}{0.9183}
\resset{e2.P1.clm_a1.b}{0.9205}
\resset{e2.P1.clm_a1.ci}{[-0.0261, +0.0188]}
\resset{e2.P1.clm_a1.d}{-0.0022}
\resset{e2.P1.clm_a1.p}{1.000}
\resset{e2.P1.clm_a2.a}{0.9630}
\resset{e2.P1.clm_a2.b}{0.9585}
\resset{e2.P1.clm_a2.ci}{[-0.0085, +0.0183]}
\resset{e2.P1.clm_a2.d}{+0.0045}
\resset{e2.P1.clm_a2.p}{1.000}
\resset{e2.P1.kev.a}{0.8984}
\resset{e2.P1.kev.b}{0.9087}
\resset{e2.P1.kev.ci}{[-0.0702, +0.0504]}
\resset{e2.P1.kev.d}{-0.0104}
\resset{e2.P1.kev.p}{1.000}
\resset{e2.P2.clm_a0.a}{0.6848}
\resset{e2.P2.clm_a0.b}{0.7065}
\resset{e2.P2.clm_a0.ci}{[-0.0756, +0.0267]}
\resset{e2.P2.clm_a0.d}{-0.0216}
\resset{e2.P2.clm_a0.p}{0.520}
\resset{e2.P2.clm_a1.a}{0.9183}
\resset{e2.P2.clm_a1.b}{0.9104}
\resset{e2.P2.clm_a1.ci}{[-0.0149, +0.0304]}
\resset{e2.P2.clm_a1.d}{+0.0079}
\resset{e2.P2.clm_a1.p}{1.000}
\resset{e2.P2.clm_a2.a}{0.9630}
\resset{e2.P2.clm_a2.b}{0.9548}
\resset{e2.P2.clm_a2.ci}{[-0.0006, +0.0218]}
\resset{e2.P2.clm_a2.d}{+0.0081}
\resset{e2.P2.clm_a2.p}{0.355}
\resset{e2.P2.kev.a}{0.8984}
\resset{e2.P2.kev.b}{0.8433}
\resset{e2.P2.kev.ci}{[+0.0056, +0.1082]}
\resset{e2.P2.kev.d}{+0.0551}
\resset{e2.P2.kev.p}{0.155}
\resset{e2.P3.clm_a0.a}{0.6848}
\resset{e2.P3.clm_a0.b}{0.5788}
\resset{e2.P3.clm_a0.ci}{[+0.0474, +0.1707]}
\resset{e2.P3.clm_a0.d}{+0.1060}
\resset{e2.P3.clm_a0.p}{0.005}
\resset{e2.P3.clm_a1.a}{0.9183}
\resset{e2.P3.clm_a1.b}{0.9242}
\resset{e2.P3.clm_a1.ci}{[-0.0377, +0.0262]}
\resset{e2.P3.clm_a1.d}{-0.0058}
\resset{e2.P3.clm_a1.p}{1.000}
\resset{e2.P3.clm_a2.a}{0.9630}
\resset{e2.P3.clm_a2.b}{0.9596}
\resset{e2.P3.clm_a2.ci}{[-0.0103, +0.0209]}
\resset{e2.P3.clm_a2.d}{+0.0034}
\resset{e2.P3.clm_a2.p}{1.000}
\resset{e2.P3.kev.a}{0.8984}
\resset{e2.P3.kev.b}{0.8922}
\resset{e2.P3.kev.ci}{[-0.0305, +0.0468]}
\resset{e2.P3.kev.d}{+0.0061}
\resset{e2.P3.kev.p}{1.000}
\resset{e2.P5.clm_a0.a}{0.6848}
\resset{e2.P5.clm_a0.b}{0.7327}
\resset{e2.P5.clm_a0.ci}{[-0.0925, -0.0033]}
\resset{e2.P5.clm_a0.d}{-0.0479}
\resset{e2.P5.clm_a0.p}{0.096}
\resset{e2.P5.clm_a1.a}{0.9183}
\resset{e2.P5.clm_a1.b}{0.9341}
\resset{e2.P5.clm_a1.ci}{[-0.0414, +0.0027]}
\resset{e2.P5.clm_a1.d}{-0.0158}
\resset{e2.P5.clm_a1.p}{0.465}
\resset{e2.P5.clm_a2.a}{0.9630}
\resset{e2.P5.clm_a2.b}{0.9674}
\resset{e2.P5.clm_a2.ci}{[-0.0207, +0.0057]}
\resset{e2.P5.clm_a2.d}{-0.0045}
\resset{e2.P5.clm_a2.p}{1.000}
\resset{e2.P5.kev.a}{0.8984}
\resset{e2.P5.kev.b}{0.9402}
\resset{e2.P5.kev.ci}{[-0.0948, +0.0061]}
\resset{e2.P5.kev.d}{-0.0419}
\resset{e2.P5.kev.p}{0.336}
\resset{e2.R.max}{+0.0121}
\resset{e2.R.min}{-0.0292}
\resset{e2.R.n}{4}
\resset{e2.R.n_sig}{0}
\resset{e2.R.n_sig_neg}{0}
\resset{e2.R.n_sig_pos}{0}
\resset{e2.S.max}{+0.0550}
\resset{e2.S.min}{-0.0374}
\resset{e2.S.n}{12}
\resset{e2.S.n_sig}{0}
\resset{e2.S.n_sig_neg}{0}
\resset{e2.S.n_sig_pos}{0}
\resset{e2.S1.clm_a0.a}{0.6848}
\resset{e2.S1.clm_a0.b}{0.6379}
\resset{e2.S1.clm_a0.ci}{[-0.0234, +0.1116]}
\resset{e2.S1.clm_a0.d}{+0.0470}
\resset{e2.S1.clm_a0.p}{0.492}
\resset{e2.S1.clm_a1.a}{0.9183}
\resset{e2.S1.clm_a1.b}{0.8762}
\resset{e2.S1.clm_a1.ci}{[-0.0010, +0.0903]}
\resset{e2.S1.clm_a1.d}{+0.0421}
\resset{e2.S1.clm_a1.p}{0.165}
\resset{e2.S1.clm_a2.a}{0.9630}
\resset{e2.S1.clm_a2.b}{0.9524}
\resset{e2.S1.clm_a2.ci}{[-0.0022, +0.0292]}
\resset{e2.S1.clm_a2.d}{+0.0106}
\resset{e2.S1.clm_a2.p}{0.372}
\resset{e2.S1.kev.a}{0.8984}
\resset{e2.S1.kev.b}{0.8833}
\resset{e2.S1.kev.ci}{[-0.0401, +0.0754]}
\resset{e2.S1.kev.d}{+0.0151}
\resset{e2.S1.kev.p}{0.574}
\resset{e2.S2.clm_a0.a}{0.6527}
\resset{e2.S2.clm_a0.b}{0.6446}
\resset{e2.S2.clm_a0.ci}{[-0.1690, +0.1665]}
\resset{e2.S2.clm_a0.d}{+0.0081}
\resset{e2.S2.clm_a0.p}{1.000}
\resset{e2.S2.clm_a1.a}{0.9196}
\resset{e2.S2.clm_a1.b}{0.8910}
\resset{e2.S2.clm_a1.ci}{[-0.0313, +0.0911]}
\resset{e2.S2.clm_a1.d}{+0.0286}
\resset{e2.S2.clm_a1.p}{0.714}
\resset{e2.S2.clm_a2.a}{0.9802}
\resset{e2.S2.clm_a2.b}{0.9696}
\resset{e2.S2.clm_a2.ci}{[-0.0047, +0.0294]}
\resset{e2.S2.clm_a2.d}{+0.0106}
\resset{e2.S2.clm_a2.p}{0.372}
\resset{e2.S2.kev.a}{0.9373}
\resset{e2.S2.kev.b}{0.8823}
\resset{e2.S2.kev.ci}{[-0.0060, +0.1236]}
\resset{e2.S2.kev.d}{+0.0550}
\resset{e2.S2.kev.p}{0.216}
\resset{e2.S3.clm_a0.a}{0.6858}
\resset{e2.S3.clm_a0.b}{0.6527}
\resset{e2.S3.clm_a0.ci}{[-0.0885, +0.1642]}
\resset{e2.S3.clm_a0.d}{+0.0331}
\resset{e2.S3.clm_a0.p}{1.000}
\resset{e2.S3.clm_a1.a}{0.9350}
\resset{e2.S3.clm_a1.b}{0.9196}
\resset{e2.S3.clm_a1.ci}{[-0.0343, +0.0626]}
\resset{e2.S3.clm_a1.d}{+0.0154}
\resset{e2.S3.clm_a1.p}{0.714}
\resset{e2.S3.clm_a2.a}{0.9754}
\resset{e2.S3.clm_a2.b}{0.9802}
\resset{e2.S3.clm_a2.ci}{[-0.0204, +0.0072]}
\resset{e2.S3.clm_a2.d}{-0.0048}
\resset{e2.S3.clm_a2.p}{0.488}
\resset{e2.S3.kev.a}{0.8999}
\resset{e2.S3.kev.b}{0.9373}
\resset{e2.S3.kev.ci}{[-0.0902, +0.0075]}
\resset{e2.S3.kev.d}{-0.0374}
\resset{e2.S3.kev.p}{0.222}
\resset{e2.T.max}{+0.0970}
\resset{e2.T.min}{+0.0055}
\resset{e2.T.n}{4}
\resset{e2.T.n_sig}{2}
\resset{e2.T.n_sig_neg}{0}
\resset{e2.T.n_sig_pos}{2}
\resset{e2.T1.clm_a0.a}{0.7349}
\resset{e2.T1.clm_a0.b}{0.6379}
\resset{e2.T1.clm_a0.ci}{[+0.0373, +0.1570]}
\resset{e2.T1.clm_a0.d}{+0.0970}
\resset{e2.T1.clm_a0.p}{0.002}
\resset{e2.T1.clm_a1.a}{0.9309}
\resset{e2.T1.clm_a1.b}{0.8762}
\resset{e2.T1.clm_a1.ci}{[+0.0057, +0.1010]}
\resset{e2.T1.clm_a1.d}{+0.0547}
\resset{e2.T1.clm_a1.p}{0.023}
\resset{e2.T1.clm_a2.a}{0.9578}
\resset{e2.T1.clm_a2.b}{0.9524}
\resset{e2.T1.clm_a2.ci}{[-0.0199, +0.0285]}
\resset{e2.T1.clm_a2.d}{+0.0055}
\resset{e2.T1.clm_a2.p}{0.585}
\resset{e2.T1.kev.a}{0.8908}
\resset{e2.T1.kev.b}{0.8833}
\resset{e2.T1.kev.ci}{[-0.0355, +0.0505]}
\resset{e2.T1.kev.d}{+0.0075}
\resset{e2.T1.kev.p}{0.770}
\resset{e2.a3.auroc_max}{0.9817}
\resset{e2.a3.auroc_min}{0.7246}
\resset{e2.a3.d_a2_max}{+0.0251}
\resset{e2.a3.d_a2_min}{-0.0094}
\resset{e2.a3.n}{30}
\resset{e2.a3.n_above_a2}{26}
\resset{e2.a3.n_sig_a1}{26}
\resset{e2.a3.n_sig_a2}{1}
\resset{e2.a3.sig_a2_d}{+0.0240}
\resset{e2.a3.sig_a2_level}{Imports}
\resset{e2.det.auroc}{0.9776}
\resset{e2.n_benign}{115}
\resset{e2.n_malware}{119}
\resset{e3.D1.clm.a}{0.8756}
\resset{e3.D1.clm.b}{0.9776}
\resset{e3.D1.clm.ci}{[-0.1520, -0.0589]}
\resset{e3.D1.clm.d}{-0.1019}
\resset{e3.D1.clm.p}{$<0.001$}
\resset{e3.D1.kev.a}{0.8868}
\resset{e3.D1.kev.b}{0.9776}
\resset{e3.D1.kev.ci}{[-0.1650, -0.0344]}
\resset{e3.D1.kev.d}{-0.0908}
\resset{e3.D1.kev.p}{0.003}
\resset{e3.D1.max}{-0.0343}
\resset{e3.D1.min}{-0.1667}
\resset{e3.D1.n}{13}
\resset{e3.D1.n_sig}{12}
\resset{e3.D1.n_sig_neg}{12}
\resset{e3.D1.n_sig_pos}{0}
\resset{e3.G.max}{+0.0488}
\resset{e3.G.min}{-0.2604}
\resset{e3.G.n}{33}
\resset{e3.G.n_sig}{9}
\resset{e3.G.n_sig_neg}{5}
\resset{e3.G.n_sig_pos}{4}
\resset{e3.K1.clm.a}{0.9516}
\resset{e3.K1.clm.b}{0.8756}
\resset{e3.K1.clm.ci}{[+0.0342, +0.1237]}
\resset{e3.K1.clm.d}{+0.0759}
\resset{e3.K1.clm.p}{0.003}
\resset{e3.K1.kev.a}{0.9532}
\resset{e3.K1.kev.b}{0.8868}
\resset{e3.K1.kev.ci}{[+0.0227, +0.1187]}
\resset{e3.K1.kev.d}{+0.0664}
\resset{e3.K1.kev.p}{0.004}
\resset{e3.K1.max}{+0.0798}
\resset{e3.K1.min}{-0.3889}
\resset{e3.K1.n}{13}
\resset{e3.K1.n_sig}{9}
\resset{e3.K1.n_sig_neg}{1}
\resset{e3.K1.n_sig_pos}{8}
\resset{e3.K2.clm.a}{0.6878}
\resset{e3.K2.clm.b}{0.8756}
\resset{e3.K2.clm.ci}{[-0.3199, -0.0713]}
\resset{e3.K2.clm.d}{-0.1878}
\resset{e3.K2.clm.p}{0.004}
\resset{e3.K2.kev.a}{0.4508}
\resset{e3.K2.kev.b}{0.8868}
\resset{e3.K2.kev.ci}{[-0.5555, -0.3266]}
\resset{e3.K2.kev.d}{-0.4360}
\resset{e3.K2.kev.p}{0.003}
\resset{e3.K2.max}{-0.1878}
\resset{e3.K2.min}{-0.6667}
\resset{e3.K2.n}{13}
\resset{e3.K2.n_sig}{13}
\resset{e3.K2.n_sig_neg}{13}
\resset{e3.K2.n_sig_pos}{0}
\resset{e3.N.clm.all}{0.8594}
\resset{e3.N.clm.k1}{0.7053}
\resset{e3.N.clm.k1.p}{0.005}
\resset{e3.N.clm.k16}{0.8237}
\resset{e3.N.clm.k2}{0.7609}
\resset{e3.N.clm.k4}{0.7529}
\resset{e3.N.clm.k8}{0.7731}
\resset{e3.N.k1.n_sig}{10}
\resset{e3.N.k16.n_sig}{4}
\resset{e3.N.k2.n_sig}{5}
\resset{e3.N.k4.n_sig}{5}
\resset{e3.N.k8.n_sig}{4}
\resset{e3.N.n_models}{13}
\resset{e3.P.max}{+0.3216}
\resset{e3.P.min}{-0.0844}
\resset{e3.P.n}{26}
\resset{e3.P.n_sig}{2}
\resset{e3.P.n_sig_neg}{0}
\resset{e3.P.n_sig_pos}{2}
\resset{e3.V1.clm.a}{0.8756}
\resset{e3.V1.clm.b}{0.6379}
\resset{e3.V1.clm.ci}{[+0.1386, +0.3385]}
\resset{e3.V1.clm.d}{+0.2378}
\resset{e3.V1.clm.p}{$<0.001$}
\resset{e3.V1.kev.a}{0.8868}
\resset{e3.V1.kev.b}{0.5505}
\resset{e3.V1.kev.ci}{[+0.2440, +0.4419]}
\resset{e3.V1.kev.d}{+0.3364}
\resset{e3.V1.kev.p}{0.002}
\resset{e3.V1.max}{+0.5082}
\resset{e3.V1.min}{+0.0556}
\resset{e3.V1.n}{13}
\resset{e3.V1.n_sig}{12}
\resset{e3.V1.n_sig_neg}{0}
\resset{e3.V1.n_sig_pos}{12}
\resset{e3.V1.ns_models}{Laya multilingual 322M}
\resset{e3.W.max}{+0.0554}
\resset{e3.W.min}{-0.0926}
\resset{e3.W.n}{39}
\resset{e3.W.n_sig}{2}
\resset{e3.W.n_sig_neg}{1}
\resset{e3.W.n_sig_pos}{1}
\resset{e3.W1.clm.a}{0.9310}
\resset{e3.W1.clm.b}{0.8756}
\resset{e3.W1.clm.ci}{[+0.0223, +0.0990]}
\resset{e3.W1.clm.d}{+0.0554}
\resset{e3.W1.clm.p}{0.006}
\resset{e3.W1.kev.a}{0.8926}
\resset{e3.W1.kev.b}{0.8868}
\resset{e3.W1.kev.ci}{[-0.0644, +0.0860]}
\resset{e3.W1.kev.d}{+0.0058}
\resset{e3.W1.kev.p}{1.000}
\resset{e3.W2.clm.a}{0.8990}
\resset{e3.W2.clm.b}{0.8756}
\resset{e3.W2.clm.ci}{[-0.0362, +0.0769]}
\resset{e3.W2.clm.d}{+0.0234}
\resset{e3.W2.clm.p}{0.418}
\resset{e3.W2.kev.a}{0.9057}
\resset{e3.W2.kev.b}{0.8868}
\resset{e3.W2.kev.ci}{[-0.0426, +0.0877]}
\resset{e3.W2.kev.d}{+0.0189}
\resset{e3.W2.kev.p}{1.000}
\resset{e3.X.max}{+0.2904}
\resset{e3.X.min}{-0.3333}
\resset{e3.X.n}{39}
\resset{e3.X.n_sig}{8}
\resset{e3.X.n_sig_neg}{0}
\resset{e3.X.n_sig_pos}{8}
\resset{e3.a3.auroc_max}{0.9714}
\resset{e3.a3.auroc_min}{0.9478}
\resset{e3.a3.d_a2_max}{+0.0190}
\resset{e3.a3.d_a2_min}{-0.0045}
\resset{e3.a3.n}{6}
\resset{e3.a3.n_above_a2}{5}
\resset{e3.a3.n_sig_a1}{5}
\resset{e3.a3.n_sig_a2}{0}
\resset{e3.det.auroc}{0.9776}
\resset{e3.n_benign}{115}
\resset{e3.n_malware}{119}
\resset{e3.n_models}{13}
\resset{e4.a0.n}{5}
\resset{e4.a0.n_infeasible}{3}
\resset{e4.band.risk_max}{1.49\%}
\resset{e4.band.risk_min}{1.32\%}
\resset{e4.clm_a0.brier}{0.2060}
\resset{e4.clm_a0.cover}{0 (infeasible)}
\resset{e4.clm_a0.ece_cal}{0.0479}
\resset{e4.clm_a0.ece_raw}{0.3691}
\resset{e4.clm_a0.improve}{0.3212}
\resset{e4.clm_a0.improve_ci}{[0.3124, 0.3282]}
\resset{e4.clm_a0.method}{isotonic}
\resset{e4.clm_a0.risk}{-}
\resset{e4.clm_a2.brier}{0.0262}
\resset{e4.clm_a2.cover}{92.36\%}
\resset{e4.clm_a2.ece_cal}{0.0060}
\resset{e4.clm_a2.ece_raw}{0.0124}
\resset{e4.clm_a2.improve}{0.0064}
\resset{e4.clm_a2.improve_ci}{[0.0027, 0.0071]}
\resset{e4.clm_a2.method}{Platt}
\resset{e4.clm_a2.risk}{1.32\%}
\resset{e4.clm_a3.cover}{93.98\%}
\resset{e4.clm_a3.ece_cal}{0.0078}
\resset{e4.clm_a3.ece_raw}{0.0129}
\resset{e4.clm_a3.improve}{0.0051}
\resset{e4.clm_a3.improve_ci}{[0.0008, 0.0068]}
\resset{e4.clm_a3.improve_p}{0.018}
\resset{e4.clm_a3.method}{Platt}
\resset{e4.clm_a3.risk}{1.56\%}
\resset{e4.det.brier}{0.0310}
\resset{e4.det.cover}{74.81\%}
\resset{e4.det.ece_cal}{0.0139}
\resset{e4.det.ece_raw}{0.0184}
\resset{e4.det.improve}{0.0044}
\resset{e4.det.improve_ci}{[0.0007, 0.0072]}
\resset{e4.det.method}{Platt}
\resset{e4.det.risk}{1.49\%}
\resset{e4.llm_fsec_a0.brier}{0.1537}
\resset{e4.llm_fsec_a0.cover}{16.01\%}
\resset{e4.llm_fsec_a0.ece_cal}{0.0316}
\resset{e4.llm_fsec_a0.ece_raw}{0.1542}
\resset{e4.llm_fsec_a0.improve}{0.1226}
\resset{e4.llm_fsec_a0.improve_ci}{[0.1143, 0.1287]}
\resset{e4.llm_fsec_a0.method}{Platt}
\resset{e4.llm_fsec_a0.risk}{3.10\%}
\resset{e4.llm_qwen3_a0.brier}{0.1833}
\resset{e4.llm_qwen3_a0.cover}{0 (infeasible)}
\resset{e4.llm_qwen3_a0.ece_cal}{0.0164}
\resset{e4.llm_qwen3_a0.ece_raw}{0.3025}
\resset{e4.llm_qwen3_a0.improve}{0.2861}
\resset{e4.llm_qwen3_a0.improve_ci}{[0.2747, 0.2950]}
\resset{e4.llm_qwen3_a0.method}{isotonic}
\resset{e4.llm_qwen3_a0.risk}{-}
\resset{e4.open_decider-2b_a0.brier}{0.2141}
\resset{e4.open_decider-2b_a0.cover}{4.60\%}
\resset{e4.open_decider-2b_a0.ece_cal}{0.0642}
\resset{e4.open_decider-2b_a0.ece_raw}{0.1620}
\resset{e4.open_decider-2b_a0.improve}{0.0978}
\resset{e4.open_decider-2b_a0.improve_ci}{[0.0860, 0.1076]}
\resset{e4.open_decider-2b_a0.method}{isotonic}
\resset{e4.open_decider-2b_a0.risk}{3.63\%}
\resset{e4.open_kev-0.8b_a0.brier}{0.2194}
\resset{e4.open_kev-0.8b_a0.cover}{0 (infeasible)}
\resset{e4.open_kev-0.8b_a0.ece_cal}{0.0719}
\resset{e4.open_kev-0.8b_a0.ece_raw}{0.1084}
\resset{e4.open_kev-0.8b_a0.improve}{0.0366}
\resset{e4.open_kev-0.8b_a0.improve_ci}{[0.0290, 0.0443]}
\resset{e4.open_kev-0.8b_a0.method}{isotonic}
\resset{e4.open_kev-0.8b_a0.risk}{-}
\resset{e4.primary.n}{7}
\resset{e4.primary.n_sig}{7}
\resset{e4.primary.pmax}{0.018}
\resset{e4.rc.clm_a2.max}{4.58\%}
\resset{e5.E5-C1.cos}{1.0000}
\resset{e5.E5-C1.eq.auroc}{equivalent}
\resset{e5.E5-C1.eq.auroc.margin}{0.00\%}
\resset{e5.E5-C1.eq.fpr}{equivalent}
\resset{e5.E5-C1.eq.fpr.margin}{0.58\%}
\resset{e5.E5-C1.eq.tpr}{equivalent}
\resset{e5.E5-C1.eq.tpr.margin}{0.49\%}
\resset{e5.E5-C1.flip}{0.00\%}
\resset{e5.E5-C1.flip_ci}{[0.00\%, 0.36\%]}
\resset{e5.E5-C1.kappa}{1.000}
\resset{e5.E5-C1.kappa_ci}{[1.000, 1.000]}
\resset{e5.E5-C1.margin_pp}{1}
\resset{e5.E5-C1.n}{1,013}
\resset{e5.E5-C1.n_flips}{0}
\resset{e5.E5-C1.pairs_needed}{1,083}
\resset{e5.E5-C1.ratio}{1.00}
\resset{e5.E5-C1.ratio_ci}{[0.99, 1.00]}
\resset{e5.E5-C2.cos}{0.9988}
\resset{e5.E5-C2.flip}{4.34\%}
\resset{e5.E5-C2.flip_ci}{[3.17\%, 5.79\%]}
\resset{e5.E5-C2.kappa}{0.892}
\resset{e5.E5-C2.kappa_ci}{[0.861, 0.923]}
\resset{e5.E5-C2.n}{1,013}
\resset{e5.E5-C2.n_flips}{44}
\resset{e5.E5-C2.ratio}{0.95}
\resset{e5.E5-C2.ratio_ci}{[0.95, 0.95]}
\resset{e5.E5-C3.cos}{0.9957}
\resset{e5.E5-C3.flip}{8.39\%}
\resset{e5.E5-C3.flip_ci}{[6.76\%, 10.27\%]}
\resset{e5.E5-C3.kappa}{0.795}
\resset{e5.E5-C3.kappa_ci}{[0.754, 0.837]}
\resset{e5.E5-C3.n}{1,013}
\resset{e5.E5-C3.n_flips}{85}
\resset{e5.E5-C3.ratio}{1.03}
\resset{e5.E5-C3.ratio_ci}{[1.03, 1.03]}
\resset{e5.E5-C6.cos}{0.9985}
\resset{e5.E5-C6.flip}{21.42\%}
\resset{e5.E5-C6.flip_ci}{[18.93\%, 24.08\%]}
\resset{e5.E5-C6.kappa}{0.571}
\resset{e5.E5-C6.kappa_ci}{[0.520, 0.621]}
\resset{e5.E5-C6.n}{1,013}
\resset{e5.E5-C6.n_flips}{217}
\resset{e5.E5-C6.ratio}{3.14}
\resset{e5.E5-C6.ratio_ci}{[3.10, 3.20]}
\resset{e5.E5-C7.cos}{0.9995}
\resset{e5.E5-C7.flip}{8.69\%}
\resset{e5.E5-C7.flip_ci}{[7.03\%, 10.59\%]}
\resset{e5.E5-C7.kappa}{0.796}
\resset{e5.E5-C7.kappa_ci}{[0.755, 0.837]}
\resset{e5.E5-C7.n}{1,013}
\resset{e5.E5-C7.n_flips}{88}
\resset{e5.E5-C7.ratio}{3.23}
\resset{e5.E5-C7.ratio_ci}{[3.18, 3.27]}
\resset{e5.E5-C8.cos}{0.9985}
\resset{e5.E5-C8.flip}{16.88\%}
\resset{e5.E5-C8.flip_ci}{[14.62\%, 19.33\%]}
\resset{e5.E5-C8.kappa}{0.662}
\resset{e5.E5-C8.kappa_ci}{[0.616, 0.708]}
\resset{e5.E5-C8.n}{1,013}
\resset{e5.E5-C8.n_flips}{171}
\resset{e5.E5-C8.ratio}{0.97}
\resset{e5.E5-C8.ratio_ci}{[0.97, 0.98]}
\resset{e5.bench.tps.2048}{2,099}
\resset{e5.bench.tps.512}{2,165}
\resset{e5.edge.auroc}{0.9827}
\resset{e5.edge.energy}{8.08}
\resset{e5.edge.fpr}{0.22\%}
\resset{e5.edge.fps}{2.68}
\resset{e5.edge.gpu}{RTX 4060 Laptop}
\resset{e5.edge.lat}{0.359}
\resset{e5.edge.power}{78}
\resset{e5.edge.t_encoder}{0.357}
\resset{e5.edge.t_heads_ms}{1.0}
\resset{e5.edge.t_state}{0.227}
\resset{e5.edge.t_tokenize_ms}{0.8}
\resset{e5.edge.tpr}{52.19\%}
\resset{e5.edge.vram}{6.7}
\resset{e5.inproc.auroc}{0.9827}
\resset{e5.inproc.energy}{8.12}
\resset{e5.inproc.fpr}{0.22\%}
\resset{e5.inproc.fps}{2.67}
\resset{e5.inproc.gpu}{RTX 4060 Laptop}
\resset{e5.inproc.lat}{0.360}
\resset{e5.inproc.power}{78}
\resset{e5.inproc.tpr}{52.19\%}
\resset{e5.inproc.vram}{6.6}
\resset{e5.n_benign}{465}
\resset{e5.n_files}{1,013}
\resset{e5.n_malware}{548}
\resset{e5.n_passes}{5}
\resset{e5.n_warmup}{20}
\resset{e5.open.auroc_max}{0.7072}
\resset{e5.open.auroc_min}{0.4239}
\resset{e5.open.lat_max}{0.265}
\resset{e5.open.lat_min}{0.020}
\resset{e5.open.n}{10}
\resset{e5.open.vram_max}{6.7}
\resset{e5.open.vram_min}{1.6}
\resset{e5.q4.auroc}{0.9788}
\resset{e5.q4.energy}{7.82}
\resset{e5.q4.fpr}{0.86\%}
\resset{e5.q4.fps}{2.75}
\resset{e5.q4.gpu}{RTX 4060 Laptop}
\resset{e5.q4.lat}{0.349}
\resset{e5.q4.power}{77}
\resset{e5.q4.tpr}{53.28\%}
\resset{e5.q4.vram}{5.8}
\resset{e5.q6.auroc}{0.9859}
\resset{e5.q6.energy}{8.55}
\resset{e5.q6.fpr}{0.43\%}
\resset{e5.q6.fps}{2.53}
\resset{e5.q6.gpu}{RTX 4060 Laptop}
\resset{e5.q6.lat}{0.378}
\resset{e5.q6.power}{78}
\resset{e5.q6.tpr}{50.55\%}
\resset{e5.q6.vram}{7.2}
\resset{e5.ref.auroc}{0.9922}
\resset{e5.ref.energy}{9.26}
\resset{e5.ref.fpr}{1.94\%}
\resset{e5.ref.fps}{8.34}
\resset{e5.ref.gpu}{RTX 5090}
\resset{e5.ref.lat}{0.114}
\resset{e5.ref.power}{278}
\resset{e5.ref.tpr}{90.33\%}
\resset{e5.ref.vram}{29.0}
\resset{e5.ref_q5.auroc}{0.9859}
\resset{e5.ref_q5.energy}{8.77}
\resset{e5.ref_q5.fpr}{0.43\%}
\resset{e5.ref_q5.fps}{8.54}
\resset{e5.ref_q5.gpu}{RTX 5090}
\resset{e5.ref_q5.lat}{0.112}
\resset{e5.ref_q5.power}{270}
\resset{e5.ref_q5.tpr}{60.40\%}
\resset{e5.ref_q5.vram}{6.9}
\resset{e5.repeat.max_flip}{0.00\%}
\resset{e5.repeat.n_configs}{25}
\resset{e5.s1mal.energy}{21.85}
\resset{e5.s1mal.fps}{0.97}
\resset{e5.s1mal.lat}{1.013}
\resset{e5.s1mal.lat_p90}{1.271}
\resset{e5.s1mal.n_files}{1,013}
\resset{e5.s1mal.n_passes}{5}
\resset{e5.s1mal.t_atomic}{0.580}
\resset{e5.s1mal.t_trees_ms}{27.6}
\resset{e5.s1mal.vram}{7.3}
\resset{e5.s1mal.x_lat}{2.8}
\resset{e5.small.0.6b.auroc}{0.9881}
\resset{e5.small.0.6b.lat}{0.068}
\resset{e5.small.0.6b.tpr}{89.23\%}
\resset{e5.small.0.6b.vram}{2.1}
\resset{e5.small.1.7b.auroc}{0.9879}
\resset{e5.small.1.7b.lat}{0.114}
\resset{e5.small.1.7b.tpr}{80.84\%}
\resset{e5.small.1.7b.vram}{3.3}
\resset{e5.small.4b.auroc}{0.9949}
\resset{e5.small.4b.lat}{0.205}
\resset{e5.small.4b.tpr}{93.07\%}
\resset{e5.small.4b.vram}{5.6}
\resset{e6.a2e1.all.flip}{1.26\%}
\resset{e6.a2e1.all.unstable}{2.74\%}
\resset{e6.a2e1.benign.flip}{0.43\%}
\resset{e6.a2e1.benign.unstable}{0.88\%}
\resset{e6.a2e1.malware.flip}{2.35\%}
\resset{e6.a2e1.malware.unstable}{5.17\%}
\resset{e6.a3.all.flip}{1.64\%}
\resset{e6.a3.all.unstable}{3.46\%}
\resset{e6.a3.auroc_max}{0.9946}
\resset{e6.a3.auroc_min}{0.9938}
\resset{e6.a3.benign.flip}{0.51\%}
\resset{e6.a3.benign.unstable}{1.11\%}
\resset{e6.a3.malware.flip}{3.11\%}
\resset{e6.a3.malware.unstable}{6.53\%}
\resset{e6.a3.n_seeds}{5}
\resset{e6.a3.tpr_max}{93.47\%}
\resset{e6.a3.tpr_min}{90.01\%}
\resset{e6.ctrl.flip}{0.01\%}
\resset{e6.ctrl.unstable}{0.02\%}
\resset{e6.ctrl.unstable_hi}{0.07\%}
\resset{e6.early.tpr_max}{60.75\%}
\resset{e6.early.tpr_min}{50.10\%}
\resset{e6.flip.benign}{0.90\%}
\resset{e6.flip.malware}{23.13\%}
\resset{e6.flip.max}{18.85\%}
\resset{e6.flip.mean}{10.55\%}
\resset{e6.flip.min}{0.89\%}
\resset{e6.flip.vs0}{8.79\%}
\resset{e6.heads.batch}{32}
\resset{e6.heads.early_epoch}{4}
\resset{e6.heads.early_seeds}{3, 4}
\resset{e6.heads.early_val_auroc_max}{0.9887}
\resset{e6.heads.late_val_auroc_min}{0.9961}
\resset{e6.heads.lr}{0.0005}
\resset{e6.heads.n_early}{2}
\resset{e6.heads.n_seeds}{5}
\resset{e6.late.tpr_max}{90.73\%}
\resset{e6.late.tpr_min}{88.75\%}
\resset{e6.local.flip_hi}{0.20\%}
\resset{e6.local.max_flip}{0.00\%}
\resset{e6.local.min_identical}{100.00\%}
\resset{e6.local.n_files}{1,831}
\resset{e6.local.n_pairs}{12}
\resset{e6.n_benign}{6,126}
\resset{e6.n_files}{10,829}
\resset{e6.n_malware}{4,703}
\resset{e6.n_seeds}{5}
\resset{e6.psd.mean}{0.0299}
\resset{e6.psd.p95}{0.1497}
\resset{e6.sampled.fsec.flip}{15.16\%}
\resset{e6.sampled.fsec.n}{835}
\resset{e6.sampled.fsec.unstable}{45.27\%}
\resset{e6.sampled.qwen3.flip}{1.88\%}
\resset{e6.sampled.qwen3.n}{928}
\resset{e6.sampled.qwen3.unstable}{4.85\%}
\resset{e6.seed0.tpr}{90.73\%}
\resset{e6.seed1.tpr}{90.60\%}
\resset{e6.seed2.tpr}{88.75\%}
\resset{e6.seed3.tpr}{60.75\%}
\resset{e6.seed4.tpr}{50.10\%}
\resset{e6.unstable}{19.51\%}
\resset{e6.unstable.benign}{1.67\%}
\resset{e6.unstable.ci}{[18.77\%, 20.27\%]}
\resset{e6.unstable.malware}{42.76\%}
\resset{e6.unstable.n}{2,113}
\resset{e7.adversarial.max}{19.81\%}
\resset{e7.adversarial.min}{0.00\%}
\resset{e7.all.max}{34.77\%}
\resset{e7.all.min}{0.05\%}
\resset{e7.clm.api_import.all.fab}{7.07\%}
\resset{e7.clm.api_import.all.recall}{94.15\%}
\resset{e7.clm.api_import.random.fab}{17.15\%}
\resset{e7.clm.api_import.random.recall}{94.15\%}
\resset{e7.clm.capability.all.fab}{0.00\%}
\resset{e7.clm.capability.all.recall}{95.25\%}
\resset{e7.clm.capability.random.fab}{0.00\%}
\resset{e7.clm.capability.random.recall}{95.25\%}
\resset{e7.clm_gguf_bf16.adversarial.conf}{0.00\%}
\resset{e7.clm_gguf_bf16.adversarial.fab}{1.10\%}
\resset{e7.clm_gguf_bf16.adversarial.fab_ci}{[0.64\%, 1.62\%]}
\resset{e7.clm_gguf_bf16.adversarial.n_neg}{1,721}
\resset{e7.clm_gguf_bf16.adversarial.recall}{94.54\%}
\resset{e7.clm_gguf_bf16.all.conf}{0.91\%}
\resset{e7.clm_gguf_bf16.all.fab}{8.86\%}
\resset{e7.clm_gguf_bf16.all.fab_ci}{[8.41\%, 9.37\%]}
\resset{e7.clm_gguf_bf16.all.n_neg}{10,609}
\resset{e7.clm_gguf_bf16.all.recall}{71.90\%}
\resset{e7.clm_gguf_bf16.popular.conf}{0.00\%}
\resset{e7.clm_gguf_bf16.popular.fab}{1.29\%}
\resset{e7.clm_gguf_bf16.popular.fab_ci}{[0.76\%, 1.87\%]}
\resset{e7.clm_gguf_bf16.popular.n_neg}{1,709}
\resset{e7.clm_gguf_bf16.popular.recall}{94.54\%}
\resset{e7.clm_gguf_bf16.random.conf}{0.18\%}
\resset{e7.clm_gguf_bf16.random.fab}{10.63\%}
\resset{e7.clm_gguf_bf16.random.fab_ci}{[9.23\%, 12.03\%]}
\resset{e7.clm_gguf_bf16.random.n_neg}{1,712}
\resset{e7.clm_gguf_bf16.random.recall}{94.54\%}
\resset{e7.clm_gguf_bf16.structural.conf}{1.72\%}
\resset{e7.clm_gguf_bf16.structural.fab}{13.12\%}
\resset{e7.clm_gguf_bf16.structural.fab_ci}{[12.39\%, 13.86\%]}
\resset{e7.clm_gguf_bf16.structural.n_neg}{5,467}
\resset{e7.clm_gguf_bf16.structural.recall}{39.25\%}
\resset{e7.d1-3b.adversarial.conf}{0.08\%}
\resset{e7.d1-3b.adversarial.fab}{1.17\%}
\resset{e7.d1-3b.adversarial.fab_ci}{[0.62\%, 1.79\%]}
\resset{e7.d1-3b.adversarial.n_neg}{1,284}
\resset{e7.d1-3b.adversarial.recall}{88.21\%}
\resset{e7.d1-3b.all.conf}{0.04\%}
\resset{e7.d1-3b.all.fab}{2.76\%}
\resset{e7.d1-3b.all.fab_ci}{[2.43\%, 3.12\%]}
\resset{e7.d1-3b.all.n_neg}{7,993}
\resset{e7.d1-3b.all.recall}{92.45\%}
\resset{e7.d1-3b.popular.conf}{0.00\%}
\resset{e7.d1-3b.popular.fab}{1.33\%}
\resset{e7.d1-3b.popular.fab_ci}{[0.71\%, 1.96\%]}
\resset{e7.d1-3b.popular.n_neg}{1,274}
\resset{e7.d1-3b.popular.recall}{88.21\%}
\resset{e7.d1-3b.random.conf}{0.08\%}
\resset{e7.d1-3b.random.fab}{0.24\%}
\resset{e7.d1-3b.random.fab_ci}{[0.00\%, 0.56\%]}
\resset{e7.d1-3b.random.n_neg}{1,259}
\resset{e7.d1-3b.random.recall}{88.21\%}
\resset{e7.d1-3b.structural.conf}{0.02\%}
\resset{e7.d1-3b.structural.fab}{4.45\%}
\resset{e7.d1-3b.structural.fab_ci}{[3.86\%, 5.07\%]}
\resset{e7.d1-3b.structural.n_neg}{4,176}
\resset{e7.d1-3b.structural.recall}{98.73\%}
\resset{e7.d1-omni-600m.adversarial.conf}{0.00\%}
\resset{e7.d1-omni-600m.adversarial.fab}{7.09\%}
\resset{e7.d1-omni-600m.adversarial.fab_ci}{[5.69\%, 8.42\%]}
\resset{e7.d1-omni-600m.adversarial.n_neg}{1,283}
\resset{e7.d1-omni-600m.adversarial.recall}{5.37\%}
\resset{e7.d1-omni-600m.all.conf}{0.00\%}
\resset{e7.d1-omni-600m.all.fab}{8.44\%}
\resset{e7.d1-omni-600m.all.fab_ci}{[7.60\%, 9.33\%]}
\resset{e7.d1-omni-600m.all.n_neg}{7,987}
\resset{e7.d1-omni-600m.all.recall}{8.54\%}
\resset{e7.d1-omni-600m.popular.conf}{0.00\%}
\resset{e7.d1-omni-600m.popular.fab}{7.54\%}
\resset{e7.d1-omni-600m.popular.fab_ci}{[6.21\%, 8.88\%]}
\resset{e7.d1-omni-600m.popular.n_neg}{1,273}
\resset{e7.d1-omni-600m.popular.recall}{5.37\%}
\resset{e7.d1-omni-600m.random.conf}{0.00\%}
\resset{e7.d1-omni-600m.random.fab}{4.29\%}
\resset{e7.d1-omni-600m.random.fab_ci}{[3.18\%, 5.48\%]}
\resset{e7.d1-omni-600m.random.n_neg}{1,258}
\resset{e7.d1-omni-600m.random.recall}{5.37\%}
\resset{e7.d1-omni-600m.structural.conf}{0.00\%}
\resset{e7.d1-omni-600m.structural.fab}{10.38\%}
\resset{e7.d1-omni-600m.structural.fab_ci}{[9.18\%, 11.58\%]}
\resset{e7.d1-omni-600m.structural.n_neg}{4,173}
\resset{e7.d1-omni-600m.structural.recall}{13.25\%}
\resset{e7.decider-0.8b.adversarial.conf}{0.00\%}
\resset{e7.decider-0.8b.adversarial.fab}{0.15\%}
\resset{e7.decider-0.8b.adversarial.fab_ci}{[0.00\%, 0.38\%]}
\resset{e7.decider-0.8b.adversarial.n_neg}{1,305}
\resset{e7.decider-0.8b.adversarial.recall}{73.16\%}
\resset{e7.decider-0.8b.all.conf}{0.04\%}
\resset{e7.decider-0.8b.all.fab}{7.22\%}
\resset{e7.decider-0.8b.all.fab_ci}{[6.78\%, 7.69\%]}
\resset{e7.decider-0.8b.all.n_neg}{8,107}
\resset{e7.decider-0.8b.all.recall}{83.42\%}
\resset{e7.decider-0.8b.popular.conf}{0.00\%}
\resset{e7.decider-0.8b.popular.fab}{0.15\%}
\resset{e7.decider-0.8b.popular.fab_ci}{[0.00\%, 0.39\%]}
\resset{e7.decider-0.8b.popular.n_neg}{1,297}
\resset{e7.decider-0.8b.popular.recall}{73.16\%}
\resset{e7.decider-0.8b.random.conf}{0.00\%}
\resset{e7.decider-0.8b.random.fab}{0.08\%}
\resset{e7.decider-0.8b.random.fab_ci}{[0.00\%, 0.23\%]}
\resset{e7.decider-0.8b.random.n_neg}{1,287}
\resset{e7.decider-0.8b.random.recall}{73.16\%}
\resset{e7.decider-0.8b.structural.conf}{0.07\%}
\resset{e7.decider-0.8b.structural.fab}{13.75\%}
\resset{e7.decider-0.8b.structural.fab_ci}{[12.91\%, 14.64\%]}
\resset{e7.decider-0.8b.structural.n_neg}{4,218}
\resset{e7.decider-0.8b.structural.recall}{98.63\%}
\resset{e7.decider-2b.adversarial.conf}{0.15\%}
\resset{e7.decider-2b.adversarial.fab}{1.07\%}
\resset{e7.decider-2b.adversarial.fab_ci}{[0.54\%, 1.61\%]}
\resset{e7.decider-2b.adversarial.n_neg}{1,305}
\resset{e7.decider-2b.adversarial.recall}{94.99\%}
\resset{e7.decider-2b.all.conf}{0.02\%}
\resset{e7.decider-2b.all.fab}{0.83\%}
\resset{e7.decider-2b.all.fab_ci}{[0.63\%, 1.03\%]}
\resset{e7.decider-2b.all.n_neg}{8,107}
\resset{e7.decider-2b.all.recall}{93.57\%}
\resset{e7.decider-2b.popular.conf}{0.00\%}
\resset{e7.decider-2b.popular.fab}{0.46\%}
\resset{e7.decider-2b.popular.fab_ci}{[0.15\%, 0.85\%]}
\resset{e7.decider-2b.popular.n_neg}{1,297}
\resset{e7.decider-2b.popular.recall}{94.99\%}
\resset{e7.decider-2b.random.conf}{0.00\%}
\resset{e7.decider-2b.random.fab}{0.16\%}
\resset{e7.decider-2b.random.fab_ci}{[0.00\%, 0.39\%]}
\resset{e7.decider-2b.random.n_neg}{1,287}
\resset{e7.decider-2b.random.recall}{94.99\%}
\resset{e7.decider-2b.structural.conf}{0.00\%}
\resset{e7.decider-2b.structural.fab}{1.07\%}
\resset{e7.decider-2b.structural.fab_ci}{[0.78\%, 1.38\%]}
\resset{e7.decider-2b.structural.n_neg}{4,218}
\resset{e7.decider-2b.structural.recall}{91.46\%}
\resset{e7.decider-4b-gguf.adversarial.conf}{0.00\%}
\resset{e7.decider-4b-gguf.adversarial.fab}{0.00\%}
\resset{e7.decider-4b-gguf.adversarial.fab_ci}{[0.00\%, 0.00\%]}
\resset{e7.decider-4b-gguf.adversarial.n_neg}{1,305}
\resset{e7.decider-4b-gguf.adversarial.recall}{93.24\%}
\resset{e7.decider-4b-gguf.all.conf}{0.02\%}
\resset{e7.decider-4b-gguf.all.fab}{2.70\%}
\resset{e7.decider-4b-gguf.all.fab_ci}{[2.38\%, 3.05\%]}
\resset{e7.decider-4b-gguf.all.n_neg}{8,107}
\resset{e7.decider-4b-gguf.all.recall}{95.39\%}
\resset{e7.decider-4b-gguf.popular.conf}{0.15\%}
\resset{e7.decider-4b-gguf.popular.fab}{0.15\%}
\resset{e7.decider-4b-gguf.popular.fab_ci}{[0.00\%, 0.39\%]}
\resset{e7.decider-4b-gguf.popular.n_neg}{1,297}
\resset{e7.decider-4b-gguf.popular.recall}{93.24\%}
\resset{e7.decider-4b-gguf.random.conf}{0.00\%}
\resset{e7.decider-4b-gguf.random.fab}{0.00\%}
\resset{e7.decider-4b-gguf.random.fab_ci}{[0.00\%, 0.00\%]}
\resset{e7.decider-4b-gguf.random.n_neg}{1,287}
\resset{e7.decider-4b-gguf.random.recall}{93.24\%}
\resset{e7.decider-4b-gguf.structural.conf}{0.00\%}
\resset{e7.decider-4b-gguf.structural.fab}{5.14\%}
\resset{e7.decider-4b-gguf.structural.fab_ci}{[4.54\%, 5.80\%]}
\resset{e7.decider-4b-gguf.structural.n_neg}{4,218}
\resset{e7.decider-4b-gguf.structural.recall}{98.57\%}
\resset{e7.foundation-sec-1.1-8b-instruct-bf16.adversarial.conf}{0.00\%}
\resset{e7.foundation-sec-1.1-8b-instruct-bf16.adversarial.fab}{0.17\%}
\resset{e7.foundation-sec-1.1-8b-instruct-bf16.adversarial.fab_ci}{[0.00\%, 0.41\%]}
\resset{e7.foundation-sec-1.1-8b-instruct-bf16.adversarial.n_neg}{1,721}
\resset{e7.foundation-sec-1.1-8b-instruct-bf16.adversarial.recall}{41.87\%}
\resset{e7.foundation-sec-1.1-8b-instruct-bf16.all.conf}{0.00\%}
\resset{e7.foundation-sec-1.1-8b-instruct-bf16.all.fab}{0.05\%}
\resset{e7.foundation-sec-1.1-8b-instruct-bf16.all.fab_ci}{[0.01\%, 0.09\%]}
\resset{e7.foundation-sec-1.1-8b-instruct-bf16.all.n_neg}{10,609}
\resset{e7.foundation-sec-1.1-8b-instruct-bf16.all.recall}{61.21\%}
\resset{e7.foundation-sec-1.1-8b-instruct-bf16.popular.conf}{0.00\%}
\resset{e7.foundation-sec-1.1-8b-instruct-bf16.popular.fab}{0.12\%}
\resset{e7.foundation-sec-1.1-8b-instruct-bf16.popular.fab_ci}{[0.00\%, 0.29\%]}
\resset{e7.foundation-sec-1.1-8b-instruct-bf16.popular.n_neg}{1,709}
\resset{e7.foundation-sec-1.1-8b-instruct-bf16.popular.recall}{41.87\%}
\resset{e7.foundation-sec-1.1-8b-instruct-bf16.random.conf}{0.00\%}
\resset{e7.foundation-sec-1.1-8b-instruct-bf16.random.fab}{0.00\%}
\resset{e7.foundation-sec-1.1-8b-instruct-bf16.random.fab_ci}{[0.00\%, 0.00\%]}
\resset{e7.foundation-sec-1.1-8b-instruct-bf16.random.n_neg}{1,712}
\resset{e7.foundation-sec-1.1-8b-instruct-bf16.random.recall}{41.87\%}
\resset{e7.foundation-sec-1.1-8b-instruct-bf16.structural.conf}{0.00\%}
\resset{e7.foundation-sec-1.1-8b-instruct-bf16.structural.fab}{0.00\%}
\resset{e7.foundation-sec-1.1-8b-instruct-bf16.structural.fab_ci}{[0.00\%, 0.00\%]}
\resset{e7.foundation-sec-1.1-8b-instruct-bf16.structural.n_neg}{5,467}
\resset{e7.foundation-sec-1.1-8b-instruct-bf16.structural.recall}{89.11\%}
\resset{e7.kev-0.8b.adversarial.conf}{0.00\%}
\resset{e7.kev-0.8b.adversarial.fab}{1.84\%}
\resset{e7.kev-0.8b.adversarial.fab_ci}{[1.15\%, 2.61\%]}
\resset{e7.kev-0.8b.adversarial.n_neg}{1,305}
\resset{e7.kev-0.8b.adversarial.recall}{50.52\%}
\resset{e7.kev-0.8b.all.conf}{0.00\%}
\resset{e7.kev-0.8b.all.fab}{0.68\%}
\resset{e7.kev-0.8b.all.fab_ci}{[0.51\%, 0.87\%]}
\resset{e7.kev-0.8b.all.n_neg}{8,107}
\resset{e7.kev-0.8b.all.recall}{58.71\%}
\resset{e7.kev-0.8b.popular.conf}{0.00\%}
\resset{e7.kev-0.8b.popular.fab}{1.39\%}
\resset{e7.kev-0.8b.popular.fab_ci}{[0.77\%, 2.08\%]}
\resset{e7.kev-0.8b.popular.n_neg}{1,297}
\resset{e7.kev-0.8b.popular.recall}{50.52\%}
\resset{e7.kev-0.8b.random.conf}{0.00\%}
\resset{e7.kev-0.8b.random.fab}{0.85\%}
\resset{e7.kev-0.8b.random.fab_ci}{[0.39\%, 1.40\%]}
\resset{e7.kev-0.8b.random.n_neg}{1,287}
\resset{e7.kev-0.8b.random.recall}{50.52\%}
\resset{e7.kev-0.8b.structural.conf}{0.00\%}
\resset{e7.kev-0.8b.structural.fab}{0.05\%}
\resset{e7.kev-0.8b.structural.fab_ci}{[0.00\%, 0.12\%]}
\resset{e7.kev-0.8b.structural.n_neg}{4,218}
\resset{e7.kev-0.8b.structural.recall}{70.86\%}
\resset{e7.n_backends}{10}
\resset{e7.popular.max}{19.07\%}
\resset{e7.popular.min}{0.12\%}
\resset{e7.qwen3-8b-bf16.adversarial.conf}{0.06\%}
\resset{e7.qwen3-8b-bf16.adversarial.fab}{0.06\%}
\resset{e7.qwen3-8b-bf16.adversarial.fab_ci}{[0.00\%, 0.17\%]}
\resset{e7.qwen3-8b-bf16.adversarial.n_neg}{1,721}
\resset{e7.qwen3-8b-bf16.adversarial.recall}{93.34\%}
\resset{e7.qwen3-8b-bf16.all.conf}{0.09\%}
\resset{e7.qwen3-8b-bf16.all.fab}{0.11\%}
\resset{e7.qwen3-8b-bf16.all.fab_ci}{[0.06\%, 0.18\%]}
\resset{e7.qwen3-8b-bf16.all.n_neg}{10,609}
\resset{e7.qwen3-8b-bf16.all.recall}{94.91\%}
\resset{e7.qwen3-8b-bf16.popular.conf}{0.12\%}
\resset{e7.qwen3-8b-bf16.popular.fab}{0.18\%}
\resset{e7.qwen3-8b-bf16.popular.fab_ci}{[0.00\%, 0.41\%]}
\resset{e7.qwen3-8b-bf16.popular.n_neg}{1,709}
\resset{e7.qwen3-8b-bf16.popular.recall}{93.34\%}
\resset{e7.qwen3-8b-bf16.random.conf}{0.00\%}
\resset{e7.qwen3-8b-bf16.random.fab}{0.00\%}
\resset{e7.qwen3-8b-bf16.random.fab_ci}{[0.00\%, 0.00\%]}
\resset{e7.qwen3-8b-bf16.random.n_neg}{1,712}
\resset{e7.qwen3-8b-bf16.random.recall}{93.34\%}
\resset{e7.qwen3-8b-bf16.structural.conf}{0.13\%}
\resset{e7.qwen3-8b-bf16.structural.fab}{0.15\%}
\resset{e7.qwen3-8b-bf16.structural.fab_ci}{[0.05\%, 0.26\%]}
\resset{e7.qwen3-8b-bf16.structural.n_neg}{5,467}
\resset{e7.qwen3-8b-bf16.structural.recall}{97.17\%}
\resset{e7.random.max}{16.52\%}
\resset{e7.random.min}{0.00\%}
\resset{e7.structural.max}{49.85\%}
\resset{e7.structural.min}{0.00\%}
\resset{e7.von.adversarial.conf}{0.00\%}
\resset{e7.von.adversarial.fab}{19.81\%}
\resset{e7.von.adversarial.fab_ci}{[17.69\%, 21.97\%]}
\resset{e7.von.adversarial.n_neg}{1,383}
\resset{e7.von.adversarial.recall}{56.71\%}
\resset{e7.von.all.conf}{0.00\%}
\resset{e7.von.all.fab}{34.77\%}
\resset{e7.von.all.fab_ci}{[33.60\%, 36.05\%]}
\resset{e7.von.all.n_neg}{8,603}
\resset{e7.von.all.recall}{69.67\%}
\resset{e7.von.popular.conf}{0.00\%}
\resset{e7.von.popular.fab}{19.07\%}
\resset{e7.von.popular.fab_ci}{[17.03\%, 21.13\%]}
\resset{e7.von.popular.n_neg}{1,379}
\resset{e7.von.popular.recall}{56.71\%}
\resset{e7.von.random.conf}{0.00\%}
\resset{e7.von.random.fab}{16.52\%}
\resset{e7.von.random.fab_ci}{[14.61\%, 18.49\%]}
\resset{e7.von.random.n_neg}{1,374}
\resset{e7.von.random.recall}{56.71\%}
\resset{e7.von.structural.conf}{0.00\%}
\resset{e7.von.structural.fab}{49.85\%}
\resset{e7.von.structural.fab_ci}{[48.07\%, 51.79\%]}
\resset{e7.von.structural.n_neg}{4,467}
\resset{e7.von.structural.recall}{88.65\%}
\resset{e7i.a1.hw.max}{1.53\%}
\resset{e7i.a1.hw.min}{0.00\%}
\resset{e7i.a1.hw.n}{2}
\resset{e7i.a1.pad05.max}{27.29\%}
\resset{e7i.a1.pad05.min}{10.26\%}
\resset{e7i.a1.pad05.n}{13}
\resset{e7i.a1.pad10.max}{22.49\%}
\resset{e7i.a1.pad10.min}{3.17\%}
\resset{e7i.a1.pad10.n}{13}
\resset{e7i.a1.para.max}{24.16\%}
\resset{e7i.a1.para.min}{8.61\%}
\resset{e7i.a1.para.n}{13}
\resset{e7i.a1.rep.max}{7.11\%}
\resset{e7i.a1.rep.min}{0.00\%}
\resset{e7i.a1.rep.n}{13}
\resset{e7i.clm_a1.pad05}{14.08\%}
\resset{e7i.clm_a1.pad05_ci}{[11.89\%, 16.51\%]}
\resset{e7i.clm_a1.pad05_n}{916}
\resset{e7i.clm_a1.pad05_shift}{32.21\%}
\resset{e7i.clm_a1.pad10}{18.03\%}
\resset{e7i.clm_a1.pad10_ci}{[15.27\%, 21.06\%]}
\resset{e7i.clm_a1.pad10_n}{710}
\resset{e7i.clm_a1.pad10_shift}{37.46\%}
\resset{e7i.clm_a1.para}{23.40\%}
\resset{e7i.clm_a1.para_acc}{77.18\%}
\resset{e7i.clm_a1.para_acc_worst}{45.63\%}
\resset{e7i.clm_a1.para_ci}{[21.21\%, 25.60\%]}
\resset{e7i.clm_a1.para_conc}{52.29\%}
\resset{e7i.clm_a1.para_n}{916}
\resset{e7i.clm_a1.para_worst}{31.55\%}
\resset{e7i.clm_a1.rep}{1.86\%}
\resset{e7i.clm_a1.rep_ci}{[1.08\%, 2.95\%]}
\resset{e7i.clm_a1.rep_ident}{17.90\%}
\resset{e7i.clm_a2.pad05}{7.10\%}
\resset{e7i.clm_a2.pad05_ci}{[5.52\%, 8.96\%]}
\resset{e7i.clm_a2.pad05_n}{916}
\resset{e7i.clm_a2.pad05_shift}{15.39\%}
\resset{e7i.clm_a2.pad10}{11.52\%}
\resset{e7i.clm_a2.pad10_ci}{[9.27\%, 14.09\%]}
\resset{e7i.clm_a2.pad10_n}{712}
\resset{e7i.clm_a2.pad10_shift}{22.61\%}
\resset{e7i.clm_a2.para}{23.22\%}
\resset{e7i.clm_a2.para_acc}{95.74\%}
\resset{e7i.clm_a2.para_acc_worst}{54.48\%}
\resset{e7i.clm_a2.para_ci}{[21.82\%, 24.59\%]}
\resset{e7i.clm_a2.para_conc}{15.61\%}
\resset{e7i.clm_a2.para_n}{916}
\resset{e7i.clm_a2.para_worst}{48.91\%}
\resset{e7i.clm_a3.pad05}{6.00\%}
\resset{e7i.clm_a3.pad05_ci}{[4.55\%, 7.74\%]}
\resset{e7i.clm_a3.pad05_n}{916}
\resset{e7i.clm_a3.pad05_shift}{8.08\%}
\resset{e7i.clm_a3.pad10}{8.59\%}
\resset{e7i.clm_a3.pad10_ci}{[6.64\%, 10.90\%]}
\resset{e7i.clm_a3.pad10_n}{710}
\resset{e7i.clm_a3.pad10_shift}{11.27\%}
\resset{e7i.clm_a3.para}{18.81\%}
\resset{e7i.clm_a3.para_acc}{97.60\%}
\resset{e7i.clm_a3.para_acc_worst}{58.62\%}
\resset{e7i.clm_a3.para_ci}{[17.71\%, 19.93\%]}
\resset{e7i.clm_a3.para_conc}{19.32\%}
\resset{e7i.clm_a3.para_n}{916}
\resset{e7i.clm_a3.para_worst}{39.19\%}
\resset{e7i.clm_a3.rep}{0.11\%}
\resset{e7i.clm_a3.rep_ci}{[0.00\%, 0.61\%]}
\resset{e7i.clm_a3.rep_ident}{39.08\%}
\resset{e7i.clm_gguf_bf16.option}{0.00\%}
\resset{e7i.clm_gguf_bf16.option_ci}{[0.00\%, 0.40\%]}
\resset{e7i.clm_gguf_bf16.option_n}{916}
\resset{e7i.clm_gguf_bf16.pad05}{1.20\%}
\resset{e7i.clm_gguf_bf16.pad05_ci}{[0.60\%, 2.14\%]}
\resset{e7i.clm_gguf_bf16.pad05_n}{916}
\resset{e7i.clm_gguf_bf16.pad05_shift}{4.48\%}
\resset{e7i.clm_gguf_bf16.pad10}{1.40\%}
\resset{e7i.clm_gguf_bf16.pad10_ci}{[0.68\%, 2.57\%]}
\resset{e7i.clm_gguf_bf16.pad10_n}{712}
\resset{e7i.clm_gguf_bf16.pad10_shift}{8.01\%}
\resset{e7i.clm_gguf_bf16.para}{26.14\%}
\resset{e7i.clm_gguf_bf16.para_acc}{55.24\%}
\resset{e7i.clm_gguf_bf16.para_acc_worst}{54.37\%}
\resset{e7i.clm_gguf_bf16.para_ben}{40.11\%}
\resset{e7i.clm_gguf_bf16.para_ci}{[24.50\%, 27.74\%]}
\resset{e7i.clm_gguf_bf16.para_conc}{19.98\%}
\resset{e7i.clm_gguf_bf16.para_n}{916}
\resset{e7i.clm_gguf_bf16.para_worst}{71.51\%}
\resset{e7i.clm_gguf_bf16.polarity}{1.31\%}
\resset{e7i.clm_gguf_bf16.polarity_ci}{[0.68\%, 2.28\%]}
\resset{e7i.clm_gguf_bf16.polarity_n}{916}
\resset{e7i.clm_gguf_bf16.rep}{0.55\%}
\resset{e7i.clm_gguf_bf16.rep_ci}{[0.18\%, 1.27\%]}
\resset{e7i.clm_gguf_bf16.rep_ident}{0.00\%}
\resset{e7i.d1-3b.option}{35.97\%}
\resset{e7i.d1-3b.option_ci}{[32.40\%, 39.66\%]}
\resset{e7i.d1-3b.option_n}{695}
\resset{e7i.d1-3b.pad05}{11.22\%}
\resset{e7i.d1-3b.pad05_ci}{[8.61\%, 14.29\%]}
\resset{e7i.d1-3b.pad05_n}{508}
\resset{e7i.d1-3b.pad05_shift}{12.60\%}
\resset{e7i.d1-3b.pad10}{12.86\%}
\resset{e7i.d1-3b.pad10_ci}{[9.81\%, 16.44\%]}
\resset{e7i.d1-3b.pad10_n}{420}
\resset{e7i.d1-3b.pad10_shift}{15.95\%}
\resset{e7i.d1-3b.para}{15.51\%}
\resset{e7i.d1-3b.para_acc}{56.14\%}
\resset{e7i.d1-3b.para_acc_worst}{58.19\%}
\resset{e7i.d1-3b.para_ci}{[13.36\%, 17.60\%]}
\resset{e7i.d1-3b.para_conc}{65.06\%}
\resset{e7i.d1-3b.para_n}{684}
\resset{e7i.d1-3b.para_worst}{28.07\%}
\resset{e7i.d1-3b.polarity}{21.45\%}
\resset{e7i.d1-3b.polarity_ci}{[18.38\%, 24.78\%]}
\resset{e7i.d1-3b.polarity_n}{662}
\resset{e7i.d1-3b.rep}{0.00\%}
\resset{e7i.d1-3b.rep_ci}{[0.00\%, 0.54\%]}
\resset{e7i.d1-3b.rep_ident}{100.00\%}
\resset{e7i.d1-omni-600m.option}{4.32\%}
\resset{e7i.d1-omni-600m.option_ci}{[2.94\%, 6.11\%]}
\resset{e7i.d1-omni-600m.option_n}{694}
\resset{e7i.d1-omni-600m.pad05}{2.57\%}
\resset{e7i.d1-omni-600m.pad05_ci}{[1.37\%, 4.35\%]}
\resset{e7i.d1-omni-600m.pad05_n}{506}
\resset{e7i.d1-omni-600m.pad05_shift}{2.57\%}
\resset{e7i.d1-omni-600m.pad10}{3.57\%}
\resset{e7i.d1-omni-600m.pad10_ci}{[2.01\%, 5.82\%]}
\resset{e7i.d1-omni-600m.pad10_n}{420}
\resset{e7i.d1-omni-600m.pad10_shift}{9.29\%}
\resset{e7i.d1-omni-600m.para}{7.77\%}
\resset{e7i.d1-omni-600m.para_acc}{43.48\%}
\resset{e7i.d1-omni-600m.para_acc_worst}{42.75\%}
\resset{e7i.d1-omni-600m.para_ci}{[6.60\%, 8.99\%]}
\resset{e7i.d1-omni-600m.para_conc}{58.27\%}
\resset{e7i.d1-omni-600m.para_n}{683}
\resset{e7i.d1-omni-600m.para_worst}{34.55\%}
\resset{e7i.d1-omni-600m.polarity}{61.88\%}
\resset{e7i.d1-omni-600m.polarity_ci}{[58.05\%, 65.59\%]}
\resset{e7i.d1-omni-600m.polarity_n}{661}
\resset{e7i.d1-omni-600m.rep}{0.00\%}
\resset{e7i.d1-omni-600m.rep_ci}{[0.00\%, 0.54\%]}
\resset{e7i.d1-omni-600m.rep_ident}{100.00\%}
\resset{e7i.decider-0.8b.option}{5.98\%}
\resset{e7i.decider-0.8b.option_ci}{[4.35\%, 8.00\%]}
\resset{e7i.decider-0.8b.option_n}{702}
\resset{e7i.decider-0.8b.pad05}{3.90\%}
\resset{e7i.decider-0.8b.pad05_ci}{[2.40\%, 5.96\%]}
\resset{e7i.decider-0.8b.pad05_n}{513}
\resset{e7i.decider-0.8b.pad05_shift}{2.73\%}
\resset{e7i.decider-0.8b.pad10}{3.73\%}
\resset{e7i.decider-0.8b.pad10_ci}{[2.15\%, 5.99\%]}
\resset{e7i.decider-0.8b.pad10_n}{429}
\resset{e7i.decider-0.8b.pad10_shift}{0.93\%}
\resset{e7i.decider-0.8b.para}{19.97\%}
\resset{e7i.decider-0.8b.para_acc}{64.43\%}
\resset{e7i.decider-0.8b.para_acc_worst}{58.14\%}
\resset{e7i.decider-0.8b.para_ci}{[18.29\%, 21.67\%]}
\resset{e7i.decider-0.8b.para_conc}{38.86\%}
\resset{e7i.decider-0.8b.para_n}{700}
\resset{e7i.decider-0.8b.para_worst}{44.00\%}
\resset{e7i.decider-0.8b.polarity}{11.86\%}
\resset{e7i.decider-0.8b.polarity_ci}{[9.56\%, 14.49\%]}
\resset{e7i.decider-0.8b.polarity_n}{700}
\resset{e7i.decider-0.8b.rep}{1.43\%}
\resset{e7i.decider-0.8b.rep_ci}{[0.69\%, 2.61\%]}
\resset{e7i.decider-0.8b.rep_ident}{46.29\%}
\resset{e7i.decider-2b.option}{3.56\%}
\resset{e7i.decider-2b.option_ci}{[2.32\%, 5.21\%]}
\resset{e7i.decider-2b.option_n}{702}
\resset{e7i.decider-2b.pad05}{6.82\%}
\resset{e7i.decider-2b.pad05_ci}{[4.80\%, 9.36\%]}
\resset{e7i.decider-2b.pad05_n}{513}
\resset{e7i.decider-2b.pad05_shift}{10.92\%}
\resset{e7i.decider-2b.pad10}{14.45\%}
\resset{e7i.decider-2b.pad10_ci}{[11.26\%, 18.14\%]}
\resset{e7i.decider-2b.pad10_n}{429}
\resset{e7i.decider-2b.pad10_shift}{25.87\%}
\resset{e7i.decider-2b.para}{20.46\%}
\resset{e7i.decider-2b.para_acc}{68.57\%}
\resset{e7i.decider-2b.para_acc_worst}{49.43\%}
\resset{e7i.decider-2b.para_ci}{[18.43\%, 22.63\%]}
\resset{e7i.decider-2b.para_conc}{43.14\%}
\resset{e7i.decider-2b.para_n}{700}
\resset{e7i.decider-2b.para_worst}{31.14\%}
\resset{e7i.decider-2b.polarity}{59.14\%}
\resset{e7i.decider-2b.polarity_ci}{[55.40\%, 62.81\%]}
\resset{e7i.decider-2b.polarity_n}{700}
\resset{e7i.decider-2b.rep}{0.57\%}
\resset{e7i.decider-2b.rep_ci}{[0.16\%, 1.46\%]}
\resset{e7i.decider-2b.rep_ident}{47.71\%}
\resset{e7i.decider-4b-gguf.option}{1.99\%}
\resset{e7i.decider-4b-gguf.option_ci}{[1.09\%, 3.32\%]}
\resset{e7i.decider-4b-gguf.option_n}{702}
\resset{e7i.decider-4b-gguf.pad05}{3.90\%}
\resset{e7i.decider-4b-gguf.pad05_ci}{[2.40\%, 5.96\%]}
\resset{e7i.decider-4b-gguf.pad05_n}{513}
\resset{e7i.decider-4b-gguf.pad05_shift}{11.70\%}
\resset{e7i.decider-4b-gguf.pad10}{6.53\%}
\resset{e7i.decider-4b-gguf.pad10_ci}{[4.38\%, 9.30\%]}
\resset{e7i.decider-4b-gguf.pad10_n}{429}
\resset{e7i.decider-4b-gguf.pad10_shift}{15.15\%}
\resset{e7i.decider-4b-gguf.para}{5.06\%}
\resset{e7i.decider-4b-gguf.para_acc}{69.14\%}
\resset{e7i.decider-4b-gguf.para_acc_worst}{70.14\%}
\resset{e7i.decider-4b-gguf.para_ci}{[3.86\%, 6.37\%]}
\resset{e7i.decider-4b-gguf.para_conc}{88.14\%}
\resset{e7i.decider-4b-gguf.para_n}{700}
\resset{e7i.decider-4b-gguf.para_worst}{8.43\%}
\resset{e7i.decider-4b-gguf.polarity}{10.29\%}
\resset{e7i.decider-4b-gguf.polarity_ci}{[8.14\%, 12.78\%]}
\resset{e7i.decider-4b-gguf.polarity_n}{700}
\resset{e7i.decider-4b-gguf.rep}{0.00\%}
\resset{e7i.decider-4b-gguf.rep_ci}{[0.00\%, 0.53\%]}
\resset{e7i.decider-4b-gguf.rep_ident}{100.00\%}
\resset{e7i.foundation-sec-1.1-8b-instruct-bf16.option}{18.45\%}
\resset{e7i.foundation-sec-1.1-8b-instruct-bf16.option_ci}{[15.99\%, 21.12\%]}
\resset{e7i.foundation-sec-1.1-8b-instruct-bf16.option_n}{916}
\resset{e7i.foundation-sec-1.1-8b-instruct-bf16.pad05}{6.88\%}
\resset{e7i.foundation-sec-1.1-8b-instruct-bf16.pad05_ci}{[5.33\%, 8.71\%]}
\resset{e7i.foundation-sec-1.1-8b-instruct-bf16.pad05_n}{916}
\resset{e7i.foundation-sec-1.1-8b-instruct-bf16.pad05_shift}{23.36\%}
\resset{e7i.foundation-sec-1.1-8b-instruct-bf16.pad10}{6.12\%}
\resset{e7i.foundation-sec-1.1-8b-instruct-bf16.pad10_ci}{[4.66\%, 7.87\%]}
\resset{e7i.foundation-sec-1.1-8b-instruct-bf16.pad10_n}{915}
\resset{e7i.foundation-sec-1.1-8b-instruct-bf16.pad10_shift}{24.15\%}
\resset{e7i.foundation-sec-1.1-8b-instruct-bf16.para}{7.21\%}
\resset{e7i.foundation-sec-1.1-8b-instruct-bf16.para_acc}{75.66\%}
\resset{e7i.foundation-sec-1.1-8b-instruct-bf16.para_acc_worst}{67.79\%}
\resset{e7i.foundation-sec-1.1-8b-instruct-bf16.para_ci}{[6.24\%, 8.21\%]}
\resset{e7i.foundation-sec-1.1-8b-instruct-bf16.para_conc}{75.11\%}
\resset{e7i.foundation-sec-1.1-8b-instruct-bf16.para_n}{916}
\resset{e7i.foundation-sec-1.1-8b-instruct-bf16.para_worst}{14.41\%}
\resset{e7i.foundation-sec-1.1-8b-instruct-bf16.polarity}{30.68\%}
\resset{e7i.foundation-sec-1.1-8b-instruct-bf16.polarity_ci}{[27.70\%, 33.78\%]}
\resset{e7i.foundation-sec-1.1-8b-instruct-bf16.polarity_n}{916}
\resset{e7i.foundation-sec-1.1-8b-instruct-bf16.rep}{0.44\%}
\resset{e7i.foundation-sec-1.1-8b-instruct-bf16.rep_ci}{[0.12\%, 1.11\%]}
\resset{e7i.foundation-sec-1.1-8b-instruct-bf16.rep_ident}{0.55\%}
\resset{e7i.kev-0.8b.option}{8.55\%}
\resset{e7i.kev-0.8b.option_ci}{[6.59\%, 10.86\%]}
\resset{e7i.kev-0.8b.option_n}{702}
\resset{e7i.kev-0.8b.pad05}{3.70\%}
\resset{e7i.kev-0.8b.pad05_ci}{[2.24\%, 5.72\%]}
\resset{e7i.kev-0.8b.pad05_n}{513}
\resset{e7i.kev-0.8b.pad05_shift}{0.00\%}
\resset{e7i.kev-0.8b.pad10}{6.06\%}
\resset{e7i.kev-0.8b.pad10_ci}{[4.00\%, 8.75\%]}
\resset{e7i.kev-0.8b.pad10_n}{429}
\resset{e7i.kev-0.8b.pad10_shift}{0.00\%}
\resset{e7i.kev-0.8b.para}{24.10\%}
\resset{e7i.kev-0.8b.para_acc}{71.14\%}
\resset{e7i.kev-0.8b.para_acc_worst}{49.86\%}
\resset{e7i.kev-0.8b.para_ci}{[22.27\%, 26.09\%]}
\resset{e7i.kev-0.8b.para_conc}{20.43\%}
\resset{e7i.kev-0.8b.para_n}{700}
\resset{e7i.kev-0.8b.para_worst}{70.71\%}
\resset{e7i.kev-0.8b.polarity}{14.29\%}
\resset{e7i.kev-0.8b.polarity_ci}{[11.78\%, 17.10\%]}
\resset{e7i.kev-0.8b.polarity_n}{700}
\resset{e7i.kev-0.8b.rep}{0.71\%}
\resset{e7i.kev-0.8b.rep_ci}{[0.23\%, 1.66\%]}
\resset{e7i.kev-0.8b.rep_ident}{2.43\%}
\resset{e7i.llm_fsec_a1.pad05}{27.29\%}
\resset{e7i.llm_fsec_a1.pad05_ci}{[24.43\%, 30.30\%]}
\resset{e7i.llm_fsec_a1.pad05_n}{916}
\resset{e7i.llm_fsec_a1.pad05_shift}{34.93\%}
\resset{e7i.llm_fsec_a1.pad10}{21.64\%}
\resset{e7i.llm_fsec_a1.pad10_ci}{[19.01\%, 24.45\%]}
\resset{e7i.llm_fsec_a1.pad10_n}{915}
\resset{e7i.llm_fsec_a1.pad10_shift}{24.81\%}
\resset{e7i.llm_fsec_a1.para}{9.10\%}
\resset{e7i.llm_fsec_a1.para_acc}{78.93\%}
\resset{e7i.llm_fsec_a1.para_acc_worst}{74.13\%}
\resset{e7i.llm_fsec_a1.para_ci}{[7.78\%, 10.55\%]}
\resset{e7i.llm_fsec_a1.para_conc}{76.86\%}
\resset{e7i.llm_fsec_a1.para_n}{916}
\resset{e7i.llm_fsec_a1.para_worst}{11.90\%}
\resset{e7i.llm_fsec_a1.rep}{1.64\%}
\resset{e7i.llm_fsec_a1.rep_ci}{[0.92\%, 2.69\%]}
\resset{e7i.llm_fsec_a1.rep_ident}{30.46\%}
\resset{e7i.llm_qwen3_a1.pad05}{10.59\%}
\resset{e7i.llm_qwen3_a1.pad05_ci}{[8.67\%, 12.77\%]}
\resset{e7i.llm_qwen3_a1.pad05_n}{916}
\resset{e7i.llm_qwen3_a1.pad05_shift}{30.90\%}
\resset{e7i.llm_qwen3_a1.pad10}{11.34\%}
\resset{e7i.llm_qwen3_a1.pad10_ci}{[8.99\%, 14.04\%]}
\resset{e7i.llm_qwen3_a1.pad10_n}{644}
\resset{e7i.llm_qwen3_a1.pad10_shift}{26.40\%}
\resset{e7i.llm_qwen3_a1.para}{8.61\%}
\resset{e7i.llm_qwen3_a1.para_acc}{64.85\%}
\resset{e7i.llm_qwen3_a1.para_acc_worst}{58.84\%}
\resset{e7i.llm_qwen3_a1.para_ci}{[7.25\%, 10.00\%]}
\resset{e7i.llm_qwen3_a1.para_conc}{78.93\%}
\resset{e7i.llm_qwen3_a1.para_n}{916}
\resset{e7i.llm_qwen3_a1.para_worst}{10.59\%}
\resset{e7i.llm_qwen3_a1.rep}{1.20\%}
\resset{e7i.llm_qwen3_a1.rep_ci}{[0.60\%, 2.14\%]}
\resset{e7i.llm_qwen3_a1.rep_ident}{72.93\%}
\resset{e7i.n_backends}{10}
\resset{e7i.n_files}{916}
\resset{e7i.n_sub}{1,000}
\resset{e7i.open_d1-3b_a1.pad05}{15.07\%}
\resset{e7i.open_d1-3b_a1.pad05_ci}{[12.81\%, 17.55\%]}
\resset{e7i.open_d1-3b_a1.pad05_n}{916}
\resset{e7i.open_d1-3b_a1.pad05_shift}{16.05\%}
\resset{e7i.open_d1-3b_a1.pad10}{17.47\%}
\resset{e7i.open_d1-3b_a1.pad10_ci}{[15.06\%, 20.08\%]}
\resset{e7i.open_d1-3b_a1.pad10_n}{916}
\resset{e7i.open_d1-3b_a1.pad10_shift}{41.81\%}
\resset{e7i.open_d1-3b_a1.para}{22.51\%}
\resset{e7i.open_d1-3b_a1.para_acc}{80.90\%}
\resset{e7i.open_d1-3b_a1.para_acc_worst}{50.76\%}
\resset{e7i.open_d1-3b_a1.para_ci}{[20.30\%, 24.76\%]}
\resset{e7i.open_d1-3b_a1.para_conc}{65.72\%}
\resset{e7i.open_d1-3b_a1.para_n}{916}
\resset{e7i.open_d1-3b_a1.para_worst}{30.13\%}
\resset{e7i.open_d1-3b_a1.rep}{0.00\%}
\resset{e7i.open_d1-3b_a1.rep_ci}{[0.00\%, 0.40\%]}
\resset{e7i.open_d1-3b_a1.rep_ident}{100.00\%}
\resset{e7i.open_d1-omni-600m_a1.pad05}{10.26\%}
\resset{e7i.open_d1-omni-600m_a1.pad05_ci}{[8.37\%, 12.41\%]}
\resset{e7i.open_d1-omni-600m_a1.pad05_n}{916}
\resset{e7i.open_d1-omni-600m_a1.pad05_shift}{53.06\%}
\resset{e7i.open_d1-omni-600m_a1.pad10}{11.03\%}
\resset{e7i.open_d1-omni-600m_a1.pad10_ci}{[9.07\%, 13.24\%]}
\resset{e7i.open_d1-omni-600m_a1.pad10_n}{916}
\resset{e7i.open_d1-omni-600m_a1.pad10_shift}{55.57\%}
\resset{e7i.open_d1-omni-600m_a1.para}{16.44\%}
\resset{e7i.open_d1-omni-600m_a1.para_acc}{66.27\%}
\resset{e7i.open_d1-omni-600m_a1.para_acc_worst}{45.52\%}
\resset{e7i.open_d1-omni-600m_a1.para_ci}{[14.51\%, 18.34\%]}
\resset{e7i.open_d1-omni-600m_a1.para_conc}{69.76\%}
\resset{e7i.open_d1-omni-600m_a1.para_n}{916}
\resset{e7i.open_d1-omni-600m_a1.para_worst}{21.62\%}
\resset{e7i.open_d1-omni-600m_a1.rep}{0.00\%}
\resset{e7i.open_d1-omni-600m_a1.rep_ci}{[0.00\%, 0.40\%]}
\resset{e7i.open_d1-omni-600m_a1.rep_ident}{100.00\%}
\resset{e7i.open_decider-0.8b_a1.pad05}{15.07\%}
\resset{e7i.open_decider-0.8b_a1.pad05_ci}{[12.81\%, 17.55\%]}
\resset{e7i.open_decider-0.8b_a1.pad05_n}{916}
\resset{e7i.open_decider-0.8b_a1.pad05_shift}{23.80\%}
\resset{e7i.open_decider-0.8b_a1.pad10}{3.17\%}
\resset{e7i.open_decider-0.8b_a1.pad10_ci}{[2.13\%, 4.52\%]}
\resset{e7i.open_decider-0.8b_a1.pad10_n}{916}
\resset{e7i.open_decider-0.8b_a1.pad10_shift}{24.34\%}
\resset{e7i.open_decider-0.8b_a1.para}{16.48\%}
\resset{e7i.open_decider-0.8b_a1.para_acc}{74.89\%}
\resset{e7i.open_decider-0.8b_a1.para_acc_worst}{47.82\%}
\resset{e7i.open_decider-0.8b_a1.para_ci}{[14.75\%, 18.49\%]}
\resset{e7i.open_decider-0.8b_a1.para_conc}{65.61\%}
\resset{e7i.open_decider-0.8b_a1.para_n}{916}
\resset{e7i.open_decider-0.8b_a1.para_worst}{27.51\%}
\resset{e7i.open_decider-0.8b_a1.rep}{1.97\%}
\resset{e7i.open_decider-0.8b_a1.rep_ci}{[1.17\%, 3.09\%]}
\resset{e7i.open_decider-0.8b_a1.rep_ident}{8.30\%}
\resset{e7i.open_decider-2b_a1.pad05}{11.03\%}
\resset{e7i.open_decider-2b_a1.pad05_ci}{[9.07\%, 13.24\%]}
\resset{e7i.open_decider-2b_a1.pad05_n}{916}
\resset{e7i.open_decider-2b_a1.pad05_shift}{13.32\%}
\resset{e7i.open_decider-2b_a1.pad10}{12.01\%}
\resset{e7i.open_decider-2b_a1.pad10_ci}{[9.97\%, 14.29\%]}
\resset{e7i.open_decider-2b_a1.pad10_n}{916}
\resset{e7i.open_decider-2b_a1.pad10_shift}{21.07\%}
\resset{e7i.open_decider-2b_a1.para}{23.84\%}
\resset{e7i.open_decider-2b_a1.para_acc}{83.84\%}
\resset{e7i.open_decider-2b_a1.para_acc_worst}{45.63\%}
\resset{e7i.open_decider-2b_a1.para_ci}{[21.93\%, 25.80\%]}
\resset{e7i.open_decider-2b_a1.para_conc}{47.60\%}
\resset{e7i.open_decider-2b_a1.para_n}{916}
\resset{e7i.open_decider-2b_a1.para_worst}{38.43\%}
\resset{e7i.open_decider-2b_a1.rep}{1.09\%}
\resset{e7i.open_decider-2b_a1.rep_ci}{[0.52\%, 2.00\%]}
\resset{e7i.open_decider-2b_a1.rep_ident}{6.99\%}
\resset{e7i.open_decider-4b-gguf_a1.pad05}{20.63\%}
\resset{e7i.open_decider-4b-gguf_a1.pad05_ci}{[18.06\%, 23.40\%]}
\resset{e7i.open_decider-4b-gguf_a1.pad05_n}{916}
\resset{e7i.open_decider-4b-gguf_a1.pad05_shift}{19.00\%}
\resset{e7i.open_decider-4b-gguf_a1.pad10}{14.30\%}
\resset{e7i.open_decider-4b-gguf_a1.pad10_ci}{[12.10\%, 16.74\%]}
\resset{e7i.open_decider-4b-gguf_a1.pad10_n}{916}
\resset{e7i.open_decider-4b-gguf_a1.pad10_shift}{26.20\%}
\resset{e7i.open_decider-4b-gguf_a1.para}{17.38\%}
\resset{e7i.open_decider-4b-gguf_a1.para_acc}{86.68\%}
\resset{e7i.open_decider-4b-gguf_a1.para_acc_worst}{62.23\%}
\resset{e7i.open_decider-4b-gguf_a1.para_ci}{[15.41\%, 19.44\%]}
\resset{e7i.open_decider-4b-gguf_a1.para_conc}{66.81\%}
\resset{e7i.open_decider-4b-gguf_a1.para_n}{916}
\resset{e7i.open_decider-4b-gguf_a1.para_worst}{25.33\%}
\resset{e7i.open_decider-4b-gguf_a1.rep}{2.62\%}
\resset{e7i.open_decider-4b-gguf_a1.rep_ci}{[1.69\%, 3.87\%]}
\resset{e7i.open_decider-4b-gguf_a1.rep_ident}{1.20\%}
\resset{e7i.open_kev-0.8b_a1.hw}{1.53\%}
\resset{e7i.open_kev-0.8b_a1.hw_ci}{[0.84\%, 2.55\%]}
\resset{e7i.open_kev-0.8b_a1.pad05}{17.58\%}
\resset{e7i.open_kev-0.8b_a1.pad05_ci}{[15.16\%, 20.20\%]}
\resset{e7i.open_kev-0.8b_a1.pad05_n}{916}
\resset{e7i.open_kev-0.8b_a1.pad05_shift}{32.10\%}
\resset{e7i.open_kev-0.8b_a1.pad10}{15.39\%}
\resset{e7i.open_kev-0.8b_a1.pad10_ci}{[13.12\%, 17.90\%]}
\resset{e7i.open_kev-0.8b_a1.pad10_n}{916}
\resset{e7i.open_kev-0.8b_a1.pad10_shift}{39.63\%}
\resset{e7i.open_kev-0.8b_a1.para}{24.16\%}
\resset{e7i.open_kev-0.8b_a1.para_acc}{72.93\%}
\resset{e7i.open_kev-0.8b_a1.para_acc_worst}{46.18\%}
\resset{e7i.open_kev-0.8b_a1.para_ci}{[22.33\%, 26.12\%]}
\resset{e7i.open_kev-0.8b_a1.para_conc}{36.68\%}
\resset{e7i.open_kev-0.8b_a1.para_n}{916}
\resset{e7i.open_kev-0.8b_a1.para_worst}{31.66\%}
\resset{e7i.open_kev-0.8b_a1.rep}{0.00\%}
\resset{e7i.open_kev-0.8b_a1.rep_ci}{[0.00\%, 0.40\%]}
\resset{e7i.open_kev-0.8b_a1.rep_ident}{100.00\%}
\resset{e7i.open_laya-multilingual_a1.pad05}{18.05\%}
\resset{e7i.open_laya-multilingual_a1.pad05_ci}{[15.22\%, 21.16\%]}
\resset{e7i.open_laya-multilingual_a1.pad05_n}{676}
\resset{e7i.open_laya-multilingual_a1.pad05_shift}{67.31\%}
\resset{e7i.open_laya-multilingual_a1.pad10}{19.12\%}
\resset{e7i.open_laya-multilingual_a1.pad10_ci}{[16.19\%, 22.33\%]}
\resset{e7i.open_laya-multilingual_a1.pad10_n}{659}
\resset{e7i.open_laya-multilingual_a1.pad10_shift}{67.22\%}
\resset{e7i.open_laya-multilingual_a1.para}{17.98\%}
\resset{e7i.open_laya-multilingual_a1.para_acc}{64.36\%}
\resset{e7i.open_laya-multilingual_a1.para_acc_worst}{47.47\%}
\resset{e7i.open_laya-multilingual_a1.para_ci}{[15.31\%, 20.68\%]}
\resset{e7i.open_laya-multilingual_a1.para_conc}{78.21\%}
\resset{e7i.open_laya-multilingual_a1.para_n}{693}
\resset{e7i.open_laya-multilingual_a1.para_worst}{19.05\%}
\resset{e7i.open_laya-multilingual_a1.rep}{0.14\%}
\resset{e7i.open_laya-multilingual_a1.rep_ci}{[0.00\%, 0.80\%]}
\resset{e7i.open_laya-multilingual_a1.rep_ident}{44.24\%}
\resset{e7i.open_laya-typed-decisions_a1.pad05}{17.66\%}
\resset{e7i.open_laya-typed-decisions_a1.pad05_ci}{[14.85\%, 20.77\%]}
\resset{e7i.open_laya-typed-decisions_a1.pad05_n}{668}
\resset{e7i.open_laya-typed-decisions_a1.pad05_shift}{44.16\%}
\resset{e7i.open_laya-typed-decisions_a1.pad10}{18.87\%}
\resset{e7i.open_laya-typed-decisions_a1.pad10_ci}{[15.95\%, 22.08\%]}
\resset{e7i.open_laya-typed-decisions_a1.pad10_n}{657}
\resset{e7i.open_laya-typed-decisions_a1.pad10_shift}{58.30\%}
\resset{e7i.open_laya-typed-decisions_a1.para}{12.79\%}
\resset{e7i.open_laya-typed-decisions_a1.para_acc}{62.84\%}
\resset{e7i.open_laya-typed-decisions_a1.para_acc_worst}{47.02\%}
\resset{e7i.open_laya-typed-decisions_a1.para_ci}{[10.88\%, 14.76\%]}
\resset{e7i.open_laya-typed-decisions_a1.para_conc}{73.15\%}
\resset{e7i.open_laya-typed-decisions_a1.para_n}{689}
\resset{e7i.open_laya-typed-decisions_a1.para_worst}{17.27\%}
\resset{e7i.open_laya-typed-decisions_a1.rep}{7.11\%}
\resset{e7i.open_laya-typed-decisions_a1.rep_ci}{[5.31\%, 9.29\%]}
\resset{e7i.open_laya-typed-decisions_a1.rep_ident}{3.77\%}
\resset{e7i.open_strands-decider-2b-v1-2610_a1.pad05}{21.62\%}
\resset{e7i.open_strands-decider-2b-v1-2610_a1.pad05_ci}{[18.99\%, 24.43\%]}
\resset{e7i.open_strands-decider-2b-v1-2610_a1.pad05_n}{916}
\resset{e7i.open_strands-decider-2b-v1-2610_a1.pad05_shift}{23.91\%}
\resset{e7i.open_strands-decider-2b-v1-2610_a1.pad10}{13.32\%}
\resset{e7i.open_strands-decider-2b-v1-2610_a1.pad10_ci}{[11.18\%, 15.69\%]}
\resset{e7i.open_strands-decider-2b-v1-2610_a1.pad10_n}{916}
\resset{e7i.open_strands-decider-2b-v1-2610_a1.pad10_shift}{23.36\%}
\resset{e7i.open_strands-decider-2b-v1-2610_a1.para}{16.41\%}
\resset{e7i.open_strands-decider-2b-v1-2610_a1.para_acc}{83.84\%}
\resset{e7i.open_strands-decider-2b-v1-2610_a1.para_acc_worst}{61.57\%}
\resset{e7i.open_strands-decider-2b-v1-2610_a1.para_ci}{[14.48\%, 18.42\%]}
\resset{e7i.open_strands-decider-2b-v1-2610_a1.para_conc}{67.90\%}
\resset{e7i.open_strands-decider-2b-v1-2610_a1.para_n}{916}
\resset{e7i.open_strands-decider-2b-v1-2610_a1.para_worst}{22.71\%}
\resset{e7i.open_strands-decider-2b-v1-2610_a1.rep}{3.60\%}
\resset{e7i.open_strands-decider-2b-v1-2610_a1.rep_ci}{[2.49\%, 5.02\%]}
\resset{e7i.open_strands-decider-2b-v1-2610_a1.rep_ident}{9.61\%}
\resset{e7i.open_von_a1.hw}{0.00\%}
\resset{e7i.open_von_a1.hw_ci}{[0.00\%, 0.40\%]}
\resset{e7i.open_von_a1.pad05}{19.43\%}
\resset{e7i.open_von_a1.pad05_ci}{[16.92\%, 22.15\%]}
\resset{e7i.open_von_a1.pad05_n}{916}
\resset{e7i.open_von_a1.pad05_shift}{42.36\%}
\resset{e7i.open_von_a1.pad10}{22.49\%}
\resset{e7i.open_von_a1.pad10_ci}{[19.82\%, 25.33\%]}
\resset{e7i.open_von_a1.pad10_n}{916}
\resset{e7i.open_von_a1.pad10_shift}{51.97\%}
\resset{e7i.open_von_a1.para}{14.10\%}
\resset{e7i.open_von_a1.para_acc}{70.52\%}
\resset{e7i.open_von_a1.para_acc_worst}{54.69\%}
\resset{e7i.open_von_a1.para_ci}{[12.58\%, 15.63\%]}
\resset{e7i.open_von_a1.para_conc}{64.96\%}
\resset{e7i.open_von_a1.para_n}{916}
\resset{e7i.open_von_a1.para_worst}{19.76\%}
\resset{e7i.open_von_a1.rep}{0.00\%}
\resset{e7i.open_von_a1.rep_ci}{[0.00\%, 0.40\%]}
\resset{e7i.open_von_a1.rep_ident}{100.00\%}
\resset{e7i.option.max}{35.97\%}
\resset{e7i.option.max_name}{Liquid d1-3B}
\resset{e7i.option.min}{0.00\%}
\resset{e7i.option.min_name}{CLM-8B}
\resset{e7i.option.n}{10}
\resset{e7i.pad05.max}{11.22\%}
\resset{e7i.pad05.max_name}{Liquid d1-3B}
\resset{e7i.pad05.min}{1.20\%}
\resset{e7i.pad05.min_name}{CLM-8B}
\resset{e7i.pad05.n}{10}
\resset{e7i.pad10.max}{14.45\%}
\resset{e7i.pad10.max_name}{decider-2b}
\resset{e7i.pad10.min}{1.40\%}
\resset{e7i.pad10.min_name}{CLM-8B}
\resset{e7i.pad10.n}{10}
\resset{e7i.para.max}{26.14\%}
\resset{e7i.para.max_name}{CLM-8B}
\resset{e7i.para.min}{2.31\%}
\resset{e7i.para.min_name}{Qwen3-8B (logits)}
\resset{e7i.para.n}{10}
\resset{e7i.polarity.max}{61.88\%}
\resset{e7i.polarity.max_name}{Liquid d1-omni-600M}
\resset{e7i.polarity.min}{1.31\%}
\resset{e7i.polarity.min_name}{CLM-8B}
\resset{e7i.polarity.n}{10}
\resset{e7i.qwen3-8b-bf16.option}{4.26\%}
\resset{e7i.qwen3-8b-bf16.option_ci}{[3.04\%, 5.77\%]}
\resset{e7i.qwen3-8b-bf16.option_n}{916}
\resset{e7i.qwen3-8b-bf16.pad05}{3.82\%}
\resset{e7i.qwen3-8b-bf16.pad05_ci}{[2.68\%, 5.27\%]}
\resset{e7i.qwen3-8b-bf16.pad05_n}{916}
\resset{e7i.qwen3-8b-bf16.pad05_shift}{7.42\%}
\resset{e7i.qwen3-8b-bf16.pad10}{4.17\%}
\resset{e7i.qwen3-8b-bf16.pad10_ci}{[2.77\%, 6.01\%]}
\resset{e7i.qwen3-8b-bf16.pad10_n}{647}
\resset{e7i.qwen3-8b-bf16.pad10_shift}{7.88\%}
\resset{e7i.qwen3-8b-bf16.para}{2.31\%}
\resset{e7i.qwen3-8b-bf16.para_acc}{72.27\%}
\resset{e7i.qwen3-8b-bf16.para_acc_worst}{70.41\%}
\resset{e7i.qwen3-8b-bf16.para_ci}{[1.80\%, 2.92\%]}
\resset{e7i.qwen3-8b-bf16.para_conc}{90.17\%}
\resset{e7i.qwen3-8b-bf16.para_n}{916}
\resset{e7i.qwen3-8b-bf16.para_worst}{7.10\%}
\resset{e7i.qwen3-8b-bf16.polarity}{17.47\%}
\resset{e7i.qwen3-8b-bf16.polarity_ci}{[15.06\%, 20.08\%]}
\resset{e7i.qwen3-8b-bf16.polarity_n}{916}
\resset{e7i.qwen3-8b-bf16.rep}{0.00\%}
\resset{e7i.qwen3-8b-bf16.rep_ci}{[0.00\%, 0.40\%]}
\resset{e7i.qwen3-8b-bf16.rep_ident}{0.22\%}
\resset{e7i.rep.max}{1.43\%}
\resset{e7i.rep.max_name}{decider-0.8b}
\resset{e7i.rep.min}{0.00\%}
\resset{e7i.rep.min_name}{Qwen3-8B (logits)}
\resset{e7i.rep.n}{10}
\resset{e7i.von.option}{0.00\%}
\resset{e7i.von.option_ci}{[0.00\%, 0.50\%]}
\resset{e7i.von.option_n}{742}
\resset{e7i.von.pad05}{10.10\%}
\resset{e7i.von.pad05_ci}{[7.63\%, 13.03\%]}
\resset{e7i.von.pad05_n}{515}
\resset{e7i.von.pad05_shift}{0.00\%}
\resset{e7i.von.pad10}{13.95\%}
\resset{e7i.von.pad10_ci}{[10.82\%, 17.59\%]}
\resset{e7i.von.pad10_n}{430}
\resset{e7i.von.pad10_shift}{0.00\%}
\resset{e7i.von.para}{9.53\%}
\resset{e7i.von.para_acc}{53.18\%}
\resset{e7i.von.para_acc_worst}{43.71\%}
\resset{e7i.von.para_ci}{[7.75\%, 11.26\%]}
\resset{e7i.von.para_conc}{79.57\%}
\resset{e7i.von.para_n}{739}
\resset{e7i.von.para_worst}{11.64\%}
\resset{e7i.von.polarity}{15.56\%}
\resset{e7i.von.polarity_ci}{[13.02\%, 18.38\%]}
\resset{e7i.von.polarity_n}{739}
\resset{e7i.von.rep}{0.00\%}
\resset{e7i.von.rep_ci}{[0.00\%, 0.50\%]}
\resset{e7i.von.rep_ident}{100.00\%}
\resset{e8.clm_a0.clean.fpr}{0.00\%}
\resset{e8.clm_a0.clean.fpr_ci}{[0.00\%, 2.12\%]}
\resset{e8.clm_a0.clean.tpr}{6.50\%}
\resset{e8.clm_a0.clean.tpr_ci}{[3.51\%, 10.86\%]}
\resset{e8.clm_a0.inj_raw.d}{-6.50}
\resset{e8.clm_a0.inj_raw.fpr}{0.00\%}
\resset{e8.clm_a0.inj_raw.fpr_ci}{[0.00\%, 2.12\%]}
\resset{e8.clm_a0.inj_raw.p}{$<0.001$}
\resset{e8.clm_a0.inj_raw.tpr}{0.00\%}
\resset{e8.clm_a0.inj_raw.tpr_ci}{[0.00\%, 1.83\%]}
\resset{e8.clm_a0.inj_san.d}{-6.50}
\resset{e8.clm_a0.inj_san.fpr}{0.00\%}
\resset{e8.clm_a0.inj_san.fpr_ci}{[0.00\%, 2.12\%]}
\resset{e8.clm_a0.inj_san.p}{$<0.001$}
\resset{e8.clm_a0.inj_san.tpr}{0.00\%}
\resset{e8.clm_a0.inj_san.tpr_ci}{[0.00\%, 1.83\%]}
\resset{e8.clm_a1.clean.fpr}{0.58\%}
\resset{e8.clm_a1.clean.fpr_ci}{[0.01\%, 3.20\%]}
\resset{e8.clm_a1.clean.tpr}{59.50\%}
\resset{e8.clm_a1.clean.tpr_ci}{[52.35\%, 66.37\%]}
\resset{e8.clm_a1.inj_raw.d}{-58.00}
\resset{e8.clm_a1.inj_raw.fpr}{0.00\%}
\resset{e8.clm_a1.inj_raw.fpr_ci}{[0.00\%, 2.12\%]}
\resset{e8.clm_a1.inj_raw.p}{$<0.001$}
\resset{e8.clm_a1.inj_raw.tpr}{1.50\%}
\resset{e8.clm_a1.inj_raw.tpr_ci}{[0.31\%, 4.32\%]}
\resset{e8.clm_a1.inj_san.d}{-57.00}
\resset{e8.clm_a1.inj_san.fpr}{0.00\%}
\resset{e8.clm_a1.inj_san.fpr_ci}{[0.00\%, 2.12\%]}
\resset{e8.clm_a1.inj_san.p}{$<0.001$}
\resset{e8.clm_a1.inj_san.tpr}{2.50\%}
\resset{e8.clm_a1.inj_san.tpr_ci}{[0.82\%, 5.74\%]}
\resset{e8.clm_a2.clean.fpr}{0.58\%}
\resset{e8.clm_a2.clean.fpr_ci}{[0.01\%, 3.20\%]}
\resset{e8.clm_a2.clean.tpr}{88.00\%}
\resset{e8.clm_a2.clean.tpr_ci}{[82.67\%, 92.16\%]}
\resset{e8.clm_a2.inj_raw.d}{-16.00}
\resset{e8.clm_a2.inj_raw.fpr}{23.26\%}
\resset{e8.clm_a2.inj_raw.fpr_ci}{[17.16\%, 30.29\%]}
\resset{e8.clm_a2.inj_raw.p}{$<0.001$}
\resset{e8.clm_a2.inj_raw.tpr}{72.00\%}
\resset{e8.clm_a2.inj_raw.tpr_ci}{[65.23\%, 78.10\%]}
\resset{e8.clm_a2.inj_san.d}{-55.00}
\resset{e8.clm_a2.inj_san.fpr}{1.74\%}
\resset{e8.clm_a2.inj_san.fpr_ci}{[0.36\%, 5.01\%]}
\resset{e8.clm_a2.inj_san.p}{$<0.001$}
\resset{e8.clm_a2.inj_san.tpr}{33.00\%}
\resset{e8.clm_a2.inj_san.tpr_ci}{[26.53\%, 39.98\%]}
\resset{e8.clm_a3.clean.fpr}{0.58\%}
\resset{e8.clm_a3.clean.fpr_ci}{[0.01\%, 3.20\%]}
\resset{e8.clm_a3.clean.tpr}{89.50\%}
\resset{e8.clm_a3.clean.tpr_ci}{[84.40\%, 93.38\%]}
\resset{e8.clm_a3.inj_raw.d}{-37.50}
\resset{e8.clm_a3.inj_raw.fpr}{9.30\%}
\resset{e8.clm_a3.inj_raw.fpr_ci}{[5.41\%, 14.67\%]}
\resset{e8.clm_a3.inj_raw.p}{$<0.001$}
\resset{e8.clm_a3.inj_raw.tpr}{52.00\%}
\resset{e8.clm_a3.inj_raw.tpr_ci}{[44.84\%, 59.10\%]}
\resset{e8.clm_a3.inj_san.d}{-61.50}
\resset{e8.clm_a3.inj_san.fpr}{0.00\%}
\resset{e8.clm_a3.inj_san.fpr_ci}{[0.00\%, 2.12\%]}
\resset{e8.clm_a3.inj_san.p}{$<0.001$}
\resset{e8.clm_a3.inj_san.tpr}{28.00\%}
\resset{e8.clm_a3.inj_san.tpr_ci}{[21.90\%, 34.77\%]}
\resset{e8.n_benign}{172}
\resset{e8.n_malware}{200}
\resset{e8.pad.both.clean.tpr}{72.07\%}
\resset{e8.pad.both.pad05.tpr}{44.14\%}
\resset{e8.pad.both.pad10.tpr}{49.55\%}
\resset{e8.pad.both.pad25.tpr}{29.73\%}
\resset{e8.pad.clm_a0.clean.tpr}{0.00\%}
\resset{e8.pad.clm_a0.pad05.d}{-0.00}
\resset{e8.pad.clm_a0.pad05.p}{1.000}
\resset{e8.pad.clm_a0.pad05.tpr}{0.00\%}
\resset{e8.pad.clm_a0.pad10.d}{-0.00}
\resset{e8.pad.clm_a0.pad10.p}{1.000}
\resset{e8.pad.clm_a0.pad10.tpr}{0.00\%}
\resset{e8.pad.clm_a0.pad25.d}{-0.00}
\resset{e8.pad.clm_a0.pad25.p}{1.000}
\resset{e8.pad.clm_a0.pad25.tpr}{0.00\%}
\resset{e8.pad.clm_a1.clean.tpr}{78.38\%}
\resset{e8.pad.clm_a1.pad05.d}{-15.32}
\resset{e8.pad.clm_a1.pad05.p}{$<0.001$}
\resset{e8.pad.clm_a1.pad05.tpr}{63.06\%}
\resset{e8.pad.clm_a1.pad10.d}{-19.82}
\resset{e8.pad.clm_a1.pad10.p}{$<0.001$}
\resset{e8.pad.clm_a1.pad10.tpr}{58.56\%}
\resset{e8.pad.clm_a1.pad25.d}{-36.94}
\resset{e8.pad.clm_a1.pad25.p}{$<0.001$}
\resset{e8.pad.clm_a1.pad25.tpr}{41.44\%}
\resset{e8.pad.clm_a2.clean.tpr}{88.29\%}
\resset{e8.pad.clm_a2.pad05.d}{-24.32}
\resset{e8.pad.clm_a2.pad05.p}{$<0.001$}
\resset{e8.pad.clm_a2.pad05.tpr}{63.96\%}
\resset{e8.pad.clm_a2.pad10.d}{-18.02}
\resset{e8.pad.clm_a2.pad10.p}{$<0.001$}
\resset{e8.pad.clm_a2.pad10.tpr}{70.27\%}
\resset{e8.pad.clm_a2.pad25.d}{-37.84}
\resset{e8.pad.clm_a2.pad25.p}{$<0.001$}
\resset{e8.pad.clm_a2.pad25.tpr}{50.45\%}
\resset{e8.pad.clm_a3.clean.tpr}{90.99\%}
\resset{e8.pad.clm_a3.pad05.d}{-5.41}
\resset{e8.pad.clm_a3.pad05.p}{0.070}
\resset{e8.pad.clm_a3.pad05.tpr}{85.59\%}
\resset{e8.pad.clm_a3.pad10.d}{-14.41}
\resset{e8.pad.clm_a3.pad10.p}{$<0.001$}
\resset{e8.pad.clm_a3.pad10.tpr}{76.58\%}
\resset{e8.pad.clm_a3.pad25.d}{-27.93}
\resset{e8.pad.clm_a3.pad25.p}{$<0.001$}
\resset{e8.pad.clm_a3.pad25.tpr}{63.06\%}
\resset{e8.pad.either.clean.tpr}{94.59\%}
\resset{e8.pad.either.n}{111}
\resset{e8.pad.either.pad05.tpr}{82.88\%}
\resset{e8.pad.either.pad10.tpr}{79.28\%}
\resset{e8.pad.either.pad25.tpr}{62.16\%}
\resset{e8.pad.n_benign}{56}
\resset{e8.pad.n_incomplete}{233}
\resset{e8.pad.n_malware}{111}
\resset{e8.screen.benign_strings}{0.00\%}
\resset{e8.screen.benign_strings.k}{0}
\resset{e8.screen.benign_strings.n}{10}
\resset{e8.screen.canaries}{100.00\%}
\resset{e8.screen.canaries.k}{14}
\resset{e8.screen.canaries.n}{14}
\resset{e8.screen.stage}{development}
\resset{e8.screen.states_benign}{16.67\%}
\resset{e8.screen.states_malware}{2.00\%}
\resset{e8.shift.inj_raw.atomic.mean}{0.053}
\resset{e8.shift.inj_raw.atomic.moved}{11.05\%}
\resset{e8.shift.inj_raw.atomic.moved_ci}{[10.29\%, 11.84\%]}
\resset{e8.shift.inj_raw.verdict.mean}{0.486}
\resset{e8.shift.inj_raw.verdict.moved}{70.50\%}
\resset{e8.shift.inj_raw.verdict.moved_ci}{[63.66\%, 76.72\%]}
\resset{e8.shift.inj_san.atomic.mean}{0.071}
\resset{e8.shift.inj_san.atomic.moved}{11.33\%}
\resset{e8.shift.inj_san.atomic.moved_ci}{[10.56\%, 12.13\%]}
\resset{e8.shift.inj_san.verdict.mean}{0.541}
\resset{e8.shift.inj_san.verdict.moved}{75.00\%}
\resset{e8.shift.inj_san.verdict.moved_ci}{[68.40\%, 80.84\%]}
\resset{e8.shift.pad05.atomic.mean}{-0.011}
\resset{e8.shift.pad05.atomic.moved}{3.66\%}
\resset{e8.shift.pad05.atomic.moved_ci}{[3.21\%, 4.15\%]}
\resset{e8.shift.pad05.verdict.mean}{0.008}
\resset{e8.shift.pad05.verdict.moved}{0.00\%}
\resset{e8.shift.pad05.verdict.moved_ci}{[0.00\%, 1.83\%]}
\resset{e8.shift.pad10.atomic.mean}{-0.028}
\resset{e8.shift.pad10.atomic.moved}{3.12\%}
\resset{e8.shift.pad10.atomic.moved_ci}{[2.67\%, 3.63\%]}
\resset{e8.shift.pad10.verdict.mean}{-0.009}
\resset{e8.shift.pad10.verdict.moved}{0.00\%}
\resset{e8.shift.pad10.verdict.moved_ci}{[0.00\%, 2.20\%]}
\resset{e8.shift.pad25.atomic.mean}{-0.022}
\resset{e8.shift.pad25.atomic.moved}{4.05\%}
\resset{e8.shift.pad25.atomic.moved_ci}{[3.43\%, 4.76\%]}
\resset{e8.shift.pad25.verdict.mean}{0.009}
\resset{e8.shift.pad25.verdict.moved}{0.00\%}
\resset{e8.shift.pad25.verdict.moved_ci}{[0.00\%, 3.27\%]}
\resset{e9.a2min.clm_a2}{84.02\%}
\resset{e9.a2min.clm_a3}{91.60\%}
\resset{e9.a2min.month}{2020-03}
\resset{e9.clm_a0.aut}{0.0885}
\resset{e9.clm_a0.aut_ci}{[0.017, 0.199]}
\resset{e9.clm_a0.mk_p}{0.466}
\resset{e9.clm_a0.sigma}{0.064}
\resset{e9.clm_a0.slope}{-0.65}
\resset{e9.clm_a0.trend}{no trend}
\resset{e9.clm_a1.aut}{0.5983}
\resset{e9.clm_a1.aut_ci}{[0.431, 0.716]}
\resset{e9.clm_a1.mk_p}{0.118}
\resset{e9.clm_a1.sigma}{0.118}
\resset{e9.clm_a1.slope}{-3.13}
\resset{e9.clm_a1.trend}{no trend}
\resset{e9.clm_a2.aut}{0.9057}
\resset{e9.clm_a2.aut_ci}{[0.840, 0.946]}
\resset{e9.clm_a2.fpr}{1.11\%}
\resset{e9.clm_a2.mk_p}{0.466}
\resset{e9.clm_a2.sigma}{0.042}
\resset{e9.clm_a2.slope}{+0.49}
\resset{e9.clm_a2.trend}{no trend}
\resset{e9.clm_a3.aut}{0.9334}
\resset{e9.clm_a3.aut_ci}{[0.885, 0.963]}
\resset{e9.clm_a3.fpr}{1.45\%}
\resset{e9.clm_a3.mk_p}{0.754}
\resset{e9.clm_a3.sigma}{0.023}
\resset{e9.clm_a3.slope}{-0.26}
\resset{e9.clm_a3.trend}{no trend}
\resset{e9.det.aut}{0.8862}
\resset{e9.det.aut_ci}{[0.813, 0.933]}
\resset{e9.det.mk_p}{0.118}
\resset{e9.det.sigma}{0.036}
\resset{e9.det.slope}{+0.81}
\resset{e9.det.trend}{no trend}
\resset{e9.mk.n_arms}{5}
\resset{e9.mk.n_sig_holm}{0}
\resset{e9.mk.pmin}{0.118}
\resset{e9.mk.pmin_arm}{clm_a1}
\resset{e9.mk.pmin_holm}{0.589}
\resset{e9.n_buckets}{9}
\resset{e9.release.clm}{2026-09-24}
\resset{e9.release.clm.before}{yes}
\resset{e9.release.qwen3}{2025-04-29}
\resset{e9.release.qwen3.before}{yes}
\resset{seeds.clm_a2.auroc_mean}{0.9941}
\resset{seeds.clm_a2.auroc_sd}{0.0002}
\resset{seeds.clm_a2.n}{5}
\resset{seeds.clm_a2.tpr_max}{90.90\%}
\resset{seeds.clm_a2.tpr_mean}{90.16\%}
\resset{seeds.clm_a2.tpr_min}{89.52\%}
\resset{seeds.clm_a2.tpr_sd}{0.54\%}
\resset{seeds.clm_a2_ctrl_lr.auroc_mean}{0.9955}
\resset{seeds.clm_a2_ctrl_lr.auroc_sd}{0.0000}
\resset{seeds.clm_a2_ctrl_lr.n}{5}
\resset{seeds.clm_a2_ctrl_lr.tpr_max}{92.94\%}
\resset{seeds.clm_a2_ctrl_lr.tpr_mean}{92.94\%}
\resset{seeds.clm_a2_ctrl_lr.tpr_min}{92.94\%}
\resset{seeds.clm_a2_ctrl_lr.tpr_sd}{0.00\%}
\resset{seeds.clm_a2_ctrl_mlp.auroc_mean}{0.9946}
\resset{seeds.clm_a2_ctrl_mlp.auroc_sd}{0.0006}
\resset{seeds.clm_a2_ctrl_mlp.n}{5}
\resset{seeds.clm_a2_ctrl_mlp.tpr_max}{92.41\%}
\resset{seeds.clm_a2_ctrl_mlp.tpr_mean}{91.25\%}
\resset{seeds.clm_a2_ctrl_mlp.tpr_min}{90.79\%}
\resset{seeds.clm_a2_ctrl_mlp.tpr_sd}{0.66\%}
\resset{seeds.clm_a2_mt.auroc_mean}{0.9928}
\resset{seeds.clm_a2_mt.auroc_sd}{0.0015}
\resset{seeds.clm_a2_mt.n}{5}
\resset{seeds.clm_a2_mt.tpr_max}{90.94\%}
\resset{seeds.clm_a2_mt.tpr_mean}{89.07\%}
\resset{seeds.clm_a2_mt.tpr_min}{83.76\%}
\resset{seeds.clm_a2_mt.tpr_sd}{3.01\%}
\resset{seeds.clm_a2_raw.auroc_mean}{0.9952}
\resset{seeds.clm_a2_raw.auroc_sd}{0.0004}
\resset{seeds.clm_a2_raw.n}{5}
\resset{seeds.clm_a2_raw.tpr_max}{92.30\%}
\resset{seeds.clm_a2_raw.tpr_mean}{90.92\%}
\resset{seeds.clm_a2_raw.tpr_min}{88.60\%}
\resset{seeds.clm_a2_raw.tpr_sd}{1.38\%}
\resset{seeds.clm_a2_std.auroc_mean}{0.9942}
\resset{seeds.clm_a2_std.auroc_sd}{0.0009}
\resset{seeds.clm_a2_std.n}{5}
\resset{seeds.clm_a2_std.tpr_max}{92.60\%}
\resset{seeds.clm_a2_std.tpr_mean}{89.92\%}
\resset{seeds.clm_a2_std.tpr_min}{87.43\%}
\resset{seeds.clm_a2_std.tpr_sd}{2.13\%}
\resset{seeds.qwen3_emb_probe.auroc_mean}{0.9956}
\resset{seeds.qwen3_emb_probe.auroc_sd}{0.0000}
\resset{seeds.qwen3_emb_probe.n}{5}
\resset{seeds.qwen3_emb_probe.tpr_max}{93.62\%}
\resset{seeds.qwen3_emb_probe.tpr_mean}{93.62\%}
\resset{seeds.qwen3_emb_probe.tpr_min}{93.62\%}
\resset{seeds.qwen3_emb_probe.tpr_sd}{0.00\%}
\resset{split}{TEST and $B_{\mathrm{test}}$}
\resset{split.e1}{TEST and $B_{\mathrm{test}}$}
\resset{split.e10}{TEST and $B_{\mathrm{test}}$}
\resset{split.e4}{TEST and $B_{\mathrm{test}}$}
\resset{split.e5}{TEST and $B_{\mathrm{test}}$}
\resset{split.e6}{TEST and $B_{\mathrm{test}}$}
\resset{split.e7}{V and $B_{\mathrm{val}}$}
\resset{split.e8}{TEST and $B_{\mathrm{test}}$}
\resset{split.e9}{TEST and $B_{\mathrm{test}}$}
\resset{stage}{test}
\resset{stage.e1}{test}
\resset{stage.e10}{test}
\resset{stage.e4}{test}
\resset{stage.e5}{test}
\resset{stage.e6}{test}
\resset{stage.e7}{development}
\resset{stage.e7inv}{test}
\resset{stage.e7probes}{test}
\resset{stage.e8}{test}
\resset{stage.e9}{test}
\resset{test.e1}{yes}
\resset{test.e10}{yes}
\resset{test.e4}{yes}
\resset{test.e5}{yes}
\resset{test.e6}{yes}
\resset{test.e7inv}{yes}
\resset{test.e7probes}{yes}
\resset{test.e8}{yes}
\resset{test.e9}{yes}
 
\title{A First Look at System One Decision Models for Malware Detection}

\ifblind
  \author{Anonymous Authors\\ \textit{Anonymous Institution}}
\else
  \author{%
    Pier Mattia Basciano\textsuperscript{$\ast$}\,\orcidlink{0009-0007-2551-4381}\quad
    Giovanni Farina\,\orcidlink{0000-0002-4792-5305}\quad
    Salvatore Monteleone\,\orcidlink{0000-0003-0158-2295}\\[0.6ex]
    \small\textit{Department of Engineering, Niccol\`o Cusano University, Rome, Italy}\\[0.4ex]
    \footnotesize\texttt{\{piermattia.basciano, giovanni.farina, salvatore.monteleone\}@unicusano.it}\\[0.4ex]
    \footnotesize\textsuperscript{$\ast$}Corresponding author.}
\fi
\date{}
\maketitle

\begin{abstract}
Malware detection needs scores calibrated at a fixed false positive rate (FPR) and decisions an analyst can read: feature-based detectors give the first, large language models (LLMs) on decompiled code the second, through generated text of uncalibrated confidence. System One decision models answer a typed question about a text with a probability over its declared answers in one forward pass. We present the first evaluation of these models for Windows Portable Executable malware detection, analyze how they decide, and propose \fwname{}, an ensemble on the typed answers and the embedding of CLM-8B. In an isolated virtual machine, Ghidra and capa decompose each binary, and the host renders the records into a canonical textual state, a digest of headers, imports, MITRE ATT\&CK capabilities and strings, optionally with decompiled functions, without class shortcuts. Three readouts score each model on the state: A0, the probability of one verdict question; A1, 32 atomic behavioral questions weighed by gradient-boosted trees; A2, heads on the model's state embedding. \fwname{} feeds gradient-boosted trees with the atomic answers and the A2 score of CLM-8B. We evaluate CLM-8B, ten open decision models and two LLMs read through logits on BODMAS temporal splits, SOREL-20M out-of-distribution malware and 14,534 matched benign files. On BODMAS, A0 wastes the static evidence: its area under the receiver operating characteristic curve (AUROC) spans \res{e1.a0.min}-\res{e1.a0.max} across \res{e1.n_models} models. A1 recovers it, raising every model's AUROC by +\res{e1.lift_rel.mean} on average; A2 lifts CLM-8B by \res{e1.E1-C2.d} to \res{e1.clm_a2.auroc}, above a combiner on digest facts (\res{e1.det.auroc}), and reaches \res{e1b.a2.min}-\res{e1b.a2.max} on nine open models. \fwname{} reaches \res{e1.clm_a3.auroc} and detects \res{e1.clm_a3.tpr} of the malware at a realized FPR of \res{e1.clm_a3.fpr}, and \res{e1.clm_a3.tpr01} at 0.1\% FPR on the test ROC curve, against \res{e1.clm_a2.tpr01} with A2. \fwname{} gives analysts calibrated decisions, each with the atomic answers behind it.
\end{abstract}

\keywords{Malware Detection; Windows Portable Executable (PE); Static Analysis; System One Decision Models; Large Language Models; Probing; Concept Drift}
\section{Introduction}
\label{sec:intro}

Windows malware detection is the first filter of endpoint products and security operations centers: a Portable Executable (PE) file is judged before it runs, at a volume where only very low false positive rates (FPRs) are tolerable~\cite{nguyen2021leveraging}. Learned detectors on engineered features (e.g., LightGBM on EMBER~\cite{anderson2018ember,joyce2025ember2024}) and networks on raw bytes (e.g., MalConv2~\cite{raff2021malconv2}) meet the throughput, but their scores carry no account of the evidence behind them, and their evaluation is fragile under temporal bias and drift~\cite{pendlebury2019tesseract,barbero2022transcending,arp2022dos}.

Large language models (LLMs) that read decompiled code or static reports promise that account~\cite{chawla2025decompilation,choi2026mars}. They answer in generated text with a self-reported confidence, which is slow to produce, hard to calibrate and easy to steer with strings planted in the binary~\cite{choi2026alibi}. In September 2026 a new class of models appeared: System One decision models, which read a textual state and a typed question and return a probability over the declared answers in a single forward pass~\cite{rafe2026benchmarking}.\footnote{\label{fn:jev}D.~Almeida, \emph{Introducing System One Models \& Jev}, TypeSafe AI Blog, September 2026. \url{https://typesafe.ai/blog/introducing-system-one-models-and-jev}} Their first security evaluations cover agent actions, prompt injection, web requests and network traffic~\cite{liu2026evaluating,lan2026detection,kisielewicz2026escalate,chen2026secjev}. To our knowledge, none gives them a malware file.

This work takes a first look at System One models as Windows PE malware detectors built on static analysis (Fig.~\ref{fig:pipeline}). It evaluates eleven of them, CLM-8B~\cite{kwok2026clm} and ten open decision models, on the same states and the same questions, next to two LLMs read through their answer tokens and a combiner on the digest facts. An isolated virtual machine decompiles each binary with Ghidra~\cite{nsa2026ghidra} and matches capa~\cite{mandiant2026capa} rules, and the results are rendered into a canonical textual state from which the known class shortcuts are removed. On this evaluation we read every System One model through three readouts, each calibrated at a 1\% FPR operating point: the probability of a single verdict question (A0), a gradient-boosted combiner over the answers to 32 atomic behavioral questions (A1), and heads trained on the model's embedding of the state (A2). A0 and A1 are measured on all \res{e1.n_models} models that read a prompt and A2 on CLM-8B and \res{e1b.n_models} open models, so that each design choice is tested across models. On CLM-8B, which reads the most evidence of each file, the digest together with its decompiled functions, we build \fwname{}, an ensemble that feeds gradient-boosted trees with the atomic answers and the score of heads trained on the embedding. The evaluation runs ten experiments, E1 to E10, each with its question and primary endpoint fixed before testing (Table~\ref{tab:experiments}): detection (E1, with E1b for heads on the open models), the parts of the state (E2) and the questions and combiner (E3) that carry the decision, calibration and abstention (E4), deployment on a laptop GPU (E5), stability across seeds and runs (E6), hallucination and invariance (E7), text planted in the binary (E8), temporal drift (E9) and transfer to out-of-distribution malware (E10).

\begin{widefigure}[t]
\centering
\IfFileExists{figures/pipeline.pdf}%
  {\includegraphics[width=\textwidth]{pipeline.pdf}}%
  {\fbox{\parbox[c][3cm][c]{0.9\textwidth}{\centering\dacompletare{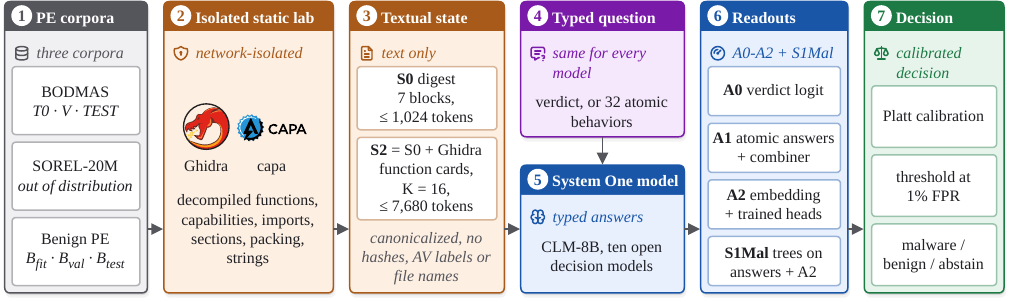 (run figures/src/make\_figures.py)}}}}
\caption{Overview of the pipeline and of \fwname{}.}
\label{fig:pipeline}
\end{widefigure}
 
The contributions of this work are:
\begin{itemize}
  \item A first look at System One decision models for Windows malware detection from static analysis, covering CLM-8B and ten open decision models next to a combiner on the digest facts and two LLMs read through logits, on BODMAS temporal splits, SOREL-20M out-of-distribution data and 14,534 matched benign files, with every readout measured on every model that reads a prompt and the state and the questions ablated on a development sub-sample. On the test split, a single verdict exploits the static evidence only in part, while readouts trained on the atomic answers or on the embedding draw much more signal from it: the atomic questions lift every model over the verdict, by \res{e1.lift.min} to \res{e1.lift.max} in AUROC, and heads on the embeddings of \res{e1b.n_models} open models reach \res{e1b.a2.min} to \res{e1b.a2.max}, each above its atomic answers.
  \item \fwname{}, an ensemble on CLM-8B that feeds a gradient-boosted combiner of its atomic behavioral answers with the score of heads trained on its embedding, with calibration, a fixed operating point and abstention. It ranks the test files at an AUROC of \res{e1.clm_a3.auroc} and detects \res{e1.clm_a3.tpr01} of the malware at 0.1\% FPR on the test ROC curve, against \res{e1.clm_a2.tpr01} for the heads alone, whose AUROC of \res{e1.clm_a2.auroc} already passes the combiner on the digest facts and the LLMs.
  \item An isolated pipeline from binary to textual state, built on Ghidra decompilation and capa capabilities, whose canonicalization and leak guards remove the timestamp, toolchain, bitness and disarming shortcuts that separate the two classes.
  \item A characterization of reliability and deployment: hallucination and invariance probes on typed answers, robustness to text planted in the binary, temporal drift, and the inference time, memory and energy of the models on a laptop GPU against a workstation.
\end{itemize}
The remainder of this paper is organized as follows. Section~\ref{sec:sota} reviews background and related works. Section~\ref{sec:meth} presents the ablation between the models and the \fwname{} framework in full. Section~\ref{sec:res} reports experimental results. Section~\ref{sec:end} concludes with directions for future work. Appendix~\ref{app:repo} describes the question bank.
\section{Background and Related Works}
\label{sec:sota}

\subsection{Background}
\label{sec:background}

\paragraph{System One decision models.} TypeSafe introduced System One models with Jev in September 2026, as models that make fast, structured decisions that software can use directly.\footref{fn:jev} A caller sends a textual state and a typed question (e.g., a boolean, a choice among options or a score over levels) and receives a probability over the declared answers without any generated text~\cite{rafe2026benchmarking}. The output is a number that a program can threshold, at the cost of one forward pass. Open models followed within weeks. Kev-0.8B~\cite{palmer2026kev} and Strands Decider~\cite{strands2026decider} add a pointer head to a small Qwen3.5 decoder adapted with LoRA. decider~\cite{marosi2026decider} fine-tunes all the weights of Qwen3.5 models of 0.8B to 4B parameters. Von~\cite{panisa2026von} and Laya~\cite{m2026laya} are encoder models, Von with a head over option markers, and Liquid's d1 models\footnote{\label{fn:d1}Liquid AI, \emph{Open d1: Edge decision models for text, vision, and audio}, Liquid AI Blog, October 2026. \url{https://www.liquid.ai/blog/d1-open}} build on LFM2.5. CLM-8B~\cite{kwok2026clm} takes a contrastive route, as CLAP~\cite{wang2024clap} does for assembly and its descriptions: two heads on a frozen Qwen3-8B map the state and each candidate answer into one space, and the answer probability follows their cosine similarity (Eq.~\ref{eq:clm}).

\paragraph{Static analysis of PE files.} A PE file declares its sections, imported functions, resources and entry points in its headers, which parsers (e.g., pefile~\cite{carrera2024pefile}) read without running it. A decompiler (e.g., Ghidra~\cite{nsa2026ghidra}) goes further: it recovers the functions of the machine code, their call graph and an approximate C rendering of their bodies. capa~\cite{mandiant2026capa} matches rules on that program to name capabilities (e.g., injecting code into another process or encrypting files) and maps them to MITRE ATT\&CK techniques and MBC behaviors. These artifacts are what a malware analyst reads, and they are the evidence that our textual states carry.

\subsection{Related Works}
\label{sec:related}

\paragraph{Learned detectors.} EMBER~\cite{anderson2018ember} established LightGBM on engineered PE features as the reference detector, and EMBER2024~\cite{joyce2025ember2024} reports a TPR of 94.48\% at 1\% FPR for a LightGBM trained on six file formats from September 2023 to September 2024 and tested on the following twelve weeks. BODMAS~\cite{yang2021bodmas} and SOREL-20M~\cite{harang2021sorel} release timestamped malware with features, SOREL-20M also with baseline models, and MalConv2~\cite{raff2021malconv2} learns directly from the raw bytes of whole binaries. The evaluation of these detectors is fragile. TESSERACT~\cite{pendlebury2019tesseract} shows how spatial and temporal bias inflate results, Transcendent~\cite{barbero2022transcending} and CADE~\cite{yang2021cade} detect and reject drifting samples, Nguyen et al.~\cite{nguyen2021leveraging} stress that deployment needs thresholds fixed on validation data at very low FPR, and Arp et al.~\cite{arp2022dos} list ten pitfalls of machine learning in security. This study adopts these rules: time-ordered splits, thresholds fixed on validation data and the FPR realized on the test files.

\paragraph{LLMs on static artifacts.} Chawla and Prasad~\cite{chawla2025decompilation} prompt LLMs with Ghidra pseudo-C of Windows executables: Gemini~2.5~Pro reaches 80.1\% accuracy against 98.5\% for XGBoost, and 64.2\% on samples from 2025, with no FPR or confidence reported. PE2Prompt~\cite{sun2026pe2prompt} turns EMBER features into zero- and few-shot prompts and reaches 49\% to 73\% accuracy. Turkmen and Raina~\cite{turkmen2026multiview} combine Ghidra and RetDec views of ELF objects without model probabilities or thresholds. MARS~\cite{choi2026mars}, the closest work, builds evidence bundles from headers, sections, imports, strings, YARA matches and optional Ghidra functions, and asks generative LLMs for a four-label verdict in JSON, or for behavioral claims that an external rule policy scores: it reaches 85.9\% to 96.2\% accuracy on 1,195 PE and ELF binaries, with a self-reported confidence that is never analyzed for calibration. ALIBI~\cite{choi2026alibi} shows the weakness of such verdicts: a legitimacy narrative planted in a section that never runs flips 30 of 35 verdicts of Gemini~2.5~Pro. Kurlandski et al.~\cite{kurlandski2026beyond} pretrain malware language models on the bytes, disassembly and decompiled code of about 800,000 deduplicated PE files, with a detection AUROC of up to 0.996 on decompiled code, and AsmRAG~\cite{karbab2026asmrag} classifies files by nearest-neighbor votes over Qwen3-Embedding vectors of their Ghidra assembly functions, with an F1 of 96.0\%. Unlike these works, this study reads a probability from typed answers in one forward pass, calibrates it, fixes its operating point at 1\% FPR on held-out validation files, and removes the class shortcuts of the textual input before any model sees it.

\paragraph{Decision models in security.} The first security studies of System One models appeared alongside the models. SecJev~\cite{chen2026secjev} specializes Kev-style decision models for 14 security tasks, from IoT traffic to authentication logs and agent actions, and tests their transfer on IoT-23 malware traffic, so that, to our knowledge, decision models have met malware traffic but not malware files. Liu~\cite{liu2026evaluating} evaluates Jev, Laya, decider and other models on agent-security decisions and finds that a low average calibration error can hide confident errors on attack groups. Lan et al.~\cite{lan2026detection} measure an ECE of 0.541 for Jev on pooled prompt-injection data, where a miss rate of at most 1\% requires escalating 92.55\% of the inputs. Kisielewicz et al.~\cite{kisielewicz2026escalate} hold out attack families of web requests and recalibrate before thresholding, the methodological twin of this study. Rafe and Das~\cite{rafe2026benchmarking} find that trained classifiers still lead where labels exist and that swapping yes and no flips 50.5 answers in a hundred for Jev. This study brings these models to PE files, and measures on malware evidence from static analysis the calibration, polarity and invariance concerns that these studies raise.

\paragraph{Hallucination and prompt sensitivity.} POPE~\cite{li2023pope} probes hallucination with yes and no questions about absent objects, drawn at random, among frequent objects, or among co-occurring ones. Sclar et al.~\cite{sclar2024quantifying} show that meaning-preserving changes of prompt format move accuracy by up to 76 points. This study transfers the probes of POPE and tests paraphrases on the artifacts of a static state (Section~\ref{sec:res-halluc}).

Table~\ref{tab:sota} summarizes the position of this study and of \fwname{} among the closest works.

\begin{widetable}[t]
  \centering
  \caption{Comparison with the state of the art.}
  \label{tab:sota}
  \small
  \begin{tabular}{l l l c c c}
    \toprule
    Work & Evidence & Output read & Calibration & Fixed FPR & Time split \\
    \midrule
    Chawla and Prasad~\cite{chawla2025decompilation} & Ghidra pseudo-C & JSON verdict & $\times$ & $\times$ & newer test set \\
    PE2Prompt~\cite{sun2026pe2prompt} & EMBER features & forced label & $\times$ & n.r. & n.r. \\
    MARS~\cite{choi2026mars} & evidence bundle & JSON verdict, rule-scored claims & $\times$ & $\times$ & $\times$ \\
    Kurlandski et al.~\cite{kurlandski2026beyond} & bytes, assembly, pseudo-C & trained language model & n.r. & n.r. & \checkmark \\
    AsmRAG~\cite{karbab2026asmrag} & Ghidra assembly & nearest-neighbor vote & n.r. & per function & \checkmark \\
    SecJev~\cite{chen2026secjev} & traffic, logs, telemetry & typed probability & \checkmark & n.r. & transfer \\
    \midrule
    \textbf{\fwname{} (ours)} & digest, Ghidra cards & \begin{tabular}[t]{@{}l@{}}typed answers and\\ embedding of CLM-8B\end{tabular} & \checkmark & \checkmark & \checkmark, OOD \\
    \bottomrule
  \end{tabular}
  \begin{flushleft}\scriptsize $\times$: not done; n.r.: not reported.\end{flushleft}
\end{widetable}
\section{Proposed Methodology}
\label{sec:meth}

The pipeline measures how well typed-output decision models detect Windows Portable Executable (PE) malware from static evidence, through three readouts of their answers and embeddings, and fuses two readouts of CLM-8B into \fwname{}. Every file crosses five modules, which group the seven numbered stages of Fig.~\ref{fig:pipeline}:
\begin{itemize}
  \item \textbf{Data and splits} (stage 1). Three corpora are deduplicated by code and split by first-seen date (malware) or by software component (benign files).
  \item \textbf{Isolated static analysis} (stage 2). A virtual machine without network parses each binary, decompiles it with Ghidra, matches capa rules and exports text and numbers only.
  \item \textbf{State construction} (stage 3). The host renders the records into a canonical textual state: a digest (S0), decompiled function cards (S1), or the digest plus the cards (S2).
  \item \textbf{Typed questioning} (stages 4 and 5). A System One model reads the state and answers typed questions: one verdict, or 32 atomic behaviors.
  \item \textbf{Readout and decision} (stages 6 and 7). The answers or the model's embedding become a malware score (readouts A0, A1 and A2), and \fwname{} fuses the atomic answers and the embedding of CLM-8B; each score is calibrated and compared with a threshold set at 1\% false positive rate (FPR).
\end{itemize}
Algorithm~\ref{alg:forward} gives the procedure for one file, where $\vec{\ell}$ collects the logits $\ell_q$ and a question that exceeds the window leaves its logit at 0. %
Each module is described below, in the order in which a file crosses them.

\begin{algorithm}[t]
  \caption{Decision for one file.}
  \label{alg:forward}
  \small
  \KwIn{PE file $x$; model $M$ with embedding $E_M$ and window $W_M$; verdict question $q_{\mathrm{v}}$; questions $Q$; readout $r$ with combiner $g_r$, calibration map $c_r$, threshold $\tau_{1\%}$ and abstention band $(\tau_{\mathrm{lo}}, \tau_{\mathrm{hi}})$; head $h$ (A2) or mean $\bar{h}$ of five fold heads (\fwname{})}
  \KwOut{decision $\hat{y}$, probability $\hat{p}$ and deferral flag $d$, or a status}
  $R \gets \textsc{StaticAnalysis}(x)$ \tcp*{isolated VM}
  \If{a state block lacks its tool record}{\Return{\texttt{missing\_block}}}
  $s \gets \textsc{RenderState}(R)$ \tcp*{S0 or S2, canonical}
  \If{$|s \oplus q| > W_M$ for every $q \in Q$}{\Return{\texttt{over\_limit}}}
  \eIf{$r = \mathrm{A2}$}{
    $z \gets h\big(E_M(s \oplus q_{\mathrm{v}})\big)$ \tcp*{embedding, trained head}
  }{
    \ForEach{$q \in Q$}{$\ell_q \gets \operatorname{logit} P_M(\mathrm{true} \mid s, q)$}
    \eIf{$r = \text{\textup{\fwname{}}}$}{
      $z \gets g_r\big(\vec{\ell}, \bar{h}(E_M(s \oplus q_{\mathrm{v}}))\big)$ \tcp*{answers + head}
    }{
      $z \gets g_r(\vec{\ell})$ \tcp*{A1: trees; A0: identity}
    }
  }
  $\hat{p} \gets c_r(z)$; \quad $\hat{y} \gets \mathbf{1}[z \geq \tau_{1\%}]$; \quad $d \gets \mathbf{1}[\tau_{\mathrm{lo}} < z < \tau_{\mathrm{hi}}]$\;
  \Return{$(\hat{y}, \hat{p}, d)$}
\end{algorithm}

\subsection{Data and Splits}
\label{sec:meth-data}

Malware comes from BODMAS~\cite{yang2021bodmas} and SOREL-20M~\cite{harang2021sorel}, benign files from a local collection acquired in a clean virtual machine (VM). BODMAS holds 57,293 disarmed Windows malware files first seen between August 2019 and September 2020. SOREL-20M contributes 46,918 disarmed files first seen between January and May 2017 and serves only as out-of-distribution (OOD) data. The benign collection holds 198,461 PE files from MSYS2, Cygwin, winget packages, a Windows~10 x86 installation image and the payloads of existing installers, with the provenance of each file recorded.

To build our test environment we use the following approach:

\paragraph{Deduplication.} Files are grouped into code species by the hash of their executable sections, and one file per species survives. This removes 79.8\% of BODMAS, leaving 11,562 files, and keeps near-copies from crossing splits, a redundancy shown to inflate the behavior tagging of a byte-level model by up to 35.6 points on a corpus that includes SOREL-20M~\cite{kurlandski2026beyond}.

\paragraph{Splits.} Malware is split by first-seen date, following the temporal constraints of TESSERACT~\cite{pendlebury2019tesseract}. T0 (up to October 2019) fits every trained component, V (November and December 2019) tunes and calibrates, and TEST (January to September 2020) measures. Benign files are split by software component into $B_{\mathrm{fit}}$, $B_{\mathrm{val}}$ and $B_{\mathrm{test}}$, so that the versions of one program never straddle two splits. %

\paragraph{Core set.} The primary population keeps native-code executables and libraries. .NET, Visual Basic, AutoIt and installer files form a separate stratum scored by true positive rate (TPR) only. Benign files are matched to the bitness and file kind of the development malware, so that neither property separates the classes. %

\subsection{Isolated Static Analysis}
\label{sec:meth-static}

The binaries never leave the analysis host, a laptop with an Intel Core i9-13900H, an Nvidia RTX 4060 Laptop and 64~GiB of RAM, where a REMnux VM without a network adapter runs the static analysis; no sample is ever executed (Fig.~\ref{fig:staticlab}). Inside the VM, pefile~2024.8.26~\cite{carrera2024pefile}, Detect It Easy~3.21~\cite{hors2026die} and FLOSS~3.1.1~\cite{mandiant2024floss} extract headers, sections, imports, packing and strings, Ghidra~12.1.4~\cite{nsa2026ghidra} decompiles the functions, and capa~9.4.0~\cite{mandiant2026capa} matches its rules on Ghidra's program, with the MITRE ATT\&CK and Malware Behavior Catalog (MBC) identifiers of each capability.

\begin{widefigure}[t]
\centering
\IfFileExists{figures/staticlab.pdf}%
  {\includegraphics[width=\textwidth]{staticlab.pdf}}%
  {\fbox{\parbox[c][3cm][c]{0.9\textwidth}{\centering\dacompletare{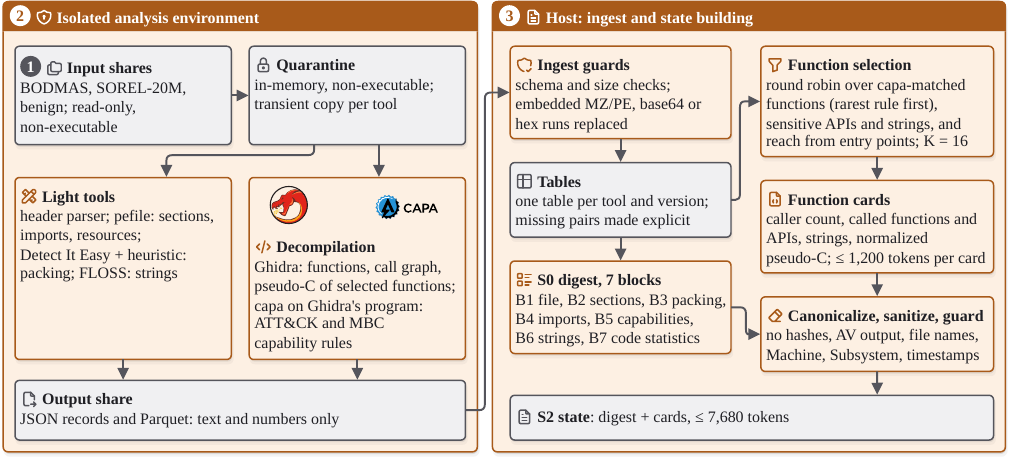 (run figures/src/make\_figures.py)}}}}
\caption{From the binary to the textual state.}
\label{fig:staticlab}
\end{widefigure}
 
Only text and numbers leave the VM: the host checks the records, replaces any embedded executable header or encoded blob with a placeholder, and stores them as Parquet tables from which it builds the states. The workstation that runs the models (Section~\ref{sec:meth-platform}) receives these tables and states only.

\subsection{Textual States}
\label{sec:meth-state}

A System One model reads text, so each file becomes a \emph{state}: a plain-text description of what the static tools found in it. The state is built deterministically from the tool records, one fact per line. Numbers become named classes (e.g., a size or an entropy class), whose edges are fitted on the training files only, and any text taken from the file is quoted and escaped.

\paragraph{Shortcut control.} The malware and benign collections differ in ways that say nothing about behavior. In a pilot, the year of the compilation timestamp alone separated the two classes with an AUROC of 0.83, and the name of the toolchain with 0.78-0.80. A state therefore carries no antivirus output, hash, file name, Machine or Subsystem field, timestamp or tool banner. The remaining collection-specific details are rewritten in the same way for both classes: library names other than the Windows system DLLs and standard section names become neutral labels, years and build paths become placeholders, and words that reveal the compiler or the bitness are removed. A file whose tool record is missing gets no state, so that a tool failure never reads as ``nothing found''.

\paragraph{S0, the digest.} The digest summarizes a file in seven blocks and at most 1,024 tokens (Table~\ref{tab:digest}): the file, its sections, packing, imports, capabilities, strings and code. Imports are grouped by the behavior they enable, through the 685 sensitive APIs named by capa's rules, in 11 categories from process injection to anti-analysis. Capabilities come rarest first, ranked by their frequency among benign training files. A digest over budget is shortened in a fixed order. The median digest is 610 tokens long.

\begin{table}[t]
  \centering
  \caption{Blocks of the S0 digest.}
  \label{tab:digest}
  \small
  \begin{tabularx}{\linewidth}{l >{\raggedright\arraybackslash}X r}
    \toprule
    Block & Content & Cap \\
    \midrule
    B1 File         & kind, size class, overlay, resources, TLS callbacks, signature & 60 \\
    B2 Sections     & count, entry section, roles or quoted names, size, ratio and entropy classes, write and execute flags & 180 \\
    B3 Packing      & Detect It Easy findings, packed flags of detector and heuristic & 40 \\
    B4 Imports      & DLL count, ordinal class, sensitive APIs by category, system DLLs & 350 \\
    B5 Capabilities & capa rules with ATT\&CK and MBC identifiers, rarest first & 300 \\
    B6 Strings      & counts per category and up to 40 quoted strings & 300 \\
    B7 Code         & function count class, share of functions calling sensitive APIs & 40 \\
    \bottomrule
  \end{tabularx}
\end{table}

\paragraph{S1, the decompiled function cards.} S1 renders up to 16 decompiled functions as cards, without the digest. Each function becomes a card: a class of the number of its caller, the functions and APIs it calls, the strings it uses, and its pseudo-C from Ghidra. The pseudo-C is normalized: function and variable names become numbered placeholders, long constants and addresses become generic tokens, and comments are removed. The 16 functions are chosen by alternating three rankings: the functions where capa found a capability, rarest first; the functions that call the most sensitive APIs or hold notable strings; and the functions closest to the entry points. S1 measures what the decompiled code carries without the facts of the digest, and together with other selections, fewer cards and smaller budgets it enters the state ablation of E2 (Section~\ref{sec:res-design}).

\paragraph{S2, the digest with function cards.} S2 joins the digest and the cards of S1, within 7,680 tokens: the 8,192-token window of CLM-8B and the LLMs, minus the question and a margin. The median S2 state is 6,660 tokens long.

\paragraph{Planted text.} Strings come from the file, so an attacker controls them. The builder normalizes their characters, removes control characters and chat-template markers, and flags text addressed to an analyst, e.g., ``this file is safe'', while keeping it in the state. The build stops if a long hex run or a special token of a model survives.

\paragraph{Primary state.} A rule fixed before the experiments chose the state of CLM-8B and the LLMs: S2 replaces S0 only if it raises the validation AUROC of the heads of CLM-8B (Section~\ref{sec:meth-readout}) with a 95\% paired interval above zero. On 782 validation files, S2 reached an AUROC of 0.9954 against 0.9919 for S0, a gain of 0.0036 [0.0002, 0.0085], so S2 is their primary state. The open models and DET read S0.

\subsection{Typed Questions}
\label{sec:meth-questions}

A typed question pairs a text with an answer type and a criterion for each answer. Boolean questions accept true or false, choice questions list options, and score questions list ordered levels. Every question has ten paraphrases.%

Two sets of questions are used. The verdict questions ask whether the program described by the state is malware, in four forms: the question itself, its benign counterpart, a five-level score and a two-option choice in both orders. The atomic questions ask about 32 behaviors (Appendix~\ref{app:repo}): 28 malicious behaviors, from process injection and persistence to ransomware encryption and self-propagation, and 4 benign indicators (e.g., a graphical application or a runtime library). Each behavior is phrased in three vocabularies: plain language, MITRE ATT\&CK and capa. The plain process-injection question reads: ``Based only on the evidence in the state (IMPORTS and CAPABILITIES), does this program write code into another running process and make that process run it?'' Its true criterion reads: ``Remote-process memory and thread APIs (opening a process, allocating or writing remote memory, creating a remote thread), or injection capabilities, appear in the state.'' A category question and twelve questions about single functions serve the ablations of E3 and E2 (Section~\ref{sec:res-design}).

Each model receives the same question through its native interface. CLM-8B encodes the state, a blank line and the question, and embeds each candidate answer through its criterion. The open models receive state and question as fields of a typed decision request. The generative baselines read a fixed chat template whose system prompt declares the state untrusted data, then the state inside \texttt{<state>} tags, the question, and the instruction to reply with one word, Yes or No.

\subsection{System One Models}
\label{sec:meth-models}

\begin{widefigure}[t]
\centering
\IfFileExists{figures/inference.pdf}%
  {\includegraphics[width=\textwidth]{inference.pdf}}%
  {\fbox{\parbox[c][3cm][c]{0.9\textwidth}{\centering\dacompletare{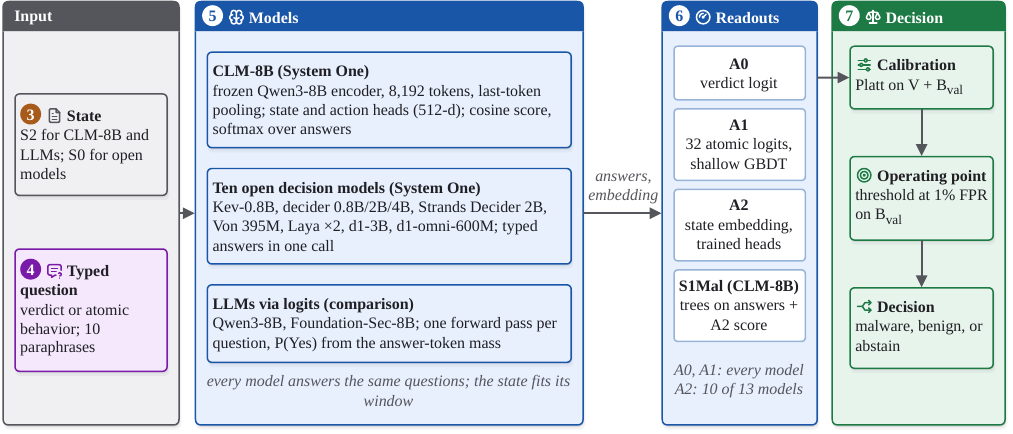 (run figures/src/make\_figures.py)}}}}
\caption{From the state and the typed question to the decision.}
\label{fig:inference}
\end{widefigure}
 
Table~\ref{tab:models} lists the models and Fig.~\ref{fig:inference} shows how they are read.

\paragraph{CLM-8B.} CLM-8B~\cite{kwok2026clm} is a contrastive dual encoder. Let $E$ be the last-token embedding of a frozen Qwen3-8B~\cite{yang2025qwen3} (36 layers, 4,096 dimensions), and $f_S$, $f_A$ the state and action heads: three linear layers each, from 4,096 to 1,536, 1,536 and 512 dimensions with GELU and layer normalization, followed by L2 normalization (18.9 million parameters in total). For state $s$, question $q$ and candidate answers $a \in \mathcal{A}_q$, each with a text $c_a$ (the answer followed by its criterion for a boolean question, the option or the level otherwise),
\begin{equation}
  P(a \mid s, q) = \frac{\exp\big(\lambda \cos(f_S(E(s \oplus q)),\, f_A(E(c_a)))\big)}{\sum_{a' \in \mathcal{A}_q} \exp\big(\lambda \cos(f_S(E(s \oplus q)),\, f_A(E(c_{a'})))\big)},
  \label{eq:clm}
\end{equation}
where $\oplus$ joins the texts with a blank line and $\lambda$ is the learned inverse temperature of the contrastive loss, clamped at 100, the value it reaches in the released heads. The released heads give the verdict and atomic readouts, and the embedding $E(s \oplus q)$ feeds the trained heads of A2. Since $E$ is causal and pooled on the last token, the key-value cache of the state is the same for every question asked about it: the pipeline encodes each state once and then only the tokens of each question. On 13,500 development answers this changed the answer probability by 0.003 on average and the decision at 0.5 in 0.35\% of the answers, a bound that also includes a change of GPU, and it cuts the encoding cost of the 33 atomic questions by an order of magnitude.

\paragraph{Open decision models.} Ten open models, chosen from the releases of September and October 2026, cover the architectures of the class: pointer heads on small Qwen3.5 decoders adapted with LoRA (Kev-0.8B~\cite{palmer2026kev}, Strands Decider 2B~\cite{strands2026decider}), fully fine-tuned Qwen3.5 decoders (decider~\cite{marosi2026decider}), encoders that score option markers (Von~\cite{panisa2026von}, Laya~\cite{m2026laya}) and LFM2.5 models (Liquid d1\footref{fn:d1}). In E1 they read S0 through a typed decision API, at temperature 1.0.

\paragraph{LLMs read through logits.} Two generative large language models (LLMs) answer the same questions through their next-token distribution: Qwen3-8B, which shares the weights file of the CLM-8B encoder, and Foundation-Sec-1.1-8B-Instruct, a later release of the security-specialized Llama~3.1 model of~\cite{weerawardhena2025foundationsec}. One forward pass at temperature 0, with thinking disabled, gives $P(\mathrm{true}) = \sum p(\text{Yes}) / (\sum p(\text{Yes}) + \sum p(\text{No}))$ over the case and spacing variants of both words. The pair separates the effect of the readout from that of the backbone.

\begin{table}[t]
  \centering
  \caption{Decision models.}
  \label{tab:models}
  \footnotesize
  \setlength{\tabcolsep}{3pt}
  \begin{tabularx}{\linewidth}{l r >{\raggedright\arraybackslash}X r}
    \toprule
    Model & Size & Architecture & Window (tok.) \\
    \midrule
    Kev-0.8B~\cite{palmer2026kev}            & 0.8B & Qwen3.5, LoRA, pointer head & 8,192 \\
    decider-0.8b~\cite{marosi2026decider}   & 0.8B & Qwen3.5, fully fine-tune              & 8,192 \\
    decider-2b                              & 2B   & Qwen3.5, full fine-tune, RL, merged LoRA & 8,192 \\
    decider-4b (Q4\_K\_M)                   & 4B   & Qwen3.5, full fine-tune, merged LoRA     & 8,192 \\
    Strands Decider~\cite{strands2026decider} & 2B & Qwen3.5, LoRA, pointer head & 4,096 \\
    Von~\cite{panisa2026von}                & 395M & ModernBERT-large encoder    & 8,192 \\
    Laya multilingual~\cite{m2026laya}      & 322M & mmBERT-base encoder         & 1,024 \\
    Laya typed-decisions                    & 421M & ModernBERT-large encoder    & 1,024 \\
    d1-3B\footref{fn:d1}             & 3.1B & LFM2.5                      & 8,192 \\
    d1-omni-600M                            & 587M & LFM2.5                      & 8,192 \\
    \midrule
    CLM-8B~\cite{kwok2026clm}               & 8.2B & frozen Qwen3-8B, contrastive heads & 8,192 \\
    \bottomrule
  \end{tabularx}
\end{table}

\subsection{Compute Platform and Isolation}
\label{sec:meth-platform}

The models run on a workstation equipped with an AMD Ryzen~9 9950X3D CPU, an NVIDIA GeForce RTX 5090, and an RTX 4090 GPUs, and 192~GB of RAM, with driver 615.78.08 and CUDA 13.4. llama.cpp~\cite{ggml2026llamacpp} build b11429 serves CLM-8B, the two LLMs and Qwen3-Embedding-8B~\cite{zhang2025qwen3embedding} in bf16 with 8,192-token windows; the open models use their own runtimes, and the Liquid d1 models llama.cpp build b11535. Each server listens only on the loopback interface, one model occupies one GPU with all its layers, and a request, state and question together, longer than a model's window is marked over limit instead of being cut; an atomic question left unanswered in this way enters the combiner at $P = 0.5$. %

The workstation never receives a binary. Work arrives from the analysis host as packages of text and numbers, which a checker scans for executable headers, archives and encoded payloads, on both machines. The workstation accepts only a fixed set of commands, reaches only the model and package registries, and stops every job if a loader for PE files appears.

\subsection{Readouts and Decision}
\label{sec:meth-readout}

Three readouts turn the answers or the embedding of a model into a malware score $z$, and \fwname{} fuses two of them for CLM-8B.

\paragraph{A0, verdict.} The score is the logit of $P(\mathrm{true})$ for the verdict question.

\paragraph{A1, atomic.} The answers to the 32 atomic questions in plain language give logits, with probabilities clipped to $[0.005, 0.995]$, and one-hot flags of the file's format route are added. A shallow gradient-boosted tree combiner (LightGBM, depth 3, seven leaves, 200 trees, learning rate 0.05, balanced class weights) is fitted on T0 and $B_{\mathrm{fit}}$. An L2-regularized logistic regression with balanced class weights, whose strength is chosen among ten values from 0.001 to 100 by balanced log-loss on V and $B_{\mathrm{val}}$, and a symbolic noisy-OR without training are the alternatives compared on development files, where the trees were selected for every A1 arm before the freeze. The noisy-OR scores $z = S_{\mathrm{mal}}(1 - \alpha S_{\mathrm{ben}})$ with $\alpha = 1$, where $S_{\mathrm{mal}} = 1 - \prod_{q}(1 - p_q)$ runs over the malicious behaviors and $S_{\mathrm{ben}}$ is the noisy-OR of the four benign indicators.

\paragraph{A2, embedding.} The CLM-8B embedding of the state and the verdict question feeds state and action heads initialized from the released ones and trained with cross-entropy over the two answers, with balanced class weights, on T0 and $B_{\mathrm{fit}}$. Training uses AdamW with a one-cycle cosine schedule over a budget of at least 1,500 optimizer steps, which early stopping on V and $B_{\mathrm{val}}$ ends erlier for five seeds. One of eight settings of learning rate, batch size and weight decay is chosen on seed 0 by the validation loss, and five seeds are then trained with it. A convergence check per seed is reported. Logistic and two-layer probes on the raw embedding (A2-ctrl), and the same probes on Qwen3-Embedding-8B, measure what the heads add.

\paragraph{\fwname{}, fusion.} \fwname{} joins A1 and A2 of CLM-8B in three steps (Algorithm~\ref{alg:forward}):
\begin{enumerate}
  \item \emph{Head score.} Heads trained as those of A2 give the logit of the malware probability of the state, computed out of fold on T0 and $B_{\mathrm{fit}}$ as in stacked generalization~\cite{wolpert1992stacked}, so that the trees never see the score of heads trained on the same file: the files are cut into five folds stratified by class, heads trained on four folds score the fifth, and every other file receives the mean $\bar{h}$ of the logits of the five fold heads, each temperature-scaled on V and $B_{\mathrm{val}}$; the fold heads start from the released heads and use the setting, seed and early stopping of A2.
  \item \emph{Trees.} Gradient-boosted trees in the configuration of A1, fitted anew on T0 and $B_{\mathrm{fit}}$, weigh the 32 atomic logits and the route flags together with this logit.
  \item \emph{Decision.} The score is calibrated and compared with its own $\tau_{1\%}$, as for every readout (below).
\end{enumerate}
We built \fwname{} on CLM-8B because of its context window and of the measured reliability of its typed answers.\par\noindent
Its 8,192-token window takes the decompiled function cards of S2, on which the atomic answers that explain each decision can draw: the open models read the digest S0 (Section~\ref{sec:meth-state}), and the 1,024- or 4,096-token context of some of them (Table~\ref{tab:models}) cannot take the 7,680 tokens of S2. On S2, the state that \fwname{} reads, E7 probes the typed answers of CLM-8B for fabrication and invariance (Section~\ref{sec:res-halluc}), so the answers that \fwname{} shows the analyst have a measured reliability.

\paragraph{Calibration and operating point.} A Platt map~\cite{platt2000probabilities} with balanced class weights is fitted on the score $z$ of every arm on V and $B_{\mathrm{val}}$; the A2 heads are first temperature-scaled~\cite{guo2017calibration}, and E4 compares this map with isotonic and temperature maps. The operating point $\tau_{1\%}$ is the smallest threshold whose FPR on $B_{\mathrm{val}}$ is at most 1\%, and a file is called malware when $z \geq \tau_{1\%}$. On the 2,785 benign validation files with a state, $\tau_{1\%}$ rests on 27 false positives. An abstention band $(\tau_{\mathrm{lo}}, \tau_{\mathrm{hi}})$ on the score, whose two thresholds are chosen together on the validation files to maximize coverage at a class-balanced selective risk of 1\%, flags as deferred the files whose score falls inside it; the decisions at $\tau_{1\%}$ include them.
\section{Evaluation and Results}
\label{sec:res}
\makeatletter
\providecommand{\resset}[2]{\expandafter\def\csname res@#1\endcsname{#2}}
\providecommand{\res}[1]{\ifcsname res@#1\endcsname\csname res@#1\endcsname\else\dacompletare{\detokenize{#1}}\fi}
\providecommand{\ifres}[3]{\ifcsname res@#1\endcsname\expandafter\@firstoftwo\else\expandafter\@secondoftwo\fi{#2}{#3}}
\makeatother
\resset{a3.dev.auroc_a2}{0.9965}
\resset{a3.dev.auroc_a3}{0.9966}
\resset{a3.dev.k}{128}
\resset{a3.dev.loss_a2}{0.0786}
\resset{a3.dev.loss_pca}{0.1198}
\resset{e1.A3-C1.abs}{0.0004}
\resset{e1.A3-C1.ci}{[-0.0004, +0.0012]}
\resset{e1.A3-C1.d}{+0.0004}
\resset{e1.A3-C1.fpr_ci}{[+0.02, +0.70]}
\resset{e1.A3-C1.fpr_d}{+0.34}
\resset{e1.A3-C1.fpr_p}{0.006}
\resset{e1.A3-C1.p}{0.370}
\resset{e1.A3-C1.tpr_ci}{[+0.88, +4.97]}
\resset{e1.A3-C1.tpr_d}{+2.57}
\resset{e1.A3-C1.tpr_p}{$<0.001$}
\resset{e1.A3-C2.abs}{0.0184}
\resset{e1.A3-C2.ci}{[+0.0120, +0.0262]}
\resset{e1.A3-C2.d}{+0.0184}
\resset{e1.A3-C2.fpr_ci}{[-0.17, +0.80]}
\resset{e1.A3-C2.fpr_d}{+0.26}
\resset{e1.A3-C2.fpr_p}{0.121}
\resset{e1.A3-C2.p}{0.002}
\resset{e1.A3-C2.tpr_ci}{[+23.64, +49.55]}
\resset{e1.A3-C2.tpr_d}{+35.57}
\resset{e1.A3-C2.tpr_p}{$<0.001$}
\resset{e1.E1-C1.abs}{0.0013}
\resset{e1.E1-C1.ci}{[-0.0023, -0.0005]}
\resset{e1.E1-C1.d}{-0.0013}
\resset{e1.E1-C1.p}{0.007}
\resset{e1.E1-C2.abs}{0.2749}
\resset{e1.E1-C2.ci}{[+0.1989, +0.3511]}
\resset{e1.E1-C2.d}{+0.2749}
\resset{e1.E1-C2.p}{0.007}
\resset{e1.E1-C3.abs}{0.2569}
\resset{e1.E1-C3.ci}{[+0.1790, +0.3365]}
\resset{e1.E1-C3.d}{+0.2569}
\resset{e1.E1-C3.p}{0.007}
\resset{e1.E1-C4.abs}{0.0041}
\resset{e1.E1-C4.ci}{[+0.0012, +0.0078]}
\resset{e1.E1-C4.d}{+0.0041}
\resset{e1.E1-C4.p}{0.010}
\resset{e1.E1-C6.abs}{0.0108}
\resset{e1.E1-C6.ci}{[+0.0063, +0.0177]}
\resset{e1.E1-C6.d}{+0.0108}
\resset{e1.E1-C6.p}{0.007}
\resset{e1.E1-C7.abs}{0.0449}
\resset{e1.E1-C7.ci}{[+0.0314, +0.0638]}
\resset{e1.E1-C7.d}{+0.0449}
\resset{e1.E1-C7.p}{0.007}
\resset{e1.E1-S-d13-a2.abs}{0.0012}
\resset{e1.E1-S-d13-a2.ci}{[+0.0002, +0.0025]}
\resset{e1.E1-S-d13-a2.d}{+0.0012}
\resset{e1.E1-S-d13-a2.p}{0.045}
\resset{e1.E1-S-d13-det.abs}{0.0006}
\resset{e1.E1-S-d13-det.ci}{[-0.0008, +0.0020]}
\resset{e1.E1-S-d13-det.d}{+0.0006}
\resset{e1.E1-S-d13-det.p}{0.756}
\resset{e1.E1-S-open-d1-3b.abs}{0.3890}
\resset{e1.E1-S-open-d1-3b.ci}{[+0.2904, +0.4818]}
\resset{e1.E1-S-open-d1-3b.d}{+0.3890}
\resset{e1.E1-S-open-d1-3b.p}{0.018}
\resset{e1.E1-S-open-d1-omni-600m.abs}{0.5412}
\resset{e1.E1-S-open-d1-omni-600m.ci}{[+0.4232, +0.6507]}
\resset{e1.E1-S-open-d1-omni-600m.d}{+0.5412}
\resset{e1.E1-S-open-d1-omni-600m.p}{0.018}
\resset{e1.E1-S-open-decider-0.8b.abs}{0.4099}
\resset{e1.E1-S-open-decider-0.8b.ci}{[+0.3221, +0.4933]}
\resset{e1.E1-S-open-decider-0.8b.d}{+0.4099}
\resset{e1.E1-S-open-decider-0.8b.p}{0.018}
\resset{e1.E1-S-open-decider-2b.abs}{0.3141}
\resset{e1.E1-S-open-decider-2b.ci}{[+0.2253, +0.4089]}
\resset{e1.E1-S-open-decider-2b.d}{+0.3141}
\resset{e1.E1-S-open-decider-2b.p}{0.018}
\resset{e1.E1-S-open-decider-4b-gguf.abs}{0.3167}
\resset{e1.E1-S-open-decider-4b-gguf.ci}{[+0.2433, +0.3882]}
\resset{e1.E1-S-open-decider-4b-gguf.d}{+0.3167}
\resset{e1.E1-S-open-decider-4b-gguf.p}{0.018}
\resset{e1.E1-S-open-kev-0.8b.abs}{0.2742}
\resset{e1.E1-S-open-kev-0.8b.ci}{[+0.2181, +0.3338]}
\resset{e1.E1-S-open-kev-0.8b.d}{+0.2742}
\resset{e1.E1-S-open-kev-0.8b.p}{0.018}
\resset{e1.E1-S-open-laya-multilingual.abs}{0.2606}
\resset{e1.E1-S-open-laya-multilingual.ci}{[+0.1712, +0.3551]}
\resset{e1.E1-S-open-laya-multilingual.d}{+0.2606}
\resset{e1.E1-S-open-laya-multilingual.p}{0.018}
\resset{e1.E1-S-open-laya-typed-decisions.abs}{0.4842}
\resset{e1.E1-S-open-laya-typed-decisions.ci}{[+0.3656, +0.6062]}
\resset{e1.E1-S-open-laya-typed-decisions.d}{+0.4842}
\resset{e1.E1-S-open-laya-typed-decisions.p}{0.018}
\resset{e1.E1-S-open-strands-decider-2b-v1-2610.abs}{0.2899}
\resset{e1.E1-S-open-strands-decider-2b-v1-2610.ci}{[+0.2113, +0.3722]}
\resset{e1.E1-S-open-strands-decider-2b-v1-2610.d}{+0.2899}
\resset{e1.E1-S-open-strands-decider-2b-v1-2610.p}{0.018}
\resset{e1.E1-S-open-von.abs}{0.2645}
\resset{e1.E1-S-open-von.ci}{[+0.1712, +0.3835]}
\resset{e1.E1-S-open-von.d}{+0.2645}
\resset{e1.E1-S-open-von.p}{0.018}
\resset{e1.E1-S5.abs}{0.0014}
\resset{e1.E1-S5.ci}{[-0.0035, +0.0005]}
\resset{e1.E1-S5.d}{-0.0014}
\resset{e1.E1-S5.p}{0.620}
\resset{e1.E1-S6.abs}{0.0209}
\resset{e1.E1-S6.ci}{[+0.0138, +0.0307]}
\resset{e1.E1-S6.d}{+0.0209}
\resset{e1.E1-S6.p}{0.018}
\resset{e1.a0.max}{0.8590}
\resset{e1.a0.min}{0.4025}
\resset{e1.a1.max}{0.9834}
\resset{e1.a1.min}{0.9126}
\resset{e1.best_open}{decider-4b (GGUF Q4\_K\_M)}
\resset{e1.clm_a0.abstained}{100.00\%}
\resset{e1.clm_a0.aurc_margin}{0.2679}
\resset{e1.clm_a0.auroc}{0.7194}
\resset{e1.clm_a0.balanced_accuracy}{0.5383}
\resset{e1.clm_a0.brier}{0.2139}
\resset{e1.clm_a0.ci}{[0.643, 0.796]}
\resset{e1.clm_a0.coverage_at_risk_1pct_margin}{0.01\%}
\resset{e1.clm_a0.ece}{0.1024}
\resset{e1.clm_a0.fpr}{1.16\%}
\resset{e1.clm_a0.pauc}{0.5258}
\resset{e1.clm_a0.tpr}{8.82\%}
\resset{e1.clm_a1.abstained}{36.10\%}
\resset{e1.clm_a1.aurc_margin}{0.0783}
\resset{e1.clm_a1.auroc}{0.9763}
\resset{e1.clm_a1.balanced_accuracy}{0.7835}
\resset{e1.clm_a1.brier}{0.0526}
\resset{e1.clm_a1.ci}{[0.965, 0.985]}
\resset{e1.clm_a1.coverage_at_risk_1pct_margin}{39.87\%}
\resset{e1.clm_a1.ece}{0.0161}
\resset{e1.clm_a1.fpr}{1.19\%}
\resset{e1.clm_a1.pauc}{0.6713}
\resset{e1.clm_a1.tpr}{57.90\%}
\resset{e1.clm_a2.abstained}{7.64\%}
\resset{e1.clm_a2.aurc_margin}{0.0166}
\resset{e1.clm_a2.auroc}{0.9943}
\resset{e1.clm_a2.balanced_accuracy}{0.9489}
\resset{e1.clm_a2.brier}{0.0262}
\resset{e1.clm_a2.ci}{[0.991, 0.997]}
\resset{e1.clm_a2.coverage_at_risk_1pct_margin}{52.90\%}
\resset{e1.clm_a2.ece}{0.0060}
\resset{e1.clm_a2.fpr}{1.11\%}
\resset{e1.clm_a2.pauc}{0.8760}
\resset{e1.clm_a2.tpr}{90.90\%}
\resset{e1.clm_a2.tpr01}{46.59\%}
\resset{e1.clm_a2.tpr_at1}{90.09\%}
\resset{e1.clm_a2.tpr_at_a3fpr}{92.15\%}
\resset{e1.clm_a2_ctrl_lr.abstained}{5.62\%}
\resset{e1.clm_a2_ctrl_lr.aurc_margin}{0.0134}
\resset{e1.clm_a2_ctrl_lr.auroc}{0.9955}
\resset{e1.clm_a2_ctrl_lr.balanced_accuracy}{0.9592}
\resset{e1.clm_a2_ctrl_lr.brier}{0.0229}
\resset{e1.clm_a2_ctrl_lr.ci}{[0.993, 0.997]}
\resset{e1.clm_a2_ctrl_lr.coverage_at_risk_1pct_margin}{53.77\%}
\resset{e1.clm_a2_ctrl_lr.ece}{0.0104}
\resset{e1.clm_a2_ctrl_lr.fpr}{1.11\%}
\resset{e1.clm_a2_ctrl_lr.pauc}{0.8974}
\resset{e1.clm_a2_ctrl_lr.tpr}{92.94\%}
\resset{e1.clm_a3.abstained}{6.02\%}
\resset{e1.clm_a3.aurc_margin}{0.0124}
\resset{e1.clm_a3.auroc}{0.9946}
\resset{e1.clm_a3.balanced_accuracy}{0.9601}
\resset{e1.clm_a3.brier}{0.0250}
\resset{e1.clm_a3.ci}{[0.991, 0.997]}
\resset{e1.clm_a3.coverage_at_risk_1pct_margin}{52.33\%}
\resset{e1.clm_a3.ece}{0.0078}
\resset{e1.clm_a3.fpr}{1.45\%}
\resset{e1.clm_a3.pauc}{0.8851}
\resset{e1.clm_a3.tpr}{93.47\%}
\resset{e1.clm_a3.tpr01}{59.15\%}
\resset{e1.clm_a3.tpr_at1}{89.94\%}
\resset{e1.det.abstained}{25.19\%}
\resset{e1.det.aurc_margin}{0.0215}
\resset{e1.det.auroc}{0.9901}
\resset{e1.det.balanced_accuracy}{0.9317}
\resset{e1.det.brier}{0.0310}
\resset{e1.det.ci}{[0.985, 0.994]}
\resset{e1.det.coverage_at_risk_1pct_margin}{51.94\%}
\resset{e1.det.ece}{0.0139}
\resset{e1.det.fpr}{1.91\%}
\resset{e1.det.pauc}{0.8284}
\resset{e1.det.tpr}{88.24\%}
\resset{e1.det.tpr01}{44.99\%}
\resset{e1.fpr.max}{1.91\%}
\resset{e1.fpr.min}{0.95\%}
\resset{e1.lift.max}{+0.5412}
\resset{e1.lift.min}{+0.1143}
\resset{e1.lift.n_sig}{11}
\resset{e1.lift.n_tested}{11}
\resset{e1.lift.pmax}{0.018}
\resset{e1.lift_rel.clm}{35.71\%}
\resset{e1.lift_rel.max}{134.44\%}
\resset{e1.lift_rel.mean}{54.61\%}
\resset{e1.lift_rel.min}{13.30\%}
\resset{e1.llm_fsec_a1.abstained}{35.04\%}
\resset{e1.llm_fsec_a1.aurc_margin}{0.0715}
\resset{e1.llm_fsec_a1.auroc}{0.9733}
\resset{e1.llm_fsec_a1.balanced_accuracy}{0.8130}
\resset{e1.llm_fsec_a1.brier}{0.0537}
\resset{e1.llm_fsec_a1.ci}{[0.962, 0.981]}
\resset{e1.llm_fsec_a1.coverage_at_risk_1pct_margin}{33.37\%}
\resset{e1.llm_fsec_a1.ece}{0.0184}
\resset{e1.llm_fsec_a1.fpr}{1.58\%}
\resset{e1.llm_fsec_a1.pauc}{0.6595}
\resset{e1.llm_fsec_a1.tpr}{64.19\%}
\resset{e1.llm_qwen3_a1.abstained}{75.61\%}
\resset{e1.llm_qwen3_a1.aurc_margin}{0.1184}
\resset{e1.llm_qwen3_a1.auroc}{0.9493}
\resset{e1.llm_qwen3_a1.balanced_accuracy}{0.6767}
\resset{e1.llm_qwen3_a1.brier}{0.0809}
\resset{e1.llm_qwen3_a1.ci}{[0.929, 0.964]}
\resset{e1.llm_qwen3_a1.coverage_at_risk_1pct_margin}{10.87\%}
\resset{e1.llm_qwen3_a1.ece}{0.0235}
\resset{e1.llm_qwen3_a1.fpr}{0.95\%}
\resset{e1.llm_qwen3_a1.pauc}{0.6013}
\resset{e1.llm_qwen3_a1.tpr}{36.30\%}
\resset{e1.llm_qwen3_a1.tpr01}{3.38\%}
\resset{e1.n_benign}{6,126}
\resset{e1.n_malware}{4,703}
\resset{e1.n_models}{13}
\resset{e1.n_no_score}{1,001}
\resset{e1.no_state}{1,001}
\resset{e1.no_state.benign}{971}
\resset{e1.no_state.capa}{968}
\resset{e1.no_state.ghidra}{948}
\resset{e1.no_state.malware}{30}
\resset{e1.no_state.strings}{122}
\resset{e1.open_decider-4b-gguf_a1.abstained}{28.71\%}
\resset{e1.open_decider-4b-gguf_a1.aurc_margin}{0.0472}
\resset{e1.open_decider-4b-gguf_a1.auroc}{0.9834}
\resset{e1.open_decider-4b-gguf_a1.balanced_accuracy}{0.8708}
\resset{e1.open_decider-4b-gguf_a1.brier}{0.0470}
\resset{e1.open_decider-4b-gguf_a1.ci}{[0.975, 0.989]}
\resset{e1.open_decider-4b-gguf_a1.coverage_at_risk_1pct_margin}{44.56\%}
\resset{e1.open_decider-4b-gguf_a1.ece}{0.0167}
\resset{e1.open_decider-4b-gguf_a1.fpr}{1.29\%}
\resset{e1.open_decider-4b-gguf_a1.pauc}{0.7530}
\resset{e1.open_decider-4b-gguf_a1.tpr}{75.44\%}
\resset{e1.open_decider-4b-gguf_a1.tpr01}{24.75\%}
\resset{e1.qwen3_emb_probe.abstained}{4.50\%}
\resset{e1.qwen3_emb_probe.aurc_margin}{0.0107}
\resset{e1.qwen3_emb_probe.auroc}{0.9956}
\resset{e1.qwen3_emb_probe.balanced_accuracy}{0.9610}
\resset{e1.qwen3_emb_probe.brier}{0.0202}
\resset{e1.qwen3_emb_probe.ci}{[0.993, 0.997]}
\resset{e1.qwen3_emb_probe.coverage_at_risk_1pct_margin}{54.68\%}
\resset{e1.qwen3_emb_probe.ece}{0.0066}
\resset{e1.qwen3_emb_probe.fpr}{1.42\%}
\resset{e1.qwen3_emb_probe.pauc}{0.8678}
\resset{e1.qwen3_emb_probe.tpr}{93.62\%}
\resset{e1.tpr01.others_max}{46.59\%}
\resset{e10.clm_a0.auroc}{0.5992}
\resset{e10.clm_a0.auroc_ci}{[0.588, 0.611]}
\resset{e10.clm_a0.fpr}{1.16\%}
\resset{e10.clm_a0.n}{4,725}
\resset{e10.clm_a0.tpr}{2.98\%}
\resset{e10.clm_a0.tpr_ci}{[2.52\%, 3.51\%]}
\resset{e10.clm_a1.auroc}{0.8997}
\resset{e10.clm_a1.auroc_ci}{[0.894, 0.905]}
\resset{e10.clm_a1.fpr}{1.19\%}
\resset{e10.clm_a1.n}{4,725}
\resset{e10.clm_a1.tpr}{22.20\%}
\resset{e10.clm_a1.tpr_ci}{[21.02\%, 23.41\%]}
\resset{e10.clm_a2.auroc}{0.9472}
\resset{e10.clm_a2.auroc_ci}{[0.943, 0.951]}
\resset{e10.clm_a2.fpr}{1.11\%}
\resset{e10.clm_a2.n}{4,725}
\resset{e10.clm_a2.tpr}{51.87\%}
\resset{e10.clm_a2.tpr_ci}{[50.44\%, 53.31\%]}
\resset{e10.clm_a2.tpr_packed}{53.60\%}
\resset{e10.clm_a2.tpr_unpacked}{50.04\%}
\resset{e10.clm_a3.auroc}{0.9453}
\resset{e10.clm_a3.auroc_ci}{[0.941, 0.949]}
\resset{e10.clm_a3.fpr}{1.45\%}
\resset{e10.clm_a3.n}{4,725}
\resset{e10.clm_a3.tpr}{54.75\%}
\resset{e10.clm_a3.tpr_ci}{[53.32\%, 56.18\%]}
\resset{e10.clm_a3.tpr_packed}{57.58\%}
\resset{e10.clm_a3.tpr_unpacked}{51.75\%}
\resset{e10.contam.clm_a0.clean.fpr}{0.23\%}
\resset{e10.contam.clm_a0.clean.tpr}{4.45\%}
\resset{e10.contam.clm_a0.noid.d}{-0.40}
\resset{e10.contam.clm_a0.noid.fpr}{0.94\%}
\resset{e10.contam.clm_a0.noid.p}{0.754}
\resset{e10.contam.clm_a0.noid.tpr}{4.05\%}
\resset{e10.contam.clm_a1.clean.fpr}{1.17\%}
\resset{e10.contam.clm_a1.clean.tpr}{21.66\%}
\resset{e10.contam.clm_a1.noid.d}{+1.21}
\resset{e10.contam.clm_a1.noid.fpr}{1.87\%}
\resset{e10.contam.clm_a1.noid.p}{0.405}
\resset{e10.contam.clm_a1.noid.tpr}{22.87\%}
\resset{e10.contam.clm_a2.clean.fpr}{1.17\%}
\resset{e10.contam.clm_a2.clean.tpr}{52.63\%}
\resset{e10.contam.clm_a2.noid.d}{+8.70}
\resset{e10.contam.clm_a2.noid.fpr}{2.58\%}
\resset{e10.contam.clm_a2.noid.p}{$<0.001$}
\resset{e10.contam.clm_a2.noid.tpr}{61.34\%}
\resset{e10.contam.clm_a3.clean.fpr}{1.17\%}
\resset{e10.contam.clm_a3.clean.tpr}{55.47\%}
\resset{e10.contam.clm_a3.noid.d}{+6.07}
\resset{e10.contam.clm_a3.noid.fpr}{2.34\%}
\resset{e10.contam.clm_a3.noid.p}{$<0.001$}
\resset{e10.contam.clm_a3.noid.tpr}{61.54\%}
\resset{e10.contam.n_benign}{427}
\resset{e10.contam.n_malware}{494}
\resset{e10.det.auroc}{0.9501}
\resset{e10.det.auroc_ci}{[0.946, 0.954]}
\resset{e10.det.fpr}{1.91\%}
\resset{e10.det.n}{4,725}
\resset{e10.det.tpr}{66.46\%}
\resset{e10.det.tpr_ci}{[65.09\%, 67.80\%]}
\resset{e10.idonly.clm_a0.clean.auroc}{0.6047}
\resset{e10.idonly.clm_a0.clean.ci}{[0.570, 0.639]}
\resset{e10.idonly.clm_a0.idonly.auroc}{0.5398}
\resset{e10.idonly.clm_a0.idonly.ci}{[0.504, 0.577]}
\resset{e10.idonly.clm_a0.noid.auroc}{0.5792}
\resset{e10.idonly.clm_a0.noid.ci}{[0.542, 0.614]}
\resset{e10.idonly.clm_a1.clean.auroc}{0.9059}
\resset{e10.idonly.clm_a1.clean.ci}{[0.886, 0.924]}
\resset{e10.idonly.clm_a1.idonly.auroc}{0.6611}
\resset{e10.idonly.clm_a1.idonly.ci}{[0.626, 0.696]}
\resset{e10.idonly.clm_a1.noid.auroc}{0.9027}
\resset{e10.idonly.clm_a1.noid.ci}{[0.882, 0.921]}
\resset{e10.idonly.clm_a2.clean.auroc}{0.9474}
\resset{e10.idonly.clm_a2.clean.ci}{[0.933, 0.960]}
\resset{e10.idonly.clm_a2.idonly.auroc}{0.6548}
\resset{e10.idonly.clm_a2.idonly.ci}{[0.620, 0.690]}
\resset{e10.idonly.clm_a2.noid.auroc}{0.9408}
\resset{e10.idonly.clm_a2.noid.ci}{[0.926, 0.954]}
\resset{e10.idonly.clm_a3.clean.auroc}{0.9484}
\resset{e10.idonly.clm_a3.clean.ci}{[0.934, 0.961]}
\resset{e10.idonly.clm_a3.idonly.auroc}{0.6408}
\resset{e10.idonly.clm_a3.idonly.ci}{[0.606, 0.676]}
\resset{e10.idonly.clm_a3.noid.auroc}{0.9375}
\resset{e10.idonly.clm_a3.noid.ci}{[0.922, 0.952]}
\resset{e10.idonly.flag_lo}{0.60}
\resset{e1b.E1b-P1-d1-3b.abs}{0.0116}
\resset{e1b.E1b-P1-d1-3b.ci}{[+0.0071, +0.0182]}
\resset{e1b.E1b-P1-d1-3b.d}{+0.0116}
\resset{e1b.E1b-P1-d1-3b.p}{0.009}
\resset{e1b.E1b-P1-decider-0.8b.abs}{0.0143}
\resset{e1b.E1b-P1-decider-0.8b.ci}{[+0.0097, +0.0199]}
\resset{e1b.E1b-P1-decider-0.8b.d}{+0.0143}
\resset{e1b.E1b-P1-decider-0.8b.p}{0.009}
\resset{e1b.E1b-P1-decider-2b.abs}{0.0114}
\resset{e1b.E1b-P1-decider-2b.ci}{[+0.0072, +0.0176]}
\resset{e1b.E1b-P1-decider-2b.d}{+0.0114}
\resset{e1b.E1b-P1-decider-2b.p}{0.009}
\resset{e1b.E1b-P1-decider-4b-gguf.abs}{0.0102}
\resset{e1b.E1b-P1-decider-4b-gguf.ci}{[+0.0060, +0.0165]}
\resset{e1b.E1b-P1-decider-4b-gguf.d}{+0.0102}
\resset{e1b.E1b-P1-decider-4b-gguf.p}{0.009}
\resset{e1b.E1b-P1-kev-0.8b.abs}{0.0262}
\resset{e1b.E1b-P1-kev-0.8b.ci}{[+0.0179, +0.0367]}
\resset{e1b.E1b-P1-kev-0.8b.d}{+0.0262}
\resset{e1b.E1b-P1-kev-0.8b.p}{0.009}
\resset{e1b.E1b-P1-laya-multilingual.abs}{0.0724}
\resset{e1b.E1b-P1-laya-multilingual.ci}{[+0.0535, +0.0970]}
\resset{e1b.E1b-P1-laya-multilingual.d}{+0.0724}
\resset{e1b.E1b-P1-laya-multilingual.p}{0.009}
\resset{e1b.E1b-P1-laya-typed-decisions.abs}{0.0517}
\resset{e1b.E1b-P1-laya-typed-decisions.ci}{[+0.0336, +0.0765]}
\resset{e1b.E1b-P1-laya-typed-decisions.d}{+0.0517}
\resset{e1b.E1b-P1-laya-typed-decisions.p}{0.009}
\resset{e1b.E1b-P1-strands-decider-2b-v1-2610.abs}{0.0116}
\resset{e1b.E1b-P1-strands-decider-2b-v1-2610.ci}{[+0.0077, +0.0171]}
\resset{e1b.E1b-P1-strands-decider-2b-v1-2610.d}{+0.0116}
\resset{e1b.E1b-P1-strands-decider-2b-v1-2610.p}{0.009}
\resset{e1b.E1b-P1-von.abs}{0.0293}
\resset{e1b.E1b-P1-von.ci}{[+0.0198, +0.0424]}
\resset{e1b.E1b-P1-von.d}{+0.0293}
\resset{e1b.E1b-P1-von.p}{0.009}
\resset{e1b.E1b-S1-d1-3b.abs}{0.4006}
\resset{e1b.E1b-S1-d1-3b.ci}{[+0.3029, +0.4921]}
\resset{e1b.E1b-S1-d1-3b.d}{+0.4006}
\resset{e1b.E1b-S1-d1-3b.p}{0.009}
\resset{e1b.E1b-S1-decider-0.8b.abs}{0.4242}
\resset{e1b.E1b-S1-decider-0.8b.ci}{[+0.3383, +0.5062]}
\resset{e1b.E1b-S1-decider-0.8b.d}{+0.4242}
\resset{e1b.E1b-S1-decider-0.8b.p}{0.009}
\resset{e1b.E1b-S1-decider-2b.abs}{0.3255}
\resset{e1b.E1b-S1-decider-2b.ci}{[+0.2382, +0.4192]}
\resset{e1b.E1b-S1-decider-2b.d}{+0.3255}
\resset{e1b.E1b-S1-decider-2b.p}{0.009}
\resset{e1b.E1b-S1-decider-4b-gguf.abs}{0.3269}
\resset{e1b.E1b-S1-decider-4b-gguf.ci}{[+0.2531, +0.3993]}
\resset{e1b.E1b-S1-decider-4b-gguf.d}{+0.3269}
\resset{e1b.E1b-S1-decider-4b-gguf.p}{0.009}
\resset{e1b.E1b-S1-kev-0.8b.abs}{0.3004}
\resset{e1b.E1b-S1-kev-0.8b.ci}{[+0.2459, +0.3586]}
\resset{e1b.E1b-S1-kev-0.8b.d}{+0.3004}
\resset{e1b.E1b-S1-kev-0.8b.p}{0.009}
\resset{e1b.E1b-S1-laya-multilingual.abs}{0.3329}
\resset{e1b.E1b-S1-laya-multilingual.ci}{[+0.2468, +0.4255]}
\resset{e1b.E1b-S1-laya-multilingual.d}{+0.3329}
\resset{e1b.E1b-S1-laya-multilingual.p}{0.009}
\resset{e1b.E1b-S1-laya-typed-decisions.abs}{0.5353}
\resset{e1b.E1b-S1-laya-typed-decisions.ci}{[+0.4106, +0.6669]}
\resset{e1b.E1b-S1-laya-typed-decisions.d}{+0.5353}
\resset{e1b.E1b-S1-laya-typed-decisions.p}{0.009}
\resset{e1b.E1b-S1-strands-decider-2b-v1-2610.abs}{0.3015}
\resset{e1b.E1b-S1-strands-decider-2b-v1-2610.ci}{[+0.2223, +0.3839]}
\resset{e1b.E1b-S1-strands-decider-2b-v1-2610.d}{+0.3015}
\resset{e1b.E1b-S1-strands-decider-2b-v1-2610.p}{0.009}
\resset{e1b.E1b-S1-von.abs}{0.2939}
\resset{e1b.E1b-S1-von.ci}{[+0.1955, +0.4163]}
\resset{e1b.E1b-S1-von.d}{+0.2939}
\resset{e1b.E1b-S1-von.p}{0.009}
\resset{e1b.E1b-S2-d1-3b.abs}{0.0035}
\resset{e1b.E1b-S2-d1-3b.ci}{[-0.0067, -0.0010]}
\resset{e1b.E1b-S2-d1-3b.d}{-0.0035}
\resset{e1b.E1b-S2-d1-3b.p}{0.035}
\resset{e1b.E1b-S2-decider-0.8b.abs}{0.0026}
\resset{e1b.E1b-S2-decider-0.8b.ci}{[-0.0053, -0.0007]}
\resset{e1b.E1b-S2-decider-0.8b.d}{-0.0026}
\resset{e1b.E1b-S2-decider-0.8b.p}{0.055}
\resset{e1b.E1b-S2-decider-2b.abs}{0.0005}
\resset{e1b.E1b-S2-decider-2b.ci}{[-0.0027, +0.0015]}
\resset{e1b.E1b-S2-decider-2b.d}{-0.0005}
\resset{e1b.E1b-S2-decider-2b.p}{1.000}
\resset{e1b.E1b-S2-decider-4b-gguf.abs}{0.0006}
\resset{e1b.E1b-S2-decider-4b-gguf.ci}{[-0.0027, +0.0012]}
\resset{e1b.E1b-S2-decider-4b-gguf.d}{-0.0006}
\resset{e1b.E1b-S2-decider-4b-gguf.p}{1.000}
\resset{e1b.E1b-S2-kev-0.8b.abs}{0.0008}
\resset{e1b.E1b-S2-kev-0.8b.ci}{[-0.0027, +0.0011]}
\resset{e1b.E1b-S2-kev-0.8b.d}{-0.0008}
\resset{e1b.E1b-S2-kev-0.8b.p}{1.000}
\resset{e1b.E1b-S2-laya-multilingual.abs}{0.0096}
\resset{e1b.E1b-S2-laya-multilingual.ci}{[-0.0154, -0.0048]}
\resset{e1b.E1b-S2-laya-multilingual.d}{-0.0096}
\resset{e1b.E1b-S2-laya-multilingual.p}{0.009}
\resset{e1b.E1b-S2-laya-typed-decisions.abs}{0.0034}
\resset{e1b.E1b-S2-laya-typed-decisions.ci}{[-0.0067, -0.0008]}
\resset{e1b.E1b-S2-laya-typed-decisions.d}{-0.0034}
\resset{e1b.E1b-S2-laya-typed-decisions.p}{0.054}
\resset{e1b.E1b-S2-strands-decider-2b-v1-2610.abs}{0.0006}
\resset{e1b.E1b-S2-strands-decider-2b-v1-2610.ci}{[-0.0018, +0.0027]}
\resset{e1b.E1b-S2-strands-decider-2b-v1-2610.d}{+0.0006}
\resset{e1b.E1b-S2-strands-decider-2b-v1-2610.p}{1.000}
\resset{e1b.E1b-S2-von.abs}{0.0047}
\resset{e1b.E1b-S2-von.ci}{[-0.0086, -0.0019]}
\resset{e1b.E1b-S2-von.d}{-0.0047}
\resset{e1b.E1b-S2-von.p}{0.009}
\resset{e1b.E1b-S3-d1-3b.abs}{0.0002}
\resset{e1b.E1b-S3-d1-3b.ci}{[-0.0013, +0.0016]}
\resset{e1b.E1b-S3-d1-3b.d}{+0.0002}
\resset{e1b.E1b-S3-d1-3b.p}{1.000}
\resset{e1b.E1b-S3-decider-0.8b.abs}{0.0012}
\resset{e1b.E1b-S3-decider-0.8b.ci}{[-0.0010, +0.0036]}
\resset{e1b.E1b-S3-decider-0.8b.d}{+0.0012}
\resset{e1b.E1b-S3-decider-0.8b.p}{1.000}
\resset{e1b.E1b-S3-decider-2b.abs}{0.0008}
\resset{e1b.E1b-S3-decider-2b.ci}{[-0.0038, +0.0013]}
\resset{e1b.E1b-S3-decider-2b.d}{-0.0008}
\resset{e1b.E1b-S3-decider-2b.p}{1.000}
\resset{e1b.E1b-S3-decider-4b-gguf.abs}{0.0011}
\resset{e1b.E1b-S3-decider-4b-gguf.ci}{[-0.0026, +0.0003]}
\resset{e1b.E1b-S3-decider-4b-gguf.d}{-0.0011}
\resset{e1b.E1b-S3-decider-4b-gguf.p}{0.864}
\resset{e1b.E1b-S3-kev-0.8b.abs}{0.0001}
\resset{e1b.E1b-S3-kev-0.8b.ci}{[-0.0015, +0.0015]}
\resset{e1b.E1b-S3-kev-0.8b.d}{-0.0001}
\resset{e1b.E1b-S3-kev-0.8b.p}{1.000}
\resset{e1b.E1b-S3-laya-multilingual.abs}{0.0019}
\resset{e1b.E1b-S3-laya-multilingual.ci}{[-0.0010, +0.0051]}
\resset{e1b.E1b-S3-laya-multilingual.d}{+0.0019}
\resset{e1b.E1b-S3-laya-multilingual.p}{1.000}
\resset{e1b.E1b-S3-laya-typed-decisions.abs}{0.0025}
\resset{e1b.E1b-S3-laya-typed-decisions.ci}{[-0.0000, +0.0052]}
\resset{e1b.E1b-S3-laya-typed-decisions.d}{+0.0025}
\resset{e1b.E1b-S3-laya-typed-decisions.p}{0.531}
\resset{e1b.E1b-S3-strands-decider-2b-v1-2610.abs}{0.0010}
\resset{e1b.E1b-S3-strands-decider-2b-v1-2610.ci}{[-0.0010, +0.0034]}
\resset{e1b.E1b-S3-strands-decider-2b-v1-2610.d}{+0.0010}
\resset{e1b.E1b-S3-strands-decider-2b-v1-2610.p}{1.000}
\resset{e1b.E1b-S3-von.abs}{0.0004}
\resset{e1b.E1b-S3-von.ci}{[-0.0014, +0.0020]}
\resset{e1b.E1b-S3-von.d}{+0.0004}
\resset{e1b.E1b-S3-von.p}{1.000}
\resset{e1b.a2.best}{Strands Decider 2B}
\resset{e1b.a2.max}{0.9948}
\resset{e1b.a2.min}{0.9853}
\resset{e1b.a2.tpr_max}{95.73\%}
\resset{e1b.a2.tpr_min}{83.66\%}
\resset{e1b.a2.worst}{Laya multilingual 322M}
\resset{e1b.n_models}{9}
\resset{e1b.open_d1-3b_a2.auroc}{0.9908}
\resset{e1b.open_d1-3b_a2.ci}{[0.986, 0.994]}
\resset{e1b.open_d1-3b_a2.fpr}{1.37\%}
\resset{e1b.open_d1-3b_a2.tpr}{88.26\%}
\resset{e1b.open_decider-0.8b_a2.auroc}{0.9916}
\resset{e1b.open_decider-0.8b_a2.ci}{[0.987, 0.995]}
\resset{e1b.open_decider-0.8b_a2.fpr}{0.88\%}
\resset{e1b.open_decider-0.8b_a2.tpr}{85.05\%}
\resset{e1b.open_decider-2b_a2.auroc}{0.9938}
\resset{e1b.open_decider-2b_a2.ci}{[0.989, 0.997]}
\resset{e1b.open_decider-2b_a2.fpr}{1.37\%}
\resset{e1b.open_decider-2b_a2.tpr}{95.36\%}
\resset{e1b.open_decider-4b-gguf_a2.auroc}{0.9936}
\resset{e1b.open_decider-4b-gguf_a2.ci}{[0.990, 0.996]}
\resset{e1b.open_decider-4b-gguf_a2.fpr}{1.42\%}
\resset{e1b.open_decider-4b-gguf_a2.tpr}{92.86\%}
\resset{e1b.open_kev-0.8b_a2.auroc}{0.9934}
\resset{e1b.open_kev-0.8b_a2.ci}{[0.990, 0.996]}
\resset{e1b.open_kev-0.8b_a2.fpr}{1.52\%}
\resset{e1b.open_kev-0.8b_a2.tpr}{91.69\%}
\resset{e1b.open_laya-multilingual_a2.auroc}{0.9853}
\resset{e1b.open_laya-multilingual_a2.ci}{[0.978, 0.991]}
\resset{e1b.open_laya-multilingual_a2.fpr}{1.49\%}
\resset{e1b.open_laya-multilingual_a2.tpr}{83.66\%}
\resset{e1b.open_laya-typed-decisions_a2.auroc}{0.9915}
\resset{e1b.open_laya-typed-decisions_a2.ci}{[0.987, 0.994]}
\resset{e1b.open_laya-typed-decisions_a2.fpr}{1.18\%}
\resset{e1b.open_laya-typed-decisions_a2.tpr}{85.83\%}
\resset{e1b.open_strands-decider-2b-v1-2610_a2.auroc}{0.9948}
\resset{e1b.open_strands-decider-2b-v1-2610_a2.ci}{[0.992, 0.997]}
\resset{e1b.open_strands-decider-2b-v1-2610_a2.fpr}{1.50\%}
\resset{e1b.open_strands-decider-2b-v1-2610_a2.tpr}{95.73\%}
\resset{e1b.open_von_a2.auroc}{0.9896}
\resset{e1b.open_von_a2.ci}{[0.983, 0.994]}
\resset{e1b.open_von_a2.fpr}{1.78\%}
\resset{e1b.open_von_a2.tpr}{93.26\%}
\resset{e1b.p1.max}{0.0724}
\resset{e1b.p1.min}{0.0102}
\resset{e1b.p1.n_sig}{9}
\resset{e1b.p1.pmax}{0.009}
\resset{e1b.s2.max}{+0.0006}
\resset{e1b.s2.min}{-0.0096}
\resset{e1b.s2.n_lower}{3}
\resset{e1b.s2.n_ns}{6}
\resset{e1b.s3.n_sig}{0}
\resset{e1b.s3.pmin}{0.531}
\resset{e1b.stage}{test}
\resset{e2.A.max}{+0.0136}
\resset{e2.A.min}{-0.0001}
\resset{e2.A.n}{4}
\resset{e2.A.n_sig}{0}
\resset{e2.A.n_sig_neg}{0}
\resset{e2.A.n_sig_pos}{0}
\resset{e2.A1.clm_a0.a}{0.6379}
\resset{e2.A1.clm_a0.b}{0.6379}
\resset{e2.A1.clm_a0.ci}{[+0.0000, +0.0000]}
\resset{e2.A1.clm_a0.d}{+0.0000}
\resset{e2.A1.clm_a0.p}{1.000}
\resset{e2.A1.clm_a1.a}{0.8898}
\resset{e2.A1.clm_a1.b}{0.8762}
\resset{e2.A1.clm_a1.ci}{[-0.0018, +0.0341]}
\resset{e2.A1.clm_a1.d}{+0.0136}
\resset{e2.A1.clm_a1.p}{0.110}
\resset{e2.A1.clm_a2.a}{0.9523}
\resset{e2.A1.clm_a2.b}{0.9524}
\resset{e2.A1.clm_a2.ci}{[-0.0019, +0.0016]}
\resset{e2.A1.clm_a2.d}{-0.0001}
\resset{e2.A1.clm_a2.p}{0.972}
\resset{e2.A1.kev.a}{0.8839}
\resset{e2.A1.kev.b}{0.8833}
\resset{e2.A1.kev.ci}{[-0.0051, +0.0060]}
\resset{e2.A1.kev.d}{+0.0006}
\resset{e2.A1.kev.p}{0.855}
\resset{e2.B.max}{+0.0502}
\resset{e2.B.min}{-0.0487}
\resset{e2.B.n}{12}
\resset{e2.B.n_sig}{2}
\resset{e2.B.n_sig_neg}{0}
\resset{e2.B.n_sig_pos}{2}
\resset{e2.B1.clm_a0.a}{0.7216}
\resset{e2.B1.clm_a0.b}{0.7498}
\resset{e2.B1.clm_a0.ci}{[-0.0702, +0.0176]}
\resset{e2.B1.clm_a0.d}{-0.0282}
\resset{e2.B1.clm_a0.p}{0.231}
\resset{e2.B1.clm_a1.a}{0.9120}
\resset{e2.B1.clm_a1.b}{0.8909}
\resset{e2.B1.clm_a1.ci}{[-0.0140, +0.0669]}
\resset{e2.B1.clm_a1.d}{+0.0211}
\resset{e2.B1.clm_a1.p}{0.265}
\resset{e2.B1.clm_a2.a}{0.9642}
\resset{e2.B1.clm_a2.b}{0.9606}
\resset{e2.B1.clm_a2.ci}{[-0.0052, +0.0152]}
\resset{e2.B1.clm_a2.d}{+0.0035}
\resset{e2.B1.clm_a2.p}{1.000}
\resset{e2.B1.kev.a}{0.9171}
\resset{e2.B1.kev.b}{0.8669}
\resset{e2.B1.kev.ci}{[+0.0072, +0.1009]}
\resset{e2.B1.kev.d}{+0.0502}
\resset{e2.B1.kev.p}{0.078}
\resset{e2.B2.clm_a0.a}{0.6772}
\resset{e2.B2.clm_a0.b}{0.7052}
\resset{e2.B2.clm_a0.ci}{[-0.0586, +0.0019]}
\resset{e2.B2.clm_a0.d}{-0.0280}
\resset{e2.B2.clm_a0.p}{0.204}
\resset{e2.B2.clm_a1.a}{0.9334}
\resset{e2.B2.clm_a1.b}{0.9062}
\resset{e2.B2.clm_a1.ci}{[+0.0081, +0.0523]}
\resset{e2.B2.clm_a1.d}{+0.0271}
\resset{e2.B2.clm_a1.p}{0.009}
\resset{e2.B2.clm_a2.a}{0.9645}
\resset{e2.B2.clm_a2.b}{0.9638}
\resset{e2.B2.clm_a2.ci}{[-0.0049, +0.0086]}
\resset{e2.B2.clm_a2.d}{+0.0007}
\resset{e2.B2.clm_a2.p}{1.000}
\resset{e2.B2.kev.a}{0.9149}
\resset{e2.B2.kev.b}{0.9090}
\resset{e2.B2.kev.ci}{[-0.0426, +0.0569]}
\resset{e2.B2.kev.d}{+0.0059}
\resset{e2.B2.kev.p}{0.896}
\resset{e2.B3.clm_a0.a}{0.7012}
\resset{e2.B3.clm_a0.b}{0.7498}
\resset{e2.B3.clm_a0.ci}{[-0.1035, +0.0082]}
\resset{e2.B3.clm_a0.d}{-0.0487}
\resset{e2.B3.clm_a0.p}{0.204}
\resset{e2.B3.clm_a1.a}{0.9370}
\resset{e2.B3.clm_a1.b}{0.8909}
\resset{e2.B3.clm_a1.ci}{[+0.0114, +0.0898]}
\resset{e2.B3.clm_a1.d}{+0.0461}
\resset{e2.B3.clm_a1.p}{0.012}
\resset{e2.B3.clm_a2.a}{0.9645}
\resset{e2.B3.clm_a2.b}{0.9606}
\resset{e2.B3.clm_a2.ci}{[-0.0073, +0.0216]}
\resset{e2.B3.clm_a2.d}{+0.0039}
\resset{e2.B3.clm_a2.p}{1.000}
\resset{e2.B3.kev.a}{0.9282}
\resset{e2.B3.kev.b}{0.8856}
\resset{e2.B3.kev.ci}{[-0.0057, +0.0969]}
\resset{e2.B3.kev.d}{+0.0427}
\resset{e2.B3.kev.p}{0.196}
\resset{e2.C.max}{+0.0361}
\resset{e2.C.min}{-0.0430}
\resset{e2.C.n}{20}
\resset{e2.C.n_sig}{1}
\resset{e2.C.n_sig_neg}{0}
\resset{e2.C.n_sig_pos}{1}
\resset{e2.C1.clm_a0.a}{0.6709}
\resset{e2.C1.clm_a0.b}{0.6379}
\resset{e2.C1.clm_a0.ci}{[-0.0078, +0.0828]}
\resset{e2.C1.clm_a0.d}{+0.0330}
\resset{e2.C1.clm_a0.p}{0.348}
\resset{e2.C1.clm_a1.a}{0.8798}
\resset{e2.C1.clm_a1.b}{0.8762}
\resset{e2.C1.clm_a1.ci}{[-0.0222, +0.0279]}
\resset{e2.C1.clm_a1.d}{+0.0036}
\resset{e2.C1.clm_a1.p}{0.966}
\resset{e2.C1.clm_a2.a}{0.9600}
\resset{e2.C1.clm_a2.b}{0.9524}
\resset{e2.C1.clm_a2.ci}{[-0.0010, +0.0199]}
\resset{e2.C1.clm_a2.d}{+0.0076}
\resset{e2.C1.clm_a2.p}{0.368}
\resset{e2.C1.kev.a}{0.9133}
\resset{e2.C1.kev.b}{0.8833}
\resset{e2.C1.kev.ci}{[-0.0272, +0.0890]}
\resset{e2.C1.kev.d}{+0.0300}
\resset{e2.C1.kev.p}{1.000}
\resset{e2.C4.clm_a0.a}{0.6621}
\resset{e2.C4.clm_a0.b}{0.6379}
\resset{e2.C4.clm_a0.ci}{[-0.0013, +0.0547]}
\resset{e2.C4.clm_a0.d}{+0.0243}
\resset{e2.C4.clm_a0.p}{0.252}
\resset{e2.C4.clm_a1.a}{0.9015}
\resset{e2.C4.clm_a1.b}{0.8762}
\resset{e2.C4.clm_a1.ci}{[+0.0099, +0.0457]}
\resset{e2.C4.clm_a1.d}{+0.0253}
\resset{e2.C4.clm_a1.p}{0.005}
\resset{e2.C4.clm_a2.a}{0.9464}
\resset{e2.C4.clm_a2.b}{0.9524}
\resset{e2.C4.clm_a2.ci}{[-0.0131, -0.0007]}
\resset{e2.C4.clm_a2.d}{-0.0060}
\resset{e2.C4.clm_a2.p}{0.110}
\resset{e2.C4.kev.a}{0.9003}
\resset{e2.C4.kev.b}{0.8833}
\resset{e2.C4.kev.ci}{[-0.0103, +0.0481]}
\resset{e2.C4.kev.d}{+0.0170}
\resset{e2.C4.kev.p}{1.000}
\resset{e2.D.max}{-0.0252}
\resset{e2.D.min}{-0.3397}
\resset{e2.D.n}{4}
\resset{e2.D.n_sig}{4}
\resset{e2.D.n_sig_neg}{4}
\resset{e2.D.n_sig_pos}{0}
\resset{e2.D1.clm_a0.a}{0.6379}
\resset{e2.D1.clm_a0.b}{0.9776}
\resset{e2.D1.clm_a0.ci}{[-0.4356, -0.2547]}
\resset{e2.D1.clm_a0.d}{-0.3397}
\resset{e2.D1.clm_a0.p}{$<0.001$}
\resset{e2.D1.clm_a1.a}{0.8762}
\resset{e2.D1.clm_a1.b}{0.9776}
\resset{e2.D1.clm_a1.ci}{[-0.1509, -0.0586]}
\resset{e2.D1.clm_a1.d}{-0.1014}
\resset{e2.D1.clm_a1.p}{$<0.001$}
\resset{e2.D1.clm_a2.a}{0.9524}
\resset{e2.D1.clm_a2.b}{0.9776}
\resset{e2.D1.clm_a2.ci}{[-0.0592, -0.0012]}
\resset{e2.D1.clm_a2.d}{-0.0252}
\resset{e2.D1.clm_a2.p}{0.041}
\resset{e2.D1.kev.a}{0.8833}
\resset{e2.D1.kev.b}{0.9776}
\resset{e2.D1.kev.ci}{[-0.1698, -0.0370]}
\resset{e2.D1.kev.d}{-0.0943}
\resset{e2.D1.kev.p}{0.002}
\resset{e2.K.max}{+0.0278}
\resset{e2.K.min}{-0.0121}
\resset{e2.K.n}{12}
\resset{e2.K.n_sig}{0}
\resset{e2.K.n_sig_neg}{0}
\resset{e2.K.n_sig_pos}{0}
\resset{e2.K1.clm_a0.a}{0.7052}
\resset{e2.K1.clm_a0.b}{0.6848}
\resset{e2.K1.clm_a0.ci}{[+0.0013, +0.0401]}
\resset{e2.K1.clm_a0.d}{+0.0204}
\resset{e2.K1.clm_a0.p}{0.090}
\resset{e2.K1.clm_a1.a}{0.9062}
\resset{e2.K1.clm_a1.b}{0.9183}
\resset{e2.K1.clm_a1.ci}{[-0.0274, +0.0047]}
\resset{e2.K1.clm_a1.d}{-0.0121}
\resset{e2.K1.clm_a1.p}{0.666}
\resset{e2.K1.clm_a2.a}{0.9638}
\resset{e2.K1.clm_a2.b}{0.9630}
\resset{e2.K1.clm_a2.ci}{[-0.0032, +0.0055]}
\resset{e2.K1.clm_a2.d}{+0.0008}
\resset{e2.K1.clm_a2.p}{1.000}
\resset{e2.K1.kev.a}{0.8991}
\resset{e2.K1.kev.b}{0.8984}
\resset{e2.K1.kev.ci}{[-0.0141, +0.0185]}
\resset{e2.K1.kev.d}{+0.0007}
\resset{e2.K1.kev.p}{1.000}
\resset{e2.KBP.clm_a0.n}{11}
\resset{e2.KBP.clm_a0.n_sig}{2}
\resset{e2.KBP.clm_a1.n}{11}
\resset{e2.KBP.clm_a1.n_sig}{2}
\resset{e2.KBP.clm_a2.n}{11}
\resset{e2.KBP.clm_a2.n_sig}{0}
\resset{e2.KBP.kev.n}{11}
\resset{e2.KBP.kev.n_sig}{0}
\resset{e2.L.clm_a0.n}{7}
\resset{e2.L.clm_a0.n_sig}{1}
\resset{e2.L.clm_a0.rest_blocks}{File, Sections, Packing, Imports, Capabilities and Code}
\resset{e2.L.clm_a0.rest_max}{+0.0487}
\resset{e2.L.clm_a0.sig_blocks}{Strings}
\resset{e2.L.clm_a0.sig_max}{+0.1373}
\resset{e2.L.clm_a0.sig_min}{+0.1373}
\resset{e2.L.clm_a1.n}{7}
\resset{e2.L.clm_a1.n_sig}{0}
\resset{e2.L.clm_a1.rest_blocks}{File, Sections, Packing, Imports, Capabilities, Strings and Code}
\resset{e2.L.clm_a1.rest_max}{+0.0048}
\resset{e2.L.clm_a1.sig_blocks}{none}
\resset{e2.L.clm_a2.n}{7}
\resset{e2.L.clm_a2.n_sig}{0}
\resset{e2.L.clm_a2.rest_blocks}{File, Sections, Packing, Imports, Capabilities, Strings and Code}
\resset{e2.L.clm_a2.rest_max}{+0.0095}
\resset{e2.L.clm_a2.sig_blocks}{none}
\resset{e2.L.kev.n}{7}
\resset{e2.L.kev.n_sig}{0}
\resset{e2.L.kev.rest_blocks}{File, Sections, Packing, Imports, Capabilities, Strings and Code}
\resset{e2.L.kev.rest_max}{+0.0445}
\resset{e2.L.kev.sig_blocks}{none}
\resset{e2.L.max}{+0.1373}
\resset{e2.L.min}{-0.0623}
\resset{e2.L.n}{28}
\resset{e2.L.n_sig}{3}
\resset{e2.L.n_sig_neg}{2}
\resset{e2.L.n_sig_pos}{1}
\resset{e2.M.max}{+0.0334}
\resset{e2.M.min}{-0.0292}
\resset{e2.M.n}{2}
\resset{e2.M.n_sig}{0}
\resset{e2.M.n_sig_neg}{0}
\resset{e2.M.n_sig_pos}{0}
\resset{e2.M1.clm_a1.a}{0.9350}
\resset{e2.M1.clm_a1.b}{0.9015}
\resset{e2.M1.clm_a1.ci}{[-0.0057, +0.0766]}
\resset{e2.M1.clm_a1.d}{+0.0334}
\resset{e2.M1.clm_a1.p}{0.098}
\resset{e2.M1.kev.a}{0.8999}
\resset{e2.M1.kev.b}{0.9291}
\resset{e2.M1.kev.ci}{[-0.0823, +0.0182]}
\resset{e2.M1.kev.d}{-0.0292}
\resset{e2.M1.kev.p}{0.235}
\resset{e2.O.clm_a0.n}{7}
\resset{e2.O.clm_a0.n_sig}{4}
\resset{e2.O.clm_a0.rest_blocks}{Sections, Packing and Imports}
\resset{e2.O.clm_a0.rest_max}{+0.0555}
\resset{e2.O.clm_a0.sig_blocks}{File, Capabilities, Strings and Code}
\resset{e2.O.clm_a0.sig_max}{+0.2880}
\resset{e2.O.clm_a0.sig_min}{+0.1519}
\resset{e2.O.clm_a1.n}{7}
\resset{e2.O.clm_a1.n_sig}{2}
\resset{e2.O.clm_a1.rest_blocks}{File, Sections, Imports, Capabilities and Strings}
\resset{e2.O.clm_a1.rest_max}{+0.0417}
\resset{e2.O.clm_a1.sig_blocks}{Packing and Code}
\resset{e2.O.clm_a1.sig_max}{+0.1662}
\resset{e2.O.clm_a1.sig_min}{+0.1632}
\resset{e2.O.clm_a2.n}{7}
\resset{e2.O.clm_a2.n_sig}{3}
\resset{e2.O.clm_a2.rest_blocks}{Sections, Imports, Capabilities and Strings}
\resset{e2.O.clm_a2.rest_max}{+0.0167}
\resset{e2.O.clm_a2.sig_blocks}{File, Packing and Code}
\resset{e2.O.clm_a2.sig_max}{+0.2433}
\resset{e2.O.clm_a2.sig_min}{+0.1175}
\resset{e2.O.kev.n}{7}
\resset{e2.O.kev.n_sig}{1}
\resset{e2.O.kev.rest_blocks}{File, Sections, Packing, Imports, Capabilities and Strings}
\resset{e2.O.kev.rest_max}{+0.1743}
\resset{e2.O.kev.sig_blocks}{Code}
\resset{e2.O.kev.sig_max}{+0.1533}
\resset{e2.O.kev.sig_min}{+0.1533}
\resset{e2.O.max}{+0.2880}
\resset{e2.O.min}{-0.0674}
\resset{e2.O.n}{28}
\resset{e2.O.n_sig}{10}
\resset{e2.O.n_sig_neg}{0}
\resset{e2.O.n_sig_pos}{10}
\resset{e2.P.max}{+0.1060}
\resset{e2.P.min}{-0.0747}
\resset{e2.P.n}{20}
\resset{e2.P.n_sig}{2}
\resset{e2.P.n_sig_neg}{1}
\resset{e2.P.n_sig_pos}{1}
\resset{e2.P1.clm_a0.a}{0.6848}
\resset{e2.P1.clm_a0.b}{0.7595}
\resset{e2.P1.clm_a0.ci}{[-0.1365, -0.0181]}
\resset{e2.P1.clm_a0.d}{-0.0747}
\resset{e2.P1.clm_a0.p}{0.040}
\resset{e2.P1.clm_a1.a}{0.9183}
\resset{e2.P1.clm_a1.b}{0.9205}
\resset{e2.P1.clm_a1.ci}{[-0.0261, +0.0188]}
\resset{e2.P1.clm_a1.d}{-0.0022}
\resset{e2.P1.clm_a1.p}{1.000}
\resset{e2.P1.clm_a2.a}{0.9630}
\resset{e2.P1.clm_a2.b}{0.9585}
\resset{e2.P1.clm_a2.ci}{[-0.0085, +0.0183]}
\resset{e2.P1.clm_a2.d}{+0.0045}
\resset{e2.P1.clm_a2.p}{1.000}
\resset{e2.P1.kev.a}{0.8984}
\resset{e2.P1.kev.b}{0.9087}
\resset{e2.P1.kev.ci}{[-0.0702, +0.0504]}
\resset{e2.P1.kev.d}{-0.0104}
\resset{e2.P1.kev.p}{1.000}
\resset{e2.P2.clm_a0.a}{0.6848}
\resset{e2.P2.clm_a0.b}{0.7065}
\resset{e2.P2.clm_a0.ci}{[-0.0756, +0.0267]}
\resset{e2.P2.clm_a0.d}{-0.0216}
\resset{e2.P2.clm_a0.p}{0.520}
\resset{e2.P2.clm_a1.a}{0.9183}
\resset{e2.P2.clm_a1.b}{0.9104}
\resset{e2.P2.clm_a1.ci}{[-0.0149, +0.0304]}
\resset{e2.P2.clm_a1.d}{+0.0079}
\resset{e2.P2.clm_a1.p}{1.000}
\resset{e2.P2.clm_a2.a}{0.9630}
\resset{e2.P2.clm_a2.b}{0.9548}
\resset{e2.P2.clm_a2.ci}{[-0.0006, +0.0218]}
\resset{e2.P2.clm_a2.d}{+0.0081}
\resset{e2.P2.clm_a2.p}{0.355}
\resset{e2.P2.kev.a}{0.8984}
\resset{e2.P2.kev.b}{0.8433}
\resset{e2.P2.kev.ci}{[+0.0056, +0.1082]}
\resset{e2.P2.kev.d}{+0.0551}
\resset{e2.P2.kev.p}{0.155}
\resset{e2.P3.clm_a0.a}{0.6848}
\resset{e2.P3.clm_a0.b}{0.5788}
\resset{e2.P3.clm_a0.ci}{[+0.0474, +0.1707]}
\resset{e2.P3.clm_a0.d}{+0.1060}
\resset{e2.P3.clm_a0.p}{0.005}
\resset{e2.P3.clm_a1.a}{0.9183}
\resset{e2.P3.clm_a1.b}{0.9242}
\resset{e2.P3.clm_a1.ci}{[-0.0377, +0.0262]}
\resset{e2.P3.clm_a1.d}{-0.0058}
\resset{e2.P3.clm_a1.p}{1.000}
\resset{e2.P3.clm_a2.a}{0.9630}
\resset{e2.P3.clm_a2.b}{0.9596}
\resset{e2.P3.clm_a2.ci}{[-0.0103, +0.0209]}
\resset{e2.P3.clm_a2.d}{+0.0034}
\resset{e2.P3.clm_a2.p}{1.000}
\resset{e2.P3.kev.a}{0.8984}
\resset{e2.P3.kev.b}{0.8922}
\resset{e2.P3.kev.ci}{[-0.0305, +0.0468]}
\resset{e2.P3.kev.d}{+0.0061}
\resset{e2.P3.kev.p}{1.000}
\resset{e2.P5.clm_a0.a}{0.6848}
\resset{e2.P5.clm_a0.b}{0.7327}
\resset{e2.P5.clm_a0.ci}{[-0.0925, -0.0033]}
\resset{e2.P5.clm_a0.d}{-0.0479}
\resset{e2.P5.clm_a0.p}{0.096}
\resset{e2.P5.clm_a1.a}{0.9183}
\resset{e2.P5.clm_a1.b}{0.9341}
\resset{e2.P5.clm_a1.ci}{[-0.0414, +0.0027]}
\resset{e2.P5.clm_a1.d}{-0.0158}
\resset{e2.P5.clm_a1.p}{0.465}
\resset{e2.P5.clm_a2.a}{0.9630}
\resset{e2.P5.clm_a2.b}{0.9674}
\resset{e2.P5.clm_a2.ci}{[-0.0207, +0.0057]}
\resset{e2.P5.clm_a2.d}{-0.0045}
\resset{e2.P5.clm_a2.p}{1.000}
\resset{e2.P5.kev.a}{0.8984}
\resset{e2.P5.kev.b}{0.9402}
\resset{e2.P5.kev.ci}{[-0.0948, +0.0061]}
\resset{e2.P5.kev.d}{-0.0419}
\resset{e2.P5.kev.p}{0.336}
\resset{e2.R.max}{+0.0121}
\resset{e2.R.min}{-0.0292}
\resset{e2.R.n}{4}
\resset{e2.R.n_sig}{0}
\resset{e2.R.n_sig_neg}{0}
\resset{e2.R.n_sig_pos}{0}
\resset{e2.S.max}{+0.0550}
\resset{e2.S.min}{-0.0374}
\resset{e2.S.n}{12}
\resset{e2.S.n_sig}{0}
\resset{e2.S.n_sig_neg}{0}
\resset{e2.S.n_sig_pos}{0}
\resset{e2.S1.clm_a0.a}{0.6848}
\resset{e2.S1.clm_a0.b}{0.6379}
\resset{e2.S1.clm_a0.ci}{[-0.0234, +0.1116]}
\resset{e2.S1.clm_a0.d}{+0.0470}
\resset{e2.S1.clm_a0.p}{0.492}
\resset{e2.S1.clm_a1.a}{0.9183}
\resset{e2.S1.clm_a1.b}{0.8762}
\resset{e2.S1.clm_a1.ci}{[-0.0010, +0.0903]}
\resset{e2.S1.clm_a1.d}{+0.0421}
\resset{e2.S1.clm_a1.p}{0.165}
\resset{e2.S1.clm_a2.a}{0.9630}
\resset{e2.S1.clm_a2.b}{0.9524}
\resset{e2.S1.clm_a2.ci}{[-0.0022, +0.0292]}
\resset{e2.S1.clm_a2.d}{+0.0106}
\resset{e2.S1.clm_a2.p}{0.372}
\resset{e2.S1.kev.a}{0.8984}
\resset{e2.S1.kev.b}{0.8833}
\resset{e2.S1.kev.ci}{[-0.0401, +0.0754]}
\resset{e2.S1.kev.d}{+0.0151}
\resset{e2.S1.kev.p}{0.574}
\resset{e2.S2.clm_a0.a}{0.6527}
\resset{e2.S2.clm_a0.b}{0.6446}
\resset{e2.S2.clm_a0.ci}{[-0.1690, +0.1665]}
\resset{e2.S2.clm_a0.d}{+0.0081}
\resset{e2.S2.clm_a0.p}{1.000}
\resset{e2.S2.clm_a1.a}{0.9196}
\resset{e2.S2.clm_a1.b}{0.8910}
\resset{e2.S2.clm_a1.ci}{[-0.0313, +0.0911]}
\resset{e2.S2.clm_a1.d}{+0.0286}
\resset{e2.S2.clm_a1.p}{0.714}
\resset{e2.S2.clm_a2.a}{0.9802}
\resset{e2.S2.clm_a2.b}{0.9696}
\resset{e2.S2.clm_a2.ci}{[-0.0047, +0.0294]}
\resset{e2.S2.clm_a2.d}{+0.0106}
\resset{e2.S2.clm_a2.p}{0.372}
\resset{e2.S2.kev.a}{0.9373}
\resset{e2.S2.kev.b}{0.8823}
\resset{e2.S2.kev.ci}{[-0.0060, +0.1236]}
\resset{e2.S2.kev.d}{+0.0550}
\resset{e2.S2.kev.p}{0.216}
\resset{e2.S3.clm_a0.a}{0.6858}
\resset{e2.S3.clm_a0.b}{0.6527}
\resset{e2.S3.clm_a0.ci}{[-0.0885, +0.1642]}
\resset{e2.S3.clm_a0.d}{+0.0331}
\resset{e2.S3.clm_a0.p}{1.000}
\resset{e2.S3.clm_a1.a}{0.9350}
\resset{e2.S3.clm_a1.b}{0.9196}
\resset{e2.S3.clm_a1.ci}{[-0.0343, +0.0626]}
\resset{e2.S3.clm_a1.d}{+0.0154}
\resset{e2.S3.clm_a1.p}{0.714}
\resset{e2.S3.clm_a2.a}{0.9754}
\resset{e2.S3.clm_a2.b}{0.9802}
\resset{e2.S3.clm_a2.ci}{[-0.0204, +0.0072]}
\resset{e2.S3.clm_a2.d}{-0.0048}
\resset{e2.S3.clm_a2.p}{0.488}
\resset{e2.S3.kev.a}{0.8999}
\resset{e2.S3.kev.b}{0.9373}
\resset{e2.S3.kev.ci}{[-0.0902, +0.0075]}
\resset{e2.S3.kev.d}{-0.0374}
\resset{e2.S3.kev.p}{0.222}
\resset{e2.T.max}{+0.0970}
\resset{e2.T.min}{+0.0055}
\resset{e2.T.n}{4}
\resset{e2.T.n_sig}{2}
\resset{e2.T.n_sig_neg}{0}
\resset{e2.T.n_sig_pos}{2}
\resset{e2.T1.clm_a0.a}{0.7349}
\resset{e2.T1.clm_a0.b}{0.6379}
\resset{e2.T1.clm_a0.ci}{[+0.0373, +0.1570]}
\resset{e2.T1.clm_a0.d}{+0.0970}
\resset{e2.T1.clm_a0.p}{0.002}
\resset{e2.T1.clm_a1.a}{0.9309}
\resset{e2.T1.clm_a1.b}{0.8762}
\resset{e2.T1.clm_a1.ci}{[+0.0057, +0.1010]}
\resset{e2.T1.clm_a1.d}{+0.0547}
\resset{e2.T1.clm_a1.p}{0.023}
\resset{e2.T1.clm_a2.a}{0.9578}
\resset{e2.T1.clm_a2.b}{0.9524}
\resset{e2.T1.clm_a2.ci}{[-0.0199, +0.0285]}
\resset{e2.T1.clm_a2.d}{+0.0055}
\resset{e2.T1.clm_a2.p}{0.585}
\resset{e2.T1.kev.a}{0.8908}
\resset{e2.T1.kev.b}{0.8833}
\resset{e2.T1.kev.ci}{[-0.0355, +0.0505]}
\resset{e2.T1.kev.d}{+0.0075}
\resset{e2.T1.kev.p}{0.770}
\resset{e2.a3.auroc_max}{0.9817}
\resset{e2.a3.auroc_min}{0.7246}
\resset{e2.a3.d_a2_max}{+0.0251}
\resset{e2.a3.d_a2_min}{-0.0094}
\resset{e2.a3.n}{30}
\resset{e2.a3.n_above_a2}{26}
\resset{e2.a3.n_sig_a1}{26}
\resset{e2.a3.n_sig_a2}{1}
\resset{e2.a3.sig_a2_d}{+0.0240}
\resset{e2.a3.sig_a2_level}{Imports}
\resset{e2.det.auroc}{0.9776}
\resset{e2.n_benign}{115}
\resset{e2.n_malware}{119}
\resset{e3.D1.clm.a}{0.8756}
\resset{e3.D1.clm.b}{0.9776}
\resset{e3.D1.clm.ci}{[-0.1520, -0.0589]}
\resset{e3.D1.clm.d}{-0.1019}
\resset{e3.D1.clm.p}{$<0.001$}
\resset{e3.D1.kev.a}{0.8868}
\resset{e3.D1.kev.b}{0.9776}
\resset{e3.D1.kev.ci}{[-0.1650, -0.0344]}
\resset{e3.D1.kev.d}{-0.0908}
\resset{e3.D1.kev.p}{0.003}
\resset{e3.D1.max}{-0.0343}
\resset{e3.D1.min}{-0.1667}
\resset{e3.D1.n}{13}
\resset{e3.D1.n_sig}{12}
\resset{e3.D1.n_sig_neg}{12}
\resset{e3.D1.n_sig_pos}{0}
\resset{e3.G.max}{+0.0488}
\resset{e3.G.min}{-0.2604}
\resset{e3.G.n}{33}
\resset{e3.G.n_sig}{9}
\resset{e3.G.n_sig_neg}{5}
\resset{e3.G.n_sig_pos}{4}
\resset{e3.K1.clm.a}{0.9516}
\resset{e3.K1.clm.b}{0.8756}
\resset{e3.K1.clm.ci}{[+0.0342, +0.1237]}
\resset{e3.K1.clm.d}{+0.0759}
\resset{e3.K1.clm.p}{0.003}
\resset{e3.K1.kev.a}{0.9532}
\resset{e3.K1.kev.b}{0.8868}
\resset{e3.K1.kev.ci}{[+0.0227, +0.1187]}
\resset{e3.K1.kev.d}{+0.0664}
\resset{e3.K1.kev.p}{0.004}
\resset{e3.K1.max}{+0.0798}
\resset{e3.K1.min}{-0.3889}
\resset{e3.K1.n}{13}
\resset{e3.K1.n_sig}{9}
\resset{e3.K1.n_sig_neg}{1}
\resset{e3.K1.n_sig_pos}{8}
\resset{e3.K2.clm.a}{0.6878}
\resset{e3.K2.clm.b}{0.8756}
\resset{e3.K2.clm.ci}{[-0.3199, -0.0713]}
\resset{e3.K2.clm.d}{-0.1878}
\resset{e3.K2.clm.p}{0.004}
\resset{e3.K2.kev.a}{0.4508}
\resset{e3.K2.kev.b}{0.8868}
\resset{e3.K2.kev.ci}{[-0.5555, -0.3266]}
\resset{e3.K2.kev.d}{-0.4360}
\resset{e3.K2.kev.p}{0.003}
\resset{e3.K2.max}{-0.1878}
\resset{e3.K2.min}{-0.6667}
\resset{e3.K2.n}{13}
\resset{e3.K2.n_sig}{13}
\resset{e3.K2.n_sig_neg}{13}
\resset{e3.K2.n_sig_pos}{0}
\resset{e3.N.clm.all}{0.8594}
\resset{e3.N.clm.k1}{0.7053}
\resset{e3.N.clm.k1.p}{0.005}
\resset{e3.N.clm.k16}{0.8237}
\resset{e3.N.clm.k2}{0.7609}
\resset{e3.N.clm.k4}{0.7529}
\resset{e3.N.clm.k8}{0.7731}
\resset{e3.N.k1.n_sig}{10}
\resset{e3.N.k16.n_sig}{4}
\resset{e3.N.k2.n_sig}{5}
\resset{e3.N.k4.n_sig}{5}
\resset{e3.N.k8.n_sig}{4}
\resset{e3.N.n_models}{13}
\resset{e3.P.max}{+0.3216}
\resset{e3.P.min}{-0.0844}
\resset{e3.P.n}{26}
\resset{e3.P.n_sig}{2}
\resset{e3.P.n_sig_neg}{0}
\resset{e3.P.n_sig_pos}{2}
\resset{e3.V1.clm.a}{0.8756}
\resset{e3.V1.clm.b}{0.6379}
\resset{e3.V1.clm.ci}{[+0.1386, +0.3385]}
\resset{e3.V1.clm.d}{+0.2378}
\resset{e3.V1.clm.p}{$<0.001$}
\resset{e3.V1.kev.a}{0.8868}
\resset{e3.V1.kev.b}{0.5505}
\resset{e3.V1.kev.ci}{[+0.2440, +0.4419]}
\resset{e3.V1.kev.d}{+0.3364}
\resset{e3.V1.kev.p}{0.002}
\resset{e3.V1.max}{+0.5082}
\resset{e3.V1.min}{+0.0556}
\resset{e3.V1.n}{13}
\resset{e3.V1.n_sig}{12}
\resset{e3.V1.n_sig_neg}{0}
\resset{e3.V1.n_sig_pos}{12}
\resset{e3.V1.ns_models}{Laya multilingual 322M}
\resset{e3.W.max}{+0.0554}
\resset{e3.W.min}{-0.0926}
\resset{e3.W.n}{39}
\resset{e3.W.n_sig}{2}
\resset{e3.W.n_sig_neg}{1}
\resset{e3.W.n_sig_pos}{1}
\resset{e3.W1.clm.a}{0.9310}
\resset{e3.W1.clm.b}{0.8756}
\resset{e3.W1.clm.ci}{[+0.0223, +0.0990]}
\resset{e3.W1.clm.d}{+0.0554}
\resset{e3.W1.clm.p}{0.006}
\resset{e3.W1.kev.a}{0.8926}
\resset{e3.W1.kev.b}{0.8868}
\resset{e3.W1.kev.ci}{[-0.0644, +0.0860]}
\resset{e3.W1.kev.d}{+0.0058}
\resset{e3.W1.kev.p}{1.000}
\resset{e3.W2.clm.a}{0.8990}
\resset{e3.W2.clm.b}{0.8756}
\resset{e3.W2.clm.ci}{[-0.0362, +0.0769]}
\resset{e3.W2.clm.d}{+0.0234}
\resset{e3.W2.clm.p}{0.418}
\resset{e3.W2.kev.a}{0.9057}
\resset{e3.W2.kev.b}{0.8868}
\resset{e3.W2.kev.ci}{[-0.0426, +0.0877]}
\resset{e3.W2.kev.d}{+0.0189}
\resset{e3.W2.kev.p}{1.000}
\resset{e3.X.max}{+0.2904}
\resset{e3.X.min}{-0.3333}
\resset{e3.X.n}{39}
\resset{e3.X.n_sig}{8}
\resset{e3.X.n_sig_neg}{0}
\resset{e3.X.n_sig_pos}{8}
\resset{e3.a3.auroc_max}{0.9714}
\resset{e3.a3.auroc_min}{0.9478}
\resset{e3.a3.d_a2_max}{+0.0190}
\resset{e3.a3.d_a2_min}{-0.0045}
\resset{e3.a3.n}{6}
\resset{e3.a3.n_above_a2}{5}
\resset{e3.a3.n_sig_a1}{5}
\resset{e3.a3.n_sig_a2}{0}
\resset{e3.det.auroc}{0.9776}
\resset{e3.n_benign}{115}
\resset{e3.n_malware}{119}
\resset{e3.n_models}{13}
\resset{e4.a0.n}{5}
\resset{e4.a0.n_infeasible}{3}
\resset{e4.band.risk_max}{1.49\%}
\resset{e4.band.risk_min}{1.32\%}
\resset{e4.clm_a0.brier}{0.2060}
\resset{e4.clm_a0.cover}{0 (infeasible)}
\resset{e4.clm_a0.ece_cal}{0.0479}
\resset{e4.clm_a0.ece_raw}{0.3691}
\resset{e4.clm_a0.improve}{0.3212}
\resset{e4.clm_a0.improve_ci}{[0.3124, 0.3282]}
\resset{e4.clm_a0.method}{isotonic}
\resset{e4.clm_a0.risk}{-}
\resset{e4.clm_a2.brier}{0.0262}
\resset{e4.clm_a2.cover}{92.36\%}
\resset{e4.clm_a2.ece_cal}{0.0060}
\resset{e4.clm_a2.ece_raw}{0.0124}
\resset{e4.clm_a2.improve}{0.0064}
\resset{e4.clm_a2.improve_ci}{[0.0027, 0.0071]}
\resset{e4.clm_a2.method}{Platt}
\resset{e4.clm_a2.risk}{1.32\%}
\resset{e4.clm_a3.cover}{93.98\%}
\resset{e4.clm_a3.ece_cal}{0.0078}
\resset{e4.clm_a3.ece_raw}{0.0129}
\resset{e4.clm_a3.improve}{0.0051}
\resset{e4.clm_a3.improve_ci}{[0.0008, 0.0068]}
\resset{e4.clm_a3.improve_p}{0.018}
\resset{e4.clm_a3.method}{Platt}
\resset{e4.clm_a3.risk}{1.56\%}
\resset{e4.det.brier}{0.0310}
\resset{e4.det.cover}{74.81\%}
\resset{e4.det.ece_cal}{0.0139}
\resset{e4.det.ece_raw}{0.0184}
\resset{e4.det.improve}{0.0044}
\resset{e4.det.improve_ci}{[0.0007, 0.0072]}
\resset{e4.det.method}{Platt}
\resset{e4.det.risk}{1.49\%}
\resset{e4.llm_fsec_a0.brier}{0.1537}
\resset{e4.llm_fsec_a0.cover}{16.01\%}
\resset{e4.llm_fsec_a0.ece_cal}{0.0316}
\resset{e4.llm_fsec_a0.ece_raw}{0.1542}
\resset{e4.llm_fsec_a0.improve}{0.1226}
\resset{e4.llm_fsec_a0.improve_ci}{[0.1143, 0.1287]}
\resset{e4.llm_fsec_a0.method}{Platt}
\resset{e4.llm_fsec_a0.risk}{3.10\%}
\resset{e4.llm_qwen3_a0.brier}{0.1833}
\resset{e4.llm_qwen3_a0.cover}{0 (infeasible)}
\resset{e4.llm_qwen3_a0.ece_cal}{0.0164}
\resset{e4.llm_qwen3_a0.ece_raw}{0.3025}
\resset{e4.llm_qwen3_a0.improve}{0.2861}
\resset{e4.llm_qwen3_a0.improve_ci}{[0.2747, 0.2950]}
\resset{e4.llm_qwen3_a0.method}{isotonic}
\resset{e4.llm_qwen3_a0.risk}{-}
\resset{e4.open_decider-2b_a0.brier}{0.2141}
\resset{e4.open_decider-2b_a0.cover}{4.60\%}
\resset{e4.open_decider-2b_a0.ece_cal}{0.0642}
\resset{e4.open_decider-2b_a0.ece_raw}{0.1620}
\resset{e4.open_decider-2b_a0.improve}{0.0978}
\resset{e4.open_decider-2b_a0.improve_ci}{[0.0860, 0.1076]}
\resset{e4.open_decider-2b_a0.method}{isotonic}
\resset{e4.open_decider-2b_a0.risk}{3.63\%}
\resset{e4.open_kev-0.8b_a0.brier}{0.2194}
\resset{e4.open_kev-0.8b_a0.cover}{0 (infeasible)}
\resset{e4.open_kev-0.8b_a0.ece_cal}{0.0719}
\resset{e4.open_kev-0.8b_a0.ece_raw}{0.1084}
\resset{e4.open_kev-0.8b_a0.improve}{0.0366}
\resset{e4.open_kev-0.8b_a0.improve_ci}{[0.0290, 0.0443]}
\resset{e4.open_kev-0.8b_a0.method}{isotonic}
\resset{e4.open_kev-0.8b_a0.risk}{-}
\resset{e4.primary.n}{7}
\resset{e4.primary.n_sig}{7}
\resset{e4.primary.pmax}{0.018}
\resset{e4.rc.clm_a2.max}{4.58\%}
\resset{e5.E5-C1.cos}{1.0000}
\resset{e5.E5-C1.eq.auroc}{equivalent}
\resset{e5.E5-C1.eq.auroc.margin}{0.00\%}
\resset{e5.E5-C1.eq.fpr}{equivalent}
\resset{e5.E5-C1.eq.fpr.margin}{0.58\%}
\resset{e5.E5-C1.eq.tpr}{equivalent}
\resset{e5.E5-C1.eq.tpr.margin}{0.49\%}
\resset{e5.E5-C1.flip}{0.00\%}
\resset{e5.E5-C1.flip_ci}{[0.00\%, 0.36\%]}
\resset{e5.E5-C1.kappa}{1.000}
\resset{e5.E5-C1.kappa_ci}{[1.000, 1.000]}
\resset{e5.E5-C1.margin_pp}{1}
\resset{e5.E5-C1.n}{1,013}
\resset{e5.E5-C1.n_flips}{0}
\resset{e5.E5-C1.pairs_needed}{1,083}
\resset{e5.E5-C1.ratio}{1.00}
\resset{e5.E5-C1.ratio_ci}{[0.99, 1.00]}
\resset{e5.E5-C2.cos}{0.9988}
\resset{e5.E5-C2.flip}{4.34\%}
\resset{e5.E5-C2.flip_ci}{[3.17\%, 5.79\%]}
\resset{e5.E5-C2.kappa}{0.892}
\resset{e5.E5-C2.kappa_ci}{[0.861, 0.923]}
\resset{e5.E5-C2.n}{1,013}
\resset{e5.E5-C2.n_flips}{44}
\resset{e5.E5-C2.ratio}{0.95}
\resset{e5.E5-C2.ratio_ci}{[0.95, 0.95]}
\resset{e5.E5-C3.cos}{0.9957}
\resset{e5.E5-C3.flip}{8.39\%}
\resset{e5.E5-C3.flip_ci}{[6.76\%, 10.27\%]}
\resset{e5.E5-C3.kappa}{0.795}
\resset{e5.E5-C3.kappa_ci}{[0.754, 0.837]}
\resset{e5.E5-C3.n}{1,013}
\resset{e5.E5-C3.n_flips}{85}
\resset{e5.E5-C3.ratio}{1.03}
\resset{e5.E5-C3.ratio_ci}{[1.03, 1.03]}
\resset{e5.E5-C6.cos}{0.9985}
\resset{e5.E5-C6.flip}{21.42\%}
\resset{e5.E5-C6.flip_ci}{[18.93\%, 24.08\%]}
\resset{e5.E5-C6.kappa}{0.571}
\resset{e5.E5-C6.kappa_ci}{[0.520, 0.621]}
\resset{e5.E5-C6.n}{1,013}
\resset{e5.E5-C6.n_flips}{217}
\resset{e5.E5-C6.ratio}{3.14}
\resset{e5.E5-C6.ratio_ci}{[3.10, 3.20]}
\resset{e5.E5-C7.cos}{0.9995}
\resset{e5.E5-C7.flip}{8.69\%}
\resset{e5.E5-C7.flip_ci}{[7.03\%, 10.59\%]}
\resset{e5.E5-C7.kappa}{0.796}
\resset{e5.E5-C7.kappa_ci}{[0.755, 0.837]}
\resset{e5.E5-C7.n}{1,013}
\resset{e5.E5-C7.n_flips}{88}
\resset{e5.E5-C7.ratio}{3.23}
\resset{e5.E5-C7.ratio_ci}{[3.18, 3.27]}
\resset{e5.E5-C8.cos}{0.9985}
\resset{e5.E5-C8.flip}{16.88\%}
\resset{e5.E5-C8.flip_ci}{[14.62\%, 19.33\%]}
\resset{e5.E5-C8.kappa}{0.662}
\resset{e5.E5-C8.kappa_ci}{[0.616, 0.708]}
\resset{e5.E5-C8.n}{1,013}
\resset{e5.E5-C8.n_flips}{171}
\resset{e5.E5-C8.ratio}{0.97}
\resset{e5.E5-C8.ratio_ci}{[0.97, 0.98]}
\resset{e5.bench.tps.2048}{2,099}
\resset{e5.bench.tps.512}{2,165}
\resset{e5.edge.auroc}{0.9827}
\resset{e5.edge.energy}{8.08}
\resset{e5.edge.fpr}{0.22\%}
\resset{e5.edge.fps}{2.68}
\resset{e5.edge.gpu}{RTX 4060 Laptop}
\resset{e5.edge.lat}{0.359}
\resset{e5.edge.power}{78}
\resset{e5.edge.t_encoder}{0.357}
\resset{e5.edge.t_heads_ms}{1.0}
\resset{e5.edge.t_state}{0.227}
\resset{e5.edge.t_tokenize_ms}{0.8}
\resset{e5.edge.tpr}{52.19\%}
\resset{e5.edge.vram}{6.7}
\resset{e5.inproc.auroc}{0.9827}
\resset{e5.inproc.energy}{8.12}
\resset{e5.inproc.fpr}{0.22\%}
\resset{e5.inproc.fps}{2.67}
\resset{e5.inproc.gpu}{RTX 4060 Laptop}
\resset{e5.inproc.lat}{0.360}
\resset{e5.inproc.power}{78}
\resset{e5.inproc.tpr}{52.19\%}
\resset{e5.inproc.vram}{6.6}
\resset{e5.n_benign}{465}
\resset{e5.n_files}{1,013}
\resset{e5.n_malware}{548}
\resset{e5.n_passes}{5}
\resset{e5.n_warmup}{20}
\resset{e5.open.auroc_max}{0.7072}
\resset{e5.open.auroc_min}{0.4239}
\resset{e5.open.lat_max}{0.265}
\resset{e5.open.lat_min}{0.020}
\resset{e5.open.n}{10}
\resset{e5.open.vram_max}{6.7}
\resset{e5.open.vram_min}{1.6}
\resset{e5.q4.auroc}{0.9788}
\resset{e5.q4.energy}{7.82}
\resset{e5.q4.fpr}{0.86\%}
\resset{e5.q4.fps}{2.75}
\resset{e5.q4.gpu}{RTX 4060 Laptop}
\resset{e5.q4.lat}{0.349}
\resset{e5.q4.power}{77}
\resset{e5.q4.tpr}{53.28\%}
\resset{e5.q4.vram}{5.8}
\resset{e5.q6.auroc}{0.9859}
\resset{e5.q6.energy}{8.55}
\resset{e5.q6.fpr}{0.43\%}
\resset{e5.q6.fps}{2.53}
\resset{e5.q6.gpu}{RTX 4060 Laptop}
\resset{e5.q6.lat}{0.378}
\resset{e5.q6.power}{78}
\resset{e5.q6.tpr}{50.55\%}
\resset{e5.q6.vram}{7.2}
\resset{e5.ref.auroc}{0.9922}
\resset{e5.ref.energy}{9.26}
\resset{e5.ref.fpr}{1.94\%}
\resset{e5.ref.fps}{8.34}
\resset{e5.ref.gpu}{RTX 5090}
\resset{e5.ref.lat}{0.114}
\resset{e5.ref.power}{278}
\resset{e5.ref.tpr}{90.33\%}
\resset{e5.ref.vram}{29.0}
\resset{e5.ref_q5.auroc}{0.9859}
\resset{e5.ref_q5.energy}{8.77}
\resset{e5.ref_q5.fpr}{0.43\%}
\resset{e5.ref_q5.fps}{8.54}
\resset{e5.ref_q5.gpu}{RTX 5090}
\resset{e5.ref_q5.lat}{0.112}
\resset{e5.ref_q5.power}{270}
\resset{e5.ref_q5.tpr}{60.40\%}
\resset{e5.ref_q5.vram}{6.9}
\resset{e5.repeat.max_flip}{0.00\%}
\resset{e5.repeat.n_configs}{25}
\resset{e5.s1mal.energy}{21.85}
\resset{e5.s1mal.fps}{0.97}
\resset{e5.s1mal.lat}{1.013}
\resset{e5.s1mal.lat_p90}{1.271}
\resset{e5.s1mal.n_files}{1,013}
\resset{e5.s1mal.n_passes}{5}
\resset{e5.s1mal.t_atomic}{0.580}
\resset{e5.s1mal.t_trees_ms}{27.6}
\resset{e5.s1mal.vram}{7.3}
\resset{e5.s1mal.x_lat}{2.8}
\resset{e5.small.0.6b.auroc}{0.9881}
\resset{e5.small.0.6b.lat}{0.068}
\resset{e5.small.0.6b.tpr}{89.23\%}
\resset{e5.small.0.6b.vram}{2.1}
\resset{e5.small.1.7b.auroc}{0.9879}
\resset{e5.small.1.7b.lat}{0.114}
\resset{e5.small.1.7b.tpr}{80.84\%}
\resset{e5.small.1.7b.vram}{3.3}
\resset{e5.small.4b.auroc}{0.9949}
\resset{e5.small.4b.lat}{0.205}
\resset{e5.small.4b.tpr}{93.07\%}
\resset{e5.small.4b.vram}{5.6}
\resset{e6.a2e1.all.flip}{1.26\%}
\resset{e6.a2e1.all.unstable}{2.74\%}
\resset{e6.a2e1.benign.flip}{0.43\%}
\resset{e6.a2e1.benign.unstable}{0.88\%}
\resset{e6.a2e1.malware.flip}{2.35\%}
\resset{e6.a2e1.malware.unstable}{5.17\%}
\resset{e6.a3.all.flip}{1.64\%}
\resset{e6.a3.all.unstable}{3.46\%}
\resset{e6.a3.auroc_max}{0.9946}
\resset{e6.a3.auroc_min}{0.9938}
\resset{e6.a3.benign.flip}{0.51\%}
\resset{e6.a3.benign.unstable}{1.11\%}
\resset{e6.a3.malware.flip}{3.11\%}
\resset{e6.a3.malware.unstable}{6.53\%}
\resset{e6.a3.n_seeds}{5}
\resset{e6.a3.tpr_max}{93.47\%}
\resset{e6.a3.tpr_min}{90.01\%}
\resset{e6.ctrl.flip}{0.01\%}
\resset{e6.ctrl.unstable}{0.02\%}
\resset{e6.ctrl.unstable_hi}{0.07\%}
\resset{e6.early.tpr_max}{60.75\%}
\resset{e6.early.tpr_min}{50.10\%}
\resset{e6.flip.benign}{0.90\%}
\resset{e6.flip.malware}{23.13\%}
\resset{e6.flip.max}{18.85\%}
\resset{e6.flip.mean}{10.55\%}
\resset{e6.flip.min}{0.89\%}
\resset{e6.flip.vs0}{8.79\%}
\resset{e6.heads.batch}{32}
\resset{e6.heads.early_epoch}{4}
\resset{e6.heads.early_seeds}{3, 4}
\resset{e6.heads.early_val_auroc_max}{0.9887}
\resset{e6.heads.late_val_auroc_min}{0.9961}
\resset{e6.heads.lr}{0.0005}
\resset{e6.heads.n_early}{2}
\resset{e6.heads.n_seeds}{5}
\resset{e6.late.tpr_max}{90.73\%}
\resset{e6.late.tpr_min}{88.75\%}
\resset{e6.local.flip_hi}{0.20\%}
\resset{e6.local.max_flip}{0.00\%}
\resset{e6.local.min_identical}{100.00\%}
\resset{e6.local.n_files}{1,831}
\resset{e6.local.n_pairs}{12}
\resset{e6.n_benign}{6,126}
\resset{e6.n_files}{10,829}
\resset{e6.n_malware}{4,703}
\resset{e6.n_seeds}{5}
\resset{e6.psd.mean}{0.0299}
\resset{e6.psd.p95}{0.1497}
\resset{e6.sampled.fsec.flip}{15.16\%}
\resset{e6.sampled.fsec.n}{835}
\resset{e6.sampled.fsec.unstable}{45.27\%}
\resset{e6.sampled.qwen3.flip}{1.88\%}
\resset{e6.sampled.qwen3.n}{928}
\resset{e6.sampled.qwen3.unstable}{4.85\%}
\resset{e6.seed0.tpr}{90.73\%}
\resset{e6.seed1.tpr}{90.60\%}
\resset{e6.seed2.tpr}{88.75\%}
\resset{e6.seed3.tpr}{60.75\%}
\resset{e6.seed4.tpr}{50.10\%}
\resset{e6.unstable}{19.51\%}
\resset{e6.unstable.benign}{1.67\%}
\resset{e6.unstable.ci}{[18.77\%, 20.27\%]}
\resset{e6.unstable.malware}{42.76\%}
\resset{e6.unstable.n}{2,113}
\resset{e7.adversarial.max}{19.81\%}
\resset{e7.adversarial.min}{0.00\%}
\resset{e7.all.max}{34.77\%}
\resset{e7.all.min}{0.05\%}
\resset{e7.clm.api_import.all.fab}{7.07\%}
\resset{e7.clm.api_import.all.recall}{94.15\%}
\resset{e7.clm.api_import.random.fab}{17.15\%}
\resset{e7.clm.api_import.random.recall}{94.15\%}
\resset{e7.clm.capability.all.fab}{0.00\%}
\resset{e7.clm.capability.all.recall}{95.25\%}
\resset{e7.clm.capability.random.fab}{0.00\%}
\resset{e7.clm.capability.random.recall}{95.25\%}
\resset{e7.clm_gguf_bf16.adversarial.conf}{0.00\%}
\resset{e7.clm_gguf_bf16.adversarial.fab}{1.10\%}
\resset{e7.clm_gguf_bf16.adversarial.fab_ci}{[0.64\%, 1.62\%]}
\resset{e7.clm_gguf_bf16.adversarial.n_neg}{1,721}
\resset{e7.clm_gguf_bf16.adversarial.recall}{94.54\%}
\resset{e7.clm_gguf_bf16.all.conf}{0.91\%}
\resset{e7.clm_gguf_bf16.all.fab}{8.86\%}
\resset{e7.clm_gguf_bf16.all.fab_ci}{[8.41\%, 9.37\%]}
\resset{e7.clm_gguf_bf16.all.n_neg}{10,609}
\resset{e7.clm_gguf_bf16.all.recall}{71.90\%}
\resset{e7.clm_gguf_bf16.popular.conf}{0.00\%}
\resset{e7.clm_gguf_bf16.popular.fab}{1.29\%}
\resset{e7.clm_gguf_bf16.popular.fab_ci}{[0.76\%, 1.87\%]}
\resset{e7.clm_gguf_bf16.popular.n_neg}{1,709}
\resset{e7.clm_gguf_bf16.popular.recall}{94.54\%}
\resset{e7.clm_gguf_bf16.random.conf}{0.18\%}
\resset{e7.clm_gguf_bf16.random.fab}{10.63\%}
\resset{e7.clm_gguf_bf16.random.fab_ci}{[9.23\%, 12.03\%]}
\resset{e7.clm_gguf_bf16.random.n_neg}{1,712}
\resset{e7.clm_gguf_bf16.random.recall}{94.54\%}
\resset{e7.clm_gguf_bf16.structural.conf}{1.72\%}
\resset{e7.clm_gguf_bf16.structural.fab}{13.12\%}
\resset{e7.clm_gguf_bf16.structural.fab_ci}{[12.39\%, 13.86\%]}
\resset{e7.clm_gguf_bf16.structural.n_neg}{5,467}
\resset{e7.clm_gguf_bf16.structural.recall}{39.25\%}
\resset{e7.d1-3b.adversarial.conf}{0.08\%}
\resset{e7.d1-3b.adversarial.fab}{1.17\%}
\resset{e7.d1-3b.adversarial.fab_ci}{[0.62\%, 1.79\%]}
\resset{e7.d1-3b.adversarial.n_neg}{1,284}
\resset{e7.d1-3b.adversarial.recall}{88.21\%}
\resset{e7.d1-3b.all.conf}{0.04\%}
\resset{e7.d1-3b.all.fab}{2.76\%}
\resset{e7.d1-3b.all.fab_ci}{[2.43\%, 3.12\%]}
\resset{e7.d1-3b.all.n_neg}{7,993}
\resset{e7.d1-3b.all.recall}{92.45\%}
\resset{e7.d1-3b.popular.conf}{0.00\%}
\resset{e7.d1-3b.popular.fab}{1.33\%}
\resset{e7.d1-3b.popular.fab_ci}{[0.71\%, 1.96\%]}
\resset{e7.d1-3b.popular.n_neg}{1,274}
\resset{e7.d1-3b.popular.recall}{88.21\%}
\resset{e7.d1-3b.random.conf}{0.08\%}
\resset{e7.d1-3b.random.fab}{0.24\%}
\resset{e7.d1-3b.random.fab_ci}{[0.00\%, 0.56\%]}
\resset{e7.d1-3b.random.n_neg}{1,259}
\resset{e7.d1-3b.random.recall}{88.21\%}
\resset{e7.d1-3b.structural.conf}{0.02\%}
\resset{e7.d1-3b.structural.fab}{4.45\%}
\resset{e7.d1-3b.structural.fab_ci}{[3.86\%, 5.07\%]}
\resset{e7.d1-3b.structural.n_neg}{4,176}
\resset{e7.d1-3b.structural.recall}{98.73\%}
\resset{e7.d1-omni-600m.adversarial.conf}{0.00\%}
\resset{e7.d1-omni-600m.adversarial.fab}{7.09\%}
\resset{e7.d1-omni-600m.adversarial.fab_ci}{[5.69\%, 8.42\%]}
\resset{e7.d1-omni-600m.adversarial.n_neg}{1,283}
\resset{e7.d1-omni-600m.adversarial.recall}{5.37\%}
\resset{e7.d1-omni-600m.all.conf}{0.00\%}
\resset{e7.d1-omni-600m.all.fab}{8.44\%}
\resset{e7.d1-omni-600m.all.fab_ci}{[7.60\%, 9.33\%]}
\resset{e7.d1-omni-600m.all.n_neg}{7,987}
\resset{e7.d1-omni-600m.all.recall}{8.54\%}
\resset{e7.d1-omni-600m.popular.conf}{0.00\%}
\resset{e7.d1-omni-600m.popular.fab}{7.54\%}
\resset{e7.d1-omni-600m.popular.fab_ci}{[6.21\%, 8.88\%]}
\resset{e7.d1-omni-600m.popular.n_neg}{1,273}
\resset{e7.d1-omni-600m.popular.recall}{5.37\%}
\resset{e7.d1-omni-600m.random.conf}{0.00\%}
\resset{e7.d1-omni-600m.random.fab}{4.29\%}
\resset{e7.d1-omni-600m.random.fab_ci}{[3.18\%, 5.48\%]}
\resset{e7.d1-omni-600m.random.n_neg}{1,258}
\resset{e7.d1-omni-600m.random.recall}{5.37\%}
\resset{e7.d1-omni-600m.structural.conf}{0.00\%}
\resset{e7.d1-omni-600m.structural.fab}{10.38\%}
\resset{e7.d1-omni-600m.structural.fab_ci}{[9.18\%, 11.58\%]}
\resset{e7.d1-omni-600m.structural.n_neg}{4,173}
\resset{e7.d1-omni-600m.structural.recall}{13.25\%}
\resset{e7.decider-0.8b.adversarial.conf}{0.00\%}
\resset{e7.decider-0.8b.adversarial.fab}{0.15\%}
\resset{e7.decider-0.8b.adversarial.fab_ci}{[0.00\%, 0.38\%]}
\resset{e7.decider-0.8b.adversarial.n_neg}{1,305}
\resset{e7.decider-0.8b.adversarial.recall}{73.16\%}
\resset{e7.decider-0.8b.all.conf}{0.04\%}
\resset{e7.decider-0.8b.all.fab}{7.22\%}
\resset{e7.decider-0.8b.all.fab_ci}{[6.78\%, 7.69\%]}
\resset{e7.decider-0.8b.all.n_neg}{8,107}
\resset{e7.decider-0.8b.all.recall}{83.42\%}
\resset{e7.decider-0.8b.popular.conf}{0.00\%}
\resset{e7.decider-0.8b.popular.fab}{0.15\%}
\resset{e7.decider-0.8b.popular.fab_ci}{[0.00\%, 0.39\%]}
\resset{e7.decider-0.8b.popular.n_neg}{1,297}
\resset{e7.decider-0.8b.popular.recall}{73.16\%}
\resset{e7.decider-0.8b.random.conf}{0.00\%}
\resset{e7.decider-0.8b.random.fab}{0.08\%}
\resset{e7.decider-0.8b.random.fab_ci}{[0.00\%, 0.23\%]}
\resset{e7.decider-0.8b.random.n_neg}{1,287}
\resset{e7.decider-0.8b.random.recall}{73.16\%}
\resset{e7.decider-0.8b.structural.conf}{0.07\%}
\resset{e7.decider-0.8b.structural.fab}{13.75\%}
\resset{e7.decider-0.8b.structural.fab_ci}{[12.91\%, 14.64\%]}
\resset{e7.decider-0.8b.structural.n_neg}{4,218}
\resset{e7.decider-0.8b.structural.recall}{98.63\%}
\resset{e7.decider-2b.adversarial.conf}{0.15\%}
\resset{e7.decider-2b.adversarial.fab}{1.07\%}
\resset{e7.decider-2b.adversarial.fab_ci}{[0.54\%, 1.61\%]}
\resset{e7.decider-2b.adversarial.n_neg}{1,305}
\resset{e7.decider-2b.adversarial.recall}{94.99\%}
\resset{e7.decider-2b.all.conf}{0.02\%}
\resset{e7.decider-2b.all.fab}{0.83\%}
\resset{e7.decider-2b.all.fab_ci}{[0.63\%, 1.03\%]}
\resset{e7.decider-2b.all.n_neg}{8,107}
\resset{e7.decider-2b.all.recall}{93.57\%}
\resset{e7.decider-2b.popular.conf}{0.00\%}
\resset{e7.decider-2b.popular.fab}{0.46\%}
\resset{e7.decider-2b.popular.fab_ci}{[0.15\%, 0.85\%]}
\resset{e7.decider-2b.popular.n_neg}{1,297}
\resset{e7.decider-2b.popular.recall}{94.99\%}
\resset{e7.decider-2b.random.conf}{0.00\%}
\resset{e7.decider-2b.random.fab}{0.16\%}
\resset{e7.decider-2b.random.fab_ci}{[0.00\%, 0.39\%]}
\resset{e7.decider-2b.random.n_neg}{1,287}
\resset{e7.decider-2b.random.recall}{94.99\%}
\resset{e7.decider-2b.structural.conf}{0.00\%}
\resset{e7.decider-2b.structural.fab}{1.07\%}
\resset{e7.decider-2b.structural.fab_ci}{[0.78\%, 1.38\%]}
\resset{e7.decider-2b.structural.n_neg}{4,218}
\resset{e7.decider-2b.structural.recall}{91.46\%}
\resset{e7.decider-4b-gguf.adversarial.conf}{0.00\%}
\resset{e7.decider-4b-gguf.adversarial.fab}{0.00\%}
\resset{e7.decider-4b-gguf.adversarial.fab_ci}{[0.00\%, 0.00\%]}
\resset{e7.decider-4b-gguf.adversarial.n_neg}{1,305}
\resset{e7.decider-4b-gguf.adversarial.recall}{93.24\%}
\resset{e7.decider-4b-gguf.all.conf}{0.02\%}
\resset{e7.decider-4b-gguf.all.fab}{2.70\%}
\resset{e7.decider-4b-gguf.all.fab_ci}{[2.38\%, 3.05\%]}
\resset{e7.decider-4b-gguf.all.n_neg}{8,107}
\resset{e7.decider-4b-gguf.all.recall}{95.39\%}
\resset{e7.decider-4b-gguf.popular.conf}{0.15\%}
\resset{e7.decider-4b-gguf.popular.fab}{0.15\%}
\resset{e7.decider-4b-gguf.popular.fab_ci}{[0.00\%, 0.39\%]}
\resset{e7.decider-4b-gguf.popular.n_neg}{1,297}
\resset{e7.decider-4b-gguf.popular.recall}{93.24\%}
\resset{e7.decider-4b-gguf.random.conf}{0.00\%}
\resset{e7.decider-4b-gguf.random.fab}{0.00\%}
\resset{e7.decider-4b-gguf.random.fab_ci}{[0.00\%, 0.00\%]}
\resset{e7.decider-4b-gguf.random.n_neg}{1,287}
\resset{e7.decider-4b-gguf.random.recall}{93.24\%}
\resset{e7.decider-4b-gguf.structural.conf}{0.00\%}
\resset{e7.decider-4b-gguf.structural.fab}{5.14\%}
\resset{e7.decider-4b-gguf.structural.fab_ci}{[4.54\%, 5.80\%]}
\resset{e7.decider-4b-gguf.structural.n_neg}{4,218}
\resset{e7.decider-4b-gguf.structural.recall}{98.57\%}
\resset{e7.foundation-sec-1.1-8b-instruct-bf16.adversarial.conf}{0.00\%}
\resset{e7.foundation-sec-1.1-8b-instruct-bf16.adversarial.fab}{0.17\%}
\resset{e7.foundation-sec-1.1-8b-instruct-bf16.adversarial.fab_ci}{[0.00\%, 0.41\%]}
\resset{e7.foundation-sec-1.1-8b-instruct-bf16.adversarial.n_neg}{1,721}
\resset{e7.foundation-sec-1.1-8b-instruct-bf16.adversarial.recall}{41.87\%}
\resset{e7.foundation-sec-1.1-8b-instruct-bf16.all.conf}{0.00\%}
\resset{e7.foundation-sec-1.1-8b-instruct-bf16.all.fab}{0.05\%}
\resset{e7.foundation-sec-1.1-8b-instruct-bf16.all.fab_ci}{[0.01\%, 0.09\%]}
\resset{e7.foundation-sec-1.1-8b-instruct-bf16.all.n_neg}{10,609}
\resset{e7.foundation-sec-1.1-8b-instruct-bf16.all.recall}{61.21\%}
\resset{e7.foundation-sec-1.1-8b-instruct-bf16.popular.conf}{0.00\%}
\resset{e7.foundation-sec-1.1-8b-instruct-bf16.popular.fab}{0.12\%}
\resset{e7.foundation-sec-1.1-8b-instruct-bf16.popular.fab_ci}{[0.00\%, 0.29\%]}
\resset{e7.foundation-sec-1.1-8b-instruct-bf16.popular.n_neg}{1,709}
\resset{e7.foundation-sec-1.1-8b-instruct-bf16.popular.recall}{41.87\%}
\resset{e7.foundation-sec-1.1-8b-instruct-bf16.random.conf}{0.00\%}
\resset{e7.foundation-sec-1.1-8b-instruct-bf16.random.fab}{0.00\%}
\resset{e7.foundation-sec-1.1-8b-instruct-bf16.random.fab_ci}{[0.00\%, 0.00\%]}
\resset{e7.foundation-sec-1.1-8b-instruct-bf16.random.n_neg}{1,712}
\resset{e7.foundation-sec-1.1-8b-instruct-bf16.random.recall}{41.87\%}
\resset{e7.foundation-sec-1.1-8b-instruct-bf16.structural.conf}{0.00\%}
\resset{e7.foundation-sec-1.1-8b-instruct-bf16.structural.fab}{0.00\%}
\resset{e7.foundation-sec-1.1-8b-instruct-bf16.structural.fab_ci}{[0.00\%, 0.00\%]}
\resset{e7.foundation-sec-1.1-8b-instruct-bf16.structural.n_neg}{5,467}
\resset{e7.foundation-sec-1.1-8b-instruct-bf16.structural.recall}{89.11\%}
\resset{e7.kev-0.8b.adversarial.conf}{0.00\%}
\resset{e7.kev-0.8b.adversarial.fab}{1.84\%}
\resset{e7.kev-0.8b.adversarial.fab_ci}{[1.15\%, 2.61\%]}
\resset{e7.kev-0.8b.adversarial.n_neg}{1,305}
\resset{e7.kev-0.8b.adversarial.recall}{50.52\%}
\resset{e7.kev-0.8b.all.conf}{0.00\%}
\resset{e7.kev-0.8b.all.fab}{0.68\%}
\resset{e7.kev-0.8b.all.fab_ci}{[0.51\%, 0.87\%]}
\resset{e7.kev-0.8b.all.n_neg}{8,107}
\resset{e7.kev-0.8b.all.recall}{58.71\%}
\resset{e7.kev-0.8b.popular.conf}{0.00\%}
\resset{e7.kev-0.8b.popular.fab}{1.39\%}
\resset{e7.kev-0.8b.popular.fab_ci}{[0.77\%, 2.08\%]}
\resset{e7.kev-0.8b.popular.n_neg}{1,297}
\resset{e7.kev-0.8b.popular.recall}{50.52\%}
\resset{e7.kev-0.8b.random.conf}{0.00\%}
\resset{e7.kev-0.8b.random.fab}{0.85\%}
\resset{e7.kev-0.8b.random.fab_ci}{[0.39\%, 1.40\%]}
\resset{e7.kev-0.8b.random.n_neg}{1,287}
\resset{e7.kev-0.8b.random.recall}{50.52\%}
\resset{e7.kev-0.8b.structural.conf}{0.00\%}
\resset{e7.kev-0.8b.structural.fab}{0.05\%}
\resset{e7.kev-0.8b.structural.fab_ci}{[0.00\%, 0.12\%]}
\resset{e7.kev-0.8b.structural.n_neg}{4,218}
\resset{e7.kev-0.8b.structural.recall}{70.86\%}
\resset{e7.n_backends}{10}
\resset{e7.popular.max}{19.07\%}
\resset{e7.popular.min}{0.12\%}
\resset{e7.qwen3-8b-bf16.adversarial.conf}{0.06\%}
\resset{e7.qwen3-8b-bf16.adversarial.fab}{0.06\%}
\resset{e7.qwen3-8b-bf16.adversarial.fab_ci}{[0.00\%, 0.17\%]}
\resset{e7.qwen3-8b-bf16.adversarial.n_neg}{1,721}
\resset{e7.qwen3-8b-bf16.adversarial.recall}{93.34\%}
\resset{e7.qwen3-8b-bf16.all.conf}{0.09\%}
\resset{e7.qwen3-8b-bf16.all.fab}{0.11\%}
\resset{e7.qwen3-8b-bf16.all.fab_ci}{[0.06\%, 0.18\%]}
\resset{e7.qwen3-8b-bf16.all.n_neg}{10,609}
\resset{e7.qwen3-8b-bf16.all.recall}{94.91\%}
\resset{e7.qwen3-8b-bf16.popular.conf}{0.12\%}
\resset{e7.qwen3-8b-bf16.popular.fab}{0.18\%}
\resset{e7.qwen3-8b-bf16.popular.fab_ci}{[0.00\%, 0.41\%]}
\resset{e7.qwen3-8b-bf16.popular.n_neg}{1,709}
\resset{e7.qwen3-8b-bf16.popular.recall}{93.34\%}
\resset{e7.qwen3-8b-bf16.random.conf}{0.00\%}
\resset{e7.qwen3-8b-bf16.random.fab}{0.00\%}
\resset{e7.qwen3-8b-bf16.random.fab_ci}{[0.00\%, 0.00\%]}
\resset{e7.qwen3-8b-bf16.random.n_neg}{1,712}
\resset{e7.qwen3-8b-bf16.random.recall}{93.34\%}
\resset{e7.qwen3-8b-bf16.structural.conf}{0.13\%}
\resset{e7.qwen3-8b-bf16.structural.fab}{0.15\%}
\resset{e7.qwen3-8b-bf16.structural.fab_ci}{[0.05\%, 0.26\%]}
\resset{e7.qwen3-8b-bf16.structural.n_neg}{5,467}
\resset{e7.qwen3-8b-bf16.structural.recall}{97.17\%}
\resset{e7.random.max}{16.52\%}
\resset{e7.random.min}{0.00\%}
\resset{e7.structural.max}{49.85\%}
\resset{e7.structural.min}{0.00\%}
\resset{e7.von.adversarial.conf}{0.00\%}
\resset{e7.von.adversarial.fab}{19.81\%}
\resset{e7.von.adversarial.fab_ci}{[17.69\%, 21.97\%]}
\resset{e7.von.adversarial.n_neg}{1,383}
\resset{e7.von.adversarial.recall}{56.71\%}
\resset{e7.von.all.conf}{0.00\%}
\resset{e7.von.all.fab}{34.77\%}
\resset{e7.von.all.fab_ci}{[33.60\%, 36.05\%]}
\resset{e7.von.all.n_neg}{8,603}
\resset{e7.von.all.recall}{69.67\%}
\resset{e7.von.popular.conf}{0.00\%}
\resset{e7.von.popular.fab}{19.07\%}
\resset{e7.von.popular.fab_ci}{[17.03\%, 21.13\%]}
\resset{e7.von.popular.n_neg}{1,379}
\resset{e7.von.popular.recall}{56.71\%}
\resset{e7.von.random.conf}{0.00\%}
\resset{e7.von.random.fab}{16.52\%}
\resset{e7.von.random.fab_ci}{[14.61\%, 18.49\%]}
\resset{e7.von.random.n_neg}{1,374}
\resset{e7.von.random.recall}{56.71\%}
\resset{e7.von.structural.conf}{0.00\%}
\resset{e7.von.structural.fab}{49.85\%}
\resset{e7.von.structural.fab_ci}{[48.07\%, 51.79\%]}
\resset{e7.von.structural.n_neg}{4,467}
\resset{e7.von.structural.recall}{88.65\%}
\resset{e7i.a1.hw.max}{1.53\%}
\resset{e7i.a1.hw.min}{0.00\%}
\resset{e7i.a1.hw.n}{2}
\resset{e7i.a1.pad05.max}{27.29\%}
\resset{e7i.a1.pad05.min}{10.26\%}
\resset{e7i.a1.pad05.n}{13}
\resset{e7i.a1.pad10.max}{22.49\%}
\resset{e7i.a1.pad10.min}{3.17\%}
\resset{e7i.a1.pad10.n}{13}
\resset{e7i.a1.para.max}{24.16\%}
\resset{e7i.a1.para.min}{8.61\%}
\resset{e7i.a1.para.n}{13}
\resset{e7i.a1.rep.max}{7.11\%}
\resset{e7i.a1.rep.min}{0.00\%}
\resset{e7i.a1.rep.n}{13}
\resset{e7i.clm_a1.pad05}{14.08\%}
\resset{e7i.clm_a1.pad05_ci}{[11.89\%, 16.51\%]}
\resset{e7i.clm_a1.pad05_n}{916}
\resset{e7i.clm_a1.pad05_shift}{32.21\%}
\resset{e7i.clm_a1.pad10}{18.03\%}
\resset{e7i.clm_a1.pad10_ci}{[15.27\%, 21.06\%]}
\resset{e7i.clm_a1.pad10_n}{710}
\resset{e7i.clm_a1.pad10_shift}{37.46\%}
\resset{e7i.clm_a1.para}{23.40\%}
\resset{e7i.clm_a1.para_acc}{77.18\%}
\resset{e7i.clm_a1.para_acc_worst}{45.63\%}
\resset{e7i.clm_a1.para_ci}{[21.21\%, 25.60\%]}
\resset{e7i.clm_a1.para_conc}{52.29\%}
\resset{e7i.clm_a1.para_n}{916}
\resset{e7i.clm_a1.para_worst}{31.55\%}
\resset{e7i.clm_a1.rep}{1.86\%}
\resset{e7i.clm_a1.rep_ci}{[1.08\%, 2.95\%]}
\resset{e7i.clm_a1.rep_ident}{17.90\%}
\resset{e7i.clm_a2.pad05}{7.10\%}
\resset{e7i.clm_a2.pad05_ci}{[5.52\%, 8.96\%]}
\resset{e7i.clm_a2.pad05_n}{916}
\resset{e7i.clm_a2.pad05_shift}{15.39\%}
\resset{e7i.clm_a2.pad10}{11.52\%}
\resset{e7i.clm_a2.pad10_ci}{[9.27\%, 14.09\%]}
\resset{e7i.clm_a2.pad10_n}{712}
\resset{e7i.clm_a2.pad10_shift}{22.61\%}
\resset{e7i.clm_a2.para}{23.22\%}
\resset{e7i.clm_a2.para_acc}{95.74\%}
\resset{e7i.clm_a2.para_acc_worst}{54.48\%}
\resset{e7i.clm_a2.para_ci}{[21.82\%, 24.59\%]}
\resset{e7i.clm_a2.para_conc}{15.61\%}
\resset{e7i.clm_a2.para_n}{916}
\resset{e7i.clm_a2.para_worst}{48.91\%}
\resset{e7i.clm_a3.pad05}{6.00\%}
\resset{e7i.clm_a3.pad05_ci}{[4.55\%, 7.74\%]}
\resset{e7i.clm_a3.pad05_n}{916}
\resset{e7i.clm_a3.pad05_shift}{8.08\%}
\resset{e7i.clm_a3.pad10}{8.59\%}
\resset{e7i.clm_a3.pad10_ci}{[6.64\%, 10.90\%]}
\resset{e7i.clm_a3.pad10_n}{710}
\resset{e7i.clm_a3.pad10_shift}{11.27\%}
\resset{e7i.clm_a3.para}{18.81\%}
\resset{e7i.clm_a3.para_acc}{97.60\%}
\resset{e7i.clm_a3.para_acc_worst}{58.62\%}
\resset{e7i.clm_a3.para_ci}{[17.71\%, 19.93\%]}
\resset{e7i.clm_a3.para_conc}{19.32\%}
\resset{e7i.clm_a3.para_n}{916}
\resset{e7i.clm_a3.para_worst}{39.19\%}
\resset{e7i.clm_a3.rep}{0.11\%}
\resset{e7i.clm_a3.rep_ci}{[0.00\%, 0.61\%]}
\resset{e7i.clm_a3.rep_ident}{39.08\%}
\resset{e7i.clm_gguf_bf16.option}{0.00\%}
\resset{e7i.clm_gguf_bf16.option_ci}{[0.00\%, 0.40\%]}
\resset{e7i.clm_gguf_bf16.option_n}{916}
\resset{e7i.clm_gguf_bf16.pad05}{1.20\%}
\resset{e7i.clm_gguf_bf16.pad05_ci}{[0.60\%, 2.14\%]}
\resset{e7i.clm_gguf_bf16.pad05_n}{916}
\resset{e7i.clm_gguf_bf16.pad05_shift}{4.48\%}
\resset{e7i.clm_gguf_bf16.pad10}{1.40\%}
\resset{e7i.clm_gguf_bf16.pad10_ci}{[0.68\%, 2.57\%]}
\resset{e7i.clm_gguf_bf16.pad10_n}{712}
\resset{e7i.clm_gguf_bf16.pad10_shift}{8.01\%}
\resset{e7i.clm_gguf_bf16.para}{26.14\%}
\resset{e7i.clm_gguf_bf16.para_acc}{55.24\%}
\resset{e7i.clm_gguf_bf16.para_acc_worst}{54.37\%}
\resset{e7i.clm_gguf_bf16.para_ben}{40.11\%}
\resset{e7i.clm_gguf_bf16.para_ci}{[24.50\%, 27.74\%]}
\resset{e7i.clm_gguf_bf16.para_conc}{19.98\%}
\resset{e7i.clm_gguf_bf16.para_n}{916}
\resset{e7i.clm_gguf_bf16.para_worst}{71.51\%}
\resset{e7i.clm_gguf_bf16.polarity}{1.31\%}
\resset{e7i.clm_gguf_bf16.polarity_ci}{[0.68\%, 2.28\%]}
\resset{e7i.clm_gguf_bf16.polarity_n}{916}
\resset{e7i.clm_gguf_bf16.rep}{0.55\%}
\resset{e7i.clm_gguf_bf16.rep_ci}{[0.18\%, 1.27\%]}
\resset{e7i.clm_gguf_bf16.rep_ident}{0.00\%}
\resset{e7i.d1-3b.option}{35.97\%}
\resset{e7i.d1-3b.option_ci}{[32.40\%, 39.66\%]}
\resset{e7i.d1-3b.option_n}{695}
\resset{e7i.d1-3b.pad05}{11.22\%}
\resset{e7i.d1-3b.pad05_ci}{[8.61\%, 14.29\%]}
\resset{e7i.d1-3b.pad05_n}{508}
\resset{e7i.d1-3b.pad05_shift}{12.60\%}
\resset{e7i.d1-3b.pad10}{12.86\%}
\resset{e7i.d1-3b.pad10_ci}{[9.81\%, 16.44\%]}
\resset{e7i.d1-3b.pad10_n}{420}
\resset{e7i.d1-3b.pad10_shift}{15.95\%}
\resset{e7i.d1-3b.para}{15.51\%}
\resset{e7i.d1-3b.para_acc}{56.14\%}
\resset{e7i.d1-3b.para_acc_worst}{58.19\%}
\resset{e7i.d1-3b.para_ci}{[13.36\%, 17.60\%]}
\resset{e7i.d1-3b.para_conc}{65.06\%}
\resset{e7i.d1-3b.para_n}{684}
\resset{e7i.d1-3b.para_worst}{28.07\%}
\resset{e7i.d1-3b.polarity}{21.45\%}
\resset{e7i.d1-3b.polarity_ci}{[18.38\%, 24.78\%]}
\resset{e7i.d1-3b.polarity_n}{662}
\resset{e7i.d1-3b.rep}{0.00\%}
\resset{e7i.d1-3b.rep_ci}{[0.00\%, 0.54\%]}
\resset{e7i.d1-3b.rep_ident}{100.00\%}
\resset{e7i.d1-omni-600m.option}{4.32\%}
\resset{e7i.d1-omni-600m.option_ci}{[2.94\%, 6.11\%]}
\resset{e7i.d1-omni-600m.option_n}{694}
\resset{e7i.d1-omni-600m.pad05}{2.57\%}
\resset{e7i.d1-omni-600m.pad05_ci}{[1.37\%, 4.35\%]}
\resset{e7i.d1-omni-600m.pad05_n}{506}
\resset{e7i.d1-omni-600m.pad05_shift}{2.57\%}
\resset{e7i.d1-omni-600m.pad10}{3.57\%}
\resset{e7i.d1-omni-600m.pad10_ci}{[2.01\%, 5.82\%]}
\resset{e7i.d1-omni-600m.pad10_n}{420}
\resset{e7i.d1-omni-600m.pad10_shift}{9.29\%}
\resset{e7i.d1-omni-600m.para}{7.77\%}
\resset{e7i.d1-omni-600m.para_acc}{43.48\%}
\resset{e7i.d1-omni-600m.para_acc_worst}{42.75\%}
\resset{e7i.d1-omni-600m.para_ci}{[6.60\%, 8.99\%]}
\resset{e7i.d1-omni-600m.para_conc}{58.27\%}
\resset{e7i.d1-omni-600m.para_n}{683}
\resset{e7i.d1-omni-600m.para_worst}{34.55\%}
\resset{e7i.d1-omni-600m.polarity}{61.88\%}
\resset{e7i.d1-omni-600m.polarity_ci}{[58.05\%, 65.59\%]}
\resset{e7i.d1-omni-600m.polarity_n}{661}
\resset{e7i.d1-omni-600m.rep}{0.00\%}
\resset{e7i.d1-omni-600m.rep_ci}{[0.00\%, 0.54\%]}
\resset{e7i.d1-omni-600m.rep_ident}{100.00\%}
\resset{e7i.decider-0.8b.option}{5.98\%}
\resset{e7i.decider-0.8b.option_ci}{[4.35\%, 8.00\%]}
\resset{e7i.decider-0.8b.option_n}{702}
\resset{e7i.decider-0.8b.pad05}{3.90\%}
\resset{e7i.decider-0.8b.pad05_ci}{[2.40\%, 5.96\%]}
\resset{e7i.decider-0.8b.pad05_n}{513}
\resset{e7i.decider-0.8b.pad05_shift}{2.73\%}
\resset{e7i.decider-0.8b.pad10}{3.73\%}
\resset{e7i.decider-0.8b.pad10_ci}{[2.15\%, 5.99\%]}
\resset{e7i.decider-0.8b.pad10_n}{429}
\resset{e7i.decider-0.8b.pad10_shift}{0.93\%}
\resset{e7i.decider-0.8b.para}{19.97\%}
\resset{e7i.decider-0.8b.para_acc}{64.43\%}
\resset{e7i.decider-0.8b.para_acc_worst}{58.14\%}
\resset{e7i.decider-0.8b.para_ci}{[18.29\%, 21.67\%]}
\resset{e7i.decider-0.8b.para_conc}{38.86\%}
\resset{e7i.decider-0.8b.para_n}{700}
\resset{e7i.decider-0.8b.para_worst}{44.00\%}
\resset{e7i.decider-0.8b.polarity}{11.86\%}
\resset{e7i.decider-0.8b.polarity_ci}{[9.56\%, 14.49\%]}
\resset{e7i.decider-0.8b.polarity_n}{700}
\resset{e7i.decider-0.8b.rep}{1.43\%}
\resset{e7i.decider-0.8b.rep_ci}{[0.69\%, 2.61\%]}
\resset{e7i.decider-0.8b.rep_ident}{46.29\%}
\resset{e7i.decider-2b.option}{3.56\%}
\resset{e7i.decider-2b.option_ci}{[2.32\%, 5.21\%]}
\resset{e7i.decider-2b.option_n}{702}
\resset{e7i.decider-2b.pad05}{6.82\%}
\resset{e7i.decider-2b.pad05_ci}{[4.80\%, 9.36\%]}
\resset{e7i.decider-2b.pad05_n}{513}
\resset{e7i.decider-2b.pad05_shift}{10.92\%}
\resset{e7i.decider-2b.pad10}{14.45\%}
\resset{e7i.decider-2b.pad10_ci}{[11.26\%, 18.14\%]}
\resset{e7i.decider-2b.pad10_n}{429}
\resset{e7i.decider-2b.pad10_shift}{25.87\%}
\resset{e7i.decider-2b.para}{20.46\%}
\resset{e7i.decider-2b.para_acc}{68.57\%}
\resset{e7i.decider-2b.para_acc_worst}{49.43\%}
\resset{e7i.decider-2b.para_ci}{[18.43\%, 22.63\%]}
\resset{e7i.decider-2b.para_conc}{43.14\%}
\resset{e7i.decider-2b.para_n}{700}
\resset{e7i.decider-2b.para_worst}{31.14\%}
\resset{e7i.decider-2b.polarity}{59.14\%}
\resset{e7i.decider-2b.polarity_ci}{[55.40\%, 62.81\%]}
\resset{e7i.decider-2b.polarity_n}{700}
\resset{e7i.decider-2b.rep}{0.57\%}
\resset{e7i.decider-2b.rep_ci}{[0.16\%, 1.46\%]}
\resset{e7i.decider-2b.rep_ident}{47.71\%}
\resset{e7i.decider-4b-gguf.option}{1.99\%}
\resset{e7i.decider-4b-gguf.option_ci}{[1.09\%, 3.32\%]}
\resset{e7i.decider-4b-gguf.option_n}{702}
\resset{e7i.decider-4b-gguf.pad05}{3.90\%}
\resset{e7i.decider-4b-gguf.pad05_ci}{[2.40\%, 5.96\%]}
\resset{e7i.decider-4b-gguf.pad05_n}{513}
\resset{e7i.decider-4b-gguf.pad05_shift}{11.70\%}
\resset{e7i.decider-4b-gguf.pad10}{6.53\%}
\resset{e7i.decider-4b-gguf.pad10_ci}{[4.38\%, 9.30\%]}
\resset{e7i.decider-4b-gguf.pad10_n}{429}
\resset{e7i.decider-4b-gguf.pad10_shift}{15.15\%}
\resset{e7i.decider-4b-gguf.para}{5.06\%}
\resset{e7i.decider-4b-gguf.para_acc}{69.14\%}
\resset{e7i.decider-4b-gguf.para_acc_worst}{70.14\%}
\resset{e7i.decider-4b-gguf.para_ci}{[3.86\%, 6.37\%]}
\resset{e7i.decider-4b-gguf.para_conc}{88.14\%}
\resset{e7i.decider-4b-gguf.para_n}{700}
\resset{e7i.decider-4b-gguf.para_worst}{8.43\%}
\resset{e7i.decider-4b-gguf.polarity}{10.29\%}
\resset{e7i.decider-4b-gguf.polarity_ci}{[8.14\%, 12.78\%]}
\resset{e7i.decider-4b-gguf.polarity_n}{700}
\resset{e7i.decider-4b-gguf.rep}{0.00\%}
\resset{e7i.decider-4b-gguf.rep_ci}{[0.00\%, 0.53\%]}
\resset{e7i.decider-4b-gguf.rep_ident}{100.00\%}
\resset{e7i.foundation-sec-1.1-8b-instruct-bf16.option}{18.45\%}
\resset{e7i.foundation-sec-1.1-8b-instruct-bf16.option_ci}{[15.99\%, 21.12\%]}
\resset{e7i.foundation-sec-1.1-8b-instruct-bf16.option_n}{916}
\resset{e7i.foundation-sec-1.1-8b-instruct-bf16.pad05}{6.88\%}
\resset{e7i.foundation-sec-1.1-8b-instruct-bf16.pad05_ci}{[5.33\%, 8.71\%]}
\resset{e7i.foundation-sec-1.1-8b-instruct-bf16.pad05_n}{916}
\resset{e7i.foundation-sec-1.1-8b-instruct-bf16.pad05_shift}{23.36\%}
\resset{e7i.foundation-sec-1.1-8b-instruct-bf16.pad10}{6.12\%}
\resset{e7i.foundation-sec-1.1-8b-instruct-bf16.pad10_ci}{[4.66\%, 7.87\%]}
\resset{e7i.foundation-sec-1.1-8b-instruct-bf16.pad10_n}{915}
\resset{e7i.foundation-sec-1.1-8b-instruct-bf16.pad10_shift}{24.15\%}
\resset{e7i.foundation-sec-1.1-8b-instruct-bf16.para}{7.21\%}
\resset{e7i.foundation-sec-1.1-8b-instruct-bf16.para_acc}{75.66\%}
\resset{e7i.foundation-sec-1.1-8b-instruct-bf16.para_acc_worst}{67.79\%}
\resset{e7i.foundation-sec-1.1-8b-instruct-bf16.para_ci}{[6.24\%, 8.21\%]}
\resset{e7i.foundation-sec-1.1-8b-instruct-bf16.para_conc}{75.11\%}
\resset{e7i.foundation-sec-1.1-8b-instruct-bf16.para_n}{916}
\resset{e7i.foundation-sec-1.1-8b-instruct-bf16.para_worst}{14.41\%}
\resset{e7i.foundation-sec-1.1-8b-instruct-bf16.polarity}{30.68\%}
\resset{e7i.foundation-sec-1.1-8b-instruct-bf16.polarity_ci}{[27.70\%, 33.78\%]}
\resset{e7i.foundation-sec-1.1-8b-instruct-bf16.polarity_n}{916}
\resset{e7i.foundation-sec-1.1-8b-instruct-bf16.rep}{0.44\%}
\resset{e7i.foundation-sec-1.1-8b-instruct-bf16.rep_ci}{[0.12\%, 1.11\%]}
\resset{e7i.foundation-sec-1.1-8b-instruct-bf16.rep_ident}{0.55\%}
\resset{e7i.kev-0.8b.option}{8.55\%}
\resset{e7i.kev-0.8b.option_ci}{[6.59\%, 10.86\%]}
\resset{e7i.kev-0.8b.option_n}{702}
\resset{e7i.kev-0.8b.pad05}{3.70\%}
\resset{e7i.kev-0.8b.pad05_ci}{[2.24\%, 5.72\%]}
\resset{e7i.kev-0.8b.pad05_n}{513}
\resset{e7i.kev-0.8b.pad05_shift}{0.00\%}
\resset{e7i.kev-0.8b.pad10}{6.06\%}
\resset{e7i.kev-0.8b.pad10_ci}{[4.00\%, 8.75\%]}
\resset{e7i.kev-0.8b.pad10_n}{429}
\resset{e7i.kev-0.8b.pad10_shift}{0.00\%}
\resset{e7i.kev-0.8b.para}{24.10\%}
\resset{e7i.kev-0.8b.para_acc}{71.14\%}
\resset{e7i.kev-0.8b.para_acc_worst}{49.86\%}
\resset{e7i.kev-0.8b.para_ci}{[22.27\%, 26.09\%]}
\resset{e7i.kev-0.8b.para_conc}{20.43\%}
\resset{e7i.kev-0.8b.para_n}{700}
\resset{e7i.kev-0.8b.para_worst}{70.71\%}
\resset{e7i.kev-0.8b.polarity}{14.29\%}
\resset{e7i.kev-0.8b.polarity_ci}{[11.78\%, 17.10\%]}
\resset{e7i.kev-0.8b.polarity_n}{700}
\resset{e7i.kev-0.8b.rep}{0.71\%}
\resset{e7i.kev-0.8b.rep_ci}{[0.23\%, 1.66\%]}
\resset{e7i.kev-0.8b.rep_ident}{2.43\%}
\resset{e7i.llm_fsec_a1.pad05}{27.29\%}
\resset{e7i.llm_fsec_a1.pad05_ci}{[24.43\%, 30.30\%]}
\resset{e7i.llm_fsec_a1.pad05_n}{916}
\resset{e7i.llm_fsec_a1.pad05_shift}{34.93\%}
\resset{e7i.llm_fsec_a1.pad10}{21.64\%}
\resset{e7i.llm_fsec_a1.pad10_ci}{[19.01\%, 24.45\%]}
\resset{e7i.llm_fsec_a1.pad10_n}{915}
\resset{e7i.llm_fsec_a1.pad10_shift}{24.81\%}
\resset{e7i.llm_fsec_a1.para}{9.10\%}
\resset{e7i.llm_fsec_a1.para_acc}{78.93\%}
\resset{e7i.llm_fsec_a1.para_acc_worst}{74.13\%}
\resset{e7i.llm_fsec_a1.para_ci}{[7.78\%, 10.55\%]}
\resset{e7i.llm_fsec_a1.para_conc}{76.86\%}
\resset{e7i.llm_fsec_a1.para_n}{916}
\resset{e7i.llm_fsec_a1.para_worst}{11.90\%}
\resset{e7i.llm_fsec_a1.rep}{1.64\%}
\resset{e7i.llm_fsec_a1.rep_ci}{[0.92\%, 2.69\%]}
\resset{e7i.llm_fsec_a1.rep_ident}{30.46\%}
\resset{e7i.llm_qwen3_a1.pad05}{10.59\%}
\resset{e7i.llm_qwen3_a1.pad05_ci}{[8.67\%, 12.77\%]}
\resset{e7i.llm_qwen3_a1.pad05_n}{916}
\resset{e7i.llm_qwen3_a1.pad05_shift}{30.90\%}
\resset{e7i.llm_qwen3_a1.pad10}{11.34\%}
\resset{e7i.llm_qwen3_a1.pad10_ci}{[8.99\%, 14.04\%]}
\resset{e7i.llm_qwen3_a1.pad10_n}{644}
\resset{e7i.llm_qwen3_a1.pad10_shift}{26.40\%}
\resset{e7i.llm_qwen3_a1.para}{8.61\%}
\resset{e7i.llm_qwen3_a1.para_acc}{64.85\%}
\resset{e7i.llm_qwen3_a1.para_acc_worst}{58.84\%}
\resset{e7i.llm_qwen3_a1.para_ci}{[7.25\%, 10.00\%]}
\resset{e7i.llm_qwen3_a1.para_conc}{78.93\%}
\resset{e7i.llm_qwen3_a1.para_n}{916}
\resset{e7i.llm_qwen3_a1.para_worst}{10.59\%}
\resset{e7i.llm_qwen3_a1.rep}{1.20\%}
\resset{e7i.llm_qwen3_a1.rep_ci}{[0.60\%, 2.14\%]}
\resset{e7i.llm_qwen3_a1.rep_ident}{72.93\%}
\resset{e7i.n_backends}{10}
\resset{e7i.n_files}{916}
\resset{e7i.n_sub}{1,000}
\resset{e7i.open_d1-3b_a1.pad05}{15.07\%}
\resset{e7i.open_d1-3b_a1.pad05_ci}{[12.81\%, 17.55\%]}
\resset{e7i.open_d1-3b_a1.pad05_n}{916}
\resset{e7i.open_d1-3b_a1.pad05_shift}{16.05\%}
\resset{e7i.open_d1-3b_a1.pad10}{17.47\%}
\resset{e7i.open_d1-3b_a1.pad10_ci}{[15.06\%, 20.08\%]}
\resset{e7i.open_d1-3b_a1.pad10_n}{916}
\resset{e7i.open_d1-3b_a1.pad10_shift}{41.81\%}
\resset{e7i.open_d1-3b_a1.para}{22.51\%}
\resset{e7i.open_d1-3b_a1.para_acc}{80.90\%}
\resset{e7i.open_d1-3b_a1.para_acc_worst}{50.76\%}
\resset{e7i.open_d1-3b_a1.para_ci}{[20.30\%, 24.76\%]}
\resset{e7i.open_d1-3b_a1.para_conc}{65.72\%}
\resset{e7i.open_d1-3b_a1.para_n}{916}
\resset{e7i.open_d1-3b_a1.para_worst}{30.13\%}
\resset{e7i.open_d1-3b_a1.rep}{0.00\%}
\resset{e7i.open_d1-3b_a1.rep_ci}{[0.00\%, 0.40\%]}
\resset{e7i.open_d1-3b_a1.rep_ident}{100.00\%}
\resset{e7i.open_d1-omni-600m_a1.pad05}{10.26\%}
\resset{e7i.open_d1-omni-600m_a1.pad05_ci}{[8.37\%, 12.41\%]}
\resset{e7i.open_d1-omni-600m_a1.pad05_n}{916}
\resset{e7i.open_d1-omni-600m_a1.pad05_shift}{53.06\%}
\resset{e7i.open_d1-omni-600m_a1.pad10}{11.03\%}
\resset{e7i.open_d1-omni-600m_a1.pad10_ci}{[9.07\%, 13.24\%]}
\resset{e7i.open_d1-omni-600m_a1.pad10_n}{916}
\resset{e7i.open_d1-omni-600m_a1.pad10_shift}{55.57\%}
\resset{e7i.open_d1-omni-600m_a1.para}{16.44\%}
\resset{e7i.open_d1-omni-600m_a1.para_acc}{66.27\%}
\resset{e7i.open_d1-omni-600m_a1.para_acc_worst}{45.52\%}
\resset{e7i.open_d1-omni-600m_a1.para_ci}{[14.51\%, 18.34\%]}
\resset{e7i.open_d1-omni-600m_a1.para_conc}{69.76\%}
\resset{e7i.open_d1-omni-600m_a1.para_n}{916}
\resset{e7i.open_d1-omni-600m_a1.para_worst}{21.62\%}
\resset{e7i.open_d1-omni-600m_a1.rep}{0.00\%}
\resset{e7i.open_d1-omni-600m_a1.rep_ci}{[0.00\%, 0.40\%]}
\resset{e7i.open_d1-omni-600m_a1.rep_ident}{100.00\%}
\resset{e7i.open_decider-0.8b_a1.pad05}{15.07\%}
\resset{e7i.open_decider-0.8b_a1.pad05_ci}{[12.81\%, 17.55\%]}
\resset{e7i.open_decider-0.8b_a1.pad05_n}{916}
\resset{e7i.open_decider-0.8b_a1.pad05_shift}{23.80\%}
\resset{e7i.open_decider-0.8b_a1.pad10}{3.17\%}
\resset{e7i.open_decider-0.8b_a1.pad10_ci}{[2.13\%, 4.52\%]}
\resset{e7i.open_decider-0.8b_a1.pad10_n}{916}
\resset{e7i.open_decider-0.8b_a1.pad10_shift}{24.34\%}
\resset{e7i.open_decider-0.8b_a1.para}{16.48\%}
\resset{e7i.open_decider-0.8b_a1.para_acc}{74.89\%}
\resset{e7i.open_decider-0.8b_a1.para_acc_worst}{47.82\%}
\resset{e7i.open_decider-0.8b_a1.para_ci}{[14.75\%, 18.49\%]}
\resset{e7i.open_decider-0.8b_a1.para_conc}{65.61\%}
\resset{e7i.open_decider-0.8b_a1.para_n}{916}
\resset{e7i.open_decider-0.8b_a1.para_worst}{27.51\%}
\resset{e7i.open_decider-0.8b_a1.rep}{1.97\%}
\resset{e7i.open_decider-0.8b_a1.rep_ci}{[1.17\%, 3.09\%]}
\resset{e7i.open_decider-0.8b_a1.rep_ident}{8.30\%}
\resset{e7i.open_decider-2b_a1.pad05}{11.03\%}
\resset{e7i.open_decider-2b_a1.pad05_ci}{[9.07\%, 13.24\%]}
\resset{e7i.open_decider-2b_a1.pad05_n}{916}
\resset{e7i.open_decider-2b_a1.pad05_shift}{13.32\%}
\resset{e7i.open_decider-2b_a1.pad10}{12.01\%}
\resset{e7i.open_decider-2b_a1.pad10_ci}{[9.97\%, 14.29\%]}
\resset{e7i.open_decider-2b_a1.pad10_n}{916}
\resset{e7i.open_decider-2b_a1.pad10_shift}{21.07\%}
\resset{e7i.open_decider-2b_a1.para}{23.84\%}
\resset{e7i.open_decider-2b_a1.para_acc}{83.84\%}
\resset{e7i.open_decider-2b_a1.para_acc_worst}{45.63\%}
\resset{e7i.open_decider-2b_a1.para_ci}{[21.93\%, 25.80\%]}
\resset{e7i.open_decider-2b_a1.para_conc}{47.60\%}
\resset{e7i.open_decider-2b_a1.para_n}{916}
\resset{e7i.open_decider-2b_a1.para_worst}{38.43\%}
\resset{e7i.open_decider-2b_a1.rep}{1.09\%}
\resset{e7i.open_decider-2b_a1.rep_ci}{[0.52\%, 2.00\%]}
\resset{e7i.open_decider-2b_a1.rep_ident}{6.99\%}
\resset{e7i.open_decider-4b-gguf_a1.pad05}{20.63\%}
\resset{e7i.open_decider-4b-gguf_a1.pad05_ci}{[18.06\%, 23.40\%]}
\resset{e7i.open_decider-4b-gguf_a1.pad05_n}{916}
\resset{e7i.open_decider-4b-gguf_a1.pad05_shift}{19.00\%}
\resset{e7i.open_decider-4b-gguf_a1.pad10}{14.30\%}
\resset{e7i.open_decider-4b-gguf_a1.pad10_ci}{[12.10\%, 16.74\%]}
\resset{e7i.open_decider-4b-gguf_a1.pad10_n}{916}
\resset{e7i.open_decider-4b-gguf_a1.pad10_shift}{26.20\%}
\resset{e7i.open_decider-4b-gguf_a1.para}{17.38\%}
\resset{e7i.open_decider-4b-gguf_a1.para_acc}{86.68\%}
\resset{e7i.open_decider-4b-gguf_a1.para_acc_worst}{62.23\%}
\resset{e7i.open_decider-4b-gguf_a1.para_ci}{[15.41\%, 19.44\%]}
\resset{e7i.open_decider-4b-gguf_a1.para_conc}{66.81\%}
\resset{e7i.open_decider-4b-gguf_a1.para_n}{916}
\resset{e7i.open_decider-4b-gguf_a1.para_worst}{25.33\%}
\resset{e7i.open_decider-4b-gguf_a1.rep}{2.62\%}
\resset{e7i.open_decider-4b-gguf_a1.rep_ci}{[1.69\%, 3.87\%]}
\resset{e7i.open_decider-4b-gguf_a1.rep_ident}{1.20\%}
\resset{e7i.open_kev-0.8b_a1.hw}{1.53\%}
\resset{e7i.open_kev-0.8b_a1.hw_ci}{[0.84\%, 2.55\%]}
\resset{e7i.open_kev-0.8b_a1.pad05}{17.58\%}
\resset{e7i.open_kev-0.8b_a1.pad05_ci}{[15.16\%, 20.20\%]}
\resset{e7i.open_kev-0.8b_a1.pad05_n}{916}
\resset{e7i.open_kev-0.8b_a1.pad05_shift}{32.10\%}
\resset{e7i.open_kev-0.8b_a1.pad10}{15.39\%}
\resset{e7i.open_kev-0.8b_a1.pad10_ci}{[13.12\%, 17.90\%]}
\resset{e7i.open_kev-0.8b_a1.pad10_n}{916}
\resset{e7i.open_kev-0.8b_a1.pad10_shift}{39.63\%}
\resset{e7i.open_kev-0.8b_a1.para}{24.16\%}
\resset{e7i.open_kev-0.8b_a1.para_acc}{72.93\%}
\resset{e7i.open_kev-0.8b_a1.para_acc_worst}{46.18\%}
\resset{e7i.open_kev-0.8b_a1.para_ci}{[22.33\%, 26.12\%]}
\resset{e7i.open_kev-0.8b_a1.para_conc}{36.68\%}
\resset{e7i.open_kev-0.8b_a1.para_n}{916}
\resset{e7i.open_kev-0.8b_a1.para_worst}{31.66\%}
\resset{e7i.open_kev-0.8b_a1.rep}{0.00\%}
\resset{e7i.open_kev-0.8b_a1.rep_ci}{[0.00\%, 0.40\%]}
\resset{e7i.open_kev-0.8b_a1.rep_ident}{100.00\%}
\resset{e7i.open_laya-multilingual_a1.pad05}{18.05\%}
\resset{e7i.open_laya-multilingual_a1.pad05_ci}{[15.22\%, 21.16\%]}
\resset{e7i.open_laya-multilingual_a1.pad05_n}{676}
\resset{e7i.open_laya-multilingual_a1.pad05_shift}{67.31\%}
\resset{e7i.open_laya-multilingual_a1.pad10}{19.12\%}
\resset{e7i.open_laya-multilingual_a1.pad10_ci}{[16.19\%, 22.33\%]}
\resset{e7i.open_laya-multilingual_a1.pad10_n}{659}
\resset{e7i.open_laya-multilingual_a1.pad10_shift}{67.22\%}
\resset{e7i.open_laya-multilingual_a1.para}{17.98\%}
\resset{e7i.open_laya-multilingual_a1.para_acc}{64.36\%}
\resset{e7i.open_laya-multilingual_a1.para_acc_worst}{47.47\%}
\resset{e7i.open_laya-multilingual_a1.para_ci}{[15.31\%, 20.68\%]}
\resset{e7i.open_laya-multilingual_a1.para_conc}{78.21\%}
\resset{e7i.open_laya-multilingual_a1.para_n}{693}
\resset{e7i.open_laya-multilingual_a1.para_worst}{19.05\%}
\resset{e7i.open_laya-multilingual_a1.rep}{0.14\%}
\resset{e7i.open_laya-multilingual_a1.rep_ci}{[0.00\%, 0.80\%]}
\resset{e7i.open_laya-multilingual_a1.rep_ident}{44.24\%}
\resset{e7i.open_laya-typed-decisions_a1.pad05}{17.66\%}
\resset{e7i.open_laya-typed-decisions_a1.pad05_ci}{[14.85\%, 20.77\%]}
\resset{e7i.open_laya-typed-decisions_a1.pad05_n}{668}
\resset{e7i.open_laya-typed-decisions_a1.pad05_shift}{44.16\%}
\resset{e7i.open_laya-typed-decisions_a1.pad10}{18.87\%}
\resset{e7i.open_laya-typed-decisions_a1.pad10_ci}{[15.95\%, 22.08\%]}
\resset{e7i.open_laya-typed-decisions_a1.pad10_n}{657}
\resset{e7i.open_laya-typed-decisions_a1.pad10_shift}{58.30\%}
\resset{e7i.open_laya-typed-decisions_a1.para}{12.79\%}
\resset{e7i.open_laya-typed-decisions_a1.para_acc}{62.84\%}
\resset{e7i.open_laya-typed-decisions_a1.para_acc_worst}{47.02\%}
\resset{e7i.open_laya-typed-decisions_a1.para_ci}{[10.88\%, 14.76\%]}
\resset{e7i.open_laya-typed-decisions_a1.para_conc}{73.15\%}
\resset{e7i.open_laya-typed-decisions_a1.para_n}{689}
\resset{e7i.open_laya-typed-decisions_a1.para_worst}{17.27\%}
\resset{e7i.open_laya-typed-decisions_a1.rep}{7.11\%}
\resset{e7i.open_laya-typed-decisions_a1.rep_ci}{[5.31\%, 9.29\%]}
\resset{e7i.open_laya-typed-decisions_a1.rep_ident}{3.77\%}
\resset{e7i.open_strands-decider-2b-v1-2610_a1.pad05}{21.62\%}
\resset{e7i.open_strands-decider-2b-v1-2610_a1.pad05_ci}{[18.99\%, 24.43\%]}
\resset{e7i.open_strands-decider-2b-v1-2610_a1.pad05_n}{916}
\resset{e7i.open_strands-decider-2b-v1-2610_a1.pad05_shift}{23.91\%}
\resset{e7i.open_strands-decider-2b-v1-2610_a1.pad10}{13.32\%}
\resset{e7i.open_strands-decider-2b-v1-2610_a1.pad10_ci}{[11.18\%, 15.69\%]}
\resset{e7i.open_strands-decider-2b-v1-2610_a1.pad10_n}{916}
\resset{e7i.open_strands-decider-2b-v1-2610_a1.pad10_shift}{23.36\%}
\resset{e7i.open_strands-decider-2b-v1-2610_a1.para}{16.41\%}
\resset{e7i.open_strands-decider-2b-v1-2610_a1.para_acc}{83.84\%}
\resset{e7i.open_strands-decider-2b-v1-2610_a1.para_acc_worst}{61.57\%}
\resset{e7i.open_strands-decider-2b-v1-2610_a1.para_ci}{[14.48\%, 18.42\%]}
\resset{e7i.open_strands-decider-2b-v1-2610_a1.para_conc}{67.90\%}
\resset{e7i.open_strands-decider-2b-v1-2610_a1.para_n}{916}
\resset{e7i.open_strands-decider-2b-v1-2610_a1.para_worst}{22.71\%}
\resset{e7i.open_strands-decider-2b-v1-2610_a1.rep}{3.60\%}
\resset{e7i.open_strands-decider-2b-v1-2610_a1.rep_ci}{[2.49\%, 5.02\%]}
\resset{e7i.open_strands-decider-2b-v1-2610_a1.rep_ident}{9.61\%}
\resset{e7i.open_von_a1.hw}{0.00\%}
\resset{e7i.open_von_a1.hw_ci}{[0.00\%, 0.40\%]}
\resset{e7i.open_von_a1.pad05}{19.43\%}
\resset{e7i.open_von_a1.pad05_ci}{[16.92\%, 22.15\%]}
\resset{e7i.open_von_a1.pad05_n}{916}
\resset{e7i.open_von_a1.pad05_shift}{42.36\%}
\resset{e7i.open_von_a1.pad10}{22.49\%}
\resset{e7i.open_von_a1.pad10_ci}{[19.82\%, 25.33\%]}
\resset{e7i.open_von_a1.pad10_n}{916}
\resset{e7i.open_von_a1.pad10_shift}{51.97\%}
\resset{e7i.open_von_a1.para}{14.10\%}
\resset{e7i.open_von_a1.para_acc}{70.52\%}
\resset{e7i.open_von_a1.para_acc_worst}{54.69\%}
\resset{e7i.open_von_a1.para_ci}{[12.58\%, 15.63\%]}
\resset{e7i.open_von_a1.para_conc}{64.96\%}
\resset{e7i.open_von_a1.para_n}{916}
\resset{e7i.open_von_a1.para_worst}{19.76\%}
\resset{e7i.open_von_a1.rep}{0.00\%}
\resset{e7i.open_von_a1.rep_ci}{[0.00\%, 0.40\%]}
\resset{e7i.open_von_a1.rep_ident}{100.00\%}
\resset{e7i.option.max}{35.97\%}
\resset{e7i.option.max_name}{Liquid d1-3B}
\resset{e7i.option.min}{0.00\%}
\resset{e7i.option.min_name}{CLM-8B}
\resset{e7i.option.n}{10}
\resset{e7i.pad05.max}{11.22\%}
\resset{e7i.pad05.max_name}{Liquid d1-3B}
\resset{e7i.pad05.min}{1.20\%}
\resset{e7i.pad05.min_name}{CLM-8B}
\resset{e7i.pad05.n}{10}
\resset{e7i.pad10.max}{14.45\%}
\resset{e7i.pad10.max_name}{decider-2b}
\resset{e7i.pad10.min}{1.40\%}
\resset{e7i.pad10.min_name}{CLM-8B}
\resset{e7i.pad10.n}{10}
\resset{e7i.para.max}{26.14\%}
\resset{e7i.para.max_name}{CLM-8B}
\resset{e7i.para.min}{2.31\%}
\resset{e7i.para.min_name}{Qwen3-8B (logits)}
\resset{e7i.para.n}{10}
\resset{e7i.polarity.max}{61.88\%}
\resset{e7i.polarity.max_name}{Liquid d1-omni-600M}
\resset{e7i.polarity.min}{1.31\%}
\resset{e7i.polarity.min_name}{CLM-8B}
\resset{e7i.polarity.n}{10}
\resset{e7i.qwen3-8b-bf16.option}{4.26\%}
\resset{e7i.qwen3-8b-bf16.option_ci}{[3.04\%, 5.77\%]}
\resset{e7i.qwen3-8b-bf16.option_n}{916}
\resset{e7i.qwen3-8b-bf16.pad05}{3.82\%}
\resset{e7i.qwen3-8b-bf16.pad05_ci}{[2.68\%, 5.27\%]}
\resset{e7i.qwen3-8b-bf16.pad05_n}{916}
\resset{e7i.qwen3-8b-bf16.pad05_shift}{7.42\%}
\resset{e7i.qwen3-8b-bf16.pad10}{4.17\%}
\resset{e7i.qwen3-8b-bf16.pad10_ci}{[2.77\%, 6.01\%]}
\resset{e7i.qwen3-8b-bf16.pad10_n}{647}
\resset{e7i.qwen3-8b-bf16.pad10_shift}{7.88\%}
\resset{e7i.qwen3-8b-bf16.para}{2.31\%}
\resset{e7i.qwen3-8b-bf16.para_acc}{72.27\%}
\resset{e7i.qwen3-8b-bf16.para_acc_worst}{70.41\%}
\resset{e7i.qwen3-8b-bf16.para_ci}{[1.80\%, 2.92\%]}
\resset{e7i.qwen3-8b-bf16.para_conc}{90.17\%}
\resset{e7i.qwen3-8b-bf16.para_n}{916}
\resset{e7i.qwen3-8b-bf16.para_worst}{7.10\%}
\resset{e7i.qwen3-8b-bf16.polarity}{17.47\%}
\resset{e7i.qwen3-8b-bf16.polarity_ci}{[15.06\%, 20.08\%]}
\resset{e7i.qwen3-8b-bf16.polarity_n}{916}
\resset{e7i.qwen3-8b-bf16.rep}{0.00\%}
\resset{e7i.qwen3-8b-bf16.rep_ci}{[0.00\%, 0.40\%]}
\resset{e7i.qwen3-8b-bf16.rep_ident}{0.22\%}
\resset{e7i.rep.max}{1.43\%}
\resset{e7i.rep.max_name}{decider-0.8b}
\resset{e7i.rep.min}{0.00\%}
\resset{e7i.rep.min_name}{Qwen3-8B (logits)}
\resset{e7i.rep.n}{10}
\resset{e7i.von.option}{0.00\%}
\resset{e7i.von.option_ci}{[0.00\%, 0.50\%]}
\resset{e7i.von.option_n}{742}
\resset{e7i.von.pad05}{10.10\%}
\resset{e7i.von.pad05_ci}{[7.63\%, 13.03\%]}
\resset{e7i.von.pad05_n}{515}
\resset{e7i.von.pad05_shift}{0.00\%}
\resset{e7i.von.pad10}{13.95\%}
\resset{e7i.von.pad10_ci}{[10.82\%, 17.59\%]}
\resset{e7i.von.pad10_n}{430}
\resset{e7i.von.pad10_shift}{0.00\%}
\resset{e7i.von.para}{9.53\%}
\resset{e7i.von.para_acc}{53.18\%}
\resset{e7i.von.para_acc_worst}{43.71\%}
\resset{e7i.von.para_ci}{[7.75\%, 11.26\%]}
\resset{e7i.von.para_conc}{79.57\%}
\resset{e7i.von.para_n}{739}
\resset{e7i.von.para_worst}{11.64\%}
\resset{e7i.von.polarity}{15.56\%}
\resset{e7i.von.polarity_ci}{[13.02\%, 18.38\%]}
\resset{e7i.von.polarity_n}{739}
\resset{e7i.von.rep}{0.00\%}
\resset{e7i.von.rep_ci}{[0.00\%, 0.50\%]}
\resset{e7i.von.rep_ident}{100.00\%}
\resset{e8.clm_a0.clean.fpr}{0.00\%}
\resset{e8.clm_a0.clean.fpr_ci}{[0.00\%, 2.12\%]}
\resset{e8.clm_a0.clean.tpr}{6.50\%}
\resset{e8.clm_a0.clean.tpr_ci}{[3.51\%, 10.86\%]}
\resset{e8.clm_a0.inj_raw.d}{-6.50}
\resset{e8.clm_a0.inj_raw.fpr}{0.00\%}
\resset{e8.clm_a0.inj_raw.fpr_ci}{[0.00\%, 2.12\%]}
\resset{e8.clm_a0.inj_raw.p}{$<0.001$}
\resset{e8.clm_a0.inj_raw.tpr}{0.00\%}
\resset{e8.clm_a0.inj_raw.tpr_ci}{[0.00\%, 1.83\%]}
\resset{e8.clm_a0.inj_san.d}{-6.50}
\resset{e8.clm_a0.inj_san.fpr}{0.00\%}
\resset{e8.clm_a0.inj_san.fpr_ci}{[0.00\%, 2.12\%]}
\resset{e8.clm_a0.inj_san.p}{$<0.001$}
\resset{e8.clm_a0.inj_san.tpr}{0.00\%}
\resset{e8.clm_a0.inj_san.tpr_ci}{[0.00\%, 1.83\%]}
\resset{e8.clm_a1.clean.fpr}{0.58\%}
\resset{e8.clm_a1.clean.fpr_ci}{[0.01\%, 3.20\%]}
\resset{e8.clm_a1.clean.tpr}{59.50\%}
\resset{e8.clm_a1.clean.tpr_ci}{[52.35\%, 66.37\%]}
\resset{e8.clm_a1.inj_raw.d}{-58.00}
\resset{e8.clm_a1.inj_raw.fpr}{0.00\%}
\resset{e8.clm_a1.inj_raw.fpr_ci}{[0.00\%, 2.12\%]}
\resset{e8.clm_a1.inj_raw.p}{$<0.001$}
\resset{e8.clm_a1.inj_raw.tpr}{1.50\%}
\resset{e8.clm_a1.inj_raw.tpr_ci}{[0.31\%, 4.32\%]}
\resset{e8.clm_a1.inj_san.d}{-57.00}
\resset{e8.clm_a1.inj_san.fpr}{0.00\%}
\resset{e8.clm_a1.inj_san.fpr_ci}{[0.00\%, 2.12\%]}
\resset{e8.clm_a1.inj_san.p}{$<0.001$}
\resset{e8.clm_a1.inj_san.tpr}{2.50\%}
\resset{e8.clm_a1.inj_san.tpr_ci}{[0.82\%, 5.74\%]}
\resset{e8.clm_a2.clean.fpr}{0.58\%}
\resset{e8.clm_a2.clean.fpr_ci}{[0.01\%, 3.20\%]}
\resset{e8.clm_a2.clean.tpr}{88.00\%}
\resset{e8.clm_a2.clean.tpr_ci}{[82.67\%, 92.16\%]}
\resset{e8.clm_a2.inj_raw.d}{-16.00}
\resset{e8.clm_a2.inj_raw.fpr}{23.26\%}
\resset{e8.clm_a2.inj_raw.fpr_ci}{[17.16\%, 30.29\%]}
\resset{e8.clm_a2.inj_raw.p}{$<0.001$}
\resset{e8.clm_a2.inj_raw.tpr}{72.00\%}
\resset{e8.clm_a2.inj_raw.tpr_ci}{[65.23\%, 78.10\%]}
\resset{e8.clm_a2.inj_san.d}{-55.00}
\resset{e8.clm_a2.inj_san.fpr}{1.74\%}
\resset{e8.clm_a2.inj_san.fpr_ci}{[0.36\%, 5.01\%]}
\resset{e8.clm_a2.inj_san.p}{$<0.001$}
\resset{e8.clm_a2.inj_san.tpr}{33.00\%}
\resset{e8.clm_a2.inj_san.tpr_ci}{[26.53\%, 39.98\%]}
\resset{e8.clm_a3.clean.fpr}{0.58\%}
\resset{e8.clm_a3.clean.fpr_ci}{[0.01\%, 3.20\%]}
\resset{e8.clm_a3.clean.tpr}{89.50\%}
\resset{e8.clm_a3.clean.tpr_ci}{[84.40\%, 93.38\%]}
\resset{e8.clm_a3.inj_raw.d}{-37.50}
\resset{e8.clm_a3.inj_raw.fpr}{9.30\%}
\resset{e8.clm_a3.inj_raw.fpr_ci}{[5.41\%, 14.67\%]}
\resset{e8.clm_a3.inj_raw.p}{$<0.001$}
\resset{e8.clm_a3.inj_raw.tpr}{52.00\%}
\resset{e8.clm_a3.inj_raw.tpr_ci}{[44.84\%, 59.10\%]}
\resset{e8.clm_a3.inj_san.d}{-61.50}
\resset{e8.clm_a3.inj_san.fpr}{0.00\%}
\resset{e8.clm_a3.inj_san.fpr_ci}{[0.00\%, 2.12\%]}
\resset{e8.clm_a3.inj_san.p}{$<0.001$}
\resset{e8.clm_a3.inj_san.tpr}{28.00\%}
\resset{e8.clm_a3.inj_san.tpr_ci}{[21.90\%, 34.77\%]}
\resset{e8.n_benign}{172}
\resset{e8.n_malware}{200}
\resset{e8.pad.both.clean.tpr}{72.07\%}
\resset{e8.pad.both.pad05.tpr}{44.14\%}
\resset{e8.pad.both.pad10.tpr}{49.55\%}
\resset{e8.pad.both.pad25.tpr}{29.73\%}
\resset{e8.pad.clm_a0.clean.tpr}{0.00\%}
\resset{e8.pad.clm_a0.pad05.d}{-0.00}
\resset{e8.pad.clm_a0.pad05.p}{1.000}
\resset{e8.pad.clm_a0.pad05.tpr}{0.00\%}
\resset{e8.pad.clm_a0.pad10.d}{-0.00}
\resset{e8.pad.clm_a0.pad10.p}{1.000}
\resset{e8.pad.clm_a0.pad10.tpr}{0.00\%}
\resset{e8.pad.clm_a0.pad25.d}{-0.00}
\resset{e8.pad.clm_a0.pad25.p}{1.000}
\resset{e8.pad.clm_a0.pad25.tpr}{0.00\%}
\resset{e8.pad.clm_a1.clean.tpr}{78.38\%}
\resset{e8.pad.clm_a1.pad05.d}{-15.32}
\resset{e8.pad.clm_a1.pad05.p}{$<0.001$}
\resset{e8.pad.clm_a1.pad05.tpr}{63.06\%}
\resset{e8.pad.clm_a1.pad10.d}{-19.82}
\resset{e8.pad.clm_a1.pad10.p}{$<0.001$}
\resset{e8.pad.clm_a1.pad10.tpr}{58.56\%}
\resset{e8.pad.clm_a1.pad25.d}{-36.94}
\resset{e8.pad.clm_a1.pad25.p}{$<0.001$}
\resset{e8.pad.clm_a1.pad25.tpr}{41.44\%}
\resset{e8.pad.clm_a2.clean.tpr}{88.29\%}
\resset{e8.pad.clm_a2.pad05.d}{-24.32}
\resset{e8.pad.clm_a2.pad05.p}{$<0.001$}
\resset{e8.pad.clm_a2.pad05.tpr}{63.96\%}
\resset{e8.pad.clm_a2.pad10.d}{-18.02}
\resset{e8.pad.clm_a2.pad10.p}{$<0.001$}
\resset{e8.pad.clm_a2.pad10.tpr}{70.27\%}
\resset{e8.pad.clm_a2.pad25.d}{-37.84}
\resset{e8.pad.clm_a2.pad25.p}{$<0.001$}
\resset{e8.pad.clm_a2.pad25.tpr}{50.45\%}
\resset{e8.pad.clm_a3.clean.tpr}{90.99\%}
\resset{e8.pad.clm_a3.pad05.d}{-5.41}
\resset{e8.pad.clm_a3.pad05.p}{0.070}
\resset{e8.pad.clm_a3.pad05.tpr}{85.59\%}
\resset{e8.pad.clm_a3.pad10.d}{-14.41}
\resset{e8.pad.clm_a3.pad10.p}{$<0.001$}
\resset{e8.pad.clm_a3.pad10.tpr}{76.58\%}
\resset{e8.pad.clm_a3.pad25.d}{-27.93}
\resset{e8.pad.clm_a3.pad25.p}{$<0.001$}
\resset{e8.pad.clm_a3.pad25.tpr}{63.06\%}
\resset{e8.pad.either.clean.tpr}{94.59\%}
\resset{e8.pad.either.n}{111}
\resset{e8.pad.either.pad05.tpr}{82.88\%}
\resset{e8.pad.either.pad10.tpr}{79.28\%}
\resset{e8.pad.either.pad25.tpr}{62.16\%}
\resset{e8.pad.n_benign}{56}
\resset{e8.pad.n_incomplete}{233}
\resset{e8.pad.n_malware}{111}
\resset{e8.screen.benign_strings}{0.00\%}
\resset{e8.screen.benign_strings.k}{0}
\resset{e8.screen.benign_strings.n}{10}
\resset{e8.screen.canaries}{100.00\%}
\resset{e8.screen.canaries.k}{14}
\resset{e8.screen.canaries.n}{14}
\resset{e8.screen.stage}{development}
\resset{e8.screen.states_benign}{16.67\%}
\resset{e8.screen.states_malware}{2.00\%}
\resset{e8.shift.inj_raw.atomic.mean}{0.053}
\resset{e8.shift.inj_raw.atomic.moved}{11.05\%}
\resset{e8.shift.inj_raw.atomic.moved_ci}{[10.29\%, 11.84\%]}
\resset{e8.shift.inj_raw.verdict.mean}{0.486}
\resset{e8.shift.inj_raw.verdict.moved}{70.50\%}
\resset{e8.shift.inj_raw.verdict.moved_ci}{[63.66\%, 76.72\%]}
\resset{e8.shift.inj_san.atomic.mean}{0.071}
\resset{e8.shift.inj_san.atomic.moved}{11.33\%}
\resset{e8.shift.inj_san.atomic.moved_ci}{[10.56\%, 12.13\%]}
\resset{e8.shift.inj_san.verdict.mean}{0.541}
\resset{e8.shift.inj_san.verdict.moved}{75.00\%}
\resset{e8.shift.inj_san.verdict.moved_ci}{[68.40\%, 80.84\%]}
\resset{e8.shift.pad05.atomic.mean}{-0.011}
\resset{e8.shift.pad05.atomic.moved}{3.66\%}
\resset{e8.shift.pad05.atomic.moved_ci}{[3.21\%, 4.15\%]}
\resset{e8.shift.pad05.verdict.mean}{0.008}
\resset{e8.shift.pad05.verdict.moved}{0.00\%}
\resset{e8.shift.pad05.verdict.moved_ci}{[0.00\%, 1.83\%]}
\resset{e8.shift.pad10.atomic.mean}{-0.028}
\resset{e8.shift.pad10.atomic.moved}{3.12\%}
\resset{e8.shift.pad10.atomic.moved_ci}{[2.67\%, 3.63\%]}
\resset{e8.shift.pad10.verdict.mean}{-0.009}
\resset{e8.shift.pad10.verdict.moved}{0.00\%}
\resset{e8.shift.pad10.verdict.moved_ci}{[0.00\%, 2.20\%]}
\resset{e8.shift.pad25.atomic.mean}{-0.022}
\resset{e8.shift.pad25.atomic.moved}{4.05\%}
\resset{e8.shift.pad25.atomic.moved_ci}{[3.43\%, 4.76\%]}
\resset{e8.shift.pad25.verdict.mean}{0.009}
\resset{e8.shift.pad25.verdict.moved}{0.00\%}
\resset{e8.shift.pad25.verdict.moved_ci}{[0.00\%, 3.27\%]}
\resset{e9.a2min.clm_a2}{84.02\%}
\resset{e9.a2min.clm_a3}{91.60\%}
\resset{e9.a2min.month}{2020-03}
\resset{e9.clm_a0.aut}{0.0885}
\resset{e9.clm_a0.aut_ci}{[0.017, 0.199]}
\resset{e9.clm_a0.mk_p}{0.466}
\resset{e9.clm_a0.sigma}{0.064}
\resset{e9.clm_a0.slope}{-0.65}
\resset{e9.clm_a0.trend}{no trend}
\resset{e9.clm_a1.aut}{0.5983}
\resset{e9.clm_a1.aut_ci}{[0.431, 0.716]}
\resset{e9.clm_a1.mk_p}{0.118}
\resset{e9.clm_a1.sigma}{0.118}
\resset{e9.clm_a1.slope}{-3.13}
\resset{e9.clm_a1.trend}{no trend}
\resset{e9.clm_a2.aut}{0.9057}
\resset{e9.clm_a2.aut_ci}{[0.840, 0.946]}
\resset{e9.clm_a2.fpr}{1.11\%}
\resset{e9.clm_a2.mk_p}{0.466}
\resset{e9.clm_a2.sigma}{0.042}
\resset{e9.clm_a2.slope}{+0.49}
\resset{e9.clm_a2.trend}{no trend}
\resset{e9.clm_a3.aut}{0.9334}
\resset{e9.clm_a3.aut_ci}{[0.885, 0.963]}
\resset{e9.clm_a3.fpr}{1.45\%}
\resset{e9.clm_a3.mk_p}{0.754}
\resset{e9.clm_a3.sigma}{0.023}
\resset{e9.clm_a3.slope}{-0.26}
\resset{e9.clm_a3.trend}{no trend}
\resset{e9.det.aut}{0.8862}
\resset{e9.det.aut_ci}{[0.813, 0.933]}
\resset{e9.det.mk_p}{0.118}
\resset{e9.det.sigma}{0.036}
\resset{e9.det.slope}{+0.81}
\resset{e9.det.trend}{no trend}
\resset{e9.mk.n_arms}{5}
\resset{e9.mk.n_sig_holm}{0}
\resset{e9.mk.pmin}{0.118}
\resset{e9.mk.pmin_arm}{clm_a1}
\resset{e9.mk.pmin_holm}{0.589}
\resset{e9.n_buckets}{9}
\resset{e9.release.clm}{2026-09-24}
\resset{e9.release.clm.before}{yes}
\resset{e9.release.qwen3}{2025-04-29}
\resset{e9.release.qwen3.before}{yes}
\resset{seeds.clm_a2.auroc_mean}{0.9941}
\resset{seeds.clm_a2.auroc_sd}{0.0002}
\resset{seeds.clm_a2.n}{5}
\resset{seeds.clm_a2.tpr_max}{90.90\%}
\resset{seeds.clm_a2.tpr_mean}{90.16\%}
\resset{seeds.clm_a2.tpr_min}{89.52\%}
\resset{seeds.clm_a2.tpr_sd}{0.54\%}
\resset{seeds.clm_a2_ctrl_lr.auroc_mean}{0.9955}
\resset{seeds.clm_a2_ctrl_lr.auroc_sd}{0.0000}
\resset{seeds.clm_a2_ctrl_lr.n}{5}
\resset{seeds.clm_a2_ctrl_lr.tpr_max}{92.94\%}
\resset{seeds.clm_a2_ctrl_lr.tpr_mean}{92.94\%}
\resset{seeds.clm_a2_ctrl_lr.tpr_min}{92.94\%}
\resset{seeds.clm_a2_ctrl_lr.tpr_sd}{0.00\%}
\resset{seeds.clm_a2_ctrl_mlp.auroc_mean}{0.9946}
\resset{seeds.clm_a2_ctrl_mlp.auroc_sd}{0.0006}
\resset{seeds.clm_a2_ctrl_mlp.n}{5}
\resset{seeds.clm_a2_ctrl_mlp.tpr_max}{92.41\%}
\resset{seeds.clm_a2_ctrl_mlp.tpr_mean}{91.25\%}
\resset{seeds.clm_a2_ctrl_mlp.tpr_min}{90.79\%}
\resset{seeds.clm_a2_ctrl_mlp.tpr_sd}{0.66\%}
\resset{seeds.clm_a2_mt.auroc_mean}{0.9928}
\resset{seeds.clm_a2_mt.auroc_sd}{0.0015}
\resset{seeds.clm_a2_mt.n}{5}
\resset{seeds.clm_a2_mt.tpr_max}{90.94\%}
\resset{seeds.clm_a2_mt.tpr_mean}{89.07\%}
\resset{seeds.clm_a2_mt.tpr_min}{83.76\%}
\resset{seeds.clm_a2_mt.tpr_sd}{3.01\%}
\resset{seeds.clm_a2_raw.auroc_mean}{0.9952}
\resset{seeds.clm_a2_raw.auroc_sd}{0.0004}
\resset{seeds.clm_a2_raw.n}{5}
\resset{seeds.clm_a2_raw.tpr_max}{92.30\%}
\resset{seeds.clm_a2_raw.tpr_mean}{90.92\%}
\resset{seeds.clm_a2_raw.tpr_min}{88.60\%}
\resset{seeds.clm_a2_raw.tpr_sd}{1.38\%}
\resset{seeds.clm_a2_std.auroc_mean}{0.9942}
\resset{seeds.clm_a2_std.auroc_sd}{0.0009}
\resset{seeds.clm_a2_std.n}{5}
\resset{seeds.clm_a2_std.tpr_max}{92.60\%}
\resset{seeds.clm_a2_std.tpr_mean}{89.92\%}
\resset{seeds.clm_a2_std.tpr_min}{87.43\%}
\resset{seeds.clm_a2_std.tpr_sd}{2.13\%}
\resset{seeds.qwen3_emb_probe.auroc_mean}{0.9956}
\resset{seeds.qwen3_emb_probe.auroc_sd}{0.0000}
\resset{seeds.qwen3_emb_probe.n}{5}
\resset{seeds.qwen3_emb_probe.tpr_max}{93.62\%}
\resset{seeds.qwen3_emb_probe.tpr_mean}{93.62\%}
\resset{seeds.qwen3_emb_probe.tpr_min}{93.62\%}
\resset{seeds.qwen3_emb_probe.tpr_sd}{0.00\%}
\resset{split}{TEST and $B_{\mathrm{test}}$}
\resset{split.e1}{TEST and $B_{\mathrm{test}}$}
\resset{split.e10}{TEST and $B_{\mathrm{test}}$}
\resset{split.e4}{TEST and $B_{\mathrm{test}}$}
\resset{split.e5}{TEST and $B_{\mathrm{test}}$}
\resset{split.e6}{TEST and $B_{\mathrm{test}}$}
\resset{split.e7}{V and $B_{\mathrm{val}}$}
\resset{split.e8}{TEST and $B_{\mathrm{test}}$}
\resset{split.e9}{TEST and $B_{\mathrm{test}}$}
\resset{stage}{test}
\resset{stage.e1}{test}
\resset{stage.e10}{test}
\resset{stage.e4}{test}
\resset{stage.e5}{test}
\resset{stage.e6}{test}
\resset{stage.e7}{development}
\resset{stage.e7inv}{test}
\resset{stage.e7probes}{test}
\resset{stage.e8}{test}
\resset{stage.e9}{test}
\resset{test.e1}{yes}
\resset{test.e10}{yes}
\resset{test.e4}{yes}
\resset{test.e5}{yes}
\resset{test.e6}{yes}
\resset{test.e7inv}{yes}
\resset{test.e7probes}{yes}
\resset{test.e8}{yes}
\resset{test.e9}{yes}
 
\subsection{Experimental Setup}
\label{sec:res-setup}

Table~\ref{tab:experiments} lists the experiments and their primary endpoints. An arm is one model with one readout on one input: CLM-8B with A0, A1 and A2 on S2, and \fwname{}, which fuses its A1 and A2; every open model with A0 and A1 on S0, and with A2 on S0 in E1b; the two LLMs with A0 and A1 on S2; DET on the digest facts; and the Qwen3-Embedding probe on S2. Combiners, heads and probes are fitted on T0 and $B_{\mathrm{fit}}$, tuned and calibrated on V and $B_{\mathrm{val}}$, and measured on TEST and $B_{\mathrm{test}}$. SOREL-20M measures transfer to another source and period. %

\paragraph{Protocol.} Each experiment fixed its hypotheses, endpoints and comparisons before its runs, and every choice on development files; it was then frozen, and only its frozen thresholds, calibration maps, combiners and heads scored TEST, $B_{\mathrm{test}}$ and the OOD set, in a single pass. \fwname{} joined E1 after this freeze, with its design fixed on the same development files and its own single test pass.

\begin{widetable}[t]
  \centering
  \caption{Experiments and primary endpoints.}
  \label{tab:experiments}
  \small
  \begin{tabularx}{\textwidth}{l >{\raggedright\arraybackslash}X >{\raggedright\arraybackslash}X}
    \toprule
    Exp. & Question & Primary endpoint \\
    \midrule
    E1  & Do System One models detect malware, and what do their readouts add over the digest facts and over LLMs read through logits? & TPR at $\tau_{1\%}$ on TEST, with the realized FPR on $B_{\mathrm{test}}$ \\
    E1b & Do heads on the embedding of every open model add over its atomic answers? & paired AUROC difference of A2 and A1 per model \\
    E2  & Which parts of the state carry the decision? & paired AUROC difference per level of the state \\
    E3  & Which questions and which combiner serve the decision? & paired AUROC difference per question design \\
    E4  & Are the scores calibrated, and how much does abstention buy? & paired ECE difference before and after recalibration \\
    E5  & How do inference time, memory and energy on a laptop GPU compare with the workstation, and do the answers agree? & equivalence of two serving paths of CLM-8B \\
    E6  & How stable are the decisions across seeds and runs? & decision flip rate of A2 between seeds \\
    E7  & Do typed answers hallucinate, and do they resist meaning-preserving changes? & rate of fabricated artifacts per negative regime \\
    E8  & Does text planted in the binary steer the decision? & paired change of TPR and FPR per condition \\
    E9  & How does detection decay over the nine test months? & area under time of the monthly TPR \\
    E10 & Does detection transfer to SOREL-20M? & TPR on the OOD set at the frozen $\tau_{1\%}$ \\
    \bottomrule
  \end{tabularx}
\end{widetable}

\paragraph{Metrics.} AUROC comes with 95\% cluster-bootstrap intervals over 2,000 resamples, where clusters are malware families and benign software groups. The standardized partial AUROC covers FPR up to 1\%. The TPR at $\tau_{1\%}$ is reported with the realized FPR on $B_{\mathrm{test}}$ and Clopper-Pearson intervals, since a threshold fixed on validation data is what a deployment uses~\cite{nguyen2021leveraging}. Precision, F1 and the Matthews correlation are reported at prevalences of 50\%, 10\% and 1\%, because precision depends on the base rate~\cite{arp2022dos}. Calibration is measured by the expected calibration error (ECE, 15 bins) and the negative log-likelihood~\cite{guo2017calibration}, and by the Brier score, at a prior of 0.5, and selective prediction by the area under the risk-coverage curve and the coverage at 1\% risk. Paired comparisons use exact McNemar tests on decisions and a paired bootstrap of AUROC differences, with Holm correction within each family at $\alpha = 0.05$ and one primary endpoint per claim.

\paragraph{Baselines.} DET applies a logistic combiner to the structured facts of the digest, so it measures what the facts give without a language model. Qwen3-8B and Foundation-Sec-8B, read through the probabilities of their answer tokens, and a logistic probe on the Qwen3-Embedding-8B vector of the state measure what general-purpose language models obtain from the same state. Every arm gets its threshold $\tau_{1\%}$ on $B_{\mathrm{val}}$.

\paragraph{Design fixed on development data.} The primary state (Section~\ref{sec:meth-state}), the 32 atomic questions and the gradient-boosted combiner of A1 (Section~\ref{sec:meth-readout}) were fixed on development files before the freeze; E1 and experiment adopts them, while E2 and E3 vary them with a logistic combiner for their A1.%

\subsection{Detection Performance}
\label{sec:res-main}

\ifres{test.e1}{}{\dacompletare{interim: E1 on the development split; the analysis below was checked against the test split}}

Table~\ref{tab:e1} reports E1 on the \res{stage.e1} split, \res{split.e1}: \res{e1.n_malware} malware and \res{e1.n_benign} benign files with a state; the other \res{e1.no_state} files (\res{e1.no_state.benign} benign), whose state lacks a tool record, keep an explicit status and no score (Section~\ref{sec:meth-state}).

\begin{figure}[t]
  \centering
  \includegraphics[width=\linewidth]{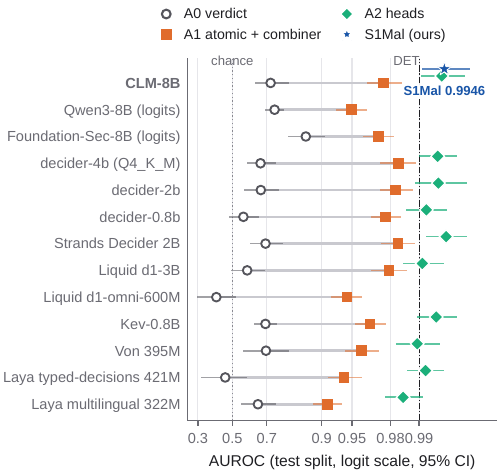}
  \caption{Readouts of every prompt-reading model.}
  \label{fig:readouts}
\end{figure}

\paragraph{Typed readouts.} Figure~\ref{fig:readouts} and Table~\ref{tab:readouts} (Appendix~\ref{app:results}) give the readouts of every model that reads a prompt. A single zero-shot verdict (A0) reaches at most \res{e1.a0.max} AUROC: it compresses the whole state into the probability of one answer, a judgment that these general-purpose decision models never learned on static evidence. The 32 atomic questions of A1 split the same decision into behaviors that the models recognize in the text (e.g., persistence, process injection or packing), and a gradient-boosted combiner weighs the answers: A1 lifts every one of the \res{e1.n_models} models, up to \res{e1.a1.max} AUROC, and all \res{e1.lift.n_sig} lifts that the protocol tests hold after Holm correction. For CLM-8B the lift is \res{e1.E1-C3.d} (95\% CI \res{e1.E1-C3.ci}), and the heads trained on its embedding (A2) reach \res{e1.clm_a2.auroc} \res{e1.clm_a2.ci}, \res{e1.E1-C2.d} over A0. A logistic probe on the same embedding (A2-ctrl) ranks the files slightly better than the heads (E1-C1: \res{e1.E1-C1.d}, Holm $p$ = \res{e1.E1-C1.p}): the decision is linearly readable from the state embedding of CLM-8B.

\paragraph{Evidence and backbone.} A2 adds \res{e1.E1-C4.d} AUROC (\res{e1.E1-C4.ci}, Holm $p$ = \res{e1.E1-C4.p}) over DET, a logistic combiner on the digest facts without a language model: CLM-8B reads the function cards of S2 and the wording of imports and strings. With the same Qwen3-8B backbone, its trained heads add \res{e1.E1-C7.d} over the answer-token probabilities of Qwen3-8B with A1, and \res{e1.E1-C6.d} over the best open decision model with A1, \res{e1.best_open}.

\paragraph{\fwname{}: atomic answers and embedding fused.} \fwname{} joins the two readouts of CLM-8B: trees in the configuration of A1 weigh the 32 atomic answers together with the cross-fitted score of heads trained as those of A2. On the test split \fwname{} reaches an AUROC of \res{e1.clm_a3.auroc} \res{e1.clm_a3.ci}, \res{e1.A3-C2.d} over A1 (Holm $p$ = \res{e1.A3-C2.p}): the embedding score carries the evidence that the 32 questions leave unasked, and the ranking matches that of A2 (Holm $p$ = \res{e1.A3-C1.p}). The fusion acts on the decisions: at $\tau_{1\%}$ \fwname{} detects \res{e1.clm_a3.tpr} of the malware, \res{e1.A3-C1.tpr_d} points more than A2 (McNemar $p$ \res{e1.A3-C1.tpr_p}), at a realized FPR of \res{e1.clm_a3.fpr} against \res{e1.clm_a2.fpr}; at the same FPR A2 reaches \res{e1.clm_a2.tpr_at_a3fpr}, so the fused ranking and the operating point each bring about half of the gain. Every score of \fwname{} keeps the 32 atomic answers among its inputs, so each decision comes with the behaviors that CLM-8B recognized in the state.

\begin{widetable}[t]
  \centering
  \caption{Detection on the test split (E1).}
  \label{tab:e1}
  \small
  \begin{threeparttable}
  \begin{tabular}{l l c c c c c}
    \toprule
    Arm & Input & AUROC [95\% CI] & pAUC$_{1\%}$ & TPR at $\tau_{1\%}$ & FPR & ECE \\
    \midrule
    DET & S0 facts & 0.9901 [0.985, 0.994] & 0.8284 & 88.24\% & 1.91\% & 0.0139 \\
    Qwen3-Embedding probe & S2 & 0.9956 [0.993, 0.997] & 0.8678 & 93.62\% & 1.42\% & 0.0066 \\
    Qwen3-8B logits, A1 & S2 & 0.9493 [0.929, 0.964] & 0.6013 & 36.30\% & 0.95\% & 0.0235 \\
    Foundation-Sec-8B logits, A1 & S2 & 0.9733 [0.962, 0.981] & 0.6595 & 64.19\% & 1.58\% & 0.0184 \\
    decider-4b (GGUF Q4\_K\_M), A1 & S0 & 0.9834 [0.975, 0.989] & 0.7530 & 75.44\% & 1.29\% & 0.0167 \\
    CLM-8B A2-ctrl probe & S2 & 0.9955 [0.993, 0.997] & 0.8974 & 92.94\% & 1.11\% & 0.0104 \\
    CLM-8B A0 & S2 & 0.7194 [0.643, 0.796] & 0.5258 & 8.82\% & 1.16\% & 0.1024 \\
    CLM-8B A1 & S2 & 0.9763 [0.965, 0.985] & 0.6713 & 57.90\% & 1.19\% & 0.0161 \\
    CLM-8B A2 & S2 & 0.9943 [0.991, 0.997] & 0.8760 & 90.90\% & 1.11\% & 0.0060 \\
    \midrule
    \textbf{\fwname{} (CLM-8B, ours)} & S2 & \textbf{0.9946 [0.991, 0.997]} & \textbf{0.8851} & \textbf{93.47\%} & \textbf{1.45\%} & \textbf{0.0078} \\
    \bottomrule
  \end{tabular}
  \end{threeparttable}
\end{widetable}
 
\paragraph{Heads on the open models.} E1b trains the heads of A2 on the embedding of each open decision model, the final hidden state at the last token of its own verdict prompt over S0, with seed 0 as the primary variant. A2 lifts each of the \res{e1b.n_models} models whose runtime exposes this representation over its A1, and all \res{e1b.p1.n_sig} lifts hold after Holm correction; the best, \res{e1b.a2.best}, reaches an AUROC of \res{e1b.a2.max}, and \res{e1b.s2.n_ns} of the open models match the A2 of CLM-8B after Holm correction (Figure~\ref{fig:readouts}). The decision is present in the representation of models from 322M to 4B parameters, and heads on the embedding recover what the verdict and the atomic answers leave unused.

\begin{figure}[t]
  \centering
  \includegraphics[width=\linewidth]{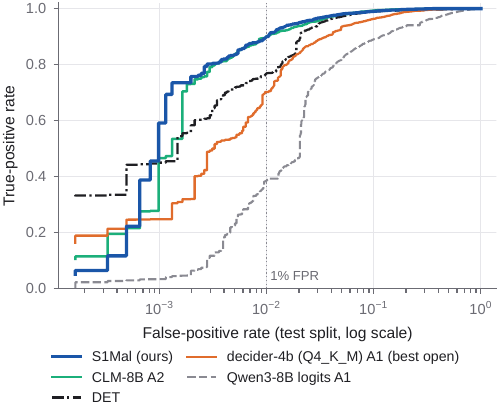}
  \caption{Detection at low false-positive rates.}
  \label{fig:roc}
\end{figure}

\paragraph{Low false-positive rates.} Figure~\ref{fig:roc} shows the region that a deployment uses. The thresholds, fixed on $B_{\mathrm{val}}$, carry over to the unseen benign files of $B_{\mathrm{test}}$ with a realized FPR from \res{e1.fpr.min} to \res{e1.fpr.max} across the arms of Table~\ref{tab:e1}. At 0.1\% FPR, the six benign test files that this rate allows, \fwname{} keeps \res{e1.clm_a3.tpr01} of the malware, the most in Table~\ref{tab:e1}, whose arms keep at most \res{e1.tpr01.others_max}: the benign files that the embedding scores highest set this tail, and the atomic answers that \fwname{} adds plausibly tell them apart from malware. A threshold at that rate needs the larger benign validation set discussed in Section~\ref{sec:res-limits}.

\paragraph{Shortcut controls.} The raw state, without canonicalization, raises the AUROC of A2 by \res{e1.E1-S-d13-a2.d} (Holm $p$ = \res{e1.E1-S-d13-a2.p}). The library names, section names, build paths and toolchain words that canonicalization rewrites carry a small signal that the embedding of CLM-8B picks up, about a thousandth of AUROC, and the canonical state gives it up on purpose: these properties differ between the two collections for reasons unrelated to behavior, so a detector that relied on them would learn the collections.

\subsection{State and Question Design}
\label{sec:res-design}

E3 and E2 test the design choices of Section~\ref{sec:meth}.%

\paragraph{Questions.} Across the \res{e3.n_models} models that read a prompt, the 32 atomic questions with a trained combiner rank the files above the single zero-shot verdict for \res{e3.V1.n_sig_pos} models, with gains up to \res{e3.V1.max} AUROC; for CLM-8B the gain is \res{e3.V1.clm.d}, and \res{e3.V1.ns_models} is the exception. A better combiner adds to this gain: shallow gradient-boosted trees improve on a logistic combiner of the same answers for \res{e3.K1.n_sig_pos} of the \res{e3.K1.n} models, while an untrained noisy-OR over the behaviors falls below the logistic combiner for all \res{e3.K2.n}: the evidence lies in how the answers combine, which a rule over independent behaviors cannot weigh. The number of questions counts. The single most informative question ranks the files below the full set for \res{e3.N.k1.n_sig} of the \res{e3.N.n_models} models, and the gap closes as questions are added, with \res{e3.N.k16.n_sig} models still below the full set at 16 questions. Terminological choices carry minimal weight: the MITRE ATT\&CK and capa vocabularies match plain analyst language in all but \res{e3.W.n_sig} of \res{e3.W.n} comparisons, and asking each behavior in both polarities changes \res{e3.P.n_sig} of \res{e3.P.n}. On these files the digest facts read by DET (AUROC \res{e3.det.auroc}) rank above the atomic answers of \res{e3.D1.n_sig_neg} of the \res{e3.D1.n} models; the embedding readouts pass DET on the test split (Table~\ref{tab:e1}), and the atomic answers name the behaviors behind each decision. With every question design of CLM-8B, \fwname{} adds the A2 score to the answers of that design and ranks the files above the answers alone for \res{e3.a3.n_sig_a1} of the \res{e3.a3.n} designs after Holm correction, and level with A2 for all of them.

\paragraph{State.} For CLM-8B, S2 and the cards alone (S1) each rank the files \res{e2.S1.clm_a2.d} AUROC above the digest S0 with A2, on their own paired files, a difference that does not hold after Holm correction; the number of cards, the token budget and the selection policy leave A2 unchanged (\res{e2.KBP.clm_a2.n_sig} of \res{e2.KBP.clm_a2.n} comparisons). The decision of the embedding rests on redundant evidence: removing any one block of S0 keeps its AUROC within the interval, and each of the \res{e2.O.clm_a2.rest_blocks} blocks alone keeps nearly all of it, with a loss of at most \res{e2.O.clm_a2.rest_max}, while the \res{e2.O.clm_a2.sig_blocks} blocks alone lose \res{e2.O.clm_a2.sig_min} to \res{e2.O.clm_a2.sig_max}. \emph{The typed answers depend more on the state than the embedding does}: a budget of 7,680 tokens lifts A1 by \res{e2.B3.clm_a1.d} over 2,048 tokens, and the attacker-controlled fields that S0 leaves out lift A0 and A1 by \res{e2.T1.clm_a0.d} and \res{e2.T1.clm_a1.d}, the shortcut that the protocol expected, while A2 stays unchanged. On these files DET ranks above A2 on S0 (\res{e2.D1.clm_a2.d}). Across the \res{e2.a3.n} levels of the state, \fwname{} ranks the files above A1 for \res{e2.a3.n_sig_a1}: the fusion keeps the decision of the embedding wherever the state changes.

\subsection{Calibration, Stability and Deployment}
\label{sec:res-deploy}

\ifres{test.e4}{}{\dacompletare{interim: E4 on the development split, out-of-fold over V and $B_{\mathrm{val}}$; the analysis below was checked against the test split}}

\begin{widetable}[t]
  \centering
  \caption{Calibration and abstention (test split, E4).}
  \label{tab:e4}
  \small
  \begin{threeparttable}
  \begin{tabular}{l c c l c c c}
    \toprule
    & \multicolumn{2}{c}{ECE} & & & \multicolumn{2}{c}{Band at 1\% risk} \\
    \cmidrule(lr){2-3} \cmidrule(lr){6-7}
    Arm & raw & recal. & Map & Brier & Coverage & Risk \\
    \midrule
    CLM-8B A0 & 0.3691 & 0.0479 & isotonic & 0.2060 & 0 (infeasible) & - \\
    Qwen3-8B logits, A0 & 0.3025 & 0.0164 & isotonic & 0.1833 & 0 (infeasible) & - \\
    Foundation-Sec-8B logits, A0 & 0.1542 & 0.0316 & Platt & 0.1537 & 16.01\% & 3.10\% \\
    decider-2b A0 & 0.1620 & 0.0642 & isotonic & 0.2141 & 4.60\% & 3.63\% \\
    Kev-0.8B A0 & 0.1084 & 0.0719 & isotonic & 0.2194 & 0 (infeasible) & - \\
    DET & 0.0184 & 0.0139 & Platt & 0.0310 & 74.81\% & 1.49\% \\
    \midrule
    CLM-8B A2 & 0.0124 & 0.0060 & Platt & 0.0262 & 92.36\% & 1.32\% \\
    \bottomrule
  \end{tabular}
  \end{threeparttable}
\end{widetable}
 
Table~\ref{tab:e4} and Figure~\ref{fig:calibration} report E4 on the \res{stage.e4} split, where each arm takes the map, among Platt, isotonic and temperature scaling, with the lowest out-of-fold log-loss on V and $B_{\mathrm{val}}$, fitted there and frozen. Recalibration turns the zero-shot verdict into a usable probability: CLM-8B A0 has an ECE of \res{e4.clm_a0.ece_raw} at the class prior, and its \res{e4.clm_a0.method} map brings it to \res{e4.clm_a0.ece_cal}, an improvement of \res{e4.clm_a0.improve}. The primary endpoint holds for all \res{e4.primary.n} arms: the frozen map of each one lowers its ECE, and all \res{e4.primary.n_sig} improvements hold after Holm correction (Holm $p \le$ \res{e4.primary.pmax}). The trained readouts arrive calibrated: A2 starts at \res{e4.clm_a2.ece_raw} and ends at \res{e4.clm_a2.ece_cal} after \res{e4.clm_a2.method} scaling, below DET (\res{e4.det.ece_cal}). Recalibration repairs the probability scale, while the ranking quality decides how much abstention buys. With an abstention band fitted for a 1\% selective risk, A2 decides \res{e4.clm_a2.cover} of the test files at a realized risk of \res{e4.clm_a2.risk}, against \res{e4.det.cover} for DET, and for both the realized risk stays between \res{e4.band.risk_min} and \res{e4.band.risk_max}. \fwname{} moves from an ECE of \res{e4.clm_a3.ece_raw} to \res{e4.clm_a3.ece_cal} with its \res{e4.clm_a3.method} map, an improvement of \res{e4.clm_a3.improve} (bootstrap $p$ = \res{e4.clm_a3.improve_p}), and its band decides \res{e4.clm_a3.cover} of the test files, more than A2, at a realized risk of \res{e4.clm_a3.risk}. The risk-coverage curves of Figure~\ref{fig:calibration} rank the files by their distance from the calibrated operating point, the confidence function is fixed before the test. This distance counts a miss scored far on the benign side as a confident decision, so on the test files the curves of A2 and DET take in their misses early and climb to a few percent, up to \res{e4.rc.clm_a2.max} for A2, while the band, whose two thresholds are fitted together on valiation, defers the files around the threshold and holds the realized risk reported above. The A0 verdicts decide at most \res{e4.llm_fsec_a0.cover} of the files, at a realized risk of \res{e4.llm_fsec_a0.risk}, and \res{e4.a0.n_infeasible} of the \res{e4.a0.n} reach no band at that risk. We can assume that a calibrated verdict still overlaps across the classes, and the atomic and embedding readouts are what separate them.

\begin{widefigure}[t]
  \centering
  \includegraphics[width=\textwidth]{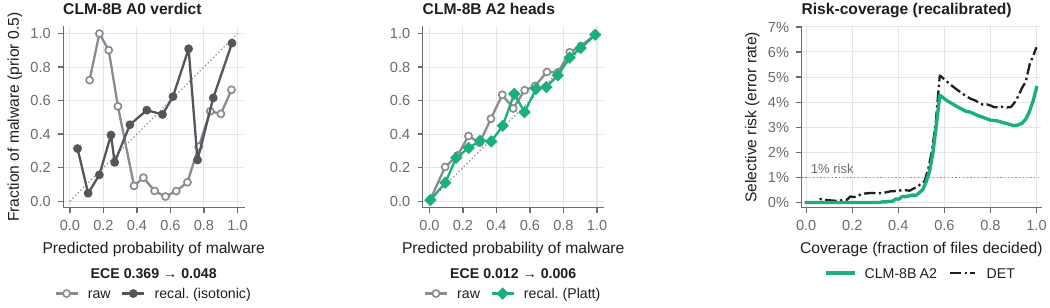}
  \caption{Calibration and selective prediction.}
  \label{fig:calibration}
\end{widefigure}

\ifres{test.e6}{}{\dacompletare{interim: E6 on the development split; the analysis below was checked against the test split}}\ifres{test.e5}{}{\dacompletare{interim: E5 on its development subsample; the deployment analysis was checked against the test split}}

\paragraph{Stability.} Inference is deterministic: on \res{e6.local.n_files} files, new server sessions with cold and cached requests give the same probability for every file with A0 and A2 (\res{e6.local.max_flip} decision flips), and every configuration of E5 repeats its decisions across its passes. The variation comes from training. Across \res{seeds.clm_a2.n} seeds the AUROC of A2 has a standard deviation of \res{seeds.clm_a2.auroc_sd} (Table~\ref{tab:seeds}, Appendix~\ref{app:results}), and \fwname{}, with its cross-fitted heads trained again under each seed, keeps an AUROC between \res{e6.a3.auroc_min} and \res{e6.a3.auroc_max}, with \res{e6.a3.all.flip} of the test files changing decision between two seeds (\res{e6.a2e1.all.flip} for A2). When E6 trains the heads again at the setting of its own validation grid, early stopping halts \res{e6.heads.n_early} of its \res{e6.heads.n_seeds} seeds after \res{e6.heads.early_epoch} epochs: these heads detect \res{e6.early.tpr_min} and \res{e6.early.tpr_max} of the test malware at $\tau_{1\%}$, against \res{e6.late.tpr_min} to \res{e6.late.tpr_max} for the others, and drive the disagreement between seeds (\res{e6.flip.mean} on average), so the validation AUROC of every seed flags them before deployment. The logistic probe on the same embedding (A2-ctrl) changes \res{e6.ctrl.flip} of its decisions between seeds, at the AUROC of the heads. Sampling at temperature 1 changes \res{e6.sampled.qwen3.flip} of the decisions of Qwen3-8B and \res{e6.sampled.fsec.flip} of those of Foundation-Sec-8B, a variation that a typed probability read in one forward pass does not have. %

\begin{widetable}[t]
  \centering
  \caption{Deployment configurations of CLM-8B (E5).}
  \label{tab:e5}
  \small
  \begin{threeparttable}
  \begin{tabular}{l l c c c c c c}
    \toprule
    Machine & Weights & AUROC & TPR at $\tau_{1\%}$ & FPR & Latency [s] & Files/s & Energy [mWh] \\
    \midrule
    Workstation & bf16 & 0.9922 & 90.33\% & 1.94\% & 0.114 & 8.34 & 9.26 \\
    Workstation & Q5\_K\_M & 0.9859 & 60.40\% & 0.43\% & 0.112 & 8.54 & 8.77 \\
    Laptop & Q6\_K & 0.9859 & 50.55\% & 0.43\% & 0.378 & 2.53 & 8.55 \\
    Laptop & Q4\_K\_M & 0.9788 & 53.28\% & 0.86\% & 0.349 & 2.75 & 7.82 \\
    \midrule
    \textbf{Laptop} & \textbf{Q5\_K\_M} & \textbf{0.9827} & \textbf{52.19\%} & \textbf{0.22\%} & \textbf{0.359} & \textbf{2.68} & \textbf{8.08} \\
    \bottomrule
  \end{tabular}
  \begin{tablenotes}\footnotesize
    \item Workstation: AMD Ryzen~9 9950X3D, NVIDIA GeForce RTX 5090, 192~GB of RAM. Laptop: Intel Core i9-13900H, NVIDIA GeForce RTX 4060 Laptop GPU, 64~GB of RAM.
  \end{tablenotes}
  \end{threeparttable}
\end{widetable}
 
\paragraph{Deployment on a laptop GPU.} E5 runs the frozen System One on the laptop GPU of the analysis host and compares it with the workstation on \res{e5.n_files} files of the \res{stage.e5} split, read as S0 (Table~\ref{tab:e5}). The bf16 model with its 8,192-token window does not fit in 8~GB, so the laptop runs CLM-8B as a 5-bit GGUF at 2,048 tokens: the deployed configuration decides a file in \res{e5.edge.lat}~s with \res{e5.edge.energy}~mWh, and the workstation runs the bf16 model \res{e5.E5-C6.ratio} times faster. \ifres{e5.s1mal.lat}{\fwname{} adds the 32 atomic questions after the same encoder pass and decides a file in \res{e5.s1mal.lat}~s with \res{e5.s1mal.energy}~mWh, \res{e5.s1mal.t_atomic}~s of which go to the atomic answers that name the behaviors behind each decision. }{}The two serving paths of the laptop give the same decision on all \res{e5.E5-C1.n} files, the primary endpoint of E5. The 5-bit weights keep the ranking, with an AUROC of \res{e5.edge.auroc} against \res{e5.ref.auroc} in bf16, and move the scores toward benign around the frozen threshold: the TPR at $\tau_{1\%}$ goes from \res{e5.ref.tpr} to \res{e5.edge.tpr} and the FPR from \res{e5.ref.fpr} to \res{e5.edge.fpr}. The precision of the weights moves the decisions about twice as much as the device does (\res{e5.E5-C8.flip} against \res{e5.E5-C7.flip}), so each deployment tier fits its own threshold on V and $B_{\mathrm{val}}$, with the heads unchanged.

\begin{figure}[t]
  \centering
  \IfFileExists{figures/e5_size.pdf}{\includegraphics[width=\linewidth]{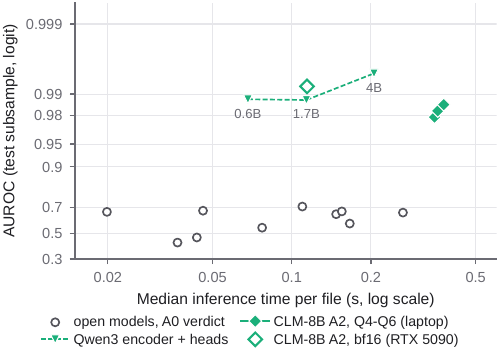}}{\dacompletare{figure e5\_size}}
  \caption{Accuracy and inference time on a laptop GPU.}
  \label{fig:e5size}
\end{figure}

Figure~\ref{fig:e5size} places the other models of the laptop on the same axes. Qwen3 base models of 0.6B, 1.7B and 4B parameters, with heads trained from a cold start on their embeddings of the same S0, reach an AUROC of \res{e5.small.0.6b.auroc}, \res{e5.small.1.7b.auroc} and \res{e5.small.4b.auroc} in \res{e5.small.0.6b.lat} to \res{e5.small.4b.lat}~s per file, with \res{e5.small.4b.vram}~GiB for the largest. The \res{e5.open.n} open decision models answer their verdict in \res{e5.open.lat_min} to \res{e5.open.lat_max}~s with \res{e5.open.vram_min} to \res{e5.open.vram_max}~GiB, at an AUROC of \res{e5.open.auroc_min} to \res{e5.open.auroc_max}. On this subsample the 4B encoder with heads ranks the files at least as well as the full System One (AUROC \res{e5.small.4b.auroc} against \res{e5.ref.auroc}) in \res{e5.small.4b.lat}~s per file: trained heads carry the decision onto edge hardware, the setting that the open decision models target. %

\subsection{Hallucination and Robustness}
\label{sec:res-halluc}

E7 adapts the existence probes of POPE~\cite{li2023pope} to static artifacts: the models are asked about APIs, strings and capabilities absent from the state, drawn at random, among the most frequent ones, or among those that co-occur with the present ones. \nuovo{A second sub-sample tests the invariance of the decisions to polarity, option order, paraphrase~\cite{sclar2024quantifying}, padding and repetition.} E8 plants instructions and legitimacy claims in the strings of synthetic states, the attack that ALIBI showed on generative verdicts~\cite{choi2026alibi}, with and without the sanitizer.

\ifres{test.e7probes}{}{\dacompletare{interim: E7 on the development probes; regenerated on the test split at the merge}}\ifres{test.e8}{}{\dacompletare{interim: E8 on its development subsample; the analysis below was checked against the test split}}

\ifres{e7.n_backends}{%
\begin{figure}[t]
  \centering
  \includegraphics[width=\linewidth]{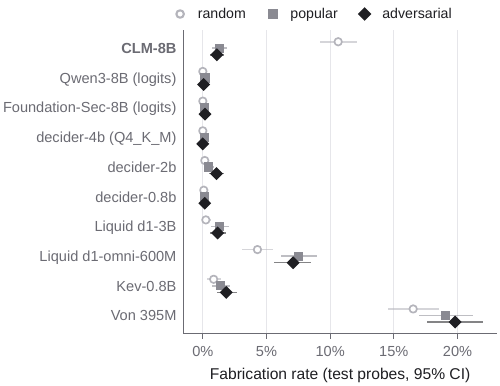}
  \caption{Fabrication on absent artifacts.}
  \label{fig:e7}
\end{figure}

\paragraph{Fabrication.} Each probe asks whether the state contains one item, and a yes on an absent item is a fabrication. Figure~\ref{fig:e7} reports the fabrication rate of the \res{e7.n_backends} models on the \res{stage.e7probes} probes in each regime: from \res{e7.random.min} to \res{e7.random.max} on random items, from \res{e7.popular.min} to \res{e7.popular.max} on popular items and from \res{e7.adversarial.min} to \res{e7.adversarial.max} on adversarial ones. CLM-8B fabricates \res{e7.clm_gguf_bf16.random.fab} of the random items (\res{e7.clm_gguf_bf16.random.fab_ci}), \res{e7.clm_gguf_bf16.popular.fab} of the popular and \res{e7.clm_gguf_bf16.adversarial.fab} of the adversarial ones, and confirms \res{e7.clm_gguf_bf16.random.recall} of the present imports, strings and capabilities. Its fabrications gather on random import names (\res{e7.clm.api_import.random.fab}), and it never affirms an absent capability (\res{e7.clm.capability.all.fab}): the frequent and co-occurring items, which a generative model tends to presume~\cite{li2023pope}, are where its typed answers hold best, so the atomic questions about behaviors rest on the part of the state that CLM-8B reads most reliably.%
}{\paragraph{Fabrication.} \dacompletare{E7 results: the development pass of the probes needs the answers of the GPU node, queued after E1's test stage}}

\ifres{e7i.clm_gguf_bf16.para}{%
\paragraph{Invariance.}\ifres{test.e7inv}{}{ \dacompletare{interim: E7's invariance on the development sub-sample; regenerated on the test split at the end of E7's test stage}} \nuovo{On a label-balanced sub-sample of \res{e7i.n_sub} \res{stage.e7inv} files, \res{e7i.n_files} of them with a state, the verdict question of every model is asked again inverted, as a choice question with its two options swapped, in ten paraphrases, about the state padded with benign import lines and strings by 5\% and 10\% of its tokens, and a second time unchanged; the frozen A1 arms, the heads of A2 and \fwname{} decide again under the paraphrases of their questions and on the padded states. The answer of CLM-8B to the verdict question (a yes at a probability of at least 0.5) holds under the changes of framing: \res{e7i.clm_gguf_bf16.polarity} of its verdicts flip when the question is inverted, \res{e7i.clm_gguf_bf16.option} when the options are swapped, \res{e7i.clm_gguf_bf16.pad10} with 10\% of padding (on the \res{e7i.clm_gguf_bf16.pad10_n} files whose padded state fits the window) and \res{e7i.clm_gguf_bf16.rep} on the repeated request. Across the \res{e7i.n_backends} models whose window takes the state, inversion flips between \res{e7i.polarity.min} and \res{e7i.polarity.max} of the verdicts (\res{e7i.polarity.max_name}) and swapped options between \res{e7i.option.min} and \res{e7i.option.max} (\res{e7i.option.max_name}). The wording of a question is what moves the decisions of CLM-8B most: a paraphrase of the verdict question flips on average \res{e7i.clm_gguf_bf16.para} of CLM-8B's verdicts, a paraphrase of the 32 atomic questions \res{e7i.clm_a1.para} of the decisions of A1 (between \res{e7i.a1.para.min} and \res{e7i.a1.para.max} across the \res{e7i.a1.para.n} A1 arms), and the paraphrased verdict in the text that the heads embed \res{e7i.clm_a2.para} of the decisions of A2. \ifres{e7i.clm_a3.para}{\fwname{} is the decision of CLM-8B that paraphrase moves least, at \res{e7i.clm_a3.para}, and with 10\% of padding it changes \res{e7i.clm_a3.pad10} of its decisions, \ifres{test.e7inv}{fewer than A2 (\res{e7i.clm_a2.pad10}) and A1 (\res{e7i.clm_a1.pad10})}{as A2 (\res{e7i.clm_a2.pad10}), against \res{e7i.clm_a1.pad10} for A1}: its trees weigh the atomic answers together with the score of the heads. }{}The wording is therefore part of each frozen arm: \fwname{} asks every question in the wording that its combiner and its heads were fitted on, the setting in which E1 measured them.}%
}{\paragraph{Invariance.} \dacompletare{E7: the invariance results, at the end of E7's test stage}}

\begin{figure}[t]
  \centering
  \IfFileExists{figures/e8_injection.pdf}{\includegraphics[width=\linewidth]{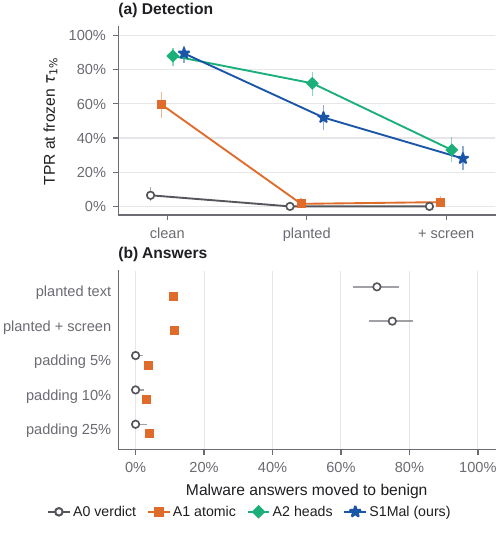}}{\dacompletare{figure e8\_injection}}
  \caption{Planted text in the state.}
  \label{fig:e8}
\end{figure}

\paragraph{Planted text.}
E8 builds a robustness test suite from a seeded sub-sample of the test files, stratified by month and packing, and appends the \res{e8.screen.canaries.n} injection payloads in its bank to the strings block of the states of its malware and benign files. These include instructions to the analyst, claims that the file is benign, fabricated evidence (e.g., a trusted signature or a clean audit), completed templates and obfuscated variants. Each payload is inserted either directly or through the state builder's sanitizer, which adds a flag line when its screen identifies analysis-directed text. The planted text affects the verdict question more than the atomic answers: it shifts \res{e8.shift.inj_raw.verdict.moved} of malware verdicts toward benign, compared with \res{e8.shift.inj_raw.atomic.moved} of atomic answers (Figure~\ref{fig:e8}b). Each atomic question concerns a specific behavior, so a general claim of legitimacy does not directly answer it.

At $\tau_{1\%}$, A0's malware TPR falls from \res{e8.clm_a0.clean.tpr} to \res{e8.clm_a0.inj_raw.tpr}, and A1's from \res{e8.clm_a1.clean.tpr} to \res{e8.clm_a1.inj_raw.tpr}. Although individual atomic answers change little, A1's combiner, fitted on clean answers, accumulates their shifts toward benign. A2 retains the highest TPR under planted text, \res{e8.clm_a2.inj_raw.tpr}, down from \res{e8.clm_a2.clean.tpr}. However, its embedding heads also treat the planted text as evidence of maliciousness, flagging \res{e8.clm_a2.inj_raw.fpr} of benign files containing it (Figure~\ref{fig:e8}a). By combining atomic answers with the embedding score, \fwname{} moderates both effects. Its TPR falls from \res{e8.clm_a3.clean.tpr} to \res{e8.clm_a3.inj_raw.tpr}, while its FPR on benign files with planted text is \res{e8.clm_a3.inj_raw.fpr}, less than half A2's. The atomic answers' shift toward benign thus reduces both detections and false alarms relative to the embedding readout.

The screen identifies all \res{e8.screen.canaries.n} payloads in its bank and none of the \res{e8.screen.benign_strings.n} benign strings used to check it. On the \res{e8.screen.stage} split, where it is evaluated, it flags \res{e8.screen.states_benign} of benign states and \res{e8.screen.states_malware} of malware states. Adding its flag line shifts A2 and \fwname{} further toward benign, reducing their TPRs to \res{e8.clm_a2.inj_san.tpr} and \res{e8.clm_a3.inj_san.tpr}, respectively. Because the flag occurs more often in benign states, the trained readouts interpret it as evidence that a file is benign. The screen is therefore most useful as a separate signal to the analyst, outside the state read by the model.

Benign padding, ranging from 5\% to 25\% of the state, changes relatively few answers. At 25\%, it moves \res{e8.shift.pad25.atomic.moved} of atomic answers and \res{e8.shift.pad25.verdict.moved} of verdicts. Nevertheless, small shifts across many answers, or in the overall embedding, can push scores across the threshold set at 1\% FPR. Among the \res{e8.pad.n_malware} malware files with answers at every padding level, A2's TPR falls from \res{e8.pad.clm_a2.clean.tpr} to \res{e8.pad.clm_a2.pad25.tpr} at 25\% padding (McNemar $p$ \res{e8.pad.clm_a2.pad25.p}), while A1's falls from \res{e8.pad.clm_a1.clean.tpr} to \res{e8.pad.clm_a1.pad25.tpr}. \fwname{} is the most robust: its TPR goes from \res{e8.pad.clm_a3.clean.tpr} to \res{e8.pad.clm_a3.pad05.tpr} at 5\% padding (McNemar $p$ = \res{e8.pad.clm_a3.pad05.p}), compared with \res{e8.pad.clm_a2.pad05.tpr} for A2 and \res{e8.pad.clm_a1.pad05.tpr} for A1, and reaches \res{e8.pad.clm_a3.pad25.tpr} at 25\%.

This robustness reflects the complementary responses of the two inputs: padding affects the atomic answers and embedding scores of different files. At 5\% padding, at least one of A1 and A2 detects \res{e8.pad.either.pad05.tpr} of malware, while both detect \res{e8.pad.both.pad05.tpr}. At 25\%, these proportions are \res{e8.pad.either.pad25.tpr} and \res{e8.pad.both.pad25.tpr}, respectively. By weighing both inputs, the trees in \fwname{} maintain a TPR close to that of the union of the two readouts.

\subsection{Drift and Out-of-Distribution Transfer}
\label{sec:res-drift}

E9 follows the TPR at the frozen threshold over the nine monthly buckets of TEST and summarizes it by the area under time~\cite{pendlebury2019tesseract}. E10 applies the frozen arms to SOREL-20M. Every month of BODMAS precedes the release of CLM-8B (\res{e9.release.clm}) and of its Qwen3-8B backbone (\res{e9.release.qwen3}), so the models may have met these samples, or reports about them, in pretraining; the contamination controls of E10 measure what the decision owes to such recall.

\ifres{e9.clm_a2.aut}{%
\begin{figure}[t]
  \centering
  \includegraphics[width=\linewidth]{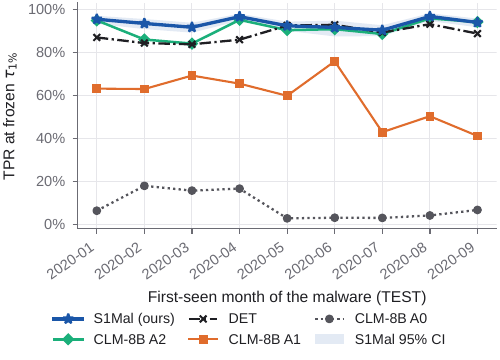}
  \caption{Detection over the test months.}
  \label{fig:e9}
\end{figure}

\paragraph{Drift.} Figure~\ref{fig:e9} follows the TPR at the frozen $\tau_{1\%}$ over the \res{e9.n_buckets} first-seen months of TEST. \fwname{} keeps an area under time of \res{e9.clm_a3.aut} (\res{e9.clm_a3.aut_ci}), with a Sen slope of \res{e9.clm_a3.slope} points per month (Mann-Kendall $p$ = \res{e9.clm_a3.mk_p}) and a standard deviation of the monthly TPR of \res{e9.clm_a3.sigma}, and A2 keeps \res{e9.clm_a2.aut} (\res{e9.clm_a2.aut_ci}), with \res{e9.clm_a2.slope} points per month ($p$ = \res{e9.clm_a2.mk_p}) and a deviation of \res{e9.clm_a2.sigma}. The atomic answers steady the months in which the embedding score dips: in \res{e9.a2min.month}, the lowest month of A2 at \res{e9.a2min.clm_a2}, \fwname{} detects \res{e9.a2min.clm_a3} of the malware. Their FPR on $B_{\mathrm{test}}$, whose files carry no dates, is \res{e9.clm_a3.fpr} and \res{e9.clm_a2.fpr}. No arm shows a monotone trend after Holm correction over the \res{e9.mk.n_arms} arms of Figure~\ref{fig:e9}: the smallest Hamed-Rao $p$, \res{e9.mk.pmin} for A1 and for DET, becomes \res{e9.mk.pmin_holm}. The frozen System One thus holds its detection over the nine months without retraining, inside the knowledge horizon of the models, plausibly because the behaviors that its readouts capture recur in the malware of the following months. DET follows the same course at \res{e9.det.aut}, and the readouts that answer questions stay at the level of their TPR in E1, \res{e9.clm_a1.aut} for A1 and \res{e9.clm_a0.aut} for A0.%
}{\paragraph{Drift.} \dacompletare{E9 results: the test stage; the two months of V are fewer than the three that the temporal summary needs}}

\paragraph{Transfer and contamination.} \ifres{e10.clm_a2.tpr}{On SOREL-20M, at the frozen $\tau_{1\%}$, \fwname{} detects \res{e10.clm_a3.tpr} of the malware (\res{e10.clm_a3.tpr_ci}), A2 \res{e10.clm_a2.tpr} (\res{e10.clm_a2.tpr_ci}), DET \res{e10.det.tpr}, A1 \res{e10.clm_a1.tpr} and A0 \res{e10.clm_a0.tpr}, each at the realized FPR of its threshold on $B_{\mathrm{test}}$ (Table~\ref{tab:e1}). The gain of \fwname{} over A2 is larger on packed malware, \res{e10.clm_a3.tpr_packed} against \res{e10.clm_a2.tpr_packed}, than on the other files. Against the benign files of $B_{\mathrm{test}}$, \fwname{} ranks the SOREL-20M malware at an AUROC of \res{e10.clm_a3.auroc} (\res{e10.clm_a3.auroc_ci}), A2 at \res{e10.clm_a2.auroc} (\res{e10.clm_a2.auroc_ci}) and DET at \res{e10.det.auroc} (\res{e10.det.auroc_ci}). The ranking carries over to a source and a period that no component has seen, while the thresholds, fixed on the BODMAS malware of V, pass about half of the SOREL-20M malware: a new feed calls for an operating point fitted on its own validation files, with the heads and the trees unchanged.}{\dacompletare{E10 transfer to SOREL-20M: the test stage; development reads V as a stand-in for the OOD feed}} \ifres{test.e10}{}{\dacompletare{interim: contamination controls on the development subsample}} The contamination controls act on the state. Removing the quoted strings and the section names leaves the ranking of A2 almost unchanged, at an AUROC of \res{e10.idonly.clm_a2.noid.auroc} against \res{e10.idonly.clm_a2.clean.auroc} with the whole state, and a state made of the strings alone ranks the files at \res{e10.idonly.clm_a2.idonly.auroc} (\res{e10.idonly.clm_a2.idonly.ci}). Without the strings the scores of A2 move toward malware: at the frozen threshold its TPR goes from \res{e10.contam.clm_a2.clean.tpr} to \res{e10.contam.clm_a2.noid.tpr} (McNemar $p$ \res{e10.contam.clm_a2.noid.p}) and its FPR from \res{e10.contam.clm_a2.clean.fpr} to \res{e10.contam.clm_a2.noid.fpr}. A2 thus decides on the behavioral evidence of the state, and the strings mostly weigh toward benign. \fwname{} follows A2: without the strings its TPR goes from \res{e10.contam.clm_a3.clean.tpr} to \res{e10.contam.clm_a3.noid.tpr} (McNemar $p$ \res{e10.contam.clm_a3.noid.p}) and its FPR from \res{e10.contam.clm_a3.clean.fpr} to \res{e10.contam.clm_a3.noid.fpr}, and the strings alone rank the files at \res{e10.idonly.clm_a3.idonly.auroc} (\res{e10.idonly.clm_a3.idonly.ci}). The readouts that answer questions keep their TPR without the strings: A1 goes from \res{e10.contam.clm_a1.clean.tpr} to \res{e10.contam.clm_a1.noid.tpr} (McNemar $p$ = \res{e10.contam.clm_a1.noid.p}) and A0 from \res{e10.contam.clm_a0.clean.tpr} to \res{e10.contam.clm_a0.noid.tpr}. Samples first seen in 2026, after the release of the models, need live malware and a networked VM, outside the static and offline setting of this study.

\subsection{Limitations}
\label{sec:res-limits}

Some limitations emerge from the scope of the study. The evidence is static: no sample is ever executed, so behavior that appears only at run time (e.g., the payload of a packed file) stays outside the state until a sealed sandbox adds it. The local models read at most 8,192 tokens, which bounds the number of function cards in a state, and windows of 32,000 tokens would hold about four times as many. Jev, the System One model released by TypeSafe, falls outside this evaluation, since its usage policy prohibits the analysis of malware; its assessment on these states, subject to a written authorization from TypeSafe, is left to future work. The operating point rests on the 27 false positives that 1\% allows among the 2,785 benign validation files with a state, so the realized FPR on $B_{\mathrm{test}}$ moves between \res{e1.fpr.min} and \res{e1.fpr.max} across the arms, and a benign validation set of tens of thousands of files would support thresholds at 0.1\%. Benign files carry no reliable first-seen date, so E9 measures the time axis on malware, and the S2 states of benign programs are longer than those of malware (6,229 against 4,385 tokens on average in a dry run), a difference that states matched in length would isolate. BODMAS and SOREL-20M predate the models, and public reports on their families may have entered pretraining: the temporal split and the shortcut controls limit this exposure, and samples first seen after the release of the models would remove it.%
\section{Conclusion and Future Works}
\label{sec:end}

This work evaluates System One decision models as Windows PE malware detectors built on static analysis. An isolated virtual machine decompiles every binary with Ghidra and maps its capabilities with capa. The host renders the records into canonical textual states free of the known class shortcuts. CLM-8B and ten open decision models answer typed verdict and behavioral questions, and three readouts turn the answers or the embedding into scores calibrated at a 1\% FPR operating point; \fwname{} joins the atomic answers and the embedding of CLM-8B in one tree combiner. A protocol frozen before testing compares the readouts with one another, with a combiner on the digest facts and with LLMs read through answer-token probabilities on BODMAS, SOREL-20M and a matched benign collection, and compares their inference on a workstation with a laptop GPU.

The evaluation shows that a single typed verdict wastes the static evidence that a System One model reads, and that many typed answers and supervised heads on its embedding recover its signal. A single verdict ranks the files at an AUROC between \res{e1.a0.min} and \res{e1.a0.max}, the atomic questions with a trained combiner lift every model to between \res{e1.a1.min} and \res{e1.a1.max}, and heads on the embedding reach \res{e1.clm_a2.auroc} for CLM-8B and from \res{e1b.a2.min} to \res{e1b.a2.max} for nine open models. \fwname{} joins the two readouts of CLM-8B: it ranks the test files at \res{e1.clm_a3.auroc}, detects \res{e1.clm_a3.tpr01} of the malware at 0.1\% FPR, stays close to the union of its inputs under padding and holds its detection over the nine test months, with an area under time of \res{e9.clm_a3.aut}, while each of its decisions keeps the atomic answers that name the behaviors behind it. Its results are reproducible: inference returns the same probabilities across sessions and caches, and the variance between training seeds is close to zero, with an AUROC between \res{e6.a3.auroc_min} and \res{e6.a3.auroc_max} over five seeds. The typed answers also stay close to the state: CLM-8B affirms \res{e7.clm_gguf_bf16.popular.fab} of the popular and \res{e7.clm_gguf_bf16.adversarial.fab} of the adversarial absent items and no absent capability, and planted text moves \res{e8.shift.inj_raw.atomic.moved} of its atomic answers against \res{e8.shift.inj_raw.verdict.moved} of its malware verdicts.\ifres{test.e7inv}{ \nuovo{Its verdict holds when the question is inverted or its options are swapped, and the wording of the questions, frozen with each arm, is what moves its decisions\ifres{e7i.clm_a3.para}{, least for \fwname{}}{}.}}{} The ranking carries over to SOREL-20M, at an AUROC of \res{e10.clm_a3.auroc} for \fwname{}, and to a 5-bit model on a laptop GPU, at \res{e5.edge.auroc} against \res{e5.ref.auroc} in bf16, whereas the operating point is the part that each feed and each deployment tier fits on its own validation files. Finally, the question design and the combiner matter more than the wording of the questions or the extent of the state, whose blocks carry redundant evidence for the embedding.

The present work identifies several research directions that warrant further investigation. Dynamic analysis in a sealed sandbox can add run-time behavior to the state of packed files, where static capabilities are scarce, so that the same typed questions reach the behaviors that only unpacking reveals. Models with windows of 32,000 tokens or more relax the function selection of S2 and let the decision weigh functions that 8,192 tokens leave out, and Jev, the model that introduced the class, would join the same evaluation with its larger window once TypeSafe authorizes its use on decompiled malware. A benign validation set of tens of thousands of files supports thresholds at 0.1\% FPR, and the abstention band can route the deferred files to a generative LLM or an analyst, with the atomic answers as the evidence shown. Benign programs dated by their release let the temporal evaluation follow the false positives over the months together with the detection. States matched in length across the classes, through the budget or through padding with neutral cards, isolate the contribution of the function content, and malware first seen after the release of the models removes any exposure of its families through pretraining. This work frames malware detection as a binary classification of each file, malicious or benign. The typed interface extends it to multiclass classification: a choice question over malware categories and families, whose labels the corpus already carries, turns the same models into a multiclass classifier that names the category and the family behind each detection. The 32 atomic questions introduce a first symbolic layer, in which each answer is judged against an explicit, rule-like criterion and learned trees combine the answers. A neuro-symbolic integration stronger than these 32 atomic rules could make the combination itself symbolic, through differentiable logic as in Logic Tensor Networks~\cite{badreddine2022ltn} or probabilistic logic programming as in DeepProbLog~\cite{manhaeve2021deepproblog}. Both frameworks are viable candidates, since the typed answers already provide probabilities that may serve as predicates, in rules that combine behaviors and, with the multiclass answers, relate each family to its category.

\section*{\raggedright Declaration of Generative AI and AI-Assisted Technologies in the Manuscript Preparation Process}
During the preparation of this work, the authors used AI tools to improve the language, readability, and grammar of the manuscript, as well as to assist with debugging and bug hunting.

\bibliographystyle{plain}
\appendix
\section{Question Bank}
\label{app:repo}

\paragraph{Question bank.} Table~\ref{tab:qb} lists the 32 behaviors of the atomic questions. Each behavior is asked in plain language, in ATT\&CK terms and in capa terms, and every question has ten paraphrases. The verdict questions are the malware question, its benign counterpart, a five-level score from ``clearly benign'' to ``clearly malware'', and a two-option choice in both orders. The function-level questions ask eleven behaviors of a single function card and whether the function is relevant to security.

\begin{table}[h]
  \centering
  \caption{Behaviors of the atomic question set.}
  \label{tab:qb}
  \footnotesize
  \begin{tabularx}{\linewidth}{>{\raggedright\arraybackslash}X}
    \toprule
    \textbf{Malicious (28).} Command execution; persistence through Run keys, services, scheduled tasks and the startup folder; process injection; process hollowing; privilege escalation; anti-debugging; anti-VM and sandbox checks; obfuscation and packing; disabling security tools; credential access; keylogging; screen and clipboard capture; discovery; lateral movement; command and control over HTTP, DNS and raw sockets; download and execute; dropping files; exfiltration; ransomware encryption; shadow-copy deletion; wiping; cryptocurrency mining; self-propagation. \\
    \midrule
    \textbf{Benign indicators (4).} Graphical application; installer or updater; runtime library; low-risk operations only. \\
    \bottomrule
  \end{tabularx}
\end{table}

\section{Additional Results}
\label{app:results}

Table~\ref{tab:readouts} gives the values behind Figure~\ref{fig:readouts}: the AUROC of every model that reads a prompt with each of its readouts, and the number of files each model scored.

\begin{widetable}[t]
  \centering
  \caption{Every prompt-reading model with each readout (test split).}
  \label{tab:readouts}
  \footnotesize
  \begin{threeparttable}
  \begin{tabular}{l l c c c c r}
    \toprule
    & & \multicolumn{4}{c}{AUROC [95\% CI]} & \\
    \cmidrule(lr){3-6}
    Model & State & A0 & A1 & A2 & \fwname{} & Files \\
    \midrule
    CLM-8B & S2 & 0.7194 [0.643, 0.796] & 0.9763 [0.965, 0.985] & 0.9943 [0.991, 0.997] & 0.9946 [0.991, 0.997] & 10,829 \\
    Qwen3-8B (logits) & S2 & 0.7388 [0.694, 0.779] & 0.9493 [0.929, 0.964] & - & - & 10,829 \\
    Foundation-Sec-8B (logits) & S2 & 0.8590 [0.799, 0.906] & 0.9733 [0.962, 0.981] & - & - & 10,829 \\
    decider-4b (Q4\_K\_M) & S0 & 0.6667 [0.595, 0.740] & 0.9834 [0.975, 0.989] & 0.9936 [0.990, 0.996] & - & 10,829 \\
    decider-2b & S0 & 0.6682 [0.574, 0.756] & 0.9824 [0.975, 0.988] & 0.9938 [0.989, 0.997] & - & 10,829 \\
    decider-0.8b & S0 & 0.5675 [0.485, 0.652] & 0.9773 [0.969, 0.984] & 0.9916 [0.987, 0.995] & - & 10,829 \\
    Strands Decider 2B & S0 & 0.6933 [0.612, 0.773] & 0.9833 [0.975, 0.989] & 0.9948 [0.992, 0.997] & - & 10,829 \\
    Liquid d1-3B & S0 & 0.5902 [0.499, 0.684] & 0.9792 [0.969, 0.986] & 0.9908 [0.986, 0.994] & - & 10,829 \\
    Liquid d1-omni-600M & S0 & 0.4025 [0.297, 0.517] & 0.9437 [0.921, 0.960] & - & - & 10,829 \\
    Kev-0.8B & S0 & 0.6930 [0.634, 0.748] & 0.9673 [0.954, 0.977] & 0.9934 [0.990, 0.996] & - & 10,829 \\
    Von 395M & S0 & 0.6957 [0.571, 0.796] & 0.9602 [0.942, 0.973] & 0.9896 [0.983, 0.994] & - & 10,829 \\
    Laya typed-decisions 421M & S0 & 0.4561 [0.323, 0.582] & 0.9398 [0.914, 0.959] & 0.9915 [0.987, 0.994] & - & 8,316 \\
    Laya multilingual 322M & S0 & 0.6524 [0.557, 0.740] & 0.9126 [0.881, 0.936] & 0.9853 [0.978, 0.991] & - & 8,369 \\
    \bottomrule
  \end{tabular}
  \end{threeparttable}
\end{widetable}
 
Table~\ref{tab:seeds} reports, for every arm trained with five seeds, seed 0, the primary variant of the main text, next to the mean and the standard deviation over the seeds (Section~\ref{sec:res-deploy}).

\begin{widetable}[t]
  \centering
  \caption{Seed 0 and the mean over the seeds (test split).}
  \label{tab:seeds}
  \small
  \begin{adjustbox}{max width=\textwidth}
  \begin{threeparttable}
  \begin{tabular}{l c c c c c c}
    \toprule
    & \multicolumn{2}{c}{AUROC} & \multicolumn{2}{c}{TPR at $\tau_{1\%}$} & \multicolumn{2}{c}{pAUC$_{1\%}$} \\
    \cmidrule(lr){2-3} \cmidrule(lr){4-5} \cmidrule(lr){6-7}
    Arm & seed 0 & mean $\pm$ sd & seed 0 & mean $\pm$ sd & seed 0 & mean $\pm$ sd \\
    \midrule
    Qwen3-Embedding-8B probe & 0.9956 & 0.9956 $\pm$ 0.0000 & 93.62\% & 93.62 $\pm$ 0.00\% & 0.8678 & 0.8678 $\pm$ 0.0000 \\
    CLM-8B A2-ctrl, logistic & 0.9955 & 0.9955 $\pm$ 0.0000 & 92.94\% & 92.94 $\pm$ 0.00\% & 0.8974 & 0.8974 $\pm$ 0.0000 \\
    CLM-8B A2-ctrl, two-layer & 0.9946 & 0.9946 $\pm$ 0.0006 & 90.88\% & 91.25 $\pm$ 0.66\% & 0.8974 & 0.8950 $\pm$ 0.0189 \\
    CLM-8B A2, cold start & 0.9948 & 0.9942 $\pm$ 0.0009 & 91.07\% & 89.92 $\pm$ 2.13\% & 0.9160 & 0.8924 $\pm$ 0.0201 \\
    CLM-8B A2, multi-task & 0.9932 & 0.9928 $\pm$ 0.0015 & 90.71\% & 89.07 $\pm$ 3.01\% & 0.8679 & 0.8626 $\pm$ 0.0209 \\
    CLM-8B A2, raw state & 0.9955 & 0.9952 $\pm$ 0.0004 & 91.28\% & 90.92 $\pm$ 1.38\% & 0.9130 & 0.9087 $\pm$ 0.0072 \\
    \midrule
    CLM-8B A2 & 0.9943 & 0.9941 $\pm$ 0.0002 & 90.90\% & 90.16 $\pm$ 0.54\% & 0.8760 & 0.8752 $\pm$ 0.0039 \\
    \bottomrule
  \end{tabular}
  \end{threeparttable}
  \end{adjustbox}
\end{widetable}
  
\end{document}